\documentclass[aps, prx, 11pt, tightenlines, amsfonts, showpacs, nofootinbib, longbibliography, notitlepage, onecolumn, superscriptaddress]{revtex4-2}

\usepackage[dvipsnames]{xcolor}
\usepackage[colorlinks=true, urlcolor=violet, linkcolor=blue, citecolor=red, hyperindex=true, linktocpage=true]{hyperref}
\usepackage{microtype} % nicer spacing and formatting
\usepackage{graphicx}
\usepackage{comment}
\usepackage{enumerate}

\usepackage{stmaryrd} % what is this??

\usepackage{xpatch}
\makeatletter
\patchcmd{\@ssect@ltx}
    {\addcontentsline{toc}{#1}{\protect\numberline{}#8}}
    {}
    {}
    {}
\makeatother

\renewcommand{\thesection}{\arabic{section}}
\renewcommand{\thesubsection}{\thesection.\arabic{subsection}}
\renewcommand{\thesubsubsection}{\thesubsection.\arabic{subsubsection}}
\makeatletter
\renewcommand{\p@subsection}{}
\renewcommand{\p@subsubsection}{}
\def\l@subsubsection#1#2{}
\makeatother

\makeatletter
\g@addto@macro\bfseries{\boldmath}
\makeatother

\newcommand{\vpd}[0]{\vphantom{\dagger}}
\newcommand{\vps}[0]{\vphantom{*}}
\newcommand{\vpp}[0]{\vphantom{\prime}}

\usepackage{amsmath,amssymb,amsfonts,mathtools,dsfont,relsize}

\usepackage{suffix} % for Ollie Hart's \matel*{}{}{} command

\usepackage{amsthm}

\usepackage{soul} % what is this?
\DeclarePairedDelimiter{\abs}{\lvert}{\rvert}
\DeclarePairedDelimiter{\expval}{\langle}{\rangle}
\DeclarePairedDelimiter{\avg}{\langle}{\rangle}
\DeclarePairedDelimiter{\ket}{\lvert}{\rangle}
\DeclarePairedDelimiter{\bra}{\langle}{\rvert}
\newcommand{\ii}[0]{\mathrm{i}}
\newcommand{\ee}[0]{\mathrm{e}}
\newcommand{\bvec}[1]{\boldsymbol{#1}}
\newcommand{\Order}[1]{\mathsf{O} ( #1 )}

\newcommand{\pd}[1]{\partial^{\vpd}_{#1}}\WithSuffix\newcommand\pd*[1]{\partial^{\,}_{#1}}
\newcommand{\pdp}[2]{\partial^{#2}_{#1}}
\newcommand{\thed}[0]{\mathrm{d}}
\newcommand{\DiracDelta}[1]{\delta \left( #1 \right)}

\newcommand{\kron}[1]{\delta^{\vpp}_{#1}}\WithSuffix\newcommand\kron*[1]{\delta_{#1}}
\newcommand{\ident}[0]{\mathds{1}}
\newcommand{\Reals}{\mathbb{R}}
\newcommand{\Comps}{\mathbb{C}}
\newcommand{\Ints}{\mathbb{Z}}
\newcommand{\Nats}{\mathbb{N}}

\newcommand{\Unitary}[1]{\mathsf{U} \hspace{-0.1em} \left( #1 \right)}\WithSuffix\newcommand\Unitary*[1]{\mathbf{U} \hspace{-0.1em} (#1)}

\renewcommand{\a}[1]{a^{\vpd}_{#1}}
\newcommand{\adag}[1]{a^{\dagger}_{#1}}

\newcommand{\B}[1]{B^{\vpd}_{#1}}
\newcommand{\Bdag}[1]{B^{\dagger}_{#1}}
\newcommand{\f}[1]{f^{\vpd}_{#1}}
\newcommand{\fdag}[1]{f^{\dagger}_{#1}}
\newcommand{\re}[1]{r^{\vpd}_{#1}}
\newcommand{\rdag}[1]{r^{\dagger}_{#1}}
\newcommand{\hilbert}[0]{\mathcal{H}}
\newcommand{\Hilbert}[1]{\hilbert^{\vpp}_{#1}}\WithSuffix\newcommand\Hilbert*[1]{\hilbert_{#1}}

\DeclareMathOperator*{\var}{var}
\newcommand{\VarOf}[1]{\var \left( #1 \right)}\WithSuffix\newcommand\VarOf*[1]{\var ( #1 ) }

\DeclareMathOperator*{\Dim}{dim}
\newcommand{\DimOf}[1]{\Dim \left( #1 \right)}\WithSuffix\newcommand\DimOf*[1]{\Dim ( #1 )}
\newcommand{\hildim}[0]{\mathcal{D}}
\newcommand{\HilDim}[1]{\hildim_{#1}}

\DeclareMathOperator*{\Spec}{spec}
\newcommand{\SpecOf}[1]{ \Spec \left( #1 \right) }\WithSuffix\newcommand\SpecOf*[1]{\Spec (#1)}

\DeclareMathOperator*{\Prob}{Prob}
\newcommand{\ProbOf}[1]{ \Prob \left( #1 \right) }\WithSuffix\newcommand\ProbOf*[1]{\Prob (#1)}

\DeclareMathOperator*{\Span}{span}
\newcommand{\SpanOf}[1]{\Span \left( #1 \right)}\WithSuffix\newcommand\SpanOf*[1]{\Span ( #1 ) }

\DeclareMathOperator*{\Endomorphism}{End}
\newcommand{\End}[1]{\Endomorphism \left( #1 \right)}\WithSuffix\newcommand\End*[1]{\Endomorphism ( #1 )}

\DeclareMathOperator*{\automorphism}{Aut}
\newcommand{\Aut}[1]{\automorphism \left( #1 \right)}\WithSuffix\newcommand\Aut*[1]{\automorphism ( #1 )}

\newcommand{\bounded}[0]{\mathcal{B}}
\newcommand{\Bounded}[1]{\bounded \left( #1 \right)}\WithSuffix\newcommand\Bounded*[1]{\bounded ( #1 )}
\newcommand{\TrClass}[1]{\bounded_{1} \left( #1 \right)}\WithSuffix\newcommand\TrClass*[1]{\bounded_{1} ( #1 )}
\newcommand{\Positive}[1]{\bounded_{+} \left( #1 \right)}\WithSuffix\newcommand\Positive*[1]{\bounded_{+} ( #1 )}
\newcommand{\Hermitian}[1]{\bounded_{H} \left( #1 \right)}\WithSuffix\newcommand\Hermitian*[1]{\bounded_{H} ( #1 )}
\newcommand{\States}[1]{\mathcal{S} \left( #1 \right)}\WithSuffix\newcommand\States*[1]{\mathcal{S} ( #1 )}
\newcommand{\PseudoStates}[1]{\mathcal{S}' \left( #1 \right)}\WithSuffix\newcommand\PseudoStates*[1]{\mathcal{S}' ( #1 )}

\newcommand{\BKop}[2]{\ket{#1} \hspace{-0.4mm} \bra{#2}}

\newcommand{\inprod}[2]{ \left\langle #1 \middle| #2 \right\rangle}

\newcommand{\matel}[3]{\left\langle #1 \middle| #2 \middle| #3 \right\rangle}
\WithSuffix\newcommand\matel*[3]{\langle #1 | #2 | #3 \rangle}

\newcommand{\tr}[1]{{\rm tr} \left[  #1  \right]}
\DeclareMathOperator*{\trace}{tr}
\newcommand{\proj}[1]{P^{\vpp}_{#1}}\WithSuffix\newcommand\proj*[1]{P^{\,}_{#1}}
\newcommand{\Proj}[2]{P^{#2}_{#1}}
\newcommand{\comm}[2]{\left[ #1, \, #2 \right]}\WithSuffix\newcommand\comm*[2]{[#1, \, #2]}
\newcommand{\acomm}[2]{\left\{ #1, \, #2 \right\} }\WithSuffix\newcommand\acomm*[2]{\{#1, \, #2\}}

\newcommand{\observ}[0]{\mathcal{O}}
\newcommand{\eig}[0]{\lambda}
\newcommand{\Eig}[1]{\eig^{\vpp}_{#1}}\WithSuffix\newcommand\Eig*[1]{\eig^{\,}_{#1}}

\newcommand{\mobserv}[0]{\mathcal{O}}
\newcommand{\meig}[0]{O}
\newcommand{\mEig}[1]{\meig^{\vpp}_{#1}}\WithSuffix\newcommand\mEig*[1]{\meig^{\,}_{#1}}

\newcommand{\eigmult}[0]{r}
\newcommand{\EigMult}[1]{\eigmult^{\vpp}_{#1}}\WithSuffix\newcommand\EigMult*[1]{\eigmult^{\,}_{#1}}

\newcommand{\Ham}[0]{H}

\newcommand{\umeas}[0]{\mathcal{U}} %for general mentions
\newcommand{\Umeas}[1]{\umeas^{\vps}_{\left[ #1 \right]}}\WithSuffix\newcommand\Umeas*[1]{\umeas^{\,}_{[#1]}}
\newcommand{\Umeasdag}[1]{\umeas^{*}_{\left[ #1 \right]}}\WithSuffix\newcommand\Umeasdag*[1]{\umeas^{*}_{[#1]}}
\newcommand{\dmeas}[0]{\mathcal{V}} %for general mentions
\newcommand{\Dmeas}[1]{\dmeas^{\vps}_{\left[ #1 \right]}}\WithSuffix\newcommand\Dmeas*[1]{\dmeas^{\,}_{[#1]}}
\newcommand{\Dmeasdag}[1]{\dmeas^{*}_{\left[ #1 \right]}}\WithSuffix\newcommand\Dmeasdag*[1]{\dmeas^{*}_{[#1]}}

\newcommand{\Pauli}[2]{\sigma^{#1}_{#2}}
\newcommand{\PauliVec}[1]{\boldsymbol{\sigma}^{\vpp}_{#1}}\WithSuffix\newcommand\PauliVec*[1]{\boldsymbol{\sigma}_{#1}}

\newcommand{\kraus}[0]{E}
\newcommand{\KrausPow}[2]{\kraus_{#1}^{#2}}
\newcommand{\Kraus}[1]{\KrausPow{#1}{\vps}}\WithSuffix\newcommand\Kraus*[1]{\KrausPow{#1}{}}
\newcommand{\KrausDag}[1]{\KrausPow{#1}{\dagger}}

\newcommand{\shift}[0]{X}
\newcommand{\weight}[0]{Z}

\newcommand{\Shift}[1]{\shift^{\vpd}_{#1}}\WithSuffix\newcommand\Shift*[1]{\shift^{\,}_{#1}}
\newcommand{\Weight}[1]{\weight^{\vpd}_{#1}}\WithSuffix\newcommand\Weight*[1]{\weight^{\,}_{#1}}
\newcommand{\ShiftPow}[2]{\shift^{#2}_{#1}}

\newcommand{\PX}[1]{X^{\vps}_{#1}}\WithSuffix\newcommand\PX*[1]{X^{\,}_{#1}}
\newcommand{\PY}[1]{Y^{\vps}_{#1}}\WithSuffix\newcommand\PY*[1]{Y^{\,}_{#1}}
\newcommand{\PZ}[1]{Z^{\vps}_{#1}}\WithSuffix\newcommand\PZ*[1]{Z^{\,}_{#1}}

\newcommand{\SShift}[1]{\widetilde{\shift}^{\vpd}_{#1}}\WithSuffix\newcommand\SShift*[1]{\widetilde{\shift}^{\,}_{#1}}
\WithSuffix\newcommand\SWeight*[1]{\widetilde{\weight}^{\,}_{#1}}
\newcommand{\Noutcome}[0]{\mathcal{N}}
\newcommand{\NKraus}[0]{q}

\numberwithin{equation}{section}

\begin{document}

\title{Overview of projective quantum measurements}

\author{Diego Barberena}
\affiliation{Department of Physics and Center for Theory of Quantum Matter, University of Colorado, Boulder CO 80309, USA}
\affiliation{JILA and NIST, University of Colorado,  Boulder, CO 80309, USA}

\author{Aaron J. Friedman}
\email{Aaron.Friedman@colorado.edu}
\affiliation{Department of Physics and Center for Theory of Quantum Matter, University of Colorado, Boulder CO 80309, USA}

\begin{abstract}
    We provide an overview of standard ``projective'' quantum measurements with the goal of elucidating connections between theory and experiment. We make use of a unitary ``Stinespring'' representation of measurements on a dilated Hilbert space that includes both the physical degrees of freedom and those of the measurement apparatus. We explain how this unitary representation (\emph{i}) is guaranteed by the axioms of quantum mechanics, (\emph{ii}) relates to both the Kraus and von Neumann representations, and (\emph{iii}) corresponds to the physical time evolution of the system and apparatus during the measurement process. The  Stinespring representation also offers significant conceptual insight into measurements, helps connects theory and experiment, is particularly useful in describing protocols involving midcircuit measurements and outcome-dependent operations, and establishes that all quantum operations are compatible with relativistic locality, among other insights. 
\end{abstract}

\date{\today}

\maketitle

\tableofcontents

\section{Introduction}
\label{sec:intro}

Measurements are a fundamental part of quantum theory. Typically codified in axiomatic formulations of quantum mechanics \cite{DiracQuantum, vonNeumannAxioms, HardyAxioms, FuchsAxioms, MackeyAxioms, WilceAxioms, MasanesAxioms, KapustinAxioms}, the standard ``projective'' measurement is intrinsically related to the most prominent departures of quantum systems from more familiar classical physics. For example, the quantization of observable outcomes and the probabilistic nature of measurements gave some of the first historical indications of the inadequacy of classical mechanics in describing, e.g., atomic systems. Despite the successes of quantum mechanics, the nature of measurements---and their implications for quantum mechanics as a %physical
theory---has been a source of substantial debate.

The apparent ``collapse'' of the wavefunction upon measurement was one of several early %historical 
sources of confusion and concern. The probabilistic nature of measurements was seen as a problem to those who expected a deterministic theory, \`a la classical physics~\cite{nodice}. Moreover, the putative violation of relativistic locality in the measurement of spatially separated entangled degrees of freedom  was viewed as a ``paradox'' that implied that quantum theory was not yet complete~\cite{epr}. While it is now understood that no such violation exists, measurements remain one of the more ``mysterious'' aspects of quantum mechanics, and are at the heart of debates over the theory's interpretation.

We stress that addressing such foundational questions (e.g., about interpretations of quantum mechanics) is \emph{not} our goal. However, doing so would require considering the details of how measurements are implemented in real experiments. Importantly, we note that the standard theoretical descriptions of quantum measurements~\cite{WheelerZurek, MargenauMeas, PeresMeas86} obfuscate their nature. Measurements and other nonunitary quantum operations are typically modeled via (\emph{i}) stochastic updates to the wavefunction, (\emph{ii}) a Lindblad master equation~\cite{lindblad1973entropy, decoherence, brasil2013simple}, or (\emph{iii}) Kraus operators~ \cite{KrausMeas1969, KrausMeas1971, KrausMeas1981, KrausBook}, all of which describe only the degrees of freedom in the system of interest, despite the fact that all such quantum operations are exclusive to \emph{open} systems. Even setting aside matters of interpretation, consideration of the experimental details surrounding measurement is important for connecting theory to experiment and designing measurement-based protocols for near-term quantum devices.

To that end, we consider a conceptually transparent and analytically powerful representation of projective measurements in terms of a unitary operator $\umeas$ acting on an enlarged Hilbert space \cite{SpeedLimit, AaronMIPT}. This representation is logically implied by all axiomatic formulations of quantum mechanics~\cite{DiracQuantum, vonNeumannAxioms, HardyAxioms, FuchsAxioms, MackeyAxioms, WilceAxioms, MasanesAxioms, KapustinAxioms}, and describes precisely the unitary time evolution of the physical system and measurement apparatus in the experimental implementation of the prototypical examples of measurements considered herein. Although an exhaustive study of all possible implementations of projective measurements is beyond the scope of this work, we expect that $\umeas$ always has this physical interpretation. In addition to elucidating various aspects of measurements---e.g.,  related to their experimental implementation and compatibility with both relativistic locality and determinism---the unitary representation we showcase is also better suited to the investigation of measurement-based protocols involving projective measurements and outcome-dependent operations. We emphasize this because such protocols have immense utility in numerous tasks relevant to near-term quantum technologies, from quantum information processing to efficient many-body state preparation \cite{AaronMIPT, SpeedLimit, AaronTeleport, QC_book, Jozsa-intro-MBQC}.

The primary goals of this paper are to (\emph{i}) provide a useful overview of projective quantum measurements in theory and experiment; (\emph{ii}) explain how these measurements are most commonly implemented in practice; (\emph{iii}) explain the rigorous mathematical origins of the unitary ``Stinespring'' representation of projective measurements; (\emph{iv}) establish how the unitary representation corresponds to the time evolution of the system and detector during the measurement process; and (\emph{v}) explain how to apply the Stinespring formalism to generic quantum protocols, its benefits (e.g., in the context of protocols with midcircuit measurements and outcome-dependent feedback), and some of its more important implications. The remainder of this paper is organized as follows.

In Sec.~\ref{sec:Measurement formalism} we derive the unitary measurement formalism from the axioms of quantum mechanics~\cite{DiracQuantum, vonNeumannAxioms, HardyAxioms, FuchsAxioms, MackeyAxioms, WilceAxioms, MasanesAxioms, KapustinAxioms}, using powerful results from $C^*$-algebras~\cite{PaulsenBook, Takesaki, Stinespring, ChoisThm, WolfNotes, AaronJamesFuture}. In Sec.~\ref{subsec:observables}, we discuss the axiomatic properties of quantum measurements. In Sec.~\ref{subsec:Kraus}, we derive the standard \emph{Kraus representation} of measurements~\cite{KrausMeas1969, KrausMeas1971, KrausMeas1981, KrausBook} using both the Stinespring dilation theorem \cite{Stinespring} and Choi's theorem~\cite{ChoisThm} for completely positive (CP) maps. In Sec.~\ref{subsec:physicist's Stinespring}, we explain the ``physicist's Stinespring theorem''~\cite{WolfNotes, AaronJamesFuture}, in which quantum operations are captured by a \emph{unitary} operator $\umeas$ on a \emph{dilated} Hilbert space $\Hilbert*{\text{dil}}$ \eqref{eq:Dilated Hilbert}, which includes both the physical Hilbert space $\hilbert$ and a Stinespring Hilbert space $\Hilbert*{\text{ss}}$. This result does not follow straightforwardly from Stinespring's theorem~\cite{Stinespring}, but was proven by Kraus~\cite{KrausBook} for finite-dimensional Hilbert spaces; a proof for the infinite-dimensional case $\hilbert = L^2 (\Reals) = \ell^2 (\Nats) \cong \Comps^{\infty}$ is the subject of forthcoming work \cite{AaronJamesFuture}. In Sec.~\ref{subsec:von Neumann}, we connect $\umeas$ to von Neumann's description of measurements using a \emph{pointer Hamiltonian}~\cite{VonNeumann, Preskill_QI} that includes the detector. All of the above is summarized in Sec.~\ref{subsec:formalism summary}

In Secs.~\ref{sec:photons} and \ref{sec:qubits}, we discuss the experimental implementations of measurements of photons and qubits, respectively. While quantum systems can be probed in a variety of ways, the final stage of these probes is typically selected from a small set of measurements that includes the examples in Secs.~\ref{sec:photons} and \ref{sec:qubits}. We also highlight that most measurement schemes involve electromagnetic modes (e.g., photons), and %the ones we consider
typically record outcomes through particle detection, which generally involve the excitation of particles in an apparatus. In Sec.~\ref{subsec:EM modes}, we review the quantum description of electromagnetic fields. In Sec.~\ref{subsec:photon counting}, we consider measurements of photon number (or intensity); despite being ``destructive'' (rather than projective), these measurements are captured by the unitary representation. In Sec.~\ref{subsec:homodyne}, we discuss interferometry measurements involving homodyne detection, which culminate in intensity measurements. In Sec.~\ref{subsec:fluorescent}, we review fluorescence measurements of various types of qubits, which also culminate in counting spontaneously emitted photons. In Sec.~\ref{subsec:dispersive}, we discuss ``dispersive readout'' of the states of, e.g., superconducting qubits. 

The examples in Secs.~\ref{sec:photons} and \ref{sec:qubits} illustrate that the dilated Hilbert space $\Hilbert{\rm dil}$ \eqref{eq:Dilated Hilbert} is \emph{physical}. We note that the \emph{minimal} Stinespring Hilbert space $\Hilbert{\text{ss}}$ that we derive in Sec.~\ref{sec:Measurement formalism} has a basis $\{ \ket{m}^{\,}_{\text{ss}} \}$ corresponding to the $\Noutcome$ possible measurement outcomes. However, in the examples considered in  Secs.~\ref{sec:photons} and \ref{sec:qubits}, we generally identify a nonminimal dilated unitary representation $\umeas$ involving additional degrees of freedom and/or multiple states reflecting the same outcome. Importantly, appropriately ``binning'' the states of $\Hilbert{\text{ss}}$ corresponding to the same outcome leads to the minimal Stinespring representation, which is unique up to the choice of  ``default'' initial state of the apparatus.

Finally, in Sec.~\ref{sec:Using Stinespring}, we discuss how to use the unitary formalism, and some of its implications. In Sec.~\ref{subsec:extract}, we show how to recover the usual Born rule comes from and extract statistics. In Sec.~\ref{subsec:adaptive}, we explain the utility of the dilated unitary representation $\umeas$ in describing measurement-based protocols. Because the outcomes are stored in explicit degrees of freedom, it is far more straightforward in the unitary representation to describe quantum operations conditioned on the outcomes of prior measurements than it is using the Kraus representation. We emphasize that, given the utility of measurement-based protocols, this is one of the main advantages of the unitary formalism. In Sec.~\ref{subsec:collapse}, we explain the appearance of ``wavefunction collapse'' using a simple model of decoherence~\cite{decoherence} based on recent progress on many-body quantum chaos---the mechanism underlying thermalization. Although the dilated unitary representation $\umeas$ is fully \emph{deterministic}, with all outcomes occurring, the fact that the measurement apparatus is classical (so that the outcome can be read off) also ensures that it is subject to decoherence~\cite{decoherence}. However, because the outcomes are associated with a symmetry charge (e.g., particle number), superpositions of distinct outcomes rapidly decohere into a mixed state, so that only one outcome is observed. In Sec.~\ref{subsec:not spooky}, we explain the absence of ``spooky action at a distance'' in measurements of entangled states, as (\emph{i}) no information is transferred and (\emph{ii}) the intrinsic nature of collapse and absence of ``branching'' of the dilated wavefunction upon measurement means there is no ``influence'' of one particle in an entangled pair on the other. Lastly, in Sec.~\ref{subsec:locality}, we explain how the unitary representation establishes that \emph{all quantum operations involving measurements obey relativistic locality} \cite{SpeedLimit}.

\section{Mathematical formalism for measurements}
\label{sec:Measurement formalism}

We begin with a careful derivation of mathematical representations of projective and similar measurements, culminating in the unitary ``Stinespring'' formulation. The Stinespring unitary $\umeas$ connects directly to the Kraus \cite{KrausMeas1969, KrausMeas1971, KrausMeas1981, KrausBook} and von Neumann \cite{vonNeumannAxioms, VonNeumann} representations of measurements, and acts on a dilated Hilbert space $\Hilbert{\rm dil} = \hilbert \otimes \Hilbert{\rm ss}$ \eqref{eq:Dilated Hilbert}. While the degrees of freedom in $\Hilbert{\rm ss}$ are nominally just a bookkeeping tool, the examples in Secs.~\ref{sec:photons} and \ref{sec:qubits} show that they are \emph{physical}, corresponding to the measurement apparatus. Readers for whom such a unitary formulation of measurements is familiar and/or intuitive may prefer to skip to the summary in Sec.~\ref{subsec:formalism summary}. 

The unitary representation of measurements we discuss below follows from mathematical results for $C^*$-algebras \cite{KrausBook, Stinespring, Takesaki, PaulsenBook, ChoisThm}. Historically, $C^*$-algebras were developed to describe the algebraic properties of the operators associated with quantum systems. In fact, every axiomatic formulation of quantum mechanics implies a $C^*$-algebra corresponding to the ``bounded'' operators $\Bounded*{\hilbert}$ on a Hilbert space $\hilbert$~\cite{DiracQuantum, vonNeumannAxioms, HardyAxioms, FuchsAxioms, MackeyAxioms, WilceAxioms, MasanesAxioms, KapustinAxioms, Stinespring, Takesaki, PaulsenBook, ChoisThm, AaronJamesFuture, WolfNotes}; moreover, the allowed updates to a quantum system---known as \emph{quantum operations}---correspond to completely positive (CP) maps between $C^*$-algebras \cite{QC_book, KrausBook}.

In particular, the density matrix $\rho$ describing a quantum system always belongs to $\Bounded*{\hilbert}$. Density matrices are positive operators $\rho \geq 0$, which means that their spectra (i.e., eigenvalues) are real and nonnegative. They also have unit trace $\trace (\rho)=1$, so their eigenvalues encode a probability distribution.  Because a quantum operation $\Phi$ is a CP map, it sends positive operators $\rho$ to positive operators $\Phi (\rho) \geq 0$, even if $\rho$ only describes part of a larger system. An important subset of quantum operations are known as \emph{quantum channels}, which are both completely positive and trace preserving (CPTP). When $\Phi$ is a quantum channel, $\rho' = \Phi (\rho)$ is also a density matrix. However, generic quantum operations are \emph{trace decreasing}, meaning that $\Phi (\rho) = \lambda \rho'$ for some density matrix $\rho'$ and real number $0 \leq \lambda \leq 1$. We consider quantum operations to allow for, e.g., measurements resulting in a particular outcome $m$, for which $\Phi_m (\rho) = p^{\,}_m \, \rho^{\,}_m$, with $p^{\,}_m$ the probability of outcome $m$ and $\rho_m$ the corresponding (and normalized) postmeasurement density matrix. Note that the map $\Phi_m : \rho \mapsto p^{\,}_m \, \rho^{\,}_m$ is always well defined and linear in $\rho$, while a map $\rho \mapsto \rho_m$ is ill defined when $p^{\,}_m = 0$, and is always a nonlinear function of $\rho$, since $p^{\,}_m$ depends on $\rho$ \cite{KrausBook, QC_book}.

The correspondence between quantum operations and CP maps between $C^*$-algebras provides for the derivations below \cite{WolfNotes, AaronJamesFuture}, due to useful properties of $C^*$-algebras and maps \cite{Takesaki, PaulsenBook, Stinespring, ChoisThm, KrausBook, AaronMIPT, SpeedLimit, WolfNotes, AaronJamesFuture}. In particular, the Stinespring dilation theorem \cite{Stinespring} implies a representation of any quantum operation $\Phi$ on a \emph{dilated} Hilbert space $\Hilbert{\rm dil}$. In the case of finite-dimensional Hilbert spaces, Choi's theorem \cite{ChoisThm} implies a representation of $\Phi$ in terms of \emph{Kraus operators} \cite{KrausMeas1969, KrausMeas1971, KrausMeas1981, KrausBook}, which is the most common representation of measurements in the literature. However, the Kraus formulation also implies the \emph{unitary} representation of measurements on $\Hilbert{\rm dil}$ that we explore herein. The unitary ``Stinespring'' representation results from a ``physicist's version'' of the Stinespring dilation theorem \cite{Stinespring} discussed in Sec.~\ref{subsec:physicist's Stinespring}, which is actually due to Kraus \cite{KrausBook}. We refer to the unitary formalism as the ``Stinespring representation'' due to its connection to the physicist's Stinespring theorem and to avoid confusion with the Kraus representation in terms of Kraus operators. A rigorous proof for the infinite-dimensional Hilbert space $\hilbert = \ell^2 (\Nats)$ is the subject of forthcoming work \cite{AaronJamesFuture}. 

In Sec.~\ref{subsec:observables} we discuss the axiomatic properties of quantum measurements. In Sec.~\ref{subsec:Kraus}, we derive the Kraus representation \cite{KrausMeas1969, KrausMeas1971, KrausMeas1981, KrausBook} from Stinespring's dilation theorem \cite{Stinespring} and Choi's theorem \cite{ChoisThm}. In Sec.~\ref{subsec:unitary measurement}, we derive the \emph{unitary} representation of measurements on a  $\Hilbert{\rm dil}$, which we relate to the Kraus formulation. In Sec.~\ref{subsec:Hamiltonian measurement}, we discuss von Neumann's ``pointer'' Hamiltonian describing measurements \cite{VonNeumann}, which we relate to the unitary representation. Finally, for convenience, we summarize the foregoing results for measurement representations in Sec.~\ref{subsec:formalism summary}.

Although we regard the extra Stinespring degrees of freedom $\Hilbert{\text{ss}} \subset \Hilbert{\rm dil}$ as a bookkeeping tool below, in Secs.~\ref{sec:photons} and \ref{sec:qubits} we show that $\Hilbert{\text{ss}}$ is \emph{physical} in several prominent examples of measurements, and we expect that it holds for generic projective and destructive measurements. In particular, $\Hilbert{\text{ss}}$ reflects the state space of the \emph{measurement apparatus}, where the dilated measurement unitary $\umeas$ \eqref{eq:Stinespring Unitary General}---or, equivalently, the von Neumann Hamiltonian $H_{\rm vN}$ \eqref{eq:Pointer Hamiltonian}---captures time evolution of the system and apparatus during the measurement process. Again, we expect this unitary description of measurements to be intuitive to many readers, who may prefer to skip to Sec.~\ref{subsec:formalism summary}.

\subsection{Measurement and spectral decomposition of observables}
\label{subsec:observables}

Consider the standard, projective measurement of a generic observable $\mobserv$ in a system %whose state is captured 
described by the density matrix $\rho$. Suppose that $\mobserv$ has $\Noutcome$ \emph{unique} eigenvalues $\{ \mEig*{m} \}$ for $0 \leq m < \Noutcome$, and denote by $\EigMult*{m}$ the multiplicity (degeneracy) of the $m$th eigenvalue $\mEig*{m}$. We then identify a set of \emph{eigenprojectors} onto the $\Noutcome$ distinct eigenspaces of $\mobserv$, i.e.,
\begin{equation}
    \label{eq:Spec Projecc}
    \mobserv \, \proj{m} = \mEig{m} \, \proj{m} ~~~~\text{with}~~~~\trace \left( \proj{m} \right) \, = \, \EigMult{m} \,,~~
\end{equation}
where $\EigMult*{m}$ is also the ``rank'' of $\proj*{m}$, meaning that $\proj*{m} = \BKop{m}{m}$ when $\EigMult*{m}=1$, and otherwise, $\proj*{m} = \sum_{k=1}^{\EigMult*{m}} \, \BKop{m_k}{m_k}$ is a sum over projectors onto degenerate eigenstates $\ket{m_k}$ of $\mobserv$ with eigenvalue $\mEig*{m}$. The eigenstates are orthonormal and form a complete basis for $\hilbert$ \cite{DiracQuantum, vonNeumannAxioms, HardyAxioms, FuchsAxioms, MackeyAxioms, WilceAxioms, MasanesAxioms, KapustinAxioms}. The $\Noutcome$ projectors \eqref{eq:Spec Projecc} form an orthogonal and complete set, meaning that,
\begin{equation}
    \label{eq:meas proj ortho complete}
    \sum\limits_{m=0}^{\Noutcome-1} \, \proj{m} = \ident  ~~~~\text{and}~~~~ \proj{m} \, \proj{n} \, = \, \kron{m,n} \, \proj{m} \, ,~~
\end{equation}
and using these projectors, we define the \emph{spectral decomposition} of the observable $\mobserv$ via
\begin{equation}
    \label{eq:measured observable}
    \mobserv = \sum\limits_{m=0}^{\Noutcome-1} \, \mEig{m} \, \proj{m} \, , ~~
\end{equation}
which is unique, and holds even in the limit $\Noutcome \to \infty$. 

The axioms of quantum mechanics \cite{DiracQuantum, vonNeumannAxioms, HardyAxioms, FuchsAxioms, MackeyAxioms, WilceAxioms, MasanesAxioms, KapustinAxioms} dictate that the outcome of projectively measuring $\mobserv$ \eqref{eq:measured observable} is  one of its eigenvalues $\mEig*{n}$. Repeating the experiment many times, the expectation value (i.e., average) of the measurement of $\mobserv$ in the state $\rho$ is given by $\expval{\mobserv}^{\,}_{\rho} = \trace (\mobserv \, \rho )$. However, the probability-theoretic definition of the expectation value also implies that
\begin{equation}
    \label{eq:Measurement Probability}
    \expval{\mobserv}^{\,}_{\rho} \, = \, \sum\limits_{m=0}^{\Noutcome-1} \, \mEig{m} \, p^{\vpp}_m \, = \, \trace \left( \, \mobserv \, \rho \,\right) \, = \, \sum\limits_{m=0}^{\Noutcome-1} \, \mEig{m} \, \trace \left(  \, \proj{m} \, \rho \, \right) ~~\implies ~~p^{\vpp}_m\, = \, \trace \left( \, \proj{m} \,\rho \, \right)  \, ,~~
\end{equation}
where $p^{\,}_m$ is the probability of observing outcome $\mEig*{m}$ upon measuring $\mobserv$ in the state $\rho$. When $\rho = \BKop{\psi}{\psi}$ is a pure state and $\EigMult*{m}=1$, then $p^{\,}_m = \inprod{m}{\psi}^2$, reproducing the familiar Born rule. 

If $\mobserv$ \eqref{eq:measured observable} is measured in the state $\rho$ and the observed outcome is $\mEig*{m}$, then the postmeasurement state \emph{collapses} into a state $\rho^{\,}_m$  \cite{DiracQuantum, vonNeumannAxioms, HardyAxioms, FuchsAxioms, MackeyAxioms, WilceAxioms, MasanesAxioms, KapustinAxioms}; up to normalization, we have that
\begin{equation}
    \label{eq:wf collapse}
    \rho \, \mapsto \, \rho^{\vpp}_m \, \propto \, \proj{m} \, \rho \, \proj{m}~~~~\text{where}~~~~\trace \left( \, \proj{m} \, \rho \, \proj{m} \, \right) \, = \, \trace \left( \, \proj{m} \, \rho \, \right) \, = \, p^{\vpp}_m \, , ~~
\end{equation}
meaning that the normalized density matrix $\rho^{\,}_m$ can only be defined if $p^{\vpp}_m > 0$. Accordingly, we identify a quantum operation (or linear CP map) $\Phi_m$, corresponding to a measurement of $\mobserv$ in the state $\rho$ resulting in the particular outcome $\mEig*{m}$, which takes the form 
\begin{equation}
    \label{eq:measurment map m}
    \Phi_m ( \rho ) \, = \, \proj{m} \, \rho \, \proj{m} \, = \, p^{\vpp}_m \, \rho^{\vpp}_m \, , ~~
\end{equation}
where $\rho^{\vpp}_m = 0$ if $p^{\vpp}_m = 0$, and $\rho^{\vpp}_m = \proj{m} \, \rho \, \proj{m} / p^{\,}_m$ otherwise. Note that if $\mobserv$ is measured again in the postmeasurement state $\rho^{\,}_m$, the probability to obtain outcome $\mEig{m}$ again is unity, as expected.

We also separately define a quantum \emph{channel} corresponding to the measurement of $\mobserv$, given by
\begin{equation}
    \label{eq:average measurement map}
    \Phi ( \rho ) \, \equiv \, \sum\limits_{m=1}^{\Noutcome} \, \Phi_m ( \rho ) \, = \, \sum\limits_{m=1}^{\Noutcome} \, \proj{m} \, \rho \, \proj{m} \, \, = \, \sum\limits_{m=1}^{\Noutcome} \, p^{\vpp}_m \, \rho^{\vpp}_m \, \equiv \, \rho^{\vpp}_{\rm av} \, ,~~
\end{equation}
which is CP, linear, and trace preserving, since $\trace \left( \rho^{\vpp}_{\rm av} \right) = \sum_{m=1}^{\Noutcome} \tr{ \Phi_m ( \rho ) }= \sum_{m=1}^{\Noutcome}  p^{\,}_m = 1$. We comment that nonprojective measurements (e.g., weak and generalized measurements) need not satisfy the foregoing properties, and we relegate their consideration to future work. As a reminder, we expect that all of the statements above hold true even in the countably infinite case $\Noutcome \to \infty$.

\subsection{Kraus representation}
\label{subsec:Kraus}

The starting point for all representations of quantum operations is the Stinespring dilation theorem \cite{Stinespring} for CP maps between $C^*$-algebras \cite{Takesaki, PaulsenBook}. As a reminder, all density matrices $\rho$ belong to the $C^*$-algebra $\Bounded*{\hilbert}$, corresponding to the bounded operators on a Hilbert space $\hilbert$; the quantum operations of interest are CP maps from  $\Bounded*{\hilbert}$ to itself. The Stinespring dilation theorem~\cite{Stinespring} establishes that all CP maps can be represented on a \emph{dilated} Hilbert space
\begin{equation}
\label{eq:Dilated Hilbert}
    \Hilbert{\rm dil} \, = \, \hilbert \otimes \Hilbert{\text{ss}} \, , ~~
\end{equation}
using the following combination of completely positive operations,
\begin{equation}
    \label{eq:Stinespring dilation}
    \Phi (\rho) \, = \, V^\dagger \, \pi (\rho ) \, V \, , ~~
\end{equation}
where $V : \hilbert \to \Hilbert{\rm dil}$ is an \emph{isometry}---i.e., a length-preserving CP map---and $\pi : \Bounded*{\hilbert} \to \Bounded*{\Hilbert{\rm dil}}$ is a  ${}^*$-homomorphism---i.e., a structure-preserving CP map between $C^*$-algebras, whose properties are unimportant to the present discussion. Further simplification is not possible in full generality.

However, when $\hilbert$ is finite dimensional~\cite{WolfNotes, AaronJamesFuture}, useful simplifications \emph{are} possible. We also expect that this extends to the countably infinite case $\hilbert = \ell^2 (\Nats)$ via inductive limits \cite{AaronJamesFuture, Davidson96, Brown, Bratteli}. For finite-dimensional $\hilbert$, we have that $\Bounded*{\hilbert} \cong M_N (\Comps)$, where $M_N (\Comps)$ is the $C^*$-algebra of $N \times N$ matrices with complex entries, for which it is well known \cite{PaulsenBook, WolfNotes, AaronJamesFuture, Takesaki} that $\pi (\rho)$ in Eq.~\ref{eq:Stinespring dilation} can be written as  $\pi (\rho) = u^\dagger ( \rho \otimes \ident^{\,}_{\text{ss}} ) \, u$ for some unitary $u$ acting on $\Hilbert{\rm dil}$. Accordingly, we have that 
\begin{equation}
    \label{eq:Stinespring Representation}
    \Phi (\rho) \, = \, V^\dagger \, \rho \otimes \ident^{\vpp}_{\text{ss}} \, V \, , ~~
\end{equation}
where we absorb $u$ into $V$ without loss of generality. We note that the representation above also invokes Choi's theorem for CP maps between finite-dimensional $C^*$-algebras~\cite{ChoisThm}.

For CP maps between generic $C^*$-algebras, which may be infinite dimensional, we stress that Eq.~\ref{eq:Stinespring Representation} is not guaranteed to hold. This is because the ${}^*$-homomorphism $\pi$ in Eq.~\ref{eq:Stinespring dilation} is not guaranteed to have a convenient functional form beyond finite-dimensional $C^*$-algebras, which are isomorphic to $M_N (\Comps)$. However, if $\hilbert$ is a tensor product of finitely many countably infinite Hilbert spaces $\Hilbert{i} = \ell^2 (\Nats)$, we expect that the Stinespring representation \eqref{eq:Stinespring Representation} remains valid \cite{AaronJamesFuture}, as these Hilbert spaces can be realized as inductive limits of a sequence of finite-dimensional Hilbert spaces $\hilbert$. The inductive limit of $M_N (\Comps)$ is the set $\mathcal{K}(\hilbert)$ of \emph{compact} operators on $\hilbert = \ell^2 (\Nats)$---an \emph{approximately finite} $C^*$-algebra~\cite{AaronJamesFuture, Davidson96, Brown, Bratteli}  to which all density matrices $\rho$ belong. Moreover, any observable $\observ \in \operatorname{End}(\hilbert)$ as $\DimOf{\hilbert} \to \infty$ is guaranteed to be close in the weak operator topology to an element of the approximately finite $C^*$-algebra $\mathcal{K}(\hilbert)$ \cite{AaronJamesFuture, Davidson96, Brown, Bratteli, Takesaki, PaulsenBook}.

We now derive the \emph{Kraus representation} of $\Phi$. Suppose that $\Hilbert{\text{ss}}$ has dimension $\HilDim{\text{ss}} = \NKraus$ and an orthonormal basis $\{ \ket{k} \}$; resolving the identity in Eq.~\ref{eq:Stinespring Representation}, we find that
\begin{equation}
    \label{eq:Kraus Representation}
    \Phi (\rho) \, = \, \sum\limits_{k=1}^{\NKraus} \, V^{\dagger} \, \rho \otimes \BKop{k}{k}^{\vpp}_{\text{ss}} \, V \, = \, 
    \sum\limits_{k=1}^{\NKraus} \, \Kraus{k} \, \rho \, \KrausDag{k} ~~~~\text{with}~~~~\Kraus{k}  \, \equiv \, V^{\dagger} \ket{k}^{\vpp}_{\text{ss}} \, ,~~
\end{equation}
where $\Kraus{k} : \hilbert \to \hilbert$ is a Kraus operator on $\hilbert$ and $\NKraus \leq \hildim^2$ (when $\hildim$ is finite). When the map $\Phi$  \eqref{eq:Kraus Representation} is a quantum channel (i.e., trace preserving), then the Kraus operators form a \emph{complete set}, 
\begin{align}
\label{eq:Kraus Completeness}
    0 \, = \, \tr{ \rho - \Phi(\rho)} \, = \, \tr{\rho \left( \ident - \sum\limits_{k=1}^{\NKraus} \, \KrausDag{k} \Kraus{k} \right) } ~~~~ \implies ~~~~ \sum\limits_{k=1}^{\NKraus} \, \KrausDag{k} \Kraus{k} \, = \, \ident \, ,~~
\end{align}
much like the projectors $\proj{m}$ \eqref{eq:Spec Projecc} form a complete set \eqref{eq:meas proj ortho complete}. When the map $\Phi$ is a quantum operation (i.e., trace decreasing),  we instead find that the Kraus operators form an \emph{incomplete set},
\begin{align}
\label{eq:Kraus Complete-able}
    0 \, \leq \, \tr{ \rho - \Phi(\rho)} \, = \, \tr{\rho \left( \ident - \sum\limits_{k=1}^{\NKraus} \, \KrausDag{k} \Kraus{k} \right) } ~~~~ \implies ~~~~ \KrausPow{0}{2} \, \equiv \, \ident - \sum\limits_{k=1}^{\NKraus} \, \KrausDag{k} \Kraus{k} \, \geq \, 0 \, ,~~
\end{align}
where the new Kraus operator $\Kraus{0}$ on the right is well defined because it is the square root of a positive operator (i.e., its eigenvalues are real and positive). Importantly, this implies that \emph{every} incomplete set of Kraus operators corresponding to a quantum operation is part of a larger, complete set of Kraus operators, since  one can always append $\Kraus{0}$ to the incomplete set to recover
\begin{equation}
    \label{eq:Kraus Completed}
    \sum\limits_{k=0}^{\NKraus} \, \KrausDag{k} \Kraus{k} \, = \, \ident \, , ~~
\end{equation}
meaning that every trace-decreasing quantum operation $\Phi$ involves a subset of a larger, complete Kraus operators that define a related, trace-preserving quantum channel.

Lastly, we discuss the Kraus representation of \emph{measurements}. First, consider the trace-decreasing quantum operation $\Phi_m$ corresponding to a measurement of $\observ$ \eqref{eq:measured observable} resulting in the $m$th outcome $\mEig*{m}$, so that $\Phi_m(\rho) = p^{\,}_m \, \rho^{\,}_m$ \eqref{eq:measurment map m} where $p^{\,}_m = \trace \left( \rho \, \proj{m} \right)$ is the probability to recover outcome $m$ \eqref{eq:Measurement Probability} and $\rho^{\,}_m$ is the collapsed postmeasurement density matrix following observation of outcome $m$. The Kraus representation \eqref{eq:Kraus Representation} that realizes the postmeasurement state $\Phi_m(\rho) = p^{\,}_m \, \rho^{\,}_m$ \eqref{eq:measurment map m} involves a single Kraus operator $\Kraus{m}=\proj{m}$ \eqref{eq:Spec Projecc}, with $\NKraus=1$. Next, consider the trace-preserving quantum channel $\Phi_{\rm av}$ corresponding to the average over a large number of measurements of $\observ$ \eqref{eq:measured observable}, meaning that $ \Phi_{\rm av} (\rho) = \rho^{\,}_{\rm av} = \sum_{m} \, p^{\,}_m \, \rho^{\,}_m = \sum_m \Phi_m (\rho)$ \eqref{eq:average measurement map}. There are $\NKraus = \Noutcome$ Kraus operators, given by $\Kraus{m}=\proj{m}$ \eqref{eq:Spec Projecc}, corresponding to the  $\Noutcome$ possible outcomes (the unique eigenvalues of $\mobserv$). Note that the set $\{ \proj{m} \}$ \eqref{eq:Spec Projecc} of Kraus operators is complete \eqref{eq:meas proj ortho complete}, as required \eqref{eq:Kraus Completed}.

\subsection{Unitary representation}
\label{subsec:unitary measurement}

We now use the Kraus representation \eqref{eq:Kraus Representation} to derive a \emph{unitary} representation of quantum operations and channels $\Phi$ on the dilated Hilbert space $\Hilbert{\rm dil}$. This representation \eqref{eq:Stinespring Unitary TP} is sometimes misattributed to the Stinespring dilation theorem~\cite{Stinespring, Takesaki, PaulsenBook} in the literature%~\cite{Marinescu}, 
, though it was actually proven by Kraus~\cite{KrausBook} for finite-dimensional quantum systems. To avoid confusion with the Kraus representation discussed in Sec.~\ref{subsec:Kraus}, we refer to Kraus's unitary representation of CP maps $\Phi$ as the ``Stinespring representation,'' and the corresponding theorem as the  ``physicist's Stinespring theorem.'' We expect this proof to extend---using the machinery of approximately finite $C^*$-algebras~\cite{Brown, Bratteli, Davidson96} and inductive and projective limits---to the infinite-dimensional Hilbert space $\ell^2 (\Nats)$~\cite{AaronJamesFuture}.

\subsubsection{The physicist's Stinespring theorem}
\label{subsec:physicist's Stinespring}

The physicist's Stinespring dilation theorem \cite{Stinespring, AaronJamesFuture} implies that any quantum channel (i.e., CPTP map) $\Phi$ acting on a finite-dimensional Hilbert space $\hilbert$ can be written in the form
\begin{equation}
    \label{eq:Stinespring Unitary TP}
    \Phi (\rho ) \, = \, \trace_{\text{ss}} \left( \, \umeas \, \rho \otimes \BKop{i}{i}^{\,}_{\text{ss}} \, \umeas^{\dagger} \, \right) \, ,~~
\end{equation}
where $\umeas$ acts unitarily on $\Hilbert{\rm dil}$ \eqref{eq:Dilated Hilbert}, $\ket{i}^{\,}_{\text{ss}} \in \Hilbert{\text{ss}}$ is some ``default'' initial state of the Stinespring Hilbert space, and the partial trace is over $\Hilbert{\text{ss}}$. Importantly, there is always a \emph{minimal} Stinespring representation---corresponding to the smallest possible dimension of $\Hilbert{\text{ss}}$ \cite{Stinespring, Takesaki, PaulsenBook, WolfNotes}---which is \emph{unique}, up to the definition of $\ket{i}^{\,}_{\text{ss}}$. We often consider nonminimal Stinespring representations.

We next extend the unitary representation of channels \eqref{eq:Stinespring Unitary TP} to quantum operations (i.e., trace-decreasing CP maps) $\Phi_m$. The Kraus operators that define a quantum operation $\Phi_m$ are a subset of those that define an associated quantum channel $\Phi$  \eqref{eq:Stinespring Unitary TP}, identified by the $\NKraus_m <  \NKraus = \HilDim{\text{ss}}$ basis states $\{ \ket{k_1}, \dots, \ket{k_{\NKraus_m}} \}$ of $\Hilbert{\text{ss}}$ \eqref{eq:Kraus Representation}. Accordingly, the quantum operation $\Phi_m$ can be written as
\begin{equation}
    \label{eq:Stinespring Unitary TD}
    \Phi_m (\rho) \, = \, \trace_{\text{ss}} \left( \, \umeas \, \rho \otimes \BKop{i}{i}^{\,}_{\text{ss}} \, \umeas^{\dagger} \, \Proj{\rm ss}{(m)} \, \right) \,, ~~
\end{equation}
where $\Proj{\rm ss}{(m)}$ projects onto the appropriate $\NKraus_m$ basis states of $\Hilbert{\text{ss}}$ associated with $\Phi_m$,  and $\umeas$ is the \emph{same} dilated unitary that defines the corresponding quantum channel $\Phi$ \eqref{eq:Stinespring Unitary TP} \cite{WolfNotes, AaronJamesFuture}. 

The dilated unitary $\umeas$ in Eqs.~\ref{eq:Stinespring Unitary TP} and \ref{eq:Stinespring Unitary TD} relates to both the Stinespring \eqref{eq:Stinespring Representation} and Kraus \eqref{eq:Kraus Representation} representations in that it acts on an arbitrary physical state $\ket{\psi} \in \hilbert$ %according to
via
\begin{equation}
\label{eq:Stinespring Unitary General}
    \umeas \ket{\psi} \otimes \ket{i}^{\,}_{\text{ss}} \, = \, \sum\limits_{k=1}^{\NKraus} \, \left( \Kraus{k} \ket{\psi} \right) \otimes \ket{k}^{\,}_{\text{ss}} \, = \, \sum\limits_{k=1}^{\NKraus} \,  \left(  V^{\dagger} \,  \ket{\psi} \otimes \ket{k}^{\,}_{\text{ss}} \right) \otimes \ket{k}^{\,}_{\text{ss}} \, ,~~
\end{equation}
where $\ket{i}^{\vpp}_{\text{ss}}$ is the ``default'' initial state on $\Hilbert{\text{ss}}$, and it is straightforward to check that this operation is length preserving (i.e., isometric) and surjective (i.e., unitary). As with the Kraus representation \eqref{eq:Kraus Representation} of Sec.~\ref{subsec:Kraus}, we expect that the results above extend to infinite-dimensional Hilbert spaces upon taking an inductive limit of finite-dimensional Hilbert spaces \cite{AaronJamesFuture, Davidson96, Brown, Bratteli}. We now consider the particular form of this unitary in the context of projective (and similar) measurements.

\subsubsection{Finite spectra}
\label{subsec:unitary finite}

We first work out the dilated unitary $\umeas$ \eqref{eq:Stinespring Unitary General} corresponding to the projective measurement of an observable $\mobserv$ \eqref{eq:measured observable} with a finite number $\Noutcome$ of unique eigenvalues. The minimum number of Kraus operators \eqref{eq:Kraus Representation} needed to represent a measurement channel is $\NKraus = \Noutcome$. For convenience, we label the eigenvalues of $\mobserv$ by $0 \leq m < \Noutcome$, so that the Kraus operators are given by $\Kraus{m}=\proj{m}$ \eqref{eq:Spec Projecc}. These Kraus operators are associated with a \emph{basis} $\{ \ket{m} \}$ of $\Hilbert{\text{ss}}$, and without loss of generality, we initialize the Stinespring register in the default state $\ket{i}^{\,}_{\text{ss}} = \ket{0}^{\,}_{\text{ss}}$, so that Eq.~\ref{eq:Stinespring Unitary General} becomes
\begin{equation}
    \label{eq:Stinespring Unitary Finite}
    \umeas \, \ket{\psi} \otimes \ket{0}^{\vpp}_{\text{ss}} \, = \, \sum\limits_{m=0}^{\Noutcome-1} \, \proj{m} \, \ket{\psi} \otimes \ket{m}^{\vpp}_{\text{ss}} \, , ~~
\end{equation}
which, in the minimal case with $\DimOf*{\Hilbert{\text{ss}}} = \Noutcome$, is \emph{uniquely} fulfilled by the dilated unitary operator
\begin{equation}
    \label{eq:Measurement Unitary Finite}
    \Umeas{\mobserv} \, \equiv \, \sum\limits_{m=0}^{\Noutcome-1} \, \proj{m} \otimes \ShiftPow{\text{ss}}{m} \, , ~~
\end{equation}
up to the choice (and definition) of the default state $\ket{0}^{\vpp}_{\text{ss}}$. The expression above involves the unitary $\Noutcome$-state Weyl \emph{shift operator}  $\shift$ \cite{AaronMIPT}, which acts on $\Hilbert{\text{ss}}$ as
\begin{equation}
    \label{eq:Weyl shift def}
    \ShiftPow{\text{ss}}{m} \, \equiv \, \sum\limits_{k=0}^{\Noutcome-1} \, \BKop{k+m~\text{mod}~\Noutcome}{k}^{\vpp}_{\text{ss}} \, ,~~
\end{equation}
and maps the default Stinespring state $\ket{0}^{\vpp}_{\text{ss}}$ to $\ket{m}^{\vpp}_{\text{ss}}$, as required. The shift operator $\shift$ is a unitary extension of the Pauli operator $\PX{}$ to Hilbert spaces with $\Noutcome \geq 2$ \cite{AaronMIPT}.

It is straightforward to verify that the Kraus operators $\Kraus{m}=\proj{m}$ form a complete set \eqref{eq:Kraus Completed} and that $\umeas$ \eqref{eq:Measurement Unitary Finite} is unitary on $\Hilbert{\rm dil}$. We next confirm  that the quantum channel $\Phi$ \eqref{eq:Stinespring Unitary TP} corresponding to $\umeas$ \eqref{eq:Measurement Unitary Finite}  leads to the outcome-averaged density matrix $\rho^{\,}_{\rm av}$ \eqref{eq:average measurement map}, 
\begin{align}
\label{eq:Stinespring finite TP check}
    \Phi (\rho) \, &= \, \trace_{\text{ss}} \left(  \umeas \, \rho \otimes \BKop{0}{0}%^{\vpp}_{\text{ss}} 
    \, \umeas^{\dagger}   \right) \, = \, \sum\limits_{m,n=0}^{\Noutcome-1} \hspace{-0.5mm} \proj{m} \, \rho \, \proj{n} \, \trace_{\text{ss}} \left( \ShiftPow{}{-m} \, \BKop{0}{0} \, \ShiftPow{}{n}  \right) \, = \, \sum\limits_{m=0}^{\Noutcome-1}  \proj{m} \, \rho \, \proj{m} \, = \, \rho^{\vpp}_{\rm av} \, , ~
\end{align}
as required. We also confirm that the quantum operation $\Phi_{\ell}$ \eqref{eq:Stinespring Unitary TD}---corresponding to a measurement of $\observ$ resulting in the %particular 
outcome $\mEig*{\ell}$---leads to $p^{\,}_{\ell}$ times the collapsed density matrix $\rho^{\,}_{\ell}$ \eqref{eq:measurment map m},
\begin{align}
    \label{eq:Stinespring finite TD check}
    \Phi_{\ell} (\rho) \, = \, \trace_{\text{ss}} \left(  \umeas \, \rho \otimes \BKop{0}{0}%^{\vpp}_{\text{ss}} 
    \, \umeas^{\dagger} \,  \Proj{\text{ss}}{( \ell )}  \right) \, = \, \sum\limits_{m,n=0}^{\Noutcome-1} \hspace{-0.5mm} \proj{m} \, \rho \, \proj{n} \, %\trace_{\text{ss}} \left( \ShiftPow{}{-m} \, \BKop{0}{0} \, \ShiftPow{}{n} \, \BKop{\ell}{\ell}  \right) 
    \inprod{\ell}{m} \inprod{n}{\ell} \, = \,  \proj{\ell} \, \rho \, \proj{\ell} \, = \, p^{\vpp}_{\ell} \, \rho^{\vpp}_{\ell} \, ,~
\end{align}
as required. In other words, the quantum operation corresponding to a measurement resulting in the outcome $\ell$ recovers from projecting $\Hilbert{\text{ss}}$ onto its $\ell$th basis state. 

Thus far, the dilated unitary $\umeas$ \eqref{eq:Stinespring Unitary Finite} has been a bookkeeping tool. In Secs.~\ref{sec:photons} and \ref{sec:qubits}, we demonstrate how, in real experiments, $\Hilbert{\text{ss}}$ \emph{physically} represents the state of the detector and $\umeas$ represents the time evolution of the system and detector during the measurement process. Since the state of the apparatus encodes the observed outcome, it is natural that the postmeasurement state $\rho^{\,}_\ell$ given outcome $\ell$ recovers from applying $\BKop{\ell}{\ell}^{\,}_{\text{ss}}$ (see also Sec.~\ref{sec:Using Stinespring}). As a reminder, the derivations above apply to the \emph{minimal} Stinespring representation with $\DimOf*{\Hilbert{\text{ss}}} = \Noutcome$. However, in general, there exist \emph{nonminimal} Stinespring representations for which $\DimOf*{\Hilbert{\text{ss}}} > \Noutcome$. These generally correspond to the actual detector degrees of freedom in the experiment, and reduce to the minimal representation upon ``binning'' distinct states corresponding to the same outcome.

\subsubsection{Single-qubit observables}
\label{subsec:unitary qubit}

First, consider measuring $\PZ{}$ on a qubit. While different physical realizations of qubits may have different interpretations and require different implementations (see Sec.~\ref{sec:qubits}), they all share the same minimal Stinespring representation \eqref{eq:Measurement Unitary Finite}. Since $\PZ{}$ has eigenvalues $\pm 1$ and eigenprojectors $\proj{\pm} = (\ident \pm \PZ{})/2$, the measurement unitary is straightforward to work out,
\begin{subequations}
\label{eq:Qubit Z Meas}
\begin{align}
    \Umeas{\PZ{}} \, &= \, \sum\limits_{n=0,1} \frac{1}{2} \left( \ident +(-1)^n \, \PZ{} \right) \otimes \ShiftPow{\text{ss}}{n} \label{eq:Qubit Z Meas General} \\
    &= \, \BKop{0}{0} \otimes \ident^{\vpp}_{\text{ss}} + \BKop{1}{1}\otimes \Shift{\text{ss}} \, = \, \operatorname{CNOT} ( {\text{ph}} \to {\text{ss}} ) \, ,~~ \label{eq:Qubit Z Meas Unitary}
\end{align}
\end{subequations}
where $\PZ{} \ket{n} = (-1)^n \, \ket{n}$, so that the measurement unitary \eqref{eq:Qubit Z Meas} is simply a CNOT gate with the physical system the ``control'' qubit and the detector the ``target'' qubit. As we discuss in Sec.~\ref{sec:qubits}, the detectors used in experiments are almost always more complicated than a single qubit; however, in the idealized limit of the measurement (i.e., vanishing probability of readout error), binning detector states into $n=0$ versus $n=1$ results in the minimal representation above~\eqref{eq:Qubit Z Meas}.

We next generalize $\umeas$ \eqref{eq:Qubit Z Meas} to arbitrary single-qubit observables $\mobserv^{\,}_j$ with eigenvalues $\mEig*{0} > \mEig*{1}$; and decomposing this operator onto the Pauli basis for qubit $j$, we have that
\begin{equation}
    \label{eq:Single-qubit Pauli decomposition}
    \mobserv^{\vpp}_j \, \equiv \, 
    \frac{1}{2} \, \trace \left(  \, \mobserv^{\vpp}_j \, \right) \, \ident_j + \frac{1}{2} \, \sum\limits_{\nu=x,y,z} \, \trace \left( \, \Pauli{\nu}{j} \, \mobserv^{\vpp}_j \, \right) \, \Pauli{\nu}{j} %\, = \, \frac{\mEig{0}+\mEig{1}}{2} \, \ident^{\,}_j + \left( \mEig{0} - \mEig{1} \right) \, \hat{\alpha}^{\vpp}_j \cdot \vec{\sigma}^{\vpp}_j 
    \, , ~~ 
\end{equation}
so that there exists a traceless, involutory operator $\bar{\mobserv}^{\,}_j$ with the same eigenvectors \cite{AaronTeleport} given by
\begin{equation}
    \label{eq:Involutory Part}
    \bar{\mobserv}^{\vpp}_j \, \equiv \, \frac{1}{\mEig{0}-\mEig{1}} \sum\limits_{\nu=x,y,z} \, \trace \left(\, \mobserv^{\vpp}_j \, \Pauli{\nu}{j} \, \right) \, \Pauli{\nu}{j}  ~~~\text{so~that}~~~ \proj{m} \, = \, \frac{1}{2} \left( \ident^{\vpp}_j + \left( -1 \right)^m \, \bar{\mobserv}^{\vpp}_j \right) \, ,~~
\end{equation}
whose eigenvalues are $\bar{\meig}^{\,}_m = (-1)^m$, where $\proj{m}$ are the spectral projectors \eqref{eq:Spec Projecc} for $\mobserv^{\,}_j$ \eqref{eq:Single-qubit Pauli decomposition} as well. The dilated unitary \eqref{eq:Measurement Unitary Finite} that captures measurement of either $\mobserv^{\,}_j$ or its involutory part $\bar{\mobserv}^{\,}_j$ is
\begin{equation}
    \label{eq:Qubit Meas Unitary General}
    \Umeas{\mobserv_j} \, = \, \frac{1}{2} \left( \ident^{\vpp}_j + \bar{\mobserv}^{\vpp}_j \right) \otimes \ident^{\vpp}_{\text{ss}} +  \frac{1}{2} \left( \ident^{\vpp}_j - \bar{\mobserv}^{\vpp}_j \right) \otimes \Shift{\text{ss}}  \, , ~~
\end{equation}
which applies to \emph{any} single-qubit observable $\mobserv^{\,}_j$ \eqref{eq:measured observable} and any operator unitarily connected thereto.

\subsubsection{Infinite spectra}
\label{subsec:unitary countably infinite}

Now, consider an observable $\mobserv$ \eqref{eq:measured observable} with infinitely many eigenvalues ($\Noutcome \to \infty$). The Weyl operator $\shift$ \eqref{eq:Weyl shift def} is no longer defined, since ``modulo $\Noutcome$'' has no meaning. Although an alternative unitary to $\shift$ \eqref{eq:Weyl shift def} may be defined by labeling eigenvalues via $n \in \Ints$ (instead of $n \in \Nats$) to realize a \emph{minimal} Stinespring representation~\cite{AaronJamesFuture}, we instead consider a \emph{nonminimal} representation on
\begin{equation}
    \label{eq:SS Infinite Qubit Hilbert}
    \Hilbert{\text{ss}} \, = \, \bigotimes\limits_{j \in \Nats} \, \Comps^{2} ~~~~\text{with}~~~N^{\,}_{\text{ss}} \, = \, \DimOf{\Hilbert*{\text{ss}}} \, = \, \infty \, ,~~
\end{equation}
corresponding to infinitely many qubits labeled by a ``site'' $j \in \Nats$. This nonminimal Stinespring Hilbert space \eqref{eq:SS Infinite Qubit Hilbert} is not separable---its dimension is \emph{uncountably} infinite by Cantor's theorem. 

Importantly, we note that the spectrum of $\mobserv$ is \emph{countably} infinite, so that the \emph{minimal} Stinespring Hilbert space has a countable basis (i.e., is separable). It is common practice to define a separable analogue of $\Hilbert*{\text{ss}}$ \eqref{eq:SS Infinite Qubit Hilbert} in terms of finite numbers of spin flips above the reference state $\ket{\bvec{0}}_\text{ss}$ (with all qubits in the $+1$ eigenstate $\ket{0}$ of $Z$). However, measuring the boson occupation number may require exciting an infinite number of qubits; moreover, separability itself is not a concern. Instead, we note  $\Hilbert*{\text{ss}}$ \eqref{eq:SS Infinite Qubit Hilbert} can be partitioned into a countably infinite number of subspaces of configurations in which exactly $n$ qubits are in the excited state $\ket{1}$, with all others in the default state $\ket{0}$. Each subspace reflects a different outcome, and imposing the condition on $\Hilbert*{\text{ss}}$ \eqref{eq:SS Infinite Qubit Hilbert} that all qubits in the state $\ket{1}$ are to the ``left'' of all qubits in the state $\ket{0}$ ensures that there is only one state from each subspace $n$---and thus, one state per outcome. In other words, labeling Stinespring qubits by the order in which they are excited leads to a countably, minimal Stinespring representation.  

That minimal Stinespring Hilbert space is simply the bosonic Hilbert space $%L^2 (\Reals) = 
\ell^2 (\Nats) \cong \Comps^{\infty}$. Still,  $\Hilbert*{\text{ss}}$ \eqref{eq:SS Infinite Qubit Hilbert} reflects the actual detector Hilbert space in experimental measurements of observables $\mobserv$ with countably infinite spectra, as we discuss in Sec.~\ref{sec:photons}. Initializing the detector qubits in the state $\ket{i}^{\,}_{\text{ss}} = \ket{\bvec{0}}^{\,}_{\text{ss}}$, we arrive at the following extension of $\umeas$ \eqref{eq:Measurement Unitary Finite} to countably infinite spectra,
\begin{equation}
    \label{eq:Measurement Unitary Countably Infinite}
    \Umeas{\mobserv} \, \equiv \, \sum\limits_{n \in \Nats} \, \proj{n} \otimes \prod\limits_{j=1}^n \, \Shift{\text{ss},j} \, ,~~
\end{equation}
where $\Shift*{\text{ss},j}$ is a Pauli $\PX*{}$ operator acting on the $j$th Stinespring qubit. Physically, this describes, e.g., the counting of the number of photons $n$ at a given frequency by creating $n$ excitations in a detector.  Because the order of excitement is unimportant, all states with $n$ excitations reflect the outcome $\mEig*{m}$. We note that $\umeas$ \eqref{eq:Measurement Unitary Countably Infinite} is only unitary on the full $\Hilbert*{\text{ss}}$ \eqref{eq:SS Infinite Qubit Hilbert}, and not on $\ell^2 (\Nats) \subset \Hilbert*{\text{ss}}$.

\subsubsection{Destructive measurements}
\label{subsec:destructive}

We now consider \emph{destructive} measurements, which are similar to the projective measurements considered thus far. While a projective measurement of $\mobserv$ \eqref{eq:measured observable} in the state $\rho$ resulting in the outcome  $\mEig*{m}$ \emph{projects} the system into the eigenstate $\rho^{\,}_m \propto \proj*{m} \, \rho \proj*{m}$ of $\mobserv$ (with the observed eigenvalue $\mEig*{m}$), a destructive measurement \emph{destroys} the measured property of the initial state $\rho$, generally resulting in a single postmeasurement state $\rho'$ regardless of the observed eigenvalue $\mEig*{m}$.

Consider a bosonic state $\ket{\psi} = \sum_{n=0}^{\infty} c_n \ket{n}$ (e.g., a harmonic oscillator or electromagnetic mode), where a \emph{projective} measurement of the boson number  $N = a^\dagger a = \sum_{n=1}^{\infty}  n \, \BKop{n}{n}$ is captured by
\begin{equation}
    \label{eq:boson count projective action}
    \Umeas{N} \, : \, \ket{\Psi^{\,}_i} \, = \,  \sum\limits_{n=0}^{\infty} \, c^{\,}_n \, \ket{n}^{\,}_{\text{ph}} \otimes \ket{0}^{\,}_{\text{ss}} ~ \mapsto ~ \ket{\Psi^{\,}_f} \, = \, \sum\limits_{n=0}^{\infty} \, c^{\,}_n \, \ket{n}^{\,}_{\text{ph}} \otimes \ket{n}^{\,}_{\text{ss}} \, ,~~
\end{equation}
under $\Umeas{N}$ \eqref{eq:Measurement Unitary Countably Infinite}. Following the measurement, if the Stinespring register is found to be in the state $\ket{n}^{\,}_{\text{ss}}$ with $n$ excited qubits, the system is in the state $\ket{n}^{\,}_{\text{ph}}$ corresponding to exactly $n$ bosons.

However, one could imagine that counting the number of bosons $N$ requires destroying them one by one, until none remain. Such a \emph{destructive} measurement is captured by the unitary
\begin{equation}
    \label{eq:boson count destructive action}
    \Dmeas{N} \, : \, \ket{\Psi^{\,}_i} \, = \,  \sum\limits_{n=0}^{\infty} \, c^{\,}_n \, \ket{n}^{\,}_{\text{ph}} \otimes \ket{0}^{\,}_{\text{ss}} ~ \mapsto ~ \ket{\Psi^{\prime}_f} \, = \, \sum\limits_{n=0}^{\infty} \, c^{\,}_n \, \ket{0}^{\,}_{\text{ph}} \otimes \ket{n}^{\,}_{\text{ss}} \, ,~~
\end{equation}
where $\ket{n}^{\,}_{\text{ss}}$ is a shorthand for \emph{any} configuration of $\Hilbert*{\text{ss}}$ \eqref{eq:SS Infinite Qubit Hilbert} with exactly $n$ qubits in the state $\ket{1}$; the physical state following the destructive measurement is always empty. 

Noting that $V :  \ket{0}^{\,}_{\text{ph}} \otimes \ket{n}^{\,}_{\text{ss}} \mapsto \ket{n}^{\,}_{\text{ph}} \otimes \ket{0}^{\,}_{\text{ss}}$ is an isometry on $\Hilbert*{\text{ph}} \otimes \ell^2 (\Nats)$, we write
\begin{equation}
    \label{eq:Destructive Isometry}
    V^\dagger \, = \, \sum\limits_{n=0}^{\infty} \, \BKop{0}{n}^{\vpp}_{\text{ph}} \otimes \Proj{\text{ss}}{(n)} \, = \, \sum\limits_{n=0}^{\infty} \, \left[ \, a \, N^{-1/2} \, \right]^n \otimes \prod\limits_{j=1}^{n} \, %\frac{1}{2} \left( \SSid{j} - \SSZ{j} \right) \, 
    \BKop{1}{1}^{\vpp}_j \hspace{-0.5mm} \prod\limits_{k=n+1}^{\infty} \, %\frac{1}{2} \left( \SSid{k} + \SSZ{k} \right) 
    \BKop{0}{0}^{\vpp}_k \, , ~~
\end{equation}
where $a$ is the \emph{lowering operator} $a \, \ket{n} = \sqrt{n} \, \ket{n-1}$,  $a^{\dagger}$ is the raising operator, and $N = a^\dagger a$. The isometry $V$ realizes a \emph{minimal} Stinespring representation on $\ell^2 (\Nats) \subset \Hilbert*{\text{ss}}$ and satisfies
\begin{equation}
    \label{eq:Destructive Isometry Check}
    V^\dagger V\, = \, \ident^{\vpp}_{\text{ph}} \otimes \ident^{\vpp}_{\text{ss}} ~~~~\text{and}~~~~V \,  V^\dagger \, = \, \ident^{\vpp}_{\text{ph}} \otimes \ident^{\vpp}_{\text{ss}} - \sum\limits_{n=1}^{\infty} \, \sum\limits_{k=0}^{n-1} \, \BKop{k}{k}^{\vpp}_{\text{ph}} \otimes \Proj{\text{ss}}{(n)} \, ,~~
\end{equation}
as required. Moreover, $V^\dagger$ can always be embedded in a unitary $\umeas'$ acting on all of $\Hilbert*{\text{ss}}$ (see Sec.~\ref{subsec:photon counting}), so the destructive measurement of $N$ is captured by the unitary operation
\begin{equation}
    \label{eq:Destructive Unitary from Projective}
    \Dmeas{N} \, = \, \umeas' \, \Umeas{N} \, , ~~
\end{equation}
where $\Umeas*{N}$ \eqref{eq:Measurement Unitary Countably Infinite} realizes a projective measurement of $N$ and $\umeas'$ is a unitary embedding of $V^\dagger$ \eqref{eq:Destructive Isometry} in $\Hilbert*{\rm dil}$ \eqref{eq:SS Infinite Qubit Hilbert}. In this sense, projective and destructive measurements are equivalent up to some unitary $\umeas'$. In the case above, the  destructive measurement is equivalent to a projective measurement followed by an outcome-dependent unitary $\umeas'$ that maps the postmeasurement physical state to some state $\ket{0}$; a projective measurement is equivalent to a destructive measurement followed by outcome-dependent restoration of the corresponding eigenstate $\ket{n}$.

\subsection{von Neumann representation}
\label{subsec:Hamiltonian measurement}

An alternative to the unitary representation of measurements outlined in Sec.~\ref{subsec:unitary measurement} is von Neumann's formulation in terms of a Hamiltonian $\Ham$ \cite{VonNeumann, Preskill_QI}. Like the unitary $\umeas$ \eqref{eq:Stinespring Unitary General}, the Hamiltonian $\Ham$ acts on the \emph{dilated} Hilbert space $\Hilbert*{\rm dil} = \Hilbert*{\text{ph}} \otimes \Hilbert*{\text{ss}}$, where now, $\Hilbert*{\text{ss}}$ explicitly represents the state of the measurement apparatus. The Hamiltonian formulation is often more useful in the context of quantum optics \cite{SqueezeBook, JacobsBook}.  We review the von Neumann representation in Sec.~\ref{subsec:von Neumann}, consider the example of the Stern-Gerlach experiment \cite{SternGerlach1, SternGerlach2, SternGerlach3} in Sec.~\ref{subsec:SternGerlach}, and connect the pointer Hamiltonian $\Ham$ to the Stinespring unitary $\umeas$ \eqref{eq:Stinespring Unitary General} in Sec.~\ref{subsec:vN SS}.

\subsubsection{Pointer Hamiltonian}
\label{subsec:von Neumann}

We now review von Neumann's description of measurements in terms of a ``pointer particle'' \cite{VonNeumann, Preskill_QI}. This model represents an early attempt to account for the details of the measurement apparatus, albeit in terms of $\Ham$ and not the unitary $\umeas$ \eqref{eq:Measurement Unitary Finite} it generates. The pointer particle propagates freely, apart from a coupling between the observable $\mobserv$ to be measured and the momentum $p$ of the pointer particle (with coupling strength $\lambda$),  i.e.,
\begin{equation}
    \label{eq:Pointer Hamiltonian}
    \Ham \, = \, \Ham_0 \otimes \ident_{\rm pp} + \ident_{\text{ph}} \otimes \frac{p^2}{2M} + \lambda \, \mobserv \otimes p \, , ~~
\end{equation}
where $\Ham_0$ is the Hamiltonian for the physical system alone, and $M \gg 1$ is the mass of the pointer particle (``pp''). As a comment, the Wigner-Araki-Yanase Theorem \cite{WignerBusch, ArakiYanase} generally requires that $\mobserv$ commute with $\Ham_0$ or else the uncertainty is generally guaranteed to be large.

The combined system is prepared in the state $\ket{ \Phi (0) } = \ket{ \phi (0) }_{\text{ph}} \otimes \ket{ \varphi (0) }_{\rm pp}$, with
\begin{subequations}
\label{eq:Example Pointer Initial State}
\begin{align}
    \ket{\phi(0)}_{\text{ph}} \, &= \, \sum\limits_{m=0}^{\Noutcome-1} \, \sum\limits_{\ell = 1}^{\EigMult*{m}} \, \phi_{m,\ell} \, \ket{m,\ell}_{\text{ph}} \label{eq:Example Pointer Initial State Physical} \\
    \ket{\varphi (0)}_{\rm pp} \, &= \, \int_{\Reals} \thed x \, \varphi_0 (x) \, \ket{x}_{\rm pp} \,=\, \int_{\Reals} \thed x \, \frac{\exp ( - (x - x_0)^2 / 4 \sigma^2 )}{\left( 2 \, \pi \, \sigma^2 \right)^{1/4}} \, \ket{x}_{\rm pp} \label{eq:Example Pointer Initial State Pointer} 
    \, ,~~
\end{align} 
\end{subequations}
where $\mobserv \, \ket{m,\ell} = \mEig*{m} \, \ket{m,\ell}$ and $\ell$ runs over the $\EigMult*{m}$ degenerate states with  eigenvalue $\mEig*{m}$. The %initial 
state $\ket{ \Phi (0) } = \ket{ \phi (0) }_{\text{ph}} \otimes \ket{ \varphi (0) }_{\rm pp}$ satisfies $\expval{x} = x_0$, $\expval{p} = 0$, $\Delta x = \sigma$, and $\Delta p = \hbar / 2 \sigma$. 

Assuming that $\comm{\mobserv}{H_0}=0$ and $M \gg 1$, one can ignore both $\Ham_0$ and the pointer's dispersion in Eq.~\ref{eq:Pointer Hamiltonian}. In this limit, the time evolution operator is well approximated by the unitary
\begin{equation}
    \label{eq:Pointer Unitary}
    %\umeas (t) \, = \, \exp \left( - \ii \, H \, t / \hbar \right) \, = \, \sum\limits_{m=0}^{\Noutcome-1} \, \proj{m} \otimes \ee^{- \ii t \left( \lambda \, \mEig{m} \, p + p^2 / 2 \, M + H_0 \right)/ \hbar} ~~ \mapsto ~~
    \umeas^{\vpd}_{\rm eff} (t) \, = \, \sum\limits_{m=0}^{\Noutcome-1} \, \proj{m} \otimes \ee^{-\ii \, t \, \lambda \, \mEig{m} \, p / \hbar} \, ,~~
\end{equation}
under which, the initial state $\ket{ \Phi (0) }$ \eqref{eq:Example Pointer Initial State} evolves into
\begin{align}
    \ket{ \Phi (t) } \, &= \, \umeas^{\vpd}_{\rm eff} (t) \,  \ket{ \Phi (0) } \, = \, \umeas^{\vpd}_{\rm eff}  (t) \, \sum\limits_{m=0}^{\Noutcome-1} \sum\limits_{\ell=1}^{n_m} \, \int_{\Reals} \thed x \,    \phi_{m,\ell} \,\ket{m,\ell}_{\text{ph}} \otimes \varphi_0 (x) \, \ket{x}_{\rm pp} \notag \\
    &= \, \sum\limits_{m,k=0}^{\Noutcome-1} \sum\limits_{\ell=1}^{n_m} \, \int_{\Reals} \thed x \,  \left( \phi_{m,\ell} \, \proj{k} \, \ket{m,\ell}_{\text{ph}} \right) \otimes \left(  \ee^{- \lambda \, t \, \mEig{k} \, \pd{x}} \,  \varphi_0 (x) \, \right) \, \ket{x}_{\rm pp} \notag \\
    &= \, \sum\limits_{m=0}^{\Noutcome-1} \sum\limits_{\ell=1}^{n_m} \,  \int_{\Reals} \thed x \,   \phi_{m,\ell} \,\ket{m,\ell}_{\text{ph}} \otimes \varphi_0 (x - \lambda \, t \, \mEig{m} ) \, \ket{x}_{\rm pp} 
    \label{eq:Pointer State Time t}
    \, , ~~
\end{align}
which can be summarized as the following map \eqref{eq:Pointer Unitary} on the initial state \eqref{eq:Example Pointer Initial State}, 
\begin{equation}
\label{eq:Pointer Map}
    \ket{ \phi (0) }_{\text{ph}} \otimes \int_{\Reals} \thed x \, \varphi_0 (x) \, \ket{x}_{\rm pp} ~\mapsto ~ \sum\limits_{m=0}^{\Noutcome-1} \, \left( \proj{m} \, \ket{ \phi (0) } \right)_{\text{ph}} \otimes  \int_{\Reals} \thed x \,  \varphi_0 (x - \lambda \, t \, \mEig{m} ) \, \ket{x}_{\rm pp} \, ,~~
\end{equation}
meaning that, for $t>0$, the full state is a sum over outcomes $m$ of the product of the physical postmeasurement state $\proj{m} \, \ket{ \phi (0) }_{\text{ph}}$ and the pointer state $\varphi_0 (x - \lambda \, t \, \mEig*{m})$.

In other words, the position of the pointer particle indicates the measurement outcome: If the initial uncertainty $\Delta x = \sigma$ in the pointer's position $x$ satisfies $\sigma \ll \lambda \, t \, \min ( \delta \mEig*{m} )$, where $\min (\delta \mEig*{m})$ is the minimum difference between consecutive eigenvalues of $\mobserv$ (ordered by absolute value), then one can distinguish the particular eigenvalue $\mEig*{m}$ by the displacement of the pointer particle's wave packet initially centered about $x_0$ with standard deviation $\Delta x = \sigma$ \eqref{eq:Example Pointer Initial State Pointer}. 

The final state \eqref{eq:Pointer Map} is realized in the limit $\comm{H_0}{\mobserv} = 0$ and $M \to \infty$ in Eq.~\ref{eq:Pointer Hamiltonian}. When $\comm{H_0}{\mobserv} \neq 0$ or the pointer particle's mass is small, the same result \eqref{eq:Pointer Map} can be achieved by increasing the coupling to the pointer particle ($\lambda \gg 1$) and decreasing the interaction time ($t \ll 1$), so that $H_0$ and the dispersion term $p^2/2 M$ in  Eq.~\ref{eq:Pointer Hamiltonian} do not have time to introduce noise to  Eq.~\ref{eq:Pointer Map}. Finally, we comment that for any fixed $t$, the von Neumann unitary evolution operator $\umeas (t)$ \eqref{eq:Pointer Unitary} is equivalent to the Stinespring measurement unitary $\umeas$ \eqref{eq:Stinespring Unitary General}.

\subsubsection{The Stern-Gerlach experiment}
\label{subsec:SternGerlach}

One of the first examples of a quantum measurement was realized by Stern and Gerlach \cite{SternGerlach1, SternGerlach2, SternGerlach3, LeBellac}. A stream of particles with intrinsic magnetic moments (e.g., silver atoms) travels in the positive $x$ direction through an apparatus that applies a magnetic field with nonzero gradient in the $z$ direction. The particles have an internal spin magnetic moment $\bvec{\mu}$, resulting in a potential $V = - \bvec{\mu} \cdot \bvec{B}$; since the gradient is nonvanishing in the $z$ direction, we expect a nonzero force on the particle, proportional to $\mu = \abs{\bvec{\mu}}$. In this sense, the atom is \emph{its own pointer particle}: The position of the atom after exiting the apparatus effectively measures its magnetic moment $\bvec{\mu}$ in the $z$ direction. Crucially, while classical mechanics predicts a continuum of final positions of the particles, corresponding to a continuum of possible magnetic moments $\mu_z = \abs{\mu} \, \cos (\theta)$, the actual experiment shows \emph{quantized} outcomes, corresponding to the particle appearing at $z=\pm \delta z$. The value of $\delta z$ is determined by experimental details that do not vary between shots.

We first construct the Hamiltonian describing the particles inside the apparatus (i.e., from $x=0$ to $x=L$). Assuming the magnetic field is $\bvec{B} = (0, - b y, B_0 + b z)$, where $B_0 \gg b$, we have that
\begin{equation}
    \label{eq:Stern Gerlach full Hamiltonian}
    H_{\rm SG} \, = \, - \frac{\hbar^2}{2 M} \left( \pdp{x}{2} + \pdp{y}{2} + \pdp{z}{2} \right) + \mu^{\,}_B \, \left( - b \, y \, \Pauli{y}{} + (B_0 + b \, z) \, \Pauli{z}{} \right)  \, , ~~
\end{equation}
where $\Pauli{n}{}$ is the $n$th Pauli matrix, and the particles' magnetic moment has vector components $\mu_n = \mu^{\,}_B \, g \, \Pauli{n}{} /2 \approx \mu^{\,}_B \, \Pauli{n}{}$,  where $\mu^{\,}_B$ is the Bohr magneton. Instead of the time-dependent Schr\"odinger equation under \eqref{eq:Stern Gerlach full Hamiltonian}, we consider the more straightforward Heisenberg evolution of operators.

We make the standard assumption that the initial state is separable with respect to the coordinates $x,y,z$. Noting that $x$ is decoupled, we assume that $\phi_x (p_x) \propto \exp ( - (p_x - M \, v)^2 / 4 \delta_x^2 )$,  where $\delta_x$ is the $x$-momentum variance, and $v$ is a velocity. In the Heisenberg picture, we find
\begin{equation}
    \label{eq:Stern Gerlach x Heis}
     x(t) \, = \, x(0) + \frac{t}{M} \, p_x(0) ~~~ \implies ~~~ \expval{x}_t \, = \, \matel{\phi_x (p_x)}{x(t)}{\phi_x (p_x)} \, = \, v \, t \, , ~~
\end{equation}
in accordance with Ehrenfest's Theorem~\cite{Ehrenfest}. Hence, the particle traverses the apparatus in time $t = L / v$; we assume that the particle's $z$ position is measured upon exiting the apparatus at $x = L$. 

The analogous expressions to Eq.~\ref{eq:Stern Gerlach x Heis} for  $y$ and $z$ are far more complicated, and do not truncate at finite order in $t$. Expressions up to $\Order{t^5}$ appear in App.~\ref{app:SG general}, and suggest simplifications. First, since the $x$-dependent part of $\ket{\Psi(t)}$ evolves independently, it may be considered separately. Second, we find in App.~\ref{app:SG general} that $\expval{y}_t = \matel{\Psi(0)}{y(t)}{\Psi(0)}$ is identically zero to $\Order{t^5}$ in generic initial states, and we expect this holds for most or all orders, so that the dynamics in the $y$ direction can be ignored (i.e., we can fix $y=0$). As a result, $H_{\rm SG}$ \eqref{eq:Stern Gerlach full Hamiltonian} is well approximated by
\begin{equation}
    \label{eq:Stern Gerlach effective Hamiltonian}
    \Ham_z \, = \, -\frac{\hbar^2}{2M} \pdp{z}{2} + \mu^{\,}_B \, \left( B_0 + b \, z \right) \, \Pauli{z}{} \, , ~~
\end{equation}
where we ignore $y$ dynamics. Suppose that the particle is initialized in the state
\begin{equation}
    \label{eq:Stern Gerlach initial state nice}
    \ket{\Psi(0)} \, = \, \sum\limits_{s = \pm 1} \, \int_{\Reals} \, \thed z \, \Psi^{\vps}_s (z,0) \, \ket{z,s} \, = \, \sum\limits_{s = \pm 1} \, c^{\vps}_s \, \int_{\Reals} \, \thed z \, \frac{\ee^{-\left( z - z_0 \right)^2/4 \, \delta^2}}{\left( 2 \, \pi \, \delta^2 \right)^{1/4}} \, \ket{z,s} \, ,~~
\end{equation}
corresponding to a Gaussian  wave packet centered at $z = z_0$ with variance $\delta$. 

The unitary operator that generates time evolution under $H_z$  \eqref{eq:Stern Gerlach effective Hamiltonian} is given by
\begin{align}
    \umeas (t) \, &= \, \exp \left( \frac{t \, \mu^{\,}_B}{\ii \, \hbar} \left( B_0 + b \, z \right) \, \Pauli{z}{} + \frac{\ii \, \hbar \, t}{2 M} \, \pdp{z}{2} \right) \notag \\
    &= \, \ee^{\ii \, b^2 \, \mu^2_B \, t^3 /M \, \hbar} \, \exp \left( \frac{t \, \mu^{\,}_B}{\ii \, \hbar}  \left( B_0 + b \, z \right) \, \Pauli{z}{} \right) \, \exp \left( \frac{\ii \, \hbar \, t}{2 M} \, \pdp{z}{2} \right) \, \exp \left( -\frac{b \, \mu^{\,}_B}{M} \, t^2 \, \pd{z} \, \Pauli{z}{} \right)  \, ,~~ \label{eq:Stern Gerlach nice unitary} 
\end{align}
via the Zassenhaus formula~\cite{MagnusBCH}. Applying this operator to the initial state $\ket{\Psi(0)}$ \eqref{eq:Stern Gerlach initial state nice} leads to
\begin{equation}
    \label{eq:Stern Gerlach nice final wavefunction}
    \Psi^{\vps}_s (z,t) \, = \, c^{\vpp}_s \, \frac{\ee^{\ii \, b^2 \, \mu^2_B \, t^3 / \hbar \, M} \, \ee^{- \ii \, s \, t \, \mu^{\,}_B (B_0 + b z)/\hbar}}{(2 \, \pi)^{1/4} \, ( \delta + \, \frac{\ii \, \delta \, t}{2 \delta})^{1/2}} \, \exp \left[ - \frac{\left( z - z_0 - \frac{b \, \mu^{\,}_B}{M} \, s \,  t^2 \right)^2}{4 \, \delta^2} \, \frac{1}{1+ \frac{\ii \, \hbar \, t}{2 \, M \, \delta^2}} \right] \, ,~~
\end{equation}
as derived in detail in App.~\ref{app:SG easy}. The probability distribution for the particle's $z$ coordinate is
\begin{equation}
    \label{eq:Stern Gerlach nice z distribution}
    %p^{\vpp}_s ( z ) \, = \, \abs{c^{\vps}_s}^2 \, \exp \left[ - \frac{\left( z - z_0 - b \, \mu^{\,}_B\,  s \,  t^2 / M \right)^2}{2  \delta^2 \, - \hbar^2  t^2 / 2 M^2 \delta^2 }   \right] \, / \, \delta \sqrt{2 \, \pi - \frac{\pi \hbar^2 t^2}{2 M^2 \delta^4} } \, ,~~
    p^{\vpp}_s ( z, t ) \, = \, p(s) \, \mathcal{N} \left( z_0 + \frac{b \, \mu^{\,}_B}{M} \, s\,  t^2 , \, \delta^2 \, \left( 1 - \frac{\hbar^2 \, t^2}{4 \, M^2 \, \delta^4} \right) \right) \, ,~~
\end{equation}
where $p(s)$ is the probability of $\expval{\Pauli{z}{}} = s = \pm 1$ in the initial state \eqref{eq:Stern Gerlach initial state nice}, and $\mathcal{N}(z_{\text{av}},\sigma^2)$ is a normal distribution with mean $z_{\text{av}}$ and variance $\sigma^2$. We evaluate $p^{\,}_s (z,t)$ \eqref{eq:Stern Gerlach nice z distribution} at time $t=L/v$, where $v$ is the initial $x$ velocity and $L$ is the length of the apparatus in the $x$ direction.

Evolving the particle's $z$ coordinate under $\Ham_z$ \eqref{eq:Stern Gerlach effective Hamiltonian} in the Heisenberg picture leads to 
\begin{equation}
    \label{eq:Stern Gerlach effective z Heis}
    z(t) \, = \, z + \frac{t}{M} \, p_z - \frac{b \, \mu^{\,}_B}{2 \, M} \, t^2 \, \Pauli{z}{} ~~~\implies ~~~ \expval{z}_{s,t} \, = \, \matel{\Psi^{\,}_s (0)}{z(t)}{\Psi^{\,}_s(0)} \, = \, z_0 - \frac{b \, \mu_B^{\,}}{2 \, M} \, s \, t^2 \, ,~~
\end{equation}
evaluated in $\ket{\Psi(0)}$ \eqref{eq:Stern Gerlach initial state nice}, where the $x$ dynamics \eqref{eq:Stern Gerlach x Heis} determine $t = L / v$. 

Because the particle's position is classically accessible, one expects to observe the classical behavior associated with the expectation values of $x$ \eqref{eq:Stern Gerlach x Heis} and $z$ \eqref{eq:Stern Gerlach effective z Heis}, with $y$ trivial. However, the particle's magnetic moment (captured via $s$) is not classically observable; moreover, one cannot observe a superposition of the macroscopically distinct $z$-basis states---corresponding to the two spin values $s = \pm 1$---in experiment (by nature, classical objects cannot appear in macroscopically distinct superpositions). Instead, any experiment appears to result in a shift of $z$ by \emph{either} $-\alpha$ (when $s=1$) or $+\alpha$ (when $s=-1$). Importantly, there is not a continuum of favored displacements, but a \emph{single} displacement, with sign given by the particle's internal magnetic moment with eigenvalues $s = \pm 1$.

\subsubsection{Connection to the Stinespring unitary}
\label{subsec:vN SS}

We now connect von Neumann's Hamiltonian representation  to the Stinespring  unitary $\umeas$ \eqref{eq:Stinespring Unitary General} from Sec.~\ref{subsec:unitary measurement}. Both representations act on the \emph{dilated} Hilbert space $\Hilbert*{\text{dil}} = \Hilbert*{\text{ph}} \otimes \Hilbert*{\text{ss}}$ \eqref{eq:Dilated Hilbert}. In the unitary case, $\Hilbert*{\text{ss}}$ is a bookkeeping tool that stores the measurement outcome; in the Hamiltonian case, it corresponds physically to an observable property of the ``pointer'' particle \cite{VonNeumann, Preskill_QI}.

The pointer Hamiltonian $\Ham$ \eqref{eq:Pointer Hamiltonian} generates the unitary evolution operator $\umeas (t) = \exp( - \ii \, t\, H /\hbar)$ \eqref{eq:Pointer Unitary}. Applying $\umeas(t)$ entangles the system and pointer particle, resulting in a sum over measurement outcomes $m$ of the physical state $\proj{m} \, \ket{\psi}$ and the pointer state $\ket{x_0 - \alpha \, \mEig*{m}}$. In this sense, the von Neumann unitary $\umeas (t)$ \eqref{eq:Pointer Unitary} \emph{is} the Stinespring  unitary $\umeas$ \eqref{eq:Stinespring Unitary General}, though possibly nonminimal, as in the case of the Stern-Gerlach experiment \cite{SternGerlach1, SternGerlach2, SternGerlach3} discussed in Sec.~\ref{subsec:SternGerlach}. However, recording a particular outcome requires ``binning'' the states of the pointer particle into a minimal set of $\Noutcome$ ``Stinespring states.'' In the Stern-Gerlach experiment, taking $z_0=0$ and for fixed initial $x$ velocity $v$ and magnetic field $B_z = B_0 + b \, z$, there are possible final $z$ positions of the pointer, localized to $z_f = \pm b \, \mu^{\,}_B \, L^2 / M \, v^2$, up to experimental imprecision and variations encoded in the initial state \eqref{eq:Stern Gerlach initial state nice}. These are binned  easily binned into $s=\pm$ according to $s = \text{sgn} (z_f)$. 

Thus, we expect that the Stinespring unitary $\umeas$ \eqref{eq:Stinespring Unitary General} is \emph{not} merely a bookkeeping tool, but a \emph{physical} operator corresponding to the time evolution of the system and measurement apparatus (a detector, pointer particle, etc.) during the measurement process. The Stinespring Hilbert space $\Hilbert{\text{ss}}$ gives a (minimal) representation of the state of the apparatus, with binning of states already encoded. We stress that the binning is itself part of the measurement; we explore numerous examples in which the $\DimOf*{\Hilbert*{\text{ss}}} \gg \Noutcome$  in Secs.~\ref{sec:photons} and \ref{sec:qubits}. In fact, this is to be expected when $\Noutcome \to \infty$ (see  Sec.~\ref{subsec:unitary countably infinite}), and the minimal Stinespring representation implicitly encodes equivalence classes between states of $\Hilbert{\text{ss}}$ that %indicate 
reflect the same measurement outcome. The technical details of how outcomes are extracted from the state of the detector are generally of little importance from the perspectives of (\emph{i}) a theory of projective and destructive measurements and (\emph{ii}) the practical or analytical treatment of protocols involving measurements and outcome-dependent operations \cite{AaronMIPT, SpeedLimit}.

\subsection{Summary of measurement representations}
\label{subsec:formalism summary}

Before moving on to experimental implementations, we briefly review the mathematical representations of projective and destructive measurements developed thus far. The axioms of quantum mechanics \cite{DiracQuantum, vonNeumannAxioms, HardyAxioms, FuchsAxioms, MackeyAxioms, WilceAxioms, MasanesAxioms, KapustinAxioms} dictate that such a measurement of an observable $\observ$ \eqref{eq:measured observable} in a system described by a density matrix $\rho$ results in one of  the $\Noutcome$ unique eigenvalues $\mEig*{m}$ of $\mobserv$, such that
\begin{equation}
    \label{eq:Copenhagen Meas Summary}
    \text{Pr}(\mEig{m}) \, \equiv \, p^{\vpp}_m \, = \, \trace \left( \, \proj{m} \, \rho \, \right)~~~~\text{and}~~~~\rho \, \mapsto \, \Phi_m (\rho)  = p^{\vpp}_m \, \rho^{\vpp}_m \, = \proj{m} \, \rho \, \proj{m} \, ,~~
\end{equation}
where $\rho^{\,}_m$ is the postmeasurement density matrix given outcome $m$ and the eigenprojector $\proj{m}$ \eqref{eq:Spec Projecc} satisfies $\mobserv \proj{m} = \mEig*{m} \, \proj{m}$ and Eq.~\ref{eq:meas proj ortho complete}. Averaging over outcomes leads to 
\begin{equation}
    \label{eq:Copenhagen average post meas}
    \rho \, \mapsto \, \Phi (\rho ) \, = \, \rho^{\vpp}_{\rm av} \, \equiv \, \sum\limits_{m=0}^{\Noutcome-1} \, p^{\vpp}_m \, \rho^{\vpp}_m \, = \,  \sum\limits_{m=0}^{\Noutcome-1} \, \proj{m} \, \rho \, \proj{m} \, ,~~
\end{equation}
which is a quantum channel (a CPTP map) \cite{QC_book}, with $\Phi (\rho) = \sum \, \Phi_m (\rho)$. 

The \emph{Kraus representations} \eqref{eq:Kraus Representation} of the quantum operation $\Phi_m \, : \, \rho \mapsto p^{\,}_m \, \rho^{\,}_m$ \eqref{eq:Copenhagen Meas Summary} and the related quantum channel $\Phi \, : \, \rho \mapsto \rho^{\,}_{\rm av}$ \eqref{eq:Copenhagen average post meas} are given straightforwardly by \cite{KrausBook}
\begin{equation}
    \label{eq:Kraus summary}
    \Phi_m (\rho) \, = \Kraus{m} \, \rho \, \KrausDag{m}~,~~\Phi(\rho) \, = \, \sum\limits_{m=0}^{\Noutcome-1} \, \Kraus{m} \, \rho \, \KrausDag{m}~,~~~\text{where}~~~\Kraus{m} \, = \, \proj{m} \, , ~~
\end{equation}
so that the \emph{Kraus operator} $\Kraus{m}$ is simply the projector $\proj{m}$ \eqref{eq:Spec Projecc}, meaning that
\begin{equation}
    \label{eq:Kraus completeness summary}
    \sum\limits_{m=0}^{\Noutcome-1} \, \KrausDag{m} \Kraus{m} \, = \, \sum\limits_{m=0}^{\Noutcome-1} \, \proj{m} \, = \, \ident \, ,~~
\end{equation}
i.e., the Kraus operators form a complete set \eqref{eq:Kraus Complete-able}. Quantum channels are trace preserving, and involve a complete set of Kraus operators, while quantum operations are trace decreasing and involve a proper subset of the complete set of Kraus operators that define an associated channel. 

Related to the Kraus representation \eqref{eq:Kraus summary} of quantum operations and channels is a \emph{unitary representation} \eqref{eq:Stinespring Unitary General}, first proven by Kraus using the Stinespring dilation theorem for finite-dimensional systems \cite{Stinespring, KrausMeas1969,  KrausMeas1971, KrausMeas1981, KrausBook, WolfNotes}. An extension to infinite-dimensional systems is the subject of forthcoming work \cite{AaronJamesFuture}. The unitary ``Stinespring'' representation is captured by
\begin{equation}
    \label{eq:Unitary Summary}
    \Phi (\rho ) \, = \, \trace_{\text{ss}} \left( \, \umeas \, \rho \otimes \BKop{i}{i}^{\vpp}_{\text{ss}} \, \umeas^{\dagger} \,  \right)~~~~\text{and}~~~~\Phi_m (\rho ) \, = \, \trace_{\text{ss}} \left( \, \umeas \, \rho \otimes \BKop{i}{i}^{\vpp}_{\text{ss}} \, \umeas^{\dagger} \, \Proj{m}{\text{ss}} \, \right) \, ,~~
\end{equation}
where $\ket{i}^{\vpp}_{\text{ss}}$ is some ``default'' initial state on $\Hilbert{\text{ss}} = \Comps^{\Noutcome}$, $\Proj{m}{\text{ss}}$ projects onto $\ket{m}^{\,}_{\text{ss}} \in \Hilbert{\text{ss}}$, and $\umeas$ acts on $\Hilbert{\rm dil} = \hilbert \otimes \Hilbert{\text{ss}}$ \eqref{eq:Dilated Hilbert} and can be expressed in terms of the Kraus operators \eqref{eq:Kraus summary} via
\begin{equation}
    \label{eq:Unitary from Kraus summary}
    \umeas \, \ket{\psi}^{\vpp}_{\text{ph}} \otimes \ket{0}^{\vpp}_{\text{ss}} \, = \, \sum\limits_{m=0}^{\Noutcome-1} \left( \Kraus{m} \, \ket{\psi} \right)^{\vpp}_{\text{ph}} \otimes \ket{m}^{\vpp}_{\text{ss}} ~~~~\implies~~~~\umeas \, = \, \sum\limits_{m=0}^{\Noutcome-1}  \, \proj{m} \otimes \ShiftPow{\text{ss}}{m} \, ,~~
\end{equation}
for projective measurements, where $\ShiftPow{}{m} \ket{0} = \ket{m}$ is the $\Noutcome$-state Weyl shift operator \eqref{eq:Weyl shift def}, and we have taken $\ket{i}^{\,}_{\text{ss}} = \ket{0}^{\,}_{\text{ss}}$ without loss of generality. Meanwhile, \emph{destructive measurements} are represented by a dilated unitary $\dmeas$ \eqref{eq:Destructive Unitary from Projective} that is equal to  the projective-measurement unitary $\umeas$ \eqref{eq:Unitary from Kraus summary} by another dilated unitary $\umeas'$. For either type of measurement, $\umeas$ \eqref{eq:Unitary from Kraus summary} realizes a \emph{minimal} Stinespring representation \cite{Takesaki}, where $\DimOf*{\Hilbert{\text{ss}}} = \Noutcome$ is the number of unique eigenvalues of the measured observable $\mobserv$ \eqref{eq:measured observable}. However, there also exist nonminimal representations with $\DimOf*{\Hilbert{\text{ss}}}  > \Noutcome$, which reduce to Eq.~\ref{eq:Unitary from Kraus summary} upon ``binning'' different states of $\Hilbert{\text{ss}}$ corresponding to the same outcome $\mEig*{m}$. We discuss how to extract various statistics from $\umeas$ \eqref{eq:Unitary from Kraus summary} in Sec.~\ref{sec:Using Stinespring}.

Finally, the \emph{von Neumann representation} describes measurements using a Hamiltonian $\Ham_{\rm vN}$ \eqref{eq:Pointer Hamiltonian}, which acts on a dilated Hilbert space $\Hilbert{\rm dil}$ \eqref{eq:Dilated Hilbert} that includes the physical system and a ``pointer'' particle \cite{VonNeumann}. The pointer plays the role of the detector, and $\Ham_{\rm vN}$ relates to $\umeas$ \eqref{eq:Unitary Summary} via
\begin{equation}
    \label{eq:Stinespring von Neumann}
    \umeas \, = \, \exp \Big( - \frac{\ii}{\hbar}  \, \int\limits_0^t \, \thed \tau \, \Ham_{\rm vN} (\tau) \Big) \, ,~~
\end{equation}
i.e., $\umeas$ \eqref{eq:Unitary Summary} corresponds to evolution under the von Neumann Hamiltonian $\Ham_{\rm vN} (\tau)$ \eqref{eq:Pointer Hamiltonian} for time $t$, and connects the Kraus and von Neumann representations of measurements. In general, $\Ham_{\rm vN}$ \eqref{eq:Pointer Hamiltonian} leads to a \emph{nonminimal} Stinespring representation; the minimal representation \eqref{eq:Unitary Summary} recovers upon identifying subspaces of $\Hilbert{\text{ss}}$ that reflect the same outcome. In the Stern-Gerlach experiment \cite{SternGerlach1, SternGerlach2, SternGerlach3, LeBellac}, e.g., the minimal representation has $\Hilbert{\text{ss}} = \Comps^2$, corresponding to  states with states where $s = \text{sgn} [ z(t) - z(0)] = \pm 1$ for any values of $M$, $L$, $v$, and $b$. The distinct subspaces with $s=\pm 1$ reflect the (quantized) $z$ component of the atom's intrinsic spin.

We discuss particular realizations of $\umeas$ and/or $H^{\,}_{\rm vN}$ for measurements of photons and qubits in Secs.~\ref{sec:photons} and \ref{sec:qubits}, respectively. While a dilated unitary representation $\umeas$ \eqref{eq:Unitary Summary} of measurements is guaranteed for any axiomatic formulation of quantum mechanics \cite{DiracQuantum, vonNeumannAxioms, HardyAxioms, FuchsAxioms, MackeyAxioms, WilceAxioms, MasanesAxioms, KapustinAxioms}, this only implies that $\umeas$ is a valid bookkeeping tool. However, the connection \eqref{eq:Stinespring von Neumann} to von Neumann's pointer Hamiltonian suggests that $\umeas$ is, in fact, \emph{physical}. In Secs.~\ref{sec:photons} and \ref{sec:qubits}, we show how this is the case in several of the most common implementations of projective and destructive measurements, and posit that this holds in all cases. In Sec.~\ref{sec:Using Stinespring}, we discuss how to use the Stinespring representation %of measurements 
to describe generic quantum protocols, as well as implications of the unitary $\umeas$ \eqref{eq:Unitary Summary} being physical.

\section{Photon measurements}
\label{sec:photons}

We first consider the measurement of photons, which not only diagnose optical systems, but are relevant to the measurements of atomic and molecular systems that we discuss in Sec.~\ref{sec:qubits}. As noted in Sec.~\ref{subsec:destructive}, one generally expects the counting of photons to be \emph{destructive}, rather than projective---though we comment that there exist projective measurements of photon number~\cite{Nogues1999, Reiserer2013, Niemietz2021} naturally described by Eq.~\ref{eq:Measurement Unitary Countably Infinite}. Importantly, such destructive measurements are nonetheless captured by the von Neumann \eqref{eq:Pointer Hamiltonian} and Stinespring \eqref{eq:Destructive Unitary from Projective} representations. These representations require that we model the measurement apparatus explicitly, in contrast to the device-independent description of photodetection introduced by Glauber~\cite{Glauber1963}.

This section is organized as follows. In Sec.~\ref{subsec:EM modes}, we review the description of photons as oscillator-like excitations of the electromagnetic fields in matter-free regions, and then extend to optical modes in cavities. In Sec.~\ref{subsec:photon counting}, we consider the destructive measurement of photon number, which serves as a proxy for measurements of intensity and other properties related to energy or occupation number \cite{Glauber1963, Olivares_2019, Combes2022}. In Sec.~\ref{subsec:homodyne} we consider homodyne detection \cite{Yuen1978, Collett1987, Barchielli_1990, Wiseman1993,Lvovsky2009} of photon quadratures \eqref{eq:quadrature X}, which relates to general interferometric probes~\cite{JacobsBook, SqueezeBook, Yuen1978, Collett1987, Barchielli_1990, Wiseman1993}.

\subsection{Electromagnetic modes}
\label{subsec:EM modes}

Quantum mechanically, photons are the excitations of the electromagnetic fields $\bvec{E}$ and $\bvec{B}$, which in a matter-free region are described by the continuum Hamiltonian \cite{LeBellac},
\begin{equation}
    \label{eq:EM Hamiltonian}
    H^{\,}_{\rm EM} \, = \, \frac{\epsilon_0}{2} \int_V \, \thed^3 \bvec{x} \, \left( \bvec{E}^2 (\bvec{x}) + c^2 \, \bvec{B}^2 (\bvec{x})\right)  \, ,~~
\end{equation}
where $\bvec{B} = \nabla \times \bvec{A}$ and $\bvec{E} = - \pd{t} \bvec{A}$, and we use the traverse (Coulomb) gauge $\nabla \cdot \bvec{A} = 0$.  

Regarding %the electromagnetic Hamiltonian 
$H^{\,}_{\rm EM}$ \eqref{eq:EM Hamiltonian}, we interpret the electric field $\bvec{E} = - \pd{t} \bvec{A}$ as the ``momentum'' conjugate to $\bvec{A}$, and the magnetic field $\bvec{B} = \nabla \times \bvec{A}$ as a ``gradient'' of $\bvec{A}$. Then, $H^{\,}_{\rm EM}$ \eqref{eq:EM Hamiltonian} resembles a theory of bosonic excitations (e.g., in a harmonic solid) and, in a box with volume $V$, we write
\begin{equation}
    \label{eq:EM vector potential operator}
    A_j (\bvec{x}) \, \equiv \, \sqrt{\frac{\hbar}{2 \, c\,  \epsilon_0 \, V}} \, \sum\limits_{\bvec{k}} \, \abs{\bvec{k}}^{-1/2} \, \sum\limits_{s=1,2} \, \left(  \, \ee^{\ii \, \bvec{k} \cdot \bvec{x}} \, \hat{e}^{\vps}_{s,j} (\bvec{k}) \, \a{{\bvec{k}},s} + \ee^{-\ii \, \bvec{k} \cdot \bvec{x}} \, \hat{e}^{*}_{s,j} (\bvec{k}) \, \adag{{\bvec{k}},s} \right) \, , ~~
\end{equation}
where $\bvec{k}$ is a Fourier wavevector, the ladder operators obey the canonical commutation relation
\begin{equation}
    \label{eq:EM ladder operators}
    \comm{\a{\bvec{k'},s'}}{\adag{\bvec{k},s}} \, = \, \kron{s,s'} \, \DiracDelta{\bvec{k}-\bvec{k'}} \, , ~~
\end{equation}
and $\hat{e}^{\vps}_{s,i}(\bvec{k})$ is the $i$the component of the \emph{polarization} unit vector $\hat{e}^{\vps}_s$, which satisfies
\begin{equation}
    \label{eq:EM polarization vector}
    \sum\limits_{s=1,2} \, \hat{e}^{\vps}_{s,i} (\bvec{k}) \, \hat{e}^*_{s,j} (\bvec{k})\, = \, \kron{i,j} - \frac{k^{\,}_i \, k^{\,}_j}{\abs{\bvec{k}}^2} \, ,~~
\end{equation}
so that the polarization vectors span the plane perpendicular to $\bvec{k}$, and there are two modes $s=1,2$ instead of the three associated with the photon's spin 1 due to gauge redundancy. 

Using Eq.~\ref{eq:EM vector potential operator} for  $\bvec{A}$ (and thereby $\bvec{E}$), we express $H^{\,}_{\rm EM}$ in terms of ladder operators \eqref{eq:EM ladder operators},
\begin{equation}
    \label{eq:EM oscillator Hamiltonian}
    H^{\vps}_{\rm EM} \, = \, \sum\limits_{\bvec{k}} \, \sum\limits_{s=1,2} \, \hbar \, \omega (\bvec{k}) \, \left( \adag{\bvec{k},s} \a{\bvec{k},s} + \frac{1}{2} \right) \, ,~~
\end{equation}
where $N^{\,}_{\bvec{k},s} = \adag{\bvec{k},s} \a{\bvec{k},s} \in \Nats$ is the number of photons with polarization $s$ and wavevector $\bvec{k}$, with corresponding dispersion $\omega (\bvec{k}) = c \, \abs{\bvec{k}}$ \cite{LeBellac}. These occupation numbers fully characterize the electromagnetic fields in a matter-free box of volume $V$. 

In many scenarios, the spatial structure of the electromagnetic fields is modified by the presence of mirrors (as in the case of cavities) or some dense medium. In other cases, the range of relevant frequencies is sufficiently small that the factor of $k^{-1/2} \propto \omega^{-1/2}$ can be pulled out of the sum in Eq.~\ref{eq:EM vector potential operator} to good approximation~\cite{SqueezeBook, JacobsBook}. These cases are still described by  $H^{\,}_{\rm EM}$ \eqref{eq:EM oscillator Hamiltonian}, but the expansion of $\bvec{A}$ in Eq.~\ref{eq:EM vector potential operator} is neither appropriate nor convenient. Instead, one writes
\begin{equation}
    \label{eq:EM general vector potential}
    \bvec{A} (\bvec{x}) \, \equiv \, \sqrt{\frac{\hbar}{2 \,  \epsilon_0 \, c}} \, \sum\limits_{n}\, \sum\limits_{s=1,2} \, \left( \, \bvec{u}_{n,s}(\bvec{x}) \, \a{n,s} + \, \bvec{u}_{n,s}^*(\bvec{x}) \, \adag{n,s} \right) \, ,~~
\end{equation}
where the vector fields $\bvec{u}_{n,s}(\bvec{x})$ are (normalized) classical solutions to Maxwell's equations with the appropriate boundary conditions. Most of these optical ``modes'' propagate and resemble plane waves far from any mirrors or media, but some may be localized in space (e.g., within a cavity). Each mode $n,s$ is associated with the creation and annihilation operators $\adag{n,s}$ and $\a{n,s}$, respectively, as with the plane-wave photon modes that realize in free space \eqref{eq:EM ladder operators} \cite{SqueezeBook, JacobsBook}. 

We also comment that, while a full accounting of all electromagnetic modes and the precise form of  $\bvec{A}$ \eqref{eq:EM vector potential operator} is  needed to resolve the spatiotemporal profile of the electromagnetic fields, it is not necessary to the measurements of interest herein~\cite{SqueezeBook, JacobsBook}. Instead, it is generally sufficient to consider one or two electromagnetic modes, which may correspond to a particular wavevector and polarization in free space, a particular confined mode in a cavity, or similar.

\subsection{Destructive measurement of photon number}
\label{subsec:photon counting}

As alluded to in Sec.~\ref{subsec:destructive}, measuring the number of photons at wavevector $\bvec{k}$ \eqref{eq:EM oscillator Hamiltonian}---or, more generally, the occupation or intensity of an electromagnetic mode $n$ \eqref{eq:EM general vector potential}---is generally \emph{destructive}, rather than projective. The reason is that the detector counts the photons by absorbing them, destroying the system's state. However, such measurements are nonetheless captured by a von Neumann Hamiltonian $H$ \eqref{eq:Pointer Hamiltonian} or Stinespring unitary $\dmeas$ \eqref{eq:boson count destructive action} acting on $\Hilbert*{\text{dil}}$ \eqref{eq:Dilated Hilbert}, and the latter is equivalent to a projective measurement $\umeas$ \eqref{eq:Measurement Unitary Countably Infinite} followed by a dilated unitary $\umeas'$ \eqref{eq:Destructive Unitary from Projective}. 

We now consider a simplified description of the destructive measurement of occupation number (i.e., photon counting), based on a toy model of a photodetector comprising $\Noutcome$ two-level systems. This toy model schematically captures a variety of distinct physical processes by which photons are detected in actual experiments. Destructive counting of photons necessarily requires their absorption, which transfers energy and momentum to the detector in the form of an ``excitation'' ($\ket{0} \mapsto \ket{1}$). The number of excitations reflects the number of photons counted; for the result to be readable, the excitations must result in a classically discernible signal. For example, the photons may excite electrons to the conduction band, producing a detectable current; electron-hole pairs in a semiconductor, altering the material's conductivity; phonons in a lattice, altering the material's thermal properties; and other mechanisms~\cite{Dennis1986}.

Importantly, all of these experimental realizations of photon counting are captured by time evolution on a dilated Hilbert space via a Hamiltonian \eqref{eq:Pointer Hamiltonian} or unitary \eqref{eq:Stinespring Unitary General}. In the interest of clarity, we neglect many technical details related to sources of noise, ``dark counts'' (i.e., false positives), thermal effects, differences between different detector realizations, the presence of multiple electromagnetic modes, and so on. Such considerations are discussed in great detail in the literature~\cite{Dennis1986}. We also comment that, while some detectors of electromagnetic radiation are not well suited to counting individual photons, they are nonetheless suitable for coarse-grained measurements of intensity. The simplified model of photon counting we consider below is also a convenient model for intensity measurements \cite{Olivares_2019} and similar energetic measurements. While the precise details of the detector degrees of freedom, set of outcomes, binning of states into outcomes, and other details may differ, the minimal Stinespring unitary is the same.

\begin{figure}
    \centering
    \includegraphics[width=0.8\textwidth]{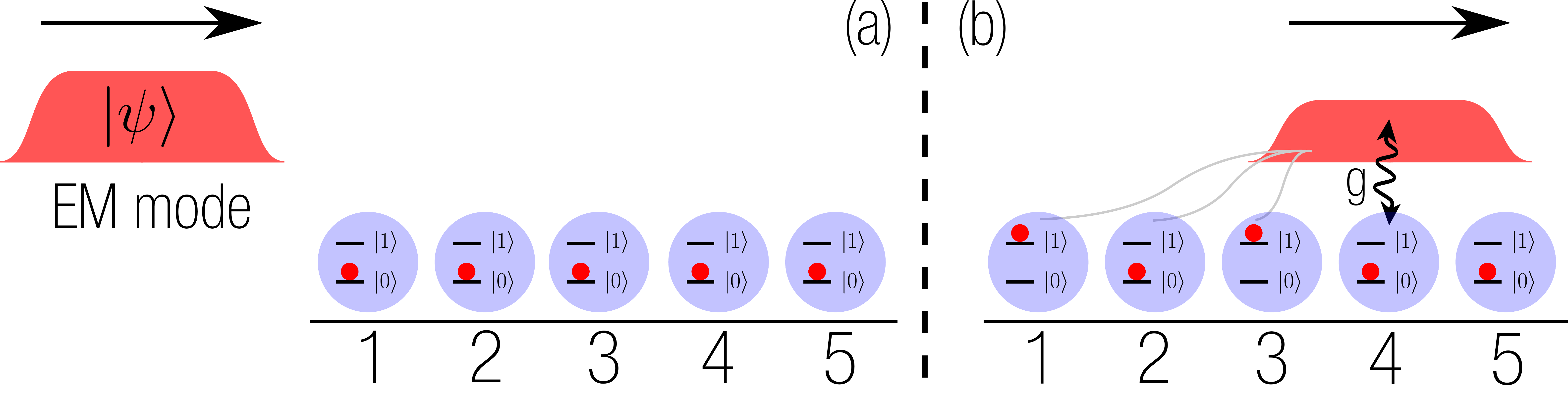}
    \caption{A simple mode of destructive measurements of photon number or intensity. (a) Prior to the measurement, the electromagnetic mode realizes a mildly localized wave packet and the electrons in the detector are represented by $\Noutcome$ two-level electrons on a line, indexed $k$. (b) During the measurement, the photons interact with the electrons (strength $g$) in sequence; each absorption exciting an electron to the state $\ket{1}$, leading to a current proportional to the number of absorbed photons.  \label{fig:DetectorModel}}  
\end{figure}

With these caveats in mind, we now consider the simplified model of destructive measurement of photon number~\cite{Mollow1968, Scully1969}. Suppose that the detector consists of electrons in the default state $\ket{0}$, which excite to the conducting state $\ket{1}$ upon absorption, giving rise to a classically discernible electrical current (though this model captures generic photodetectors). We also restrict to a \emph{single} electromagnetic mode, with ladder operators $\a{}$ and $\adag{}$. It is useful to picture this mode as a propagating wavepacket---i.e., a localized mode moving in space towards the detector [see Fig.~\ref{fig:DetectorModel}(a)]. 

The $\Noutcome$ detector electrons are described by the fermion operators $\f{s,k}$, where $s=0,1$ labels the ``orbital'' and $k \in \{1,2,\dots,\Noutcome\}$ reflects the order of excitation [see Fig.~\ref{fig:DetectorModel}(a)]. The initial states of the electromagnetic mode and detector are, respectively, $\ket{\psi} \in \ell^2 (\Nats)$ and  $\ket{i}_{\text{ss}} = \ket{\bvec{0}} = \ket{000 \cdots 000}$. The light interacts with the $k$th electron for a time $\tau$ under the Hamiltonian
\begin{equation}
\label{eq:photon counting Hamiltonian}
    \Ham_k \, = \, \ii \, g 
    \,\left(  \a{} \otimes \fdag{1,k}\f{0,k} - \adag{} \otimes \fdag{0,k}\f{1,k} \right) \, ,~~
\end{equation}
where $g$ is the strength of the electron-photon coupling. Ignoring electron-electron interactions, we replace the fermion operators with spin operators according to $\fdag{1,k}\f{0,k}\equiv \Pauli{+}{k}$ and $\fdag{0,k}\f{1,k}\equiv \Pauli{-}{k}$. The sequential interaction between the optical mode and the effective spins 1/2 leads to 
\begin{equation}
\label{eq:photon count final state def}
    \ket{\Psi} \, = \, \ee^{g\tau(\a{}\Pauli{+}{\Noutcome}-\adag{}\Pauli{-}{N})} \cdots \ee^{g\tau(\a{}\Pauli{+}{1}-\adag{}\Pauli{-}{1})} \,  \ket{\psi} \otimes \ket{ \bvec{0}}^{\vpp}_{\text{ss}} \, ,~~
\end{equation}
where $\ket{\Psi} \in \Hilbert{\rm dil}$ is  final state of the light and detector. We also define the parameter
\begin{equation}
    \label{eq:zeta def}
    \zeta \,  \equiv \, \Noutcome \, g^2 \, \tau^2 \, ,~~
\end{equation}
to be fixed and finite as we take the limits $g\tau \ll 1$ and $\Noutcome \gg 1$, so that Eq.~\ref{eq:photon count final state def} becomes
\begin{equation}
\label{eqn: photodetection final state}
    \ket{\Psi} \, =  \,\exp \bigg[ -\frac{\zeta}{2} \, \adag{}\a{} \bigg]\, \exp \bigg[ (1-\ee^{-\zeta})^{1/2} \, \a{} \otimes \Bdag{} \bigg]  \, \ket{\psi} \otimes \ket{\bvec{0}}^{\vpp}_{\text{ss}} \, .~~
\end{equation}
A careful derivation of the new ladder operator $\Bdag{}$ appears in App.~\ref{app:photonabsorption} and results in
\begin{equation}
\label{eq:big B spin operator}
    \Bdag{} \, \equiv \, \frac{1}{\sqrt{\Noutcome}}\frac{\sqrt{\zeta}}{\sqrt{\ee^{\zeta}-1}} \sum\limits_{k=1}^\Noutcome \, \ee^{\zeta(\Noutcome - k)/(2 \Noutcome)} \, \Pauli{+}{k} \, ,~~
\end{equation}
as a collective raising operator on $\Hilbert{\text{ss}}$ with $\comm*{\B{}}{\Bdag{}}=1$ in the limit $\Noutcome \to \infty$ and $\tau \to 0$, with leading corrections proportional to $g \tau$ and $\Noutcome^{-1/2}$ (see App.~\ref{app:photonabsorption} for details). 

The operator $\Bdag{}$ \eqref{eq:big B spin operator}  creates bosonic excitations in the detector above the ``vacuum'' state  $\ket{\bvec{0}}_{\text{ss}} = \ket{000 \cdots 000}$. Note that corrections to the final state $\ket{\Psi}$ \eqref{eqn: photodetection final state} vanish as $\Order{\Noutcome^{-1/2}}$ as $\Noutcome \to \infty$. Hence, the electronic state of the detector is well approximated by a bosonic mode, where $\Bdag{}\B{}$ counts the number of of electrons in the state $\ket{1}$ after interacting with the light. Those excited electrons produce a classically detectable current proportional to the number of absorbed photons. While other means of counting photons exist, they generically involve converting photons into excitations, leading to a classical signal from which the number of photons can be inferred. 

The \emph{attenuation parameter} $\zeta$ \eqref{eq:zeta def} controls the efficiency of the transduction of photons into  excitations of the detector; its name refers to the ``attenuation'' of electromagnetic energy with each absorbed photon. The transduction---and thus, the counting---of photons is most efficient when the attenuation parameter $\zeta \gg 1$ is large but finite as $\Noutcome \to \infty$ and $g\tau \to 0$. For example, if the system is initially in the Fock state $\ket{\psi^{\,}_0}^{\,}_{\text{ph}} = \ket{n}$ with exactly $n$ photons, then the measurement results in
\begin{equation}
\label{eq:photon count joint state}
    \ket{n}^{\vpp}_\text{ph} \otimes \ket{0}^{\vpp}_\text{ss} ~ \mapsto ~ \ket{\Psi} \, = \, \sum\limits_{m=0}^n  \, \ee^{\ii \theta_m} \, \binom{n}{m}^{1/2} \, (1-\ee^{-\zeta})^{m/2}  \, \ee^{-\zeta (n-m)/2} \, \ket{n-m}^{\vpp}_{\text{ph}} \otimes \ket{m}^{\vpp}_{\text{ss}} \, ,~~
\end{equation}
which includes contributions from detector states $\ket{m}^{\,}_{\text{ss}}$ with different numbers $m$ of excited electrons---and hence, counted photons. We also note the generic, $m$-dependent phase $\exp(\ii \, \theta_m) \neq 1$ above, which is sensitive to microscopic details of the apparatus, may be time dependent and vary between experimental shots, and intuitively comes from phases other than $\pm \ii$ in $\Ham_k$~\eqref{eq:photon counting Hamiltonian}. Although the phases $\{ \theta_m \}$ are not important in any single experiment shot or for detecting photons from a single mode, they are quite relevant in the more realistic settings of many electromagnetic modes (which often have spatiotemporal structure) and the extraction of statistics (which requires multiple shots and is sensitive to coherences). Moreover, the phases $\{ \theta_m \}$ provide a mechanism for \emph{decoherence}, which leads to the appearance of wavefunction collapse as we describe in Sec.~\ref{subsec:collapse}. 

For the initial Fock state $\ket{\psi^{\,}_0}^{\,}_{\text{ph}} = \ket{n}$ \eqref{eq:photon count joint state}, the probability to detect $k \leq n$ photons is
\begin{equation}
    \label{eq:photodetect prob k}
    p^{\vpp}_k \, = \, \matel*{\Psi}{\ident \otimes \Proj{\text{ss}}{(k)}}{\Psi} \, = \, \binom{n}{k} \, p^{k} \, (1-p)^{n-k}~~~~\text{with}~~~~p \, = \, 1 - \ee^{-\zeta} \, , ~~
\end{equation}
corresponding to a binomial distribution, where $p = 1 - \ee^{-\zeta}$ is the probability that an incident photon is successfully detected. For the initial Fock state $\ket{n}$, the expected number of counted photons is $n \, p$ with variance $n \, p \, (1-p)$. Hence, efficiency is greatest when the attenuation parameter $\zeta$ \eqref{eq:zeta def} is large. This also holds for more general initial states of the electromagnetic mode.

However, one can still effectively count photons even when $\zeta \lesssim 1$ is small \eqref{eq:zeta def}. Essentially, when $p \ll 1$ \eqref{eq:photodetect prob k} is small, the average number of excitations in the detector is $n p \ll n$, with variance $np(1-p)\approx np$, so that the distribution is effectively Poisson, as one might expect in the context of particle detection. If $np$ is large, the relative fluctuations with respect to the mean are small, with $\sim (np)^{-1/2} \ll 1$, so that if $m$ excitations are observed, one can infer an occupation $n = m/p$ of the electromagnetic mode with high precision, provided that $p$ is known.

Finally, we note that the simplified model of photon counting discussed above also applies to the measurement of \emph{intensity} of narrow-band sources of light. For such sources, the intensity is simply proportional to photon number flux, even if the associated measurements rely on a physical mechanism different from the photodetection mechanism considered above~\cite{Glauber1963}. We also note that Eq.~\ref{eqn: photodetection final state} is the coherent representation of the statistics of single-mode attenuated fields originally derived in Refs.~\citenum{Mollow1968}~and~\citenum{Scully1969}. In this context, the attenuation parameter $\zeta$ is replaced by $\kappa \tau$ from Ref.~\citenum{Mollow1968}. In general, the mathematical form of the associated probabilities is agnostic to the specific detector model employed, as shown in Ref.~\citenum{Srinivas1981}, which also provides formulae for multi-time correlations. Other measurements of electromagnetic modes related to photon number include irradiance and excitance. For a discussion of how the model above is modified to account for the presence of many electromagnetic modes, we refer to the  literature~\cite{Chmara1987, Fleischhauer1991, Gardiner2004-co}.

\subsection{Balanced homodyne detection}
\label{subsec:homodyne}

\begin{figure}
    \centering
    \includegraphics[width=0.8\textwidth]{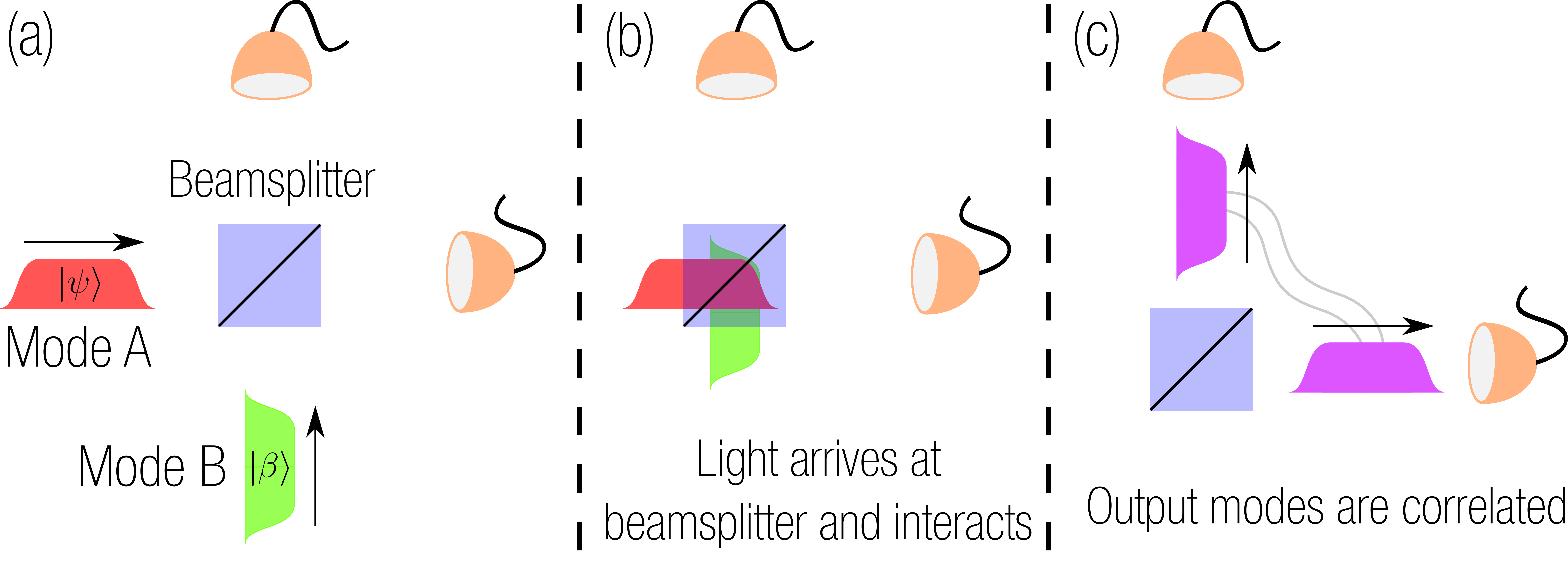}
    \caption{Schematic depiction of homodyne detection. (a) An optical mode ``A'' of interest and a strong, coherent ``B'' mode \eqref{eq:beta aux} are sent towards a beam splitter. (b) The modes interact in the beam splitter. (c) The modes exit the beam splitter with coherent correlations and propagate towards photodetectors.}
    \label{fig:homodyne detection}
\end{figure}

Another important class of optical measurements relates to \emph{interferometry}~\cite{Born2020}. When combined with other ingredients, measurements of intensity or photon number (see Sec.~\ref{subsec:photon counting}) can be used to probe ``quadratures'' of light---i.e., a family of operators of the form
\begin{equation}
    \label{eq:quadrature X}
    x^{\vpp}_{\phi} \, \equiv \, \frac{1}{\sqrt{2}} \left(  \ee^{- \ii \phi} \, \a{} + \ee^{\ii \phi} \, \adag{} \right) \,  ~~
\end{equation}
which realize canonically conjugate operators for, e.g.,  $\phi=0$ and $\phi=\pi/2$. Measuring such quadratures \eqref{eq:quadrature X} can be achieved using homodyne detection~\cite{Yuen1978,Collett1987, Barchielli_1990, Wiseman1993, Tyc_2004,Lvovsky2009}, where $\phi$ is the experimentally controllable ``homodyne angle.'' For convenience, we assume that the attenuation parameter $\zeta \gg 1$ \eqref{eq:zeta def} is large, so that the photodetectors (see Fig.~\ref{fig:homodyne detection}) efficiently count photons~\cite{Olivares_2019}.

Homodyne detection involves \emph{two} electromagnetic modes of the \emph{same} frequency, as schematically depicted in Fig.~\ref{fig:homodyne detection}. The ``A'' mode represents the physical system of interest, prepared in an arbitrary initial state $\ket{\psi}^{\,}_\text{A}$. The ``B'' mode is an auxiliary mode, prepared in the coherent state
\begin{equation}
\label{eq:beta aux}
    \ket{\beta}^{\vpp}_{\text{B}} \, = \, \ee^{-\abs{\beta}^2/2}\sum_{n\in\Nats} \frac{\beta^n}{\sqrt{n!}} \, \ket{n}^{\vpp}_{\text{B}} \, , ~~
\end{equation}
for complex $\beta \in \Comps$ with $\abs{\beta} \gg 1$. This highly populated (i.e., bright) coherent state represents the field of a laser, which is commonly termed the ``local oscillator'' (LO)~\cite{Collett1987}. Its phase ($\beta=\abs{\beta} \, \ee^{-\ii \phi}$) determines the quadrature $x^{\,}_{\phi}$ \eqref{eq:quadrature X} that is ultimately measured. For conceptual clarity, it is useful to regard both the A and B modes as propagating, localized wavepackets, as in Fig.~\ref{fig:homodyne detection}(a). With this in mind, the steps involved in homodyne measurements are as follows (see Fig.~\ref{fig:homodyne detection}): 
\begin{enumerate}
    \item The two electromagnetic modes are initially in the state $\ket{\Psi_0} = \ket{\psi}^{\,}_\text{A} %\otimes 
    \ket{\beta}^{\,}_{\text{B}}$. The modes propagate towards separate photodetectors [see Sec.~\ref{subsec:photon counting} and Fig.~\ref{fig:homodyne detection}(a)], each in the default initial state $\ket{0}^{\,}_{\text{det}}$ with no excitations. The full initial state is $\ket{\Psi_0} \otimes  \ket{0,0}^{\vpp}_{\text{ss}}$.
    \item Before reaching the detectors, the A and B modes simultaneously pass through a beam splitter [see Fig.~\ref{fig:homodyne detection}(b)], during which they evolve under the unitary~\cite{Yurke1986,PRASAD1987}
    \begin{equation}
    \label{eq:Beam Splitter U}
        U^{\vpd}_{\text{BS}} \, = \, \exp \left( \frac{\pi}{4} (\adag{} b - \a{} b^{\dagger} )  \right) \, ,~~
    \end{equation}
    where $\a{}$ and $b$ are annihilation operators for the A and B modes, respectively. Applying $U^{\,}_{\text{BS}}$ \eqref{eq:Beam Splitter U} to the initial state $\ket{\Psi_0}$ produces a complicated state  $\ket{\Psi_1} = U^{\,}_{\text{BS}} \, \ket{\Psi_0}$.
    \item The A and B modes exit the beam splitter and propagate toward their associated detector [see Fig.~\ref{fig:homodyne detection}(c)], which counts the photons in that mode as in Sec.~\ref{subsec:photon counting}, i.e., 
    \begin{equation}
    \label{eq:homodyne detector step}
        \sum\limits_{n_a,n_b \in \Nats} \Big[\inprod{n_a,n_b}{\Psi_1} \Big] \ket{n_a,n_b}\otimes\ket{0,0}^{\vpp}_{\text{ss}} ~ \mapsto ~ \sum\limits_{n_a,n_b \in \Nats} \Big[\inprod{n_a,n_b}{\Psi_1}\Big] \ket{0,0}\otimes\ket{n_a,n_b}^{\vpp}_{\text{ss}} \, , ~
    \end{equation}
    where $\ket{n_a,n_b}$ is a shorthand for $\ket{n_a}^{\,}_\text{A} %\otimes 
    \ket{n_b}^{\,}_{\text{B}}$. This process is described in Sec.~\ref{subsec:photon counting} and Fig.~\ref{fig:DetectorModel}.
    \item We express the final state in terms of the sum $N=n_a+n_b$ and difference $D=(n_a-n_b)/2$ of the counts of the two detectors, so that the final state \eqref{eq:homodyne detector step} is
    \begin{equation}
    \label{eq:homodyne post detector}
        \ket{\Psi_f} \, = \, \sum\limits_{N=0}^{\infty} \sum_{D=-N/2}^{N/2} \matel*{N,D}{U^{\vpd}_{\text{BS}}}{\Psi_0} \, \ket{0}^{\vpp}_{\text{A}} \ket{0}^{\vpp}_{\text{B}} \otimes \ket{N,D}^{\vpp}_{\text{ss}} \, .~~
    \end{equation}
\end{enumerate}
In Eq.~\ref{eq:homodyne post detector}, we introduced a new basis $\ket{N,D}$ for the physical and auxiliary modes, given by 
\begin{equation}
\label{eq:homodyne sum diff basis}
    \ket{N,D} \, = \, \frac{1}{\sqrt{\big(N/2+D\big)!}}\frac{1}{\sqrt{\big(N/2-D\big)!}}\big(\adag{}\big)^{N/2+D}\big(b^{\dagger} \big)^{N/2-D} \ket{0}^{\vpp}_\text{A} \ket{0}^{\vpp}_\text{B} \, , ~~
\end{equation}
and all that remains is to calculate the matrix element $\matel*{N,D}{U^{\vpd}_{\text{BS}}}{\Psi_0}$. As we show explicitly in App.~\ref{app:homdetect}, in the limit $\abs{\beta} \to \infty$ with $\matel*{\psi}{\adag{}\a{}}{\psi}^{\,}_\text{A} \ll \abs{\beta}$~\cite{Tyc_2004}, this matrix element takes the form~\cite{Combes2022},
\begin{equation}
\label{eqn:HomDetectMatrixElement}
    \matel*{N,D}{U^{\vpd}_{\text{BS}}}{\Psi_0} \, = \, \frac{ \ee^{\ii N \phi }}{\pi^{1/4} \abs{\beta}} \exp\bigg[-\frac{\big(N-\abs{\beta}^2\big)^2}{4 \abs{\beta}^2}\bigg]\,\psi^{\vpp}_\phi\big(D \sqrt{2} / \abs{\beta} \big) \, ,~~
\end{equation}
where $\psi^{\vpp}_\phi(x) = \inprod{x_\phi}{\psi}^{\vpp}_{\text{A}}$ is the ``quadrature wavefunction''---i.e., the overlap of the state $\ket{\psi} \in \ell^2 (\Nats)$ with the eigenbasis of the \emph{quadrature} operator $x^{\,}_{\phi}$ \eqref{eq:quadrature X}, instead of the usual $x = x_0$, i.e.,
\begin{equation}
    \label{eq:quadrature basis}
    \psi^{\vpp}_\phi (x) \, = \, \inprod{x^{\vpp}_{\phi}}{\psi} \, = \, \sum\limits_{n=0}^{\infty} \, \frac{\ee^{-\ii \, n \, \phi} \, \ee^{- x^2/2}}{\pi^{1/4} \, 2^{n/2} \, \sqrt{n!}} \, H^{\vpp}_n (x) \, \inprod{n}{\psi} \, , ~~
\end{equation}
where $H^{\vpp}_n(x)$ is the $n$th (physicist's) Hermite polynomial. Since $\phi$ is defined by the phase of the coherent state (i.e., $\beta=\abs{\beta} \, \ee^{-\ii \phi}$), the final state $\ket{\Psi_f}$ \eqref{eq:homodyne post detector} can be written 
\begin{equation}
\label{eqn:homodynefinalstate}
    \ket{\Psi_f} \, = \, \sum_{N=0}^{\infty} \sum_{D=-\infty}^{\infty} \frac{\ee^{\ii N \phi }}{\pi^{1/4} \abs{\beta}} \, \exp\bigg[-\frac{\big(N-\abs{\beta}^2\big)^2}{4 \abs{\beta}^2}\bigg]\,\psi^{\vpp}_\phi\big(D \sqrt{2} / \abs{\beta}\big) \, \ket{0,0}\otimes \ket{N,D}^{\vpp}_{\text{ss}} \, , ~~
\end{equation}
where we extend the sum over $D$ from $\pm N/2$ to $\pm \infty$ by assuming that the quadrature wavefunction $\psi^{\,}_\phi ( D \sqrt{2} / \abs{\beta} )$ vanishes for $D> N/2$. As a result, $\ket{\Psi_f}$ \eqref{eqn:homodynefinalstate} is \emph{separable} with respect to $N$ and $D$.

Several remarks about the homodyne-detection procedure described above are in order:
\begin{itemize}
    \item The total detector excitation level $N$ is uncorrelated with the difference in excitations $D$, since $\ket{\Psi_f}$ \eqref{eqn:homodynefinalstate} is a product of a function of $N$ and a function of $D$, where
    \begin{equation}
        \label{eq:homodyne det probs}
        p(N) \, = \, \mathcal{N} \left( \abs{\beta}^2 , \abs{\beta}^2 \right) ~~~~\text{and}~~~ p(D) \, = \, \left| \psi^{\vpp}_{\phi} \left( D \sqrt{2} / \abs{\beta} \right) \right|^2 \, .~~
    \end{equation}
    \item The $N$-dependent part of  $\ket{\Psi_f}$ \eqref{eqn:homodynefinalstate} reflects a very bright coherent state, where $p(N)$ \eqref{eq:homodyne det probs} realizes a normal (Gaussian) distribution whose mean and variance are both equal to $\abs{\beta}^2$. Importantly, it encodes no information about the initial physical state $\ket{\psi}$. 
    \item The coherences between different $N$ sectors are extremely sensitive to the phase $\phi$ of the LO~\cite{Combes2022}, which also appears in the factor $\ee^{\ii N\phi}$ in $\ket{\Psi_f}$ \eqref{eqn:homodynefinalstate}. Measuring coherences between macroscopically different values of $N$ thus requires very precise control (e.g., with sensitivity $1/N$) of the phase $\phi$ of the LO, or else the coherences wash out upon averaging over many measurements, which leads to an effective source of decoherence (see also Sec.~\ref{subsec:collapse}).
    \item The excitation difference $D$ samples the quadrature wavefunction $\psi_\phi$ in $\ket{\Psi_f}$ \eqref{eqn:homodynefinalstate}. The quadrature basis $x_\phi$ \eqref{eq:quadrature X}  is determined by the phase $\phi$ of the LO \eqref{eq:beta aux}, and the quadrature wavefunction $\psi_\phi ( \cdot )$ is effectively probed at points separated by the spacing $\delta x_\phi = \sqrt{2}/\abs{\beta}$, which vanishes when $\abs{\beta}$ is large so that one resolves a continuum of values of $\psi_\phi (\cdot)$. 
    \item The homodyne detection scheme described above operates in a ``balanced'' configuration, where the specific beam-splitter unitary $U_{BS}$ \eqref{eq:Beam Splitter U} transmits (and reflects) $50\%$ of the light in each mode. Once can also use unbalanced configurations of the beam splitter, where the ratio between the reflection and transmission coefficients is not $1:1$. Such setups can also be used to extract information about the state of electromagnetic modes~\cite{Wallentowitz1996, Cives2000, Kuhn2016}.
\end{itemize}
Within this framework, each experimental shot results in a single ``observed'' value of $D$, sampled from the probability distribution $p(D) = \abs{\psi_\phi}^2$ \eqref{eq:homodyne det probs}. Repeating the experiment  with the same value of the homodyne angle $\phi$ \eqref{eq:quadrature X} samples $p(D)$ with approximate spacing $\delta x_\phi = \sqrt{2}/\abs{\beta}$, allowing for the reconstruction of $p(D)$ with sensitivity $\delta x_\phi$. Moreover, repeating the experiment with \emph{different} values of $\phi$ allows for the recovery of the full quantum state $\psi_\phi$ of mode A~\cite{Smithey1993,Lvovsky2009}. 

In practice, extracting a value of $D$ from an experimental shot requires integrating the instantaneous outputs of the A and B detectors over time, and computing sums and differences. The temporal nature of such interferometric measurement processes is especially pronounced in the related context of \emph{heterodyne detection}, in which the frequencies $\omega_A$ and $\omega_B$ are different. As a result, the quadrature \eqref{eq:quadrature X} being probed changes as a function of time---i.e., the quadrature angle is  $\phi (t) = (\omega_A-\omega_B) \, t$. Moreover, extracting information from this temporal distribution requires multiplying the time-dependent output of the detectors by a sinusoidal signal that oscillates at the beat frequency $\omega_A-\omega_B$ \emph{before} integrating. Since quadratures~\eqref{eq:quadrature X} for different values of $\phi$ do not commute with one another, accommodating this technique into the Stinespring formalism---and particularly, extracting the \emph{minimal} Stinespring form---requires more careful consideration of the time-dependent processes described above, which we defer to future work. Schematically, however, heterodyne detection can be viewed as effectively measuring the non-Hermitian annihilation operator $\a{}$, whose coherent eigenstates are not mutually orthogonal and overcomplete~\cite{Shapiro2009}.

\section{Qubit measurements}
\label{sec:qubits}

We now consider the measurements of various implementations of qubits, which involve the examples of photon measurements considered in Sec.~\ref{sec:photons}. Although we restrict to two-level qubits, the ideas herein may also extend to multi-level qudits. The two levels of the qubit correspond to the computational ($Z$) basis states $\ket{0}$ and $\ket{1}$, where $Z \ket{n} = (-1)^n \ket{n}$. These states are generally associated with the internal energy levels of, e.g., atoms, ions, and superconducting circuits, where the interpretation varies depending on the underlying physical system. In atoms and ions, the levels correspond to distinct orbital and spin configurations of valence electrons; in superconducting circuits, the states describe different configurations of the superconducting current, phases, or voltages; in nitrogen-vacancy (NV) centers, the states correspond configurations of a defect and surrounding electrons; and in quantum dots, the configurations correspond to the presence of electrons in the potential wells or the spins of those electrons. 

Typically, one seeks to measure the computational ($Z$) basis state of the qubit. Other single-qubit operators can be measured by first applying a unitary change of basis---e.g., applying the Hadamard gate $H = (X+Z)/\sqrt{2}$ and measuring $Z$ is equivalent to measuring $X$. Certain multi-qubit operators can also be measured by applying entangling unitaries and then measuring $Z$. In contrast to the photon measurements of Sec.~\ref{sec:photons}, qubit measurements are generically projective---as opposed to destructive---so that the postmeasurement state is an eigenstate of the measured operator. 

In the remainder, we consider two types of measurements that are common across many different experimental systems: fluorescence state detection and dispersive readout. Since they ultimately rely on detection of light, we frequently invoke the results of Sec.~\ref{sec:photons}. While other detection mechanisms for qubits exist, they generally seem to involve particle detection in some form.

\subsection{Fluorescence measurement}
\label{subsec:fluorescent}

Detecting fluorescent photons is a common means of measuring the $Z$-basis states of qubits realized in trapped ions~\cite{Didi2003}, neutral atoms~\cite{Bakr2010, Sherson2010, Bloch2012, Ott_2016}, NV centers~\cite{DOHERTY20131, Barry2020}, and more. We first introduce a minimal model of fluorescence state detection for trapped-ion qubits, which is based on the ``electron shelving'' technique~\cite{Nagourney1986,Didi2003}. We then explain how to adapt this simple model to fluorescence state detection of other experimental realizations of qubits. 

\begin{figure}
    \centering
    \includegraphics[width=0.85\textwidth]{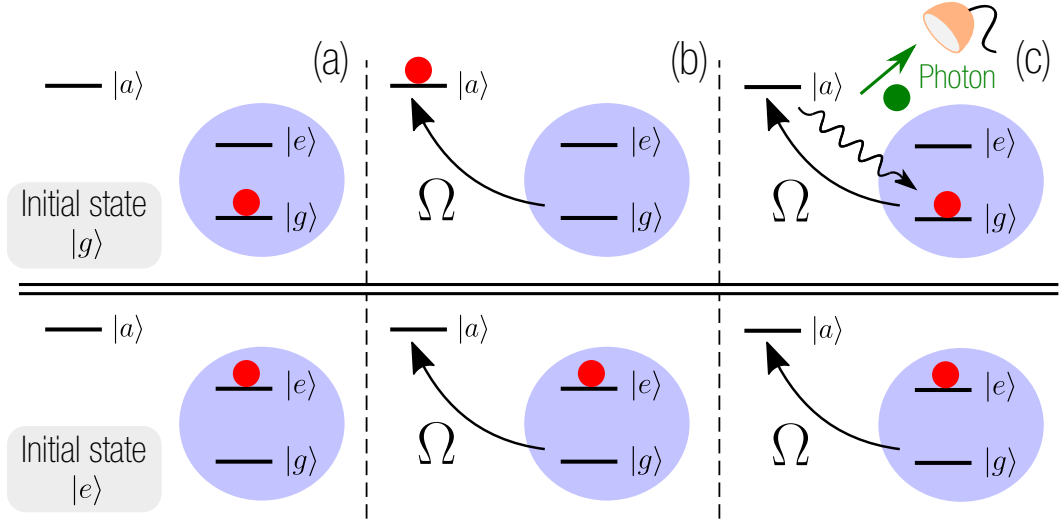}
    \caption{State detection via fluorescent photons. (a) The top (bottom) row corresponds to the initial physical state $\ket{g}$ ($\ket{e}$). (b) Shining light on the system preferentially excites the state $\ket{g}$ to $\ket{a}$. (c) Relaxation via spontaneous emission of a photon that can be detected. This process is repeated many times. }
    \label{fig:fluorescence detection}
\end{figure}

The minimal model involves three atomic energy levels: the ground state $\ket{g} = \ket{0}$ and excited state $\ket{e} = \ket{1}$ that define the computational states of the qubit, and an \emph{auxiliary} level $\ket{a}$, as depicted in Fig.~\ref{fig:fluorescence detection}. The state $\ket{a}$ is generally unstable (i.e., short lived, with a typical lifetime of a few nanoseconds). Fluorescence state detection operates as follows (see also Fig.~\ref{fig:fluorescence detection}). Incident light of the appropriate frequency $\omega_{ag} = E_a - E_g$ preferentially excites an atom from the state $\ket{g} = \ket{0}$ to the state $\ket{a}$, while leaving the state $\ket{e} = \ket{1}$ unaltered; the excited atom in the unstable state $\ket{a}$ fluoresces, spontaneously emitting a photon as it relaxes back to the state $\ket{g}$. The process is then repeated, typically in a regime of high saturation (i.e., high intensity of incident light). The physical state is inferred from the presence ($\ket{g}$) or absence ($\ket{e}$) of emitted photons~\cite{Didi2003, Bakr2010,Sherson2010, Bloch2012, Ott_2016, Nagourney1986}.

This process is modeled using a nonminimal Stinespring Hilbert space $\Hilbert{\text{ss}} = \Hilbert{\text{em}} \otimes \Hilbert{\text{det}}$, where $\Hilbert{\text{em}}$ corresponds to the electromagnetic mode containing emitted (fluorescent) photons and $\Hilbert{\text{det}}$ is the state of the photodetector. The electromagnetic mode incident on the atom is \emph{not} included in $\Hilbert{\text{em}}$; these photons merely induce the atomic transition $\ket{g} \to \ket{a}$. The state of the detector reflects the number of photons absorbed, as described in Sec.~\ref{subsec:photon counting}. The minimal representation \eqref{eq:Qubit Z Meas} corresponds to associating $n>0$ detected photons with $\ket{g}$ and $n=0$ with $\ket{e}$. 

The state of the full system (including the electromagnetic mode and detector) is initially
\begin{equation}
\label{eq:fluorescence initial}
    \ket{\Psi_0} \, = \, \left( c_g \ket{g}+c_e \ket{e} \right)_{\text{ph}} \otimes \ket{0}_{\text{em} }\ket{0}_{\text{det}} \, , ~~
\end{equation}
i.e., an arbitrary superposition of the qubit states $\ket{0}$ and $\ket{1}$, an unoccupied external electromagnetic field, and the default (unexcited) state of the photodetector. Fluorescence measurement begins with shining light on the atom, described by time evolution under the Hamiltonian
\begin{equation}
\label{eq:fluorescence Ham}
    \Ham \, = \, \frac{\omega_{ag}}{2} \big(\BKop{a}{a}-\BKop{g}{g}\big)\, +\frac{\Omega}{2} (\Pauli{}{ga} \ee^{\ii \, \omega_{ag} \, t} + \Pauli{}{ag} \ee^{-\ii \, \omega_{ag} \, t})+\Ham_\text{em}+\Ham_\text{int} \, ,~~
\end{equation}
where $\Pauli{}{ga} = \Pauli{\dagger}{ag} = \BKop{g}{a}$ and $\Omega$ is the Rabi frequency of the driving light, which is resonant only with the atomic transition $\ket{g} \leftrightarrow \ket{a}$,  $\Ham_\text{em}$ describes the evolution of the electromagnetic field, and $\Ham_\text{int}$ describes the interaction between the field and atom,  including spontaneous emission of photons via the Wigner-Weisskopf mechanism. We do not write $H_\text{int}$ explicitly, as it is quite complicated; we also stress that $\ket{e}$ does not appear in Eq.~\ref{eq:fluorescence Ham}.

Physically, the atom is repeatedly excited from $\ket{g}$ to $\ket{a}$ by the incident light via the Rabi term in $H$ with coefficient $\Omega$ in \eqref{eq:fluorescence Ham} [see Fig.~\ref{fig:fluorescence detection}(b)]. The interaction term $H_{\text{int}}$ \eqref{eq:fluorescence Ham} captures the fluorescence process, where the atomic transition $\ket{a}$ to $\ket{g}$ is accompanied by a spontaneously emitted photon [see Fig.~\ref{fig:fluorescence detection}(c)]. The emitted photons evolve under $H_{\text{em}}$ \eqref{eq:fluorescence Ham} until they are counted by the photodetector, as described in Sec.~\ref{subsec:photon counting}; the repeated excitation and emission increases the chance of detection. Supposing that this process results in $n$ photons, the intermediate state is 
\begin{equation}
\label{eq:fluorescence Psi 1}
    \ket{\Psi_1} \, = \, c_g \ket{g}^{\vpp}_{\text{ph}} \otimes\ket{n}^{\vpp}_{\text{em}} \ket{0}^{\vpp}_{\text{det}} + c_e \ket{e}^{\vpp}_{\text{ph}} \otimes\ket{0}^{\vpp}_{\text{em}} \ket{0}^{\vpp}_{\text{det}} \, ,~~
\end{equation}
where none of the photons have been absorbed by the photodetector [see Fig.~\ref{fig:fluorescence detection}(c)] for the purpose of illustration. The photons are then counted as described in Sec.~\ref{subsec:photon counting}, leading to a final state
\begin{equation}
\label{eq:fluorescence Psi final}
    \ket{\Psi_f} \, = \, c_g \ket{g}^{\vpp}_{\text{ph}} \otimes\ket{0}^{\vpp}_{\text{em}} \ket{n}^{\vpp}_{\text{det}} + c_e \ket{e}^{\vpp}_{\text{ph}} \otimes\ket{0}^{\vpp}_{\text{em}} \ket{0}^{\vpp}_{\text{det}} \, , ~~
\end{equation}
of the full system and detector, where we associate $n>0$ counted photons with outcome $g$ and $n=0$ counted photons with outcome $e$ (see also Fig.~\ref{fig:fluorescence detection}). In practice, however, the counting of emitted photons and reexcitation of the atom are concurrent, and not all emitted photons are necessarily counted. Although this may result in more complicated states than appear in  Eqs.~\ref{eq:fluorescence Psi 1} and \ref{eq:fluorescence Psi final}, we always associate $n>0$ counted photons with $\ket{g}$ and $n=0$ counted photons with $\ket{e}$.

The foregoing procedure is commonly referred to as ``fluorescence state detection''~\cite{FluorescentMeas} because it detects the $Z$-basis state. This is equivalent to measuring $Z = \BKop{g}{g} - \BKop{e}{e}$ itself.  Importantly, associating  $\ket{n=0}^{\,}_{\text{det}}$ with the outcome $\ket{1}$ ($e$) and \emph{any} state $\ket{n>0}^{\,}_{\text{det}}$ with the outcome $\ket{0}$ ($g$) defines a binning procedure that leads to the minimal Stinespring unitary \eqref{eq:Qubit Z Meas},
\begin{equation}
    \Umeas{\PZ{}}  \, = \, \BKop{g}{g}^{\vpp}_{\text{ph}} \otimes \ident^{\vpp}_{\text{ss}} + \BKop{e}{e}^{\vpp}_{\text{ph}} \otimes \Shift{\text{ss}} \, ,~~
\end{equation}
which acts on the initial state $\ket{\psi}^{\vpp}_{\text{ph}} \otimes \ket{e}^{\vpp}_{\text{ss}}$, where the default is $e$ because it is associated with no change to the state of the detector; this is the general interpretation of the default state on $\Hilbert*{\text{ss}}$.

As with photon counting in Sec.~\ref{subsec:photon counting}, successful fluorescence measurement is possible even if the photodetector only counts emitted photons with probability $p < 1$ \eqref{eq:photodetect prob k}. Intuitively, one should then shine a large number of photons on the atom, in which case the postmeasurement state is
\begin{align}
    \ket{\Psi_f (p)} \, &= \, c_g \, \ket{g} \otimes \sum\limits_{m=1}^{n} \Big[ \binom{n}{m} \, p^m(1-p)^{n-m} \Big]^{1/2}  \ket{n-m}^{\vpp}_{\text{em}} \ket{m}^{\vpp}_{\text{det}} \notag \\
    &~~~~~ + \big[c_e\ket{e}\otimes\ket{0}^{\vpp}_{\text{em}}+c_g(1-p)^{n/2} \, \ket{g}\otimes\ket{n}^{\vpp}_{\text{em}}\big]\ket{0}^{\vpp}_{\text{det}} \, ,~~\label{eq:fluorescence post meas}
\end{align}
where the two terms above correspond to $m \geq 1$ photons detected and $m=0$ photons detected, respectively. For any $p>0$, if $m>0$ photons are detected, the atom is guaranteed to be in the state $\ket{g}$, provided that sources of dark counts (i.e., false positives) are mitigated. However, for $p<1$, there is a nonzero probability to detect $m=0$ photons even if the atom is in the excitable state $\ket{g}$. If $p$ is the probability that a photon shone on the atom induces fluorescence \emph{and} the spontaneously emitted photon is absorbed by the detector, then the probability of such a ``false negative'' is $\abs{c_g}^2 \, (1-p)^{n/2}$, where $n$ is the number of photons shone on the atom. 

Hence, shining a very large number of photons on the atom exponentially reduces the probability of a false negative. In this limit, the final state can be represented schematically as
\begin{equation}
\label{eq:fluorescent final state SS}
    \ket{\Psi_f} \, = \, c_g \, \ket{g}^{\vpp}_{\text{ph}} \otimes\ket{\text{clicks}}^{\vpp}_{\text{ss}}+c_e \, \ket{e}^{\vpp}_{\text{ph}} \otimes\ket{\text{no clicks}}^{\vpp}_{\text{ss}} \, ,~~
\end{equation}
which we note is of the minimal Stinespring form now that we have binned ``clicks'' ($m>0$) versus ``no clicks'' ($m=0$). However, scattering large numbers of photons leads to unwanted heating so that, in practice, one must optimize the number of photons shone on the atom. 

We now discuss some important practical considerations involved in the experimental implementation of the foregoing simplified model of fluorescence measurements of $\PZ{}$ in atomic systems, as well as how the technique is commonly  extended to other realizations of qubits.  
\begin{itemize}
    \item The fluorescence measurement process described above is most commonly used to determine the energy levels of, e.g., trapped ions~\cite{Bohnet2016, deClercq2016, Cui2022} and neutral atoms~\cite{Kwon2017,Martinez2018b,Hines2022}. With neutral atoms, it is often preferable to induce the transition $\ket{g} \to \ket{a}$ with off-resonant light to avoid, e.g., heating-induced losses and state leakage~\cite{Martinez2018}. Alternatively, cavities can be used to enhance the collection efficiency~\cite{Bochmann2010} and ensure that light is preferentially emitted in particular directions to facilitate the collection of outgoing photons~\cite{Terraciano2009}.
    \item Fluorescence measurements are also used to determine the occupancy of ``sites'' in optical lattices~\cite{Bakr2009, Sherson2010, Bloch2012}. The states $\ket{g}$, $\ket{a}$ and $\ket{e}$ above are replaced by $\ket{1,g}$, $\ket{1,a}$ and $\ket{0}$, where  $\ket{1,g/a}$ indicates the presence of an atom in the state $g/a$ and $\ket{0}$ indicates the absence of an atom or an atom in a state other than $g$ or $a$. We also note that conservation of particle number requires that superpositions of $\ket{0}$ and $\ket{1,g/a}$ on a given ``site'' must be correlated with mixed occupations of other sites~\cite{Endres2013}. This necessitates a description of the measurement process involving Stinespring registers for each site; in many experimental implementations, the imaging technique only discriminates between even and odd occupations of several optical lattice sites, as atoms are typically lost in pairs due to light-assisted collisions~\cite{Schlosser2001}. 
    \item Fluorescence measurements can also be used as an indirect probe of which ``sites'' in an optical lattice are occupied by an atom in the state $g$. This is particularly common in tweezer arrays, where one preferentially expels all atoms in the state $e$ (e.g., via resonant heating~\cite{Evered2023, Young2020} or due to antitrapping~\cite{Barredo2018}). One then uses fluorescence measurement to obtain a ``snapshot'' of the locations of the remaining atoms, which are in the state $g$. This requires the ability to resolve light emitted from different lattice sites. We also note that this process is \emph{destructive}, as the  ejected atoms are irrevocably lost, but is useful in numerous settings. 
    \item Fluorescence state detection also applies to NV centers, whose various internal states involve configurations of, e.g., electronic spins in the vicinity of the defect~\cite{DOHERTY20131, Barry2020}. In contrast to the atomic implementation above, fluorescence measurement of NV centers require \emph{two} auxiliary internal states $\ket{a}$ and $\ket{a'}$. The light incident on the NV center generically induces transitions $\ket{g} \to \ket{a}$ and $\ket{e} \to \ket{a'}$, but the various decay pathways out of $\ket{a}$ versus $\ket{a'}$ are different. The system can return to both of the states $\ket{g}$ and $\ket{e}$ via radiative processes, and can return to the state $\ket{g}$ via nonradiative decay (via intersystem crossing). Importantly, these nonradiative processes occur with different rates, which  has two consequences: (\emph{i}) at the end of the measurement the NV center realizes the state $\ket{g}$ regardless of its initial state  and (\emph{ii}) the number of scattered photons depends on whether the initial state was $\ket{g}$ or $\ket{e}$. The final state following the measurement is given schematically by~\cite{Barry2020}
    \begin{equation}
    \label{eq:NV center fluorescence}
        \ket{\psi} \, = \, \big[c_g \, \ket{g}+c_e \, \ket{e}\big] \otimes \ket{0}^{\vpp}_{\text{em}} ~~ \mapsto ~~ \ket{g} \otimes \big[c_g \, \ket{n_g}^{\vpp}_{\text{em}} + c_e \, \ket{n_e}^{\vpp}_{\text{em}} \big] \, , ~~
    \end{equation}
    where $\ket{n_{g,e}}$ is a placeholder for some complicated state of the electromagnetic field that, on average, has $n_{e,g}$ photons. It is typically assumed to be coherent to account for Poisson statistics~\cite{Barry2020}. The outcomes corresponding to $\ket{e}$ versus and $\ket{g}$ are more subtle to differentiate than ``clicks'' versus ``no clicks,'' since the associated numbers of emitted photons are very small. However, performing the experiment many times increases the total numbers of collected photons from the states $\ket{g}$ and $\ket{e}$; it also increases their \emph{difference}, so that one can reliably distinguish the two states. Note that $\ket{\psi}$~\eqref{eq:NV center fluorescence} has the Stinespring form of a destructive measurement~\eqref{eq:Destructive Unitary from Projective}.
\end{itemize}

\subsection{Dispersive measurement}
\label{subsec:dispersive}

\begin{figure}
    \centering
    \includegraphics[width=0.8\textwidth]{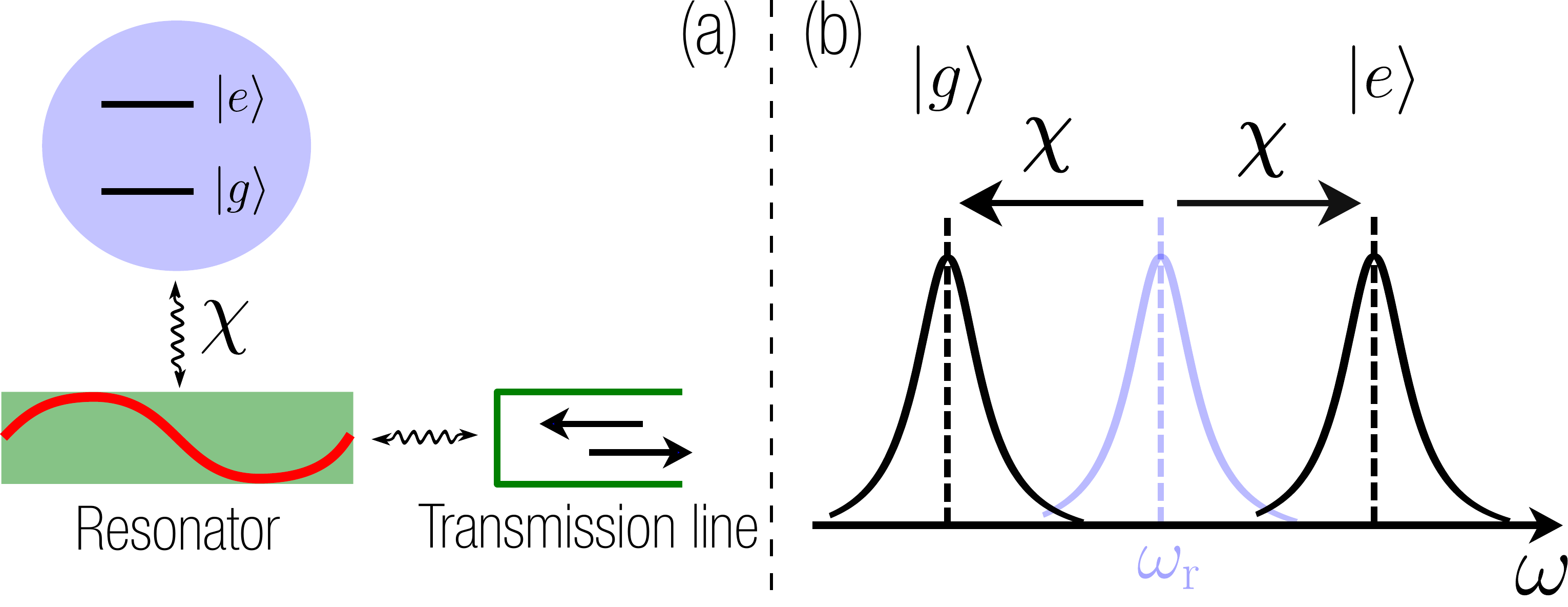}
    \caption{Left: Schematic of a superconducting qubit interacting with a resonator that is coupled to a transmission line. Right: Bare cavity frequency ($\omega_{\text{r}}$) and the qubit state-dependent frequency shifts (blue).}
    \label{fig:SchematicSQC}
\end{figure}

Here we describe dispersive state readout~\cite{Blais2021}---a common technique for determining the $\PZ{}$-basis state of superconducting qubits and atoms in cavities (microwave and optical). For concreteness, we focus on a single superconducting qubit, whose two levels generally represent physically distinct configurations of charges, phases, and/or fluxes in a superconducting circuit. This state can be measured indirectly via controllable interactions with auxiliary systems, as depicted schematically in Fig.~\ref{fig:SchematicSQC}. More concretely, one sends a pulse of light through a transmission line to probe a resonator (e.g., a cavity) that hosts confined electromagnetic modes. The interaction of the resonator modes with the qubit modify the pulse in the transmission line (see Fig.~\ref{fig:SchematicSQC}), from which the qubit's state is inferred. A minimal description consists of the qubit (with states $g$ and $e$), a single resonator mode (with photon annihilation operator $\re{}$), and the transmission line (whose continuum of propagating electromagnetic modes is represented schematically by Fock states $\ket{n}_{\text{out}}$). As in Sec.~\ref{sec:photons}, we picture these outgoing modes as being mildly localized and propagating towards a detector.

The measurement protocol involves evolution under the qubit-resonator Hamiltonian
\begin{equation}
\label{eqn:ResonantHamiltonian}
    H_{\text{QR}} \, = \, \omega_a \Pauli{z}{} + \omega_r\rdag{} \re{} + g (\rdag{}\Pauli{-}{}+\re{}\Pauli{+}{}) \, , ~~
\end{equation}
where $\omega_a$ is the frequency of transitions between the states $\ket{e}$ and $\ket{g}$ of the qubit, $\omega_r$ is the frequency of the resonator, and $g$ controls the strength of a Jaynes-Cummings interaction~\cite{Jaynes1963}, which models coherent photon absorption and emission by the qubit. In the \emph{dispersive} regime---where the large detuning $\abs{\omega_a-\omega_r} \gg g$ ensures that the atom quickly emits any absorbed photons---we perturb about $g = 0$ to good approximation, recovering the Hamiltonian
\begin{equation}
\label{eqn:DispersiveHamiltonian}
    H_{\text{QR}} \, = \, \omega_a \Pauli{z}{} + \omega_r\rdag{}\re{} + \chi(\rdag{} \re{}) \, \Pauli{z}{} \, ,~~
\end{equation}
where $\chi = g^2/(\omega_r-\omega_a)$, and this term describes a shift in the resonator frequency $\omega_r$, whose sign depends on the qubit's $\PZ{}$-basis state [see Fig.~\ref{fig:SchematicSQC}(b)]. Dispersive readout proceeds as follows:
\begin{enumerate}
    \item The qubit of interest is prepared in an arbitrary superposition $\ket{\psi} = c_g \ket{g} + c_e \ket{e}$ and the field modes (resonator and transmission line) are initially empty. The dilated initial state is
    \begin{equation}
        \label{eq:dispersive dilated initial state}
        \ket{\Psi_0} \, = \, \left( c_g \ket{g} + c_e \ket{e} \right) \otimes \ket{0}^{\vpp}_{r} \ket{0}^{\vpp}_{\text{out}} \, .~~
    \end{equation}
    \item A mode with the same frequency $\omega_r$ as the \emph{bare} resonator is sent to the resonator via the transmission line. The outgoing modes realize a coherent state $\ket{\alpha}$~\eqref{eq:beta aux}, and the average number of photons $n \sim \abs{\alpha}^2$ sent towards the resonator grows linearly in time in the late-time steady-state regime of the electromagnetic modes. The state remains separable, with
    \begin{equation}
        \label{eq:dispersive dilated second state}
        \ket{\Psi_1} \,  = \,  \ket{\psi} \otimes \ket{0}^{\,}_{r}\ket{\alpha}^{\,}_{\text{out}} \, .~~
    \end{equation}
    \item Photons sent through the transmission line enter the resonator, interact with the qubit, and escape back through the line. The resonator responds to the incoming light as a damped-driven harmonic oscillator (damped by the modes in the transmission line). Crucially, this response depends on the resonator frequency, which in turn depends on the state of the qubit via $\chi$ \eqref{eqn:DispersiveHamiltonian}. This process (with the foregoing assumptions) is captured by:
    \begin{subequations}
    \label{eq:dispersive readout g,e update}
    \begin{align}
        \ket{g} \otimes \ket{0}^{\vpp}_{\text{r}} \ket{\alpha}^{\vpp}_{\text{out}} ~ &\mapsto ~ \ket{g} \otimes\ket{\rho \, \ee^{+\ii \theta}}^{\vpp}_{\text{r}} \ket{0}^{\vpp}_{\text{out}} ~ \mapsto ~ \ket{g} \otimes\ket{0}^{\vpp}_{\text{r}} \ket{\alpha \, \ee^{+2 \ii \theta}}^{\vpp}_{\text{out}} \label{eq:dispersive readout g update}\\
        \ket{e} \otimes \ket{0}^{\vpp}_{\text{r}} \ket{\alpha}^{\vpp}_{\text{out}} ~ &\mapsto ~ \ket{e} \otimes\ket{\rho \, \ee^{-\ii \theta}}^{\vpp}_{\text{r}} \ket{0}^{\vpp}_{\text{out}} ~ \mapsto ~ \ket{e} \otimes \ket{0}^{\vpp}_{\text{r}} \ket{\alpha \, \ee^{-2 \ii \theta}}^{\vpp}_{\text{out}} \, , ~~\label{eq:dispersive readout e update}
    \end{align}
    \end{subequations}
    where $\ket{\rho \, \ee^{\pm \ii\theta}}^{\vpp}_{\text{r}}$ and $\ket{\alpha \, \ee^{\pm 2 \ii \theta}}^{\vpp}_{\text{out}}$ are coherent states \eqref{eq:beta aux}. The final state of the system is
    \begin{equation}\label{eq:dispersive readout final state}
        \ket{\Psi_f} \, = \,c_g \ket{g} \otimes \ket{0}^{\vpp}_{r} \ket{\alpha \, \ee^{2\ii \theta}}^{\vpp}_{\text{out}}+\,  c_e \ket{e}  \otimes \ket{0}^{\vpp}_{r} \ket{\alpha \, \ee^{-2\ii \theta}}^{\vpp}_{\text{out}} \, .~~
    \end{equation}
    In this schematic description, the light enters the cavity and establishes an intracavity field $\rho \, \ee^{\pm \ii \theta}$ whose phase depends on the state of the qubit ($+\ii \theta $ for $\ket{g}$ and $- \ii \theta$ for $\ket{e}$). The coherent light then escapes the resonator via the transmission line, so that its phase can be recorded. We comment that, in practice, all of these processes occur at the same time and the device is operated in steady-state conditions after transients have died off. 
    \item The phase of the returning light is extracted via homodyne detection of the momentum quadrature $p = (\a{} - \adag{})/ \ii \sqrt{2} = x_{\phi=\pi/2}$ ~\eqref{eq:beta aux}. Working in the eigenbasis of $p$,
    \begin{align}
        \ket{\alpha \,  \ee^{\pm \ii \theta}} \, = \, \pi^{-1/4}\int\limits_{-\infty}^{\infty} \thed p \, \ee^{-\ii \sqrt{2} \, \alpha p \cos \theta} \, \ee^{-(p \mp \sqrt{2} \, \alpha \sin \theta)^2/2} \, \ket{p} \, \equiv \, \int\limits_{-\infty}^{\infty} \thed p \, \psi^{\vpp}_{\pm}(p) \, \ket{p} \, , ~~
    \end{align}
    and, since the distribution of the output momentum for the initial qubit states $\ket{g}$ and $\ket{e}$ are centered at $p = \pm \sqrt{2} \alpha \sin \theta$, respectively, we associate the detection of $\ket{p>0}$ with $\ket{g}$ and of $\ket{p<0}$ with $\ket{e}$. This is procedure is not perfect: The probability to detect $p<0$ ($p>0$) when the initial state of the qubit is $g$ ($e$) is given by
    \begin{equation}
        p^{\vpp}_\text{error} \, = \, \int\limits_0^{\infty} \thed p \abs{\psi^{\,}_{-} (p)}^2 \, = \, \int\limits_{-\infty}^0 \thed p \abs{\psi^{\,}_{+} (p)}^2 \, = \, \frac{1}{2} \, \text{erfc} \left( \alpha \, \sqrt{2} \, \sin \theta \right)  \, , ~~
    \end{equation}
    which saturates to $1/2$ when $\alpha \sin \theta \to 0$, and vanishes as $\exp (-\alpha^2 \sin^2 \theta) / \alpha \sin \theta$ in the limit $\alpha \sin \theta \to \infty$. In this limit of many outgoing photons, the final state takes the form
    \begin{align}
    \label{eq:dispersive final Stinespring}
        \ket{\Psi_f} \, &= \, c_g \ket{g} \otimes \ket{0}_{\text{r}} \bigg( \int\limits_0^\infty \thed p \, \psi^{\vpp}_{+} (p) \ket{p}^{\vpp}_{\text{out}} \bigg) + c_e \ket{e} \otimes \ket{0}^{\vpp}_{\text{r}} \bigg( \int\limits_{-\infty}^0 \thed p \, \psi^{\vpp}_{-} (p) \ket{p}^{\vpp}_{\text{out}} \bigg) \, ,~~
    \end{align}
    which requires that $\theta$ is within $(0,\pi/2)$, or else one cannot discriminate between these outcomes; however, this is generally the case in experiment. The outgoing light is then measured via homodyne detection, so that correlations between the qubit and the electromagnetic mode (in the transmission line) are transferred into correlations between the qubit and the differential intensity measurements of the final detectors, as in Eq.~\ref{eqn:homodynefinalstate}. In actual experimental implementations, the phase $\theta$ in Eq.~\ref{eq:dispersive readout g,e update} is measured in a steady-state configuration in which there is a fixed \emph{flux} of photons (i.e., intensity) emanating from the resonator.  By waiting longer, transients die off and $\abs{\alpha}^2$ grows larger, linearly in time, reducing the probability of a readout error. However, separate experimental factors may degrade the measurement process over time, meaning there exists an optimal duration for the dispersive-readout process.
\end{enumerate}

\noindent We now describe the application of this detection scheme in various experimental settings:
\begin{itemize}
    \item Dispersive readout as described above is a common measurement approach for superconducting qubits~\cite{Blais2021} and detecting the presence of absence of atoms in cavities~\cite{Puppe2007}.
    \item The same scheme can also be used to measure the collective ``inversion'' of many atoms---i.e., the difference between the number of atoms in the excited versus ground state---in cavities (both optical and microwave)~\cite{SchleierSmith2010}. Instead of measuring $\Pauli{z}{}$ for a single atom, one measures $S^{\,}_z=\sum_j \Pauli{z}{j}/2$ for an ensemble of $N$ atoms. The role of the resonator is played by the cavity itself and that of the transmission-line modes is played by light leaking through the cavity mirrors, which can be measured using homodyne (or heterodyne) detection.
    
    In the ``dispersive regime,'' the atom-cavity interaction is given by Eq.~\ref{eqn:DispersiveHamiltonian}, with $\Pauli{z}{}$ replaced by $2 S^{\,}_z$. The detection scheme proceeds as described above. Note that, in this collective regime, there are $N+1$ possible values $S^{\,}_z$ instead of two. Ideally, the procedure collects enough photons to enable almost perfect discrimination between each value of inversion, yielding a postmeasurement state in Stinespring form \eqref{eq:dispersive final Stinespring}. Schematically, if $\ket{m}$ represents the different eigenstates of the inversion $S^{\,}_z$, the dispersive-readout process takes the form
    \begin{equation}
        \ket{\Psi_0}\,= \sum_{m=-N/2}^{N/2} c^{\,}_m \ket{m}\otimes\ket{0}_{\text{r}}^{\vpp}\ket{0}_{\text{out}}^{\vpp}~~ \mapsto ~~\ket{\Psi_f} \,= \sum\limits_{m=-N/2}^{N/2} c^{\,}_m \ket{m}\otimes\ket{0}_{\text{r}}^{\vpp}\ket{\alpha \, \ee^{\ii m\theta}}_{\text{out}}^{\vpp} \,,~~
    \end{equation}
    where the coherent states $\ket{\alpha \, \ee^{\ii m\theta}}$ are nearly orthogonal for different $m$-dependent phases $m \theta$. This is often not possible due to technical limitations, but the information gained by the measurement can still be used to constrain the most likely values of $S^{\,}_z$, in many cases creating entanglement between atoms in the ensemble~\cite{SchleierSmith2010, Cox2016, Hosten2016, robinson2022direct}.

    The physical mechanism underlying this collective $S^{\,}_z$ measurement---namely, the state-dependent shift of the cavity (resonator) frequency captured by Eq.~\ref{eqn:DispersiveHamiltonian}---is, strictly speaking, not restricted to the dispersive regime. It also occurs when the atoms and cavity are close to (or on) resonance---i.e.,  $\omega_a=\omega_r$ in Eq.~\ref{eqn:ResonantHamiltonian}---though the functional relationship between the atomic state and the shift in cavity frequency is different. In contrast to the dispersive scheme, three levels are typically used, where the cavity is resonant with the transition $\ket{g} \leftrightarrow \ket{a}$, rather than $\ket{e} \leftrightarrow \ket{g}$, for some auxiliary state $\ket{a}$. The objective of a measurement in this regime is not to infer the inversion $S^{\,}_z = N_e - N_g$, but simply the number $N_g$ of atoms in $\ket{g}$~\cite{Cox2016, Hosten2016}. The inversion can be determined using a similar, complementary measurement of $N_e$. These resonant measurements can also be used in the single-atom regime~\cite{Volz2011}.
    \item The dispersive response of atoms can also be used directly with atomic clouds to infer their average atomic state~\cite{Windpassinger2008}, or for spatial imaging, whereby a probe beam of light acquires spatially dependent phase shifts as it interacts with the atoms in the cloud~\cite{Andrews1996, Bradley1997, Meppelink2010}. We defer a Stinespring description of this procedure to the future, since it involves the spatial distribution of the electromagnetic field, which we have ignored thus far.  
\end{itemize}

\section{Using the unitary measurement formalism}
\label{sec:Using Stinespring}

Having discussed several prominent experimental implementations of projective and destructive measurements and their connection to the dilated unitary formalism presented in Sec.~\ref{sec:Measurement formalism}, we now discuss how to \emph{use} this formalism in practice and its various advantages, implications, and properties.

\subsection{Measurement outcomes and the Born rule}
\label{subsec:extract}

We begin by showing that the standard concepts associated with such measurements recover in the Stinespring representation of Sec.~\ref{sec:Measurement formalism}. In particular, we show how the standard Born rule may be evaluated either before or after the measurement, and is a natural consequence of Gleason's theorem~\cite{Gleason}. We also consider the extraction of expectation values. As noted in Sec.~\ref{sec:Measurement formalism}, the projective measurement of an observable $\mobserv$ \eqref{eq:measured observable} is always associated with a quantum \emph{channel}
\begin{equation*}
    \tag{\ref{eq:Stinespring Unitary TP}}
    \Phi (\rho ) \, = \, \trace_{\text{ss}} \left( \, \umeas \, \rho \otimes \BKop{i}{i}^{\,}_{\text{ss}} \, \umeas^{\dagger} \, \right) \, = \rho^{\vpp}_{\rm av} \, ,~~
\end{equation*}
where the \emph{measurement unitary} $\umeas$ acts on $\Hilbert{\rm dil} = \hilbert \otimes \Hilbert{\text{ss}}$, with $\Hilbert{\text{ss}}$ the state space of the measurement apparatus and $\ket{i}^{\,}_{\text{ss}}$ the default initial state of the apparatus. The channel $\Phi$ is CPTP, and $\rho^{\,}_{\rm av}$ is the outcome-averaged postmeasurement density matrix. 

Related to $\Phi$ \eqref{eq:Stinespring Unitary TP} is the selective operation \cite{KrausBook} corresponding to the quantum \emph{channel}
\begin{equation*}
    \tag{\ref{eq:Stinespring Unitary TD}}
    \Phi^{\vpp}_m (\rho) \, = \, \trace_{\text{ss}} \left( \, \umeas \, \rho \otimes \BKop{i}{i}^{\,}_{\text{ss}} \, \umeas^{\dagger} \, \Proj{\text{ss}}{(m)} \, \right) \, = \, p^{\vpp}_m \, \rho^{\vpp}_m \,, ~~
\end{equation*}
which is a CP map describing a projective measurement resulting in the particular outcome $m$, where  $\Proj{\text{ss}}{(m)}$ projects onto the state(s) of the apparatus associated with outcome $m$, $p^{\,}_m$ is the probability of observing outcome $m$, and $\rho^{\,}_m$ is the ``collapsed'' postmeasurement density matrix~\cite{KrausBook}. 

The foregoing expressions for $\Phi$ \eqref{eq:Stinespring Unitary TP} and $\Phi^{\,}_m$ \eqref{eq:Stinespring Unitary TD} only apply to projective measurements. The destructive measurements discussed in Sec.~\ref{subsec:destructive} generally result in the \emph{same} state of the physical system for \emph{any} outcome $m$. Apart from the postmeasurement state, however, all of the formulae below apply equally to projective and destructive measurements. 

The probability to observe outcome $m$ (i.e., the eigenvalue $\mEig*{m}$) is given by the \emph{Born rule}
\begin{equation}
    \label{eq:usual Born}
    p^{\vpp}_m \, = \, \abs{\inprod{m}{\psi}}^2 ~~~ \longleftrightarrow ~~~ p^{\vpp}_m \, = \, \trace \left( \, \proj{m} \, \rho \, \right) \, , ~~
\end{equation}
where $\rho$ is the premeasurement state of the system. The familiar expression on the left holds when $\rho = \BKop{\psi}{\psi}$ is pure and the $m$th outcome is nondegenerate \eqref{eq:measured observable}; the expression on the right is generic. Importantly, the Born rule \eqref{eq:usual Born} is generally an assumption of quantum mechanics---i.e., it is built into the axioms \cite{DiracQuantum, vonNeumannAxioms, HardyAxioms, FuchsAxioms, MackeyAxioms, WilceAxioms, MasanesAxioms, KapustinAxioms}. We note that the state $\rho$ encodes a probability distribution, which one expects can be extracted from expectation values as in Eq.~\ref{eq:usual Born}. The expectation value of an operator is a $C^*$-state---a positive linear functional. Positivity of the spectral projectors \eqref{eq:Spec Projecc} then implies that $p^{\,}_m \geq 0$, and because the projectors form a complete set \eqref{eq:meas proj ortho complete}, we have
\begin{equation}
\label{eq:Born motivation}
    0 \, \leq \, p^{\vpp}_m \, = \, \trace \left(\, \proj{m} \, \rho \, \right) \, \leq \, \sum\limits_{m=0}^{\Noutcome-1} \, \trace \left(\, \proj{m} \, \rho \, \right)  \, = \, \trace \left( \, \rho \,\right) \, = \, 1 \, ,~~
\end{equation}
so that the probabilities $p^{\,}_m$ are positive and sum to one, as required. Moreover, Gleason's theorem~\cite{Gleason} establishes that, \emph{a priori}, probabilities are always related to projectors via Eq.~\ref{eq:Born motivation}.

Using the Stinespring representation of measurements developed in Sec.~\ref{subsec:unitary measurement}, we can write
\begin{equation}
    \label{eq:Stinespring Born}
    p^{\vpp}_m \, = \, \trace \left( \, \umeas \, \rho \otimes \BKop{i}{i}^{\,}_{\text{ss}} \, \umeas^{\dagger} \, \Proj{\text{ss}}{(m)} \, \right) \, = \, \tr{ \, \Phi^{\vpp}_m (\rho) \, } \, = \, p^{\vpp}_m \, \trace \left( \rho \right) \, = \, p^{\vpp}_m \, ,~~
\end{equation}
i.e., $p^{\,}_m$ is the expectation value of the Stinespring projector $\Proj{\text{ss}}{(m)}$ in the postmeasurement state
\begin{equation}
    \label{eq:dilated postmeas state}
    \varrho (t) \, = \, \umeas \, \rho \otimes \BKop{i}{i}^{\,}_{\text{ss}} \, \umeas^{\dagger} \, = \, \sum\limits_{m=0}^{\Noutcome-1} \, p^{-1}_{m} \, \Phi^{\vpp}_m (\rho) \otimes \Proj{\text{ss}}{(m)} \, , ~~
\end{equation}
on $\Hilbert*{\text{dil}}$. As always, the Born probability corresponds to the expectation value of a projector~\eqref{eq:Born motivation}, as guaranteed by Gleason's theorem~\cite{Gleason}. The probability $p^{\,}_m$ may be extracted from the physical premeasurement state $\rho^{\,}_0$ \eqref{eq:dilated initial state} via Eq.~\ref{eq:usual Born} or the dilated postmeasurement state $\varrho (t)$ \eqref{eq:dilated postmeas state} via Eq.~\ref{eq:Stinespring Born}. The latter has a physical interpretation as the expectation value (or probability) of finding the detector in a state corresponding to outcome $m$ following evolution under $\umeas$ \eqref{eq:Stinespring Unitary General}.

Following a projective measurement resulting in outcome $m$, the system's state is
\begin{align}
\label{eq:collapsed rho}
    \rho^{\vpp}_m \, = \, p^{-1}_m \, \Phi^{\vpp}_m (\rho) \, = \, \frac{\proj{m} \, \rho \, \proj{m}}{\trace \left( \, \proj{m} \, \rho \, \right)} \, ,~~
\end{align}
provided that $p^{\,}_m > 0$. The average postmeasurement state $\rho^{\,}_{\rm av}$ is given by $\Phi(\rho)$ \eqref{eq:Stinespring Unitary TP}. However, in the dilated unitary representation, these merely correspond to particular limits of the \emph{dilated} postmeasurement density matrix $\varrho$ \eqref{eq:dilated postmeas state} of the system \emph{and} apparatus. The average over all outcomes recovers upon taking the trace over the apparatus Hilbert space $\Hilbert{\text{ss}}$ \eqref{eq:Stinespring Unitary TP}, giving the reduced postmeasurement density matrix of the physical system. The collapsed density matrix $\rho^{\,}_m$ \eqref{eq:collapsed rho} corresponds to projecting the apparatus onto the corresponding state $\ket{m}^{\,}_{\text{ss}}$, and then taking the trace  \eqref{eq:Stinespring Unitary TD}. Because the dilated unitary $\umeas$ \eqref{eq:Stinespring Unitary General}---or  Hamiltonian $\Ham$ \eqref{eq:Pointer Hamiltonian}---is both guaranteed by the axioms (see Sec.~\ref{sec:Measurement formalism}) and consistent with known experimental implementations (see Secs.~\ref{sec:photons} and \ref{sec:qubits}), we adopt the perspective that $\varrho$ is the correct postmeasurement density matrix, in which \emph{all outcomes occur}. In Sec.~\ref{subsec:collapse}, we describe why only one outcome is observed in practice.

Finally, the expectation value for measuring $\mobserv$ in some state $\rho$ can be written 
\begin{equation}
    \label{eq:usual expval}
    \expval{\mobserv}^{\vpp}_{\rho} \, = \, \sum\limits_{m=0}^{\Noutcome-1} \, p^{\vpp}_m \, \mEig{m} \, = \, \sum\limits_{m=0}^{\Noutcome-1} \, \mEig{m} \, \trace \left( \, \proj{m} \, \rho \,\right) \, = \, \trace \left( \, \mobserv \, \rho \, \right) \, ,~~
\end{equation}
in terms of the premeasurement state $\rho$, or in terms of the postmeasurement state $\varrho$ (on $\Hilbert{\rm dil}$) via
\begin{equation}
    \label{eq:expval summary}
    \expval{\mobserv}^{\vpp}_{\rho} \, = \, \sum\limits_{m=0}^{\Noutcome-1} \, p^{\vpp}_m \, \mEig{m} \, = \, \sum\limits_{m=0}^{\Noutcome-1} \, \trace_{\rm dil} \left[ \, \left( \ident^{\vpp}_{\text{ph}} \otimes \mEig{m} \, \Proj{\text{ss}}{(m)} \right) \, \varrho \, \right] \, ,~~
\end{equation}
which reproduces Eq.~\ref{eq:usual expval}, as expected. The expectation value and Born probabilities are the same for projective and destructive measurements, and may be evaluated either before or after the measurement unitary $\umeas$ is applied using $\rho$ or $\varrho(t)$, respectively. As a reminder, the fact that the Born probability takes the form of an expectation value of a projector is a consequence of Gleason's theorem~\cite{Gleason}, which extends without caveat to the dilated Hilbert space $\Hilbert{\rm dil}$ \eqref{eq:Dilated Hilbert}.

\subsection{Application to measurement-based protocols}
\label{subsec:adaptive}

Perhaps the greatest advantage of the Stinespring formulation of projective and destructive  measurements is the application to \emph{adaptive} quantum protocols. These protocols involve nonlocal quantum operations based on the outcomes of prior measurements, and are known to be faster (and sometimes less error prone) than local unitary---or even Lindblad---dynamics \cite{Lieb1972, Poulin, SpeedLimit}. However, such protocols are often cumbersome to describe in both the Kraus \eqref{eq:Kraus Representation} or von Neumann \eqref{eq:Pointer Hamiltonian} representations, especially if one is interested calculations that require evolving operators---e.g., diagnosing locality, constraining quantum protocols,  and identifying optimal strategies  \cite{SpeedLimit, AaronTeleport}.

The advantage of representing measurements unitarily on a dilated Hilbert space $\Hilbert{\rm dil} = \Hilbert{\text{ph}} \otimes \Hilbert{\text{ss}}$ \eqref{eq:Dilated Hilbert} is that the outcomes are reflected in explicit degrees of freedom in $\Hilbert{\text{ss}}$, corresponding to the states of the detectors. In experiment, distinct detectors are not necessarily required, and one can simply reset and reuse the same detector for multiple measurements and ``write down'' the outcome. However, for the purpose of calculation, it is most convenient to encode the measurement outcomes in distinct Hilbert spaces. We may also consider protocols in which the decision of \emph{whether} or not to perform a measurement (or what operator to measure) is conditioned on prior outcomes. One simply assigns Stinespring degrees of freedom to each \emph{possible} measurement, and if a measurement is not performed, one simply does not use the corresponding Stinespring register.

Consider a quantum protocol $\mathcal{W}$ involving measurements of up to $M$ observables $\{ \mobserv^{\,}_j \}$, where all aspects of $\mathcal{W}$ at any time $t$ may depend on \emph{any} prior measurement outcomes. The physical system is initialized in an arbitrary state $\rho_0$ on $\Hilbert{\text{ph}}$, and $M$ Stinespring registers labeled $j$ are initialized in the default state $\BKop{0}{0}_{{\text{ss}},j}$ of each apparatus, so that
\begin{equation}
    \label{eq:dilated initial state}
    \varrho (0) \, = \, \rho^{\vpp}_0 \otimes \BKop{\bvec{0}}{\bvec{0}}^{\vpp}_{\text{ss}} \, ,~~
\end{equation}
is the dilated initial state. The physicist's Stinespring theorem \cite{KrausBook, WolfNotes, AaronJamesFuture} ensures that $\mathcal{W}$ acts unitarily on $\Hilbert{\rm dil}$ \eqref{eq:Dilated Hilbert}, which may require enlarging $\Hilbert{\text{ss}}$ to include ``environmental'' degrees of freedom. For simplicity, we consider protocols $\mathcal{W}$ involving only measurements and unitary time evolution on $\Hilbert*{\text{ph}}$, which may be arbitrarily conditioned on prior outcomes. The unitary evolution may be generated by a discrete quantum circuit or continuous evolution under some local Hamiltonian $\Ham (t)$.  In the latter case, one must either approximate the measurement unitary $\umeas$ as instantaneous or use the Hamiltonian formulation of measurements \cite{SpeedLimit}. 

For convenience, suppose that $\mathcal{W}$ is realized by a unitary circuit on $\Hilbert{\rm dil}$ \eqref{eq:Dilated Hilbert}. We now consider the distinct types of dilated unitary ``gates'' in $\mathcal{W}$. Time evolution is realized by a unitary $U$ acting only on $\Hilbert{\text{ph}}$, while a projective measurement of an observable $\mobserv^{\,}_j$ \eqref{eq:measured observable} takes the form
\begin{equation}
\label{eq:Projective measurement gate}
    \umeas^{\vpd}_j \, = \, \sum\limits_{n=0}^{\Noutcome_j-1} \, \Proj{j}{(n)} \otimes \ShiftPow{\text{ss},j}{n} \, ,~~
\end{equation}
where $\Proj{j}{(n)}$ projects onto the $n$th eigenspace of $\mobserv^{\vpp}_j$ \eqref{eq:measured observable} associated with Stinespring label $j$. When $\Noutcome_{j} \to \infty$, $\umeas^{\,}_j$ \eqref{eq:Projective measurement gate} is replaced with a dilated Hamiltonian \eqref{eq:Pointer Hamiltonian} or with a modified unitary \eqref{eq:Measurement Unitary Countably Infinite}. For destructive measurements, one modifies the physical part of Eq.~\ref{eq:Projective measurement gate} to match, e.g., Eq.~\ref{eq:Destructive Unitary from Projective}. 

The measurement of $\mobserv^{\,}_j$ \eqref{eq:measured observable} via $\umeas^{\,}_j$ \eqref{eq:Projective measurement gate} entangles the system with the $j$th Stinespring register, which reflects the observed outcome. Importantly, we do \emph{not} immediately trace over $\Hilbert*{{\text{ss},j}}$ to realize the quantum channel $\Phi$ \eqref{eq:Stinespring Unitary TP} or the quantum operation $\Phi^{\,}_m$ \eqref{eq:Stinespring Unitary TD}. Instead, we view $\umeas^{\,}_j$ \eqref{eq:Projective measurement gate} as the time evolution of the system and the $j$th detector during the measurement of $\mobserv^{\,}_j$ \eqref{eq:measured observable}, and defer \emph{all} traces until the end of the \emph{full} calculation. The relevant quantities are given by
\begin{subequations}
\label{eq:adaptive trace at end}
    \begin{align}
        \rho^{\vpp}_{\rm av} \, &= \, \trace_{\text{ss}} \, \left( \, \mathcal{W} \, \varrho (0) \, \mathcal{W}^{\dagger} \, \right) \label{eq:final rho trajectory average} \\
        p^{\vpp}_{\bvec{n}} \, \rho^{\vpp}_{\bvec{n}} \, &= \, \trace_{\text{ss}} \, \left( \, \mathcal{W} \, \varrho (0) \, \mathcal{W}^{\dagger} \, \Proj{\text{ss}}{(\bvec{n})} \, \right) \label{eq:final rho trajectory n}  \, , ~~
    \end{align}
\end{subequations}
where $\Proj{\text{ss}}{(\bvec{n})} = \BKop{n_1 , \dots, n_M}{n_1 , \dots, n_M}^{\vpp}_{\text{ss}}$ projects (i.e., collapses) the dilated state onto a particular sequence $\bvec{n}=\{n_1,\dots,n_M\}$ of measurement outcomes \eqref{eq:final rho trajectory n}, and taking the trace averages over all outcomes \eqref{eq:final rho trajectory average}. Importantly, these traces are only taken \emph{after} $\mathcal{W}$ has been applied in its entirety, to allow for subsequent outcome-dependent operations. As a bookkeeping tool, one can view the unitary description in Eq.~\ref{eq:adaptive trace at end} as an algebraic means of accounting for all Kraus operators---including outcome-dependent operations---in a convenient manner. However, we associate $\mathcal{W}$ with the \emph{physical} time-evolution operator for the system \emph{and} detectors throughout the protocol. 

Because the trace is only taken at the end in Eq.~\ref{eq:adaptive trace at end}, we can express a unitary \emph{feedback} operation on the physical degrees of freedom in $\Hilbert{\text{ph}}$ that is conditioned on the $j$th measurement outcome via
\begin{equation}
    \label{eq:outcome-dependent gate}
    \mathcal{R}^{\vpd}_j \, = \, \sum\limits_{n=0}^{\Noutcome_j-1} \, U^{\vpd}_n \otimes \Proj{\text{ss}}{(n)} \, ,~~
\end{equation}
which applies the physical unitary $U^{\,}_n$ if the $j$th outcome was $n$, where the trivial case corresponds to $U^{\,}_n = \ident$ (i.e., do nothing). Such outcome-dependent unitaries are essential to measurement-based protocols, as without these operations, the outcome-averaged final state $\rho(t)$ is some maximally mixed state \cite{SpeedLimit, AaronMIPT, AaronTeleport}. We note that Eq.~\ref{eq:outcome-dependent gate} is easily extended to include physical unitaries that are conditioned on multiple outcomes---e.g., the parity of several qubit measurements, in the context of quantum teleportation \cite{AaronTeleport}. One can also describe \emph{conditional measurements} via
\begin{equation}
    \label{eq:outcome-dependent measurement}
    \mathcal{M}^{\vpd}_{j,k} \, = \, \sum\limits_{n=0}^{\Noutcome_j-1} \, \Umeas{\mobserv_n} \otimes \Proj{\text{ss}}{(n)} \,  ,~~
\end{equation}
which realizes a projective measurement of an observable $\mobserv_n$ if the $j$th measurement resulted in outcome $n$. The measurement unitary $\umeas^{\,}_n$ acts on $\Hilbert{\text{ss}}$ and the $k$th Stinespring register---in which the measurement outcome is encoded---conditioned on the outcome of the $j$th measurement. As with outcome-dependent feedback \eqref{eq:outcome-dependent gate}, one can also extend Eq.~\ref{eq:outcome-dependent measurement} to measurements conditioned on multiple outcomes, and the case in which no measurement is performed if the $j$th measurement resulted in outcome $m$ is captured by replacing the $n$th measurement unitary with $\ident$.

\subsection{Decoherence and ``wavefunction collapse''}
\label{subsec:collapse}

For the dilated unitary $\umeas$ \eqref{eq:Projective measurement gate} to be considered \emph{physical}, we must explain why only one outcome is \emph{observed} in any experimental measurement. It is widely understood that quantum \emph{decoherence}---the mechanism by which classical physics emerges in quantum systems---explains the appearance of wavefunction collapse~\cite{decoherence}. Here, we combine recent results on quantum chaos and thermalization with the Stinespring representation of measurements outlined in Sec.~\ref{sec:Measurement formalism}. 

Importantly, the physicist's Stinespring theorem \eqref{eq:Unitary Summary} implies a bookkeeping representation of measurements in which all outcomes occur \emph{deterministically} on $\Hilbert{\rm dil}$ \eqref{eq:Dilated Hilbert}---i.e., the dilated postmeasurement state $\varrho(t)$ \eqref{eq:dilated postmeas state} is a \emph{coherent} superposition of states $\propto \ket{\psi^{\,}_m} \otimes \ket{m}^{\,}_{\text{ss}}$, where the state $\ket{m}$ of the apparatus indicates that the system is in the state $\ket{\psi^{\,}_m}$. Although the \emph{quantum} probability distributions encoded by such coherent superpositions may exhibit interference, this is not observed in any particular experimental instance of a measurement.

Instead, the fact that only one outcome is observed indicates that $\varrho(t)$ \eqref{eq:dilated postmeas state} encodes an \emph{incoherent} (i.e., classical) probability distribution. For the ensemble of states encoded in the postmeasurement state $\varrho(t)$ \eqref{eq:dilated postmeas state} to appear classical, $\varrho(t)$ must realize a \emph{mixed state}, i.e.,
\begin{equation}
\label{eq:general mixed state}
    \varrho^{\vpp}_{\text{mix}} \, = \, \sum\limits_{m=0}^{\Noutcome-1} \, p^{\vpp}_m \, \BKop{\psi^{\,}_m}{\psi^{\,}_m} \, ,~~
\end{equation}
where $\rho^{\,}_m = \BKop{\psi^{\,}_m}{\psi^{\,}_m}$ is  pure and $p^{\,}_m$ is the classical probability for outcome $m$. While the coefficients of pure states  correspond to coherent superpositions in a particular basis and reflect \emph{quantum} probabilities, the coefficients of the different pure states in a mixed state \eqref{eq:general mixed state} reflect incoherent \emph{classical} probabilities. Mixed states have always had the physical interpretation of a classical ensemble of pure quantum states $\rho^{\,}_m$, only one of which is observed classically. A given experiment realizes a single pure state $\rho^{\,}_m$, which is sampled from $\varrho^{\,}_{\text{mix}}$ with probability  $p^{\,}_m$. 

It remains to identify a mechanism by which $\varrho (t)$ \eqref{eq:dilated postmeas state} realizes a mixed state $\varrho^{\,}_{\text{mix}}$ \eqref{eq:general mixed state}, i.e.,
\begin{align}
    \varrho(t) \, %= \, \umeas \, \varrho(0) \, \umeas^{\dagger} \, 
    = \, \sum\limits_{m,n=0}^{\Noutcome-1} \, \proj{m} \, \rho^{\vpp}_0 \, \proj{n} \otimes \BKop{m}{n}^{\vpp}_{\text{ss}} %\notag \\
    %
    %&
   ~~ \mapsto ~~ \sum\limits_{m=0}^{\Noutcome-1} \proj{m} \, \rho^{\vpp}_0 \, \proj{m} \otimes \Proj{\text{ss}}{(m)} \,  = \, \sum\limits_{m=0}^{\Noutcome-1} \, p^{\vpp}_m ~\rho^{\vpp}_m \otimes \Proj{\text{ss}}{(m)} \, , ~~ \label{eq:postmeasurement mixing}
\end{align}
where we used Eqs.~\ref{eq:dilated initial state} and \ref{eq:Projective measurement gate}, and $\rho^{\vpp}_m \otimes \Proj{\text{ss}}{(m)}$ realize pure states. Understanding how the process in Eq.~\ref{eq:postmeasurement mixing} comes about in actual experiments involves two key observations. 

The first observation is that the state of the apparatus is classically observable and in thermal equilibrium with its environment. The eigenstate thermalization hypothesis (ETH) \cite{ETH1, ETH2, ETH3}---which has since been proven~\cite{DoyonETH}---states that entanglement between a system and its environment is the mechanism for quantum thermalization. Because the apparatus is thermal, it is guaranteed to be highly (if not maximally) entangled with the environment. Studies of the chaotic quantum dynamics leading to thermalization have shown that the dynamics is well described by evolution under an ensemble of random matrices with the appropriate symmetries~\cite{BohigasChaos, YoshidaCBD, NahumOperator, RUCconTibor, U1FRUC, ConstrainedRUC, AaronMIPT}.

The second observation is that the postmeasurement state of the apparatus is stable to perturbations. In other words, the chaotic quantum dynamics that lead to the classical behavior of the apparatus must preserve the recorded outcome. Since chaotic quantum dynamics scramble as much  information as symmetries allow~\cite{NahumOperator, RUCconTibor, U1FRUC, ConstrainedRUC, AaronMIPT}, the states of $\Hilbert*{\text{ss}}$ associated with each outcome $m$ must belong to distinct symmetry sectors. Together with the first observation, this implies that the chaotic evolution of the apparatus is well approximated by a random unitary of the form
\begin{equation}
    \label{eq:random unitary}
    U \, = \, \sum\limits_{m} \, \Proj{\text{ss}}{(m)} \, U^{\vpd}_m \, \Proj{\text{ss}}{(m)} ~~~~\text{for}~~~U^{\vpd}_m \in \Unitary{\EigMult{m}} \, ,~~
\end{equation}
where the  $\EigMult*{m} \times \EigMult*{m}$ unitary  $U^{\vpd}_m$ acts in the $\EigMult*{m}$-dimensional subspace of $\Hilbert{\text{ss}}$ corresponding to the $m$th outcome $\mEig*{m}$ \eqref{eq:measured observable} and is drawn at random from $\Unitary{\EigMult*{m}}$ with uniform (Haar) measure. In the minimal Stinespring representation, each outcome has a unique state, so $\EigMult*{m}=1$ and $U^{\,}_m = \ee^{\ii \theta_m}$ is simply a complex phase. In the nonminimal representations relevant to experiments, $U^{\,}_m$ generically mixes states that reflect the same outcome $\mEig*{m}$. For example, in the case of photon counting (see Sec.~\ref{subsec:photon counting},  every state of $\Hilbert{\text{ss}}$ with $n$ conduction electrons reflects the same outcome $n$, so that the Haar-random unitary $U$ \eqref{eq:random unitary} respects a $\Unitary{1}$ symmetry corresponding to particle number.

We next show that the chaotic, symmetry-preserving unitary evolution of the apparatus---captured by Haar-averaged evolution under Eq.~\ref{eq:random unitary}---leads to wavefunction collapse. Following the measurement process, the full dilated state is generically given by
\begin{equation}
    \varrho(t) \, = \, \sum\limits_{m,n=0}^{\Noutcome-1} \proj{m} \, \rho^{\vpp}_0 \, \proj{n} \otimes \sum\limits_{j=1}^{\EigMult*{m}} \, \sum\limits_{k=1}^{\EigMult*{n}} c^*_{m,j} c^{\vps}_{n,k} \, \BKop{m,j}{n,k}^{\vpp}_{\text{ss}} \, , ~~
    \label{eq:generic postmeas state}
\end{equation}
where $\ket{m,j}$ and $\ket{n,k}$ label all states corresponding to outcomes $m$ and $n$, respectively, and $\sum_{j=1}^{\EigMult*{m}} \abs{c^{\,}_{m,j}}^2 = 1$. Evolving under $U$ \eqref{eq:random unitary} and ensemble averaging leads to 
\begin{align}
    \varrho' (t) \, &= \, \mathbb{E} \left[ \, U \, \varrho(t) \, U^\dagger \, \right] \notag \\
    &= \, \sum\limits_{m,n=0}^{\Noutcome-1} \proj{m} \, \rho^{\vpp}_0 \, \proj{n} \otimes \sum\limits_{j,k}  c^*_{m,j} c^{\vps}_{n,k} \, \sum\limits_{\ell,\ell'=0}^{\Noutcome-1} \, \Proj{\text{ss}}{(\ell)} \, \mathbb{E} \left[ U^{\vpd}_{\ell} \, \proj{\ell} \, \BKop{m,j}{n,k} \, \proj{\ell'} \, U^{\dagger}_{\ell'}  \right] \Proj{\text{ss}}{(\ell')} \notag \\
    &=  \, \sum\limits_{m,n=0}^{\Noutcome-1} \proj{m} \, \rho^{\vpp}_0 \, \proj{n} \otimes \sum\limits_{j,k}  c^*_{m,j} c^{\vps}_{n,k} \, \sum\limits_{\ell=0}^{\Noutcome-1} \, \Proj{\text{ss}}{(\ell)} \, \Phi^{(1)}_{\text{Haar},\ell} \left(  \proj{\ell} \, \BKop{m,j}{n,k} \, \proj{\ell} \right) \, ,~~\label{eq:postmeas decohere step 1}
\end{align}
where $\Phi^{(1)}_{\text{Haar},\ell} (A) = \trace \left( A \right) / \EigMult*{\ell}$ is the onefold Haar channel for the $\EigMult*{\ell} \times \EigMult*{\ell}$ unitary $U^{\,}_{\ell}$ \cite{YoshidaCBD}, which vanishes unless $\ell=\ell'$ and is proportional to the identity. In particular, we have that
\begin{equation}
\label{eq:onefold Haar}
   \Phi^{(1)}_{\text{Haar},\ell} \left(  \proj{\ell} \, \BKop{m,j}{n,k} \, \proj{\ell} \right) \,  = \, \frac{1}{\EigMult*{\ell}} \, \trace \left( \Proj{\ell} \, \BKop{m,j}{n,k} \right) \, = \, \frac{1}{\EigMult*{\ell}} \, \kron{m,\ell} \, \kron{n,\ell} \, \kron{j,k} \, ,~~
\end{equation}
and inserting this result back into Eq.~\ref{eq:postmeas decohere step 1} leads to
\begin{align}
    \label{eq:RMT decoherence}
    \varrho' (t) \, = \, \sum\limits_{m=0}^{\Noutcome-1} \proj{m} \, \rho^{\vpp}_0 \, \proj{m} \otimes \sum\limits_{j=1}^{\EigMult*{m}}  \frac{\abs{c^{\vps}_{m,j}}^2}{\EigMult*{m}} \, \Proj{\text{ss}}{(m)} \, = \,  \sum\limits_{m=0}^{\Noutcome-1} \proj{m} \, \rho^{\vpp}_0 \, \proj{m} \otimes \frac{1}{\EigMult*{m}} \, \Proj{\text{ss}}{(m)} \, ,~~
\end{align}
which has precisely the desired form \eqref{eq:postmeasurement mixing}. The final state \eqref{eq:RMT decoherence} is a sum over outcomes $m$ of the probability $p^{\,}_m$ \eqref{eq:Measurement Probability} of observing outcome $m$ times the collapsed density matrix $\rho^{\,}_m$ \eqref{eq:collapsed rho} times the maximally mixed state $\proj{m}/\EigMult*{m}$ on the subspace of $\Hilbert*{\text{ss}}$ corresponding to outcome $m$. Finally, in the \emph{minimal} Stinespring representation, each subspace has dimension $\EigMult*{m}=1$, so that
\begin{equation}
    \label{eq:minimal RMT decoherence}
    \varrho' (t) \, = \, \mathbb{E} \left[ \, U \, \umeas \, \varrho (0) \, \umeas^{\dagger} \, U^{\dagger} \, \right] \, = \, \sum\limits_{m=0}^{\Noutcome-1} \, \proj{m} \, \rho^{\vpp}_0 \, \proj{m} \otimes \BKop{m}{m}^{\vpp}_{\text{ss}} \, ,~~
\end{equation}
which is a classical mixture of $\Noutcome$ pure quantum states corresponding to the collapsed state $\rho^{\,}_m \propto \proj{m} \, \rho^{\,}_0 \, \proj{m}$ \eqref{eq:postmeasurement mixing} tensored with the projector onto the $m$th Stinespring state. In the context of chaos, the process \eqref{eq:minimal RMT decoherence} reflects relaxation  to the \emph{digaonal ensemble}, due to dephasing \cite{ETH1, ETH2, ETH3, DoyonETH}. It is ubiquitous to quantum systems that thermalize (or behave classically).

While the averaging over an ensemble of statistically similar evolution operators $U$ \eqref{eq:random unitary} on $\Hilbert*{\text{ss}}$ may appear unnatural, it is merely a useful---and highly successful---approximation of the underlying chaotic dynamics. Indeed, the onefold Haar channel \eqref{eq:onefold Haar} is \emph{nonunitary}, though still a valid quantum channel (i.e., CPTP map). One might think that such a nonunitary channel could not possibly capture the actual unitary evolution. In practice, however, we note that the chaotic evolution of the measurement apparatus also involves interactions with an environment, which may be similarly nonunitary in its action on the system after tracing out the environment. Moreover, the ensemble-averaged evolution reproduces the decohering effects of complicated interactions with the environment, and even successfully predicts the universal behaviors of isolated, thermodynamically large systems. We also note that other random-matrix ensembles common to the literature on, e.g., quantum complexity~\cite{YoshidaCBD} lead to the same result~\eqref{eq:RMT decoherence}. For these reasons, we expect the model of decoherence in terms of the random unitary $U$ \eqref{eq:random unitary} to provide a useful description of the chaotic dynamics associated with the classical (and thermal) nature of the apparatus  \cite{ETH1, ETH2, ETH3, DoyonETH, BohigasChaos, YoshidaCBD, NahumOperator,  RUCconTibor, U1FRUC, ConstrainedRUC, AaronMIPT}. 

Summarizing the discussion thus far, the axioms of quantum mechanics \cite{DiracQuantum, vonNeumannAxioms, HardyAxioms, FuchsAxioms, MackeyAxioms, WilceAxioms, MasanesAxioms, KapustinAxioms} imply that measurements are equivalent to unitary evolution on a \emph{dilated} Hilbert space $\Hilbert{\rm dil} = \Hilbert{\text{ph}} \otimes \Hilbert{\text{ss}}$ \eqref{eq:Dilated Hilbert}, where the unitary $\umeas$ \eqref{eq:Projective measurement gate} codifies the time evolution of the system ($\Hilbert{\text{ph}}$) and measurement apparatus ($\Hilbert{\text{ss}}$) during the measurement process, and the resulting density matrix \eqref{eq:postmeasurement mixing} encodes all possible outcomes. We note that this formulation of measurements is entirely \emph{deterministic}---i.e., at the level of the equations of motion, there is nothing stochastic, and all outcomes occur. 

The appearance of wavefunction collapse is merely a byproduct of our classical nature---and one we engineer. For a measurement outcome to be discernible to classical beings, it must be stored in a classical thermal state---meaning a state that is stable to the underlying and chaotic quantum dynamics. The same mechanism that leads to the classical behavior of the apparatus in the first place---and hence, its utility as a detector---also guarantee that the the postmeasurement state rapidly decoheres into a classical mixture of states \eqref{eq:RMT decoherence}.  In the context of measurements, we associate this with wavefunction collapse---in any given experiment, we experience only one pure state $\rho^{\,}_m \otimes \BKop{m}{m}^{\,}_{\text{ss}}$, sampled from $\varrho_\text{mixed}$ \eqref{eq:general mixed state} with probability $p^{\,}_m$ \eqref{eq:Stinespring Born}. 

The primary constraint this places on the apparatus itself is that distinct outcomes correspond to different ``charges'' under some symmetry; in most implementations of measurements, that symmetry corresponds to particle number. Importantly, the connection to symmetries naturally explains the emergence of a ``preferred basis.'' This also underlies the probabilistic appearance of radioactive decay---as classical beings, we cannot observe superpositions of different numbers of particles (symmetry sectors), as they rapidly decohere into mixed states, as in Eq.~\ref{eq:postmeasurement mixing}. 

A related point is that any ``strangeness'' of quantum are by design. Without assigning distinct outcomes to different symmetry charges, the state of the apparatus would not reliably indicate the outcome. Without using a classical apparatus, there would be no notion of collapse, nor an ``outcome'' to discern, without a second measurement involving a classical apparatus (e.g., in the thought experiment known as ``Wigner's friend''). Moreover, any nonclassical apparatus could not safely store information about the outcome, which could easily be lost to decoherence. 

This understanding of wavefunction collapse appears to resolve several ``paradoxes'' surrounding measurements. First, there is no paradox surrounding Schr\"odinger's cat: A cat is placed in a box with a radioactive isotope, whose decay triggers the release of poison; the paradox is that, until the cat is observed, its state is a superposition of alive and dead. However, a cat is classical---in this case, it acts as the detector for the spontaneous decay of the isotope (though we comment that far more ethical and efficient detectors exist). While the release of the poison creates a coherent superposition of the cat being dead (radioactive decay occurs) and alive (no decay occurs), because the cat is classical, this state rapidly decoheres into a mixed state in which the cat is alive and no decay occurred and the cat is dead and decay occurred. This collapse happens extremely rapidly as unitary dynamics increase the probability that the isotope decayed. As a result, there is no possibility of observing the cat to be in a coherent superposition of alive and dead, and thus no paradox \cite{WheelerZurek, decoherence}. Second, there is no concern surrounding Wigner's friend: The apparatus should be identified with the first classical system that is entangled with the quantum system of interest, since that apparatus is sufficient for collapse, and no other ingredients are required.

Finally, we note that the physical details surrounding decoherence of the apparatus beyond the toy model considered above do \emph{not} depend directly on the measured observable, but on microscopic and nonuniveral properties of the apparatus. In contrast to the minimal Stinespring representation \eqref{eq:Projective measurement gate}, there can be no general theory of, e.g., the precise time scale over which decoherence (and collapse) occurs, as this depends entirely on microscopic details that vary depending on how the measurement is implemented. However, it is likely that the time scale of decoherence is small compared to the time scale of the measurement overall, or else the apparatus would exhibit quantum effects, and would not reliably indicate the measurement outcome. Hence, we expect that  decoherence-induced collapse~\eqref{eq:RMT decoherence} occurs quite rapidly, and is merely part of the process by which the outcome becomes classically available. The crucial observations are that (\emph{i}) the measurement process itself is described by a \emph{deterministic} unitary \eqref{eq:Projective measurement gate} corresponding to time evolution of the system and apparatus; (\emph{ii}) the same criteria that guarantee that $\umeas$ realizes a classically readable measurement also guarantee the appearance of wavefunction collapse via decoherence \eqref{eq:postmeasurement mixing}; (\emph{iii}) it also explains the emergence of a preferred classical basis for the different ``branches'' of the universal wavefunction $\ket{\Psi} \in \Hilbert*{\text{dil}}$; and (\emph{iv}) that collapse is unrelated to consciousness or observation.

\subsection{Absence of ``spooky action at a distance''}
\label{subsec:not spooky}

Here we discuss how the Stinespring representation \eqref{eq:Projective measurement gate} naturally resolves the Einstein-Podolsky-Rosen (EPR) ``paradox'' \cite{epr}, in a manner consistent with results long known to the literature. Suppose that two qubits are prepared in the Bell state
\begin{equation}
    \label{eq:Bell main}
    \ket{\text{Bell}}^{\vpp}_{a,b} \, = \, \frac{1}{\sqrt{2}} \, \left( \ket{0}^{\vpp}_a \otimes \ket{0}^{\vpp}_b + \ket{1}^{\vpp}_a \otimes \ket{1}^{\vpp}_b \right) \, ,~~
\end{equation}
and we then send qubits $a$ and $b$ to Alice and Bob, respectively---who are separated by some large distance $r$---while preserving the maximally entangled state \eqref{eq:Bell main} of the qubits \cite{epr, Bell}. 

Suppose that Alice and Bob each measure $\PZ{}$ on their  qubits. From the measurement axiom~\cite{DiracQuantum, vonNeumannAxioms, HardyAxioms, FuchsAxioms, MackeyAxioms, WilceAxioms, MasanesAxioms, KapustinAxioms} 
, it would appear that whoever measures first collapses the Bell state \eqref{eq:Bell main} onto \emph{either} $\ket{00}$ or $\ket{11}$, so that whoever measures second is guaranteed to observe the same outcome. The EPR paradox centers on the putative violation of relativistic locality---e.g., how does Bob's qubit $b$ ``know'' to collapse to the same outcome observed by Alice upon her measurement of qubit $a$~\cite{epr}? The idea was that information about Alice's outcome must propagate the distance $r$ to Bob's qubit \emph{instantaneously}, which would violate causality. Even if one accepts that all outcomes occur in the universal (dilated) state $\varrho$ \eqref{eq:minimal RMT decoherence}, since the \emph{observed} outcomes correspond to distinct classical ``branches'' of $\varrho$ (i.e., Stinespring states $\ket{m}$), one might still worry that Alice's measurement determines this classical branch and  instantaneously communicates that information to Bob's qubit. 

It has been well established for decades now that there is no violation of relativistic locality (i.e., causality) in such Bell measurements, and hence, no ``paradox'' to resolve. This is particularly clear using the Stinespring representation, as we now show. The dilated description reveals that there is no need for instantaneous communication between entangled qubits following a measurement of one---i.e., there is no ``spooky action at a distance'' in the context of measuring entangled states. 

We first introduce two Stinespring qubits $A$ and $B$, corresponding to the measurement apparati used by Alice and Bob, respectively. Alice's $\PZ{a}$ measurement, e.g, is captured by the dilated unitary
\begin{align}
\label{eq:Alice Z unitary}
    \umeas^{\vpd}_A \, = \, \BKop{0}{0}^{\vpp}_a \otimes \ident^{\vpp}_{A} + \BKop{1}{1}^{\vpp}_a \otimes \Shift{A} \, ,~~
\end{align}
as in Eq.~\ref{eq:Qubit Z Meas}, and the analogous $\umeas^{\vpd}_B$ for Bob is the same up to replacing $(a,A) \to (b,B)$. The two apparati are initialized in the $\PZ{}$-basis state $\ket{0}$, so the dilated initial state is given by
\begin{equation}
    \label{eq:Bell dilated initial}
    \ket{\Psi^{\vpp}_i} \, = \, \frac{1}{\sqrt{2}} \left( \ket{00} + \ket{11} \right)^{\vpp}_{a,b} \otimes \ket{00}^{\vpp}_{A,B} \, .~~
\end{equation}
Assuming Alice measures first, we apply $\umeas^{\vpd}_A$ \eqref{eq:Alice Z unitary} to $\ket{\Psi^{\vpp}_i}$ \eqref{eq:Bell dilated initial}, resulting in 
\begin{equation}
    \label{eq:Bell dilated post Alice}
    \ket{\Psi'} \, = \, \frac{1}{\sqrt{2}} \left( \ket{000} + \ket{111} \right)^{\vpp}_{a,b,A} \otimes \ket{0}^{\vpp}_{B} \, ,~~
\end{equation}
and applying $\umeas^{\vpd}_B$ \eqref{eq:Alice Z unitary} for Bob's subsequent measurement leads to the final state
\begin{equation}
    \label{eq:Bell dilated post meas}
    \ket{\Psi^{\vpp}_f} \, = \, \frac{1}{\sqrt{2}} \left( \ket{0000} + \ket{1111} \right)^{\vpp}_{a,b,A,B}  \, ,~~
\end{equation}
from which it is clear that Alice and Bob's $\PZ{}$ measurements are \emph{guaranteed} to realize the same outcome, simply due to the structure of the Bell state. Because the two measurement unitaries act on disjoint Hilbert spaces, they commute, and their order is unimportant. The correlation between the measurement outcomes is a consequence of the correlations between the physical qubits; each measurement merely entangles the corresponding apparatus into the Bell state \eqref{eq:Bell main}, forming a Greenberger-Horne-Zeilinger (GHZ) state \cite{GHZ89} of the two physical qubits and two detector qubits. 

Essentially, the measurements merely reveal information already encoded in the state. Regarding the postmeasurement state $\ket{\Psi^{\,}_f}$ \eqref{eq:Bell dilated post meas}, we see that the \emph{physical} part of the state is  \emph{unaltered} compared to the initial state $\ket{\Psi^{\,}_i}$ \eqref{eq:Bell dilated initial}. In fact, projective measurements never alter the physical part of the state when written the basis of the measured observable $\mobserv$ \eqref{eq:measured observable}; they merely entangle the state of the apparatus into the physical state in a correlated manner. 

For example, if Alice measures $\PZ{a}$ and Bob measures $\PX{b}$, the initial state is
\begin{align}
\label{eq:Bell Z-X initial}
    \ket{\Psi^{\vpp}_i} \, = \, \frac{1}{2} \left( \ket{0,+} + \ket{0,-} + \ket{1,+} - \ket{1,-} \right)^{\vpp}_{a,b} \otimes \ket{0,0}^{\vpp}_{A,B}\, ,~~
\end{align}
in the measurement basis $\PZ{a} \otimes \PX{b}$, where $\PX{} \ket{\pm} = \pm \ket{\pm}$. The postmeasurement state is
\begin{align}
    \label{eq:Bell Z-X final}
    \ket{\Psi^{\vpp}_f} \, = \,\frac{1}{2} \left( \ket{0,+,0,+} + \ket{0,-,0,-} + \ket{1,+,1,+} - \ket{1,-,1,-} \right)^{\vpp}_{a,b,A,B} \, ,~~
\end{align}
where the physical part is unaltered compared to $\ket{\Psi^{\vpp}_i}$ \eqref{eq:Bell Z-X initial}. Again, the measurements commute. 

Comparing Eqs.~\ref{eq:Bell Z-X initial} and \ref{eq:Bell Z-X final} also shows that measurements do \emph{not} create ``branches'' of the universal wavefunction $\ket{\Psi} \in \Hilbert*{\text{dil}}$, but merely reveal superpositions \emph{already present} in $\ket{\Psi}$. %While destructive measurements alter the physical state, they only \emph{remove} branches (superpositions). 
By writing $\ket{\Psi}$ in the basis of the observables to be measured, we see that the measurement simply entangles the state of the apparatus into $\ket{\Psi}$ without altering the latter, as in Eqs.~\ref{eq:Bell dilated post meas} and \ref{eq:Bell Z-X final}. Classical information about the observed outcome emanates from the locations of \emph{each} measurement, traveling no faster than $c$. The correlations between the outcomes of measurements of entangled states $\ket{\Psi}$ are guaranteed by the structure of $\ket{\Psi}$ itself,  without the need for action at a distance.

Not only is there no \emph{need} for action at a distance---``spooky'' or otherwise---but we now explain why there cannot be any such action. Returning to the well separated Bell state \eqref{eq:Bell main}, the expectation value of any Pauli measurement by either Alice or Bob is given by
\begin{equation}
    \label{eq:Bell meas prob}
    \expval{\Pauli{\nu}{a/b}} \, = \, \trace \left( \Pauli{\nu}{a/b} \, \rho^{\vpp}_{\text{Bell}} \right) \, = \, \frac{1}{4} \, \trace \left[ \Pauli{\nu}{a/b} \left( \ident + \PX{a}\PX{b} + \PZ{a}\PZ{b} -\PY{a}\PY{b} \right) \right] \, = \, 0 \, ,~~
\end{equation}
so that, if they repeat the experiment many times without communicating, Alice and Bob each conclude that their qubit is in the maximally mixed state $\rho^{\,}_{a/b} = \ident /2$, so that all measurement outcomes are equally likely. They can only deduce that their qubits are maximally entangled with \emph{something else}---e.g., a thermal bath---but not that they are part of a Bell state.

In fact, without communicating, Alice and Bob cannot learn anything about the other party. If they both know that they share a Bell state, they cannot determine which of the four Bell states they share from local measurements. Even knowing which Bell state they share, they cannot determine \emph{whether} the other party has measured yet, even knowing what they plan to measure. Alternatively, even knowing that the other party has measured, they cannot determine which operator was measured. The \emph{only} way that Alice and Bob can see correlations between their measurements is by sending classical signals, which obey relativistic causality. The \emph{only} way Alice and Bob can use their entangled state to transmit quantum information is again by sending a classical signal containing details of the measurements and their outcomes~\cite{SpeedLimit, AaronTeleport}. In other words, there is no ``paradox'' because measuring one party in an entangled state does not transmit information or influence of any kind to the other parties---there is no action at a distance, as is now well known.

\subsection{Quantum mechanics is local}
\label{subsec:locality}

We now explain that not only measurements of Bell states \cite{epr, Bell}, but \emph{all} of quantum mechanics, is compatible with relativistic locality~\cite{SpeedLimit}. In the process, we show how the Stinespring Stinespring representation of Sec.~\ref{sec:Measurement formalism} implies that large swaths of quantum dynamics---including those with measurements and nonlocal feedback---obey a much stronger, \emph{nonrelativistic} notion of locality \cite{SpeedLimit} in the sense of the Lieb-Robinson theorem~\cite{Lieb1972}. Such notions of locality have profound implications for quantum dynamics, phases of matter, and more~\cite{SpeedLimit, Lieb1972, hastings_rev}. 

We emphasize that we use the term ``locality'' exclusively to refer to the idea that objects are only influenced by events within their causal light cone. We do \emph{not} refer to the notion of ``quantum nonlocality'' associated with Bell states \cite{Bell}, which instead refers to the fact any \emph{classical} hidden-variables model describing quantum phenomena \emph{must} violate relativistic locality~\cite{NonlocalityAxiom, BellNonloc}. Instead, we use the term ``locality'' to refer to the idea that information and influence of any kind can always be traced dynamically, and never exceed the speed of light $c$. Moreover, in most nonrelativistic scenarios, there exists an \emph{emergent} and stronger notion of locality. 

We first review locality in the context of unitary time evolution of isolated quantum systems. Consider a lattice spin system described by a local Hamiltonian $\Ham$ (e.g., interactions between neighboring spins) with energy scale $J$ and average spacing $a$ between spins. The Lieb-Robinson bound establishes that the distance $x$ over which quantum information can be transferred (or correlations and/or entanglement generated, etc.)  in time $t$ obeys the inequality
\begin{equation}
    \label{eq:LR bound}
    x \,  \leq \, 2 v t \, ,~~
\end{equation}
where $v \sim a J/ \hbar \ll c$ is an \emph{emergent} speed limit on quantum information~\cite{Lieb1972}. The bound comes from the fact that correlations, entanglement, and quantum information can spread out from some point no faster than $v$, leading to a region of size $2 v t$ in time $t$. Formally, the bound \eqref{eq:LR bound} derives from the fact that generic commutators satisfy $| [ A(x,t), B(0,0)] | \lesssim \exp(2 v t - x)$ as $x \to \infty$.  

The Lieb-Robinson bound \eqref{eq:LR bound} also applies to systems with interactions in finite local regions, exponentially decaying interactions $V(r) \sim \ee^{-r/\xi}$~\cite{hastings_rev}, and even power-law interactions $V(r) \sim r^{-\alpha}$ for $\alpha \geq 2d+1$ in $d$ spatial dimensions~\cite{LRfossfeig, chen2019finite, Tran:2020xpc, CLYreview}. A modified Lieb-Robinson bound with a nonlinear ``light cone'' recovers for $d < \alpha < 2d+1$ \cite{CLYreview}. The foregoing Hamiltonians all have an emergent notion of \emph{nonrelativistic} locality that is even stronger (i.e., more constraining) than standard, relativistic locality. Importantly, \emph{all} Hamiltonians $\Ham$ obey \emph{relativistic} locality: When $\alpha \leq d$, $V(r)$ is mediated by a field whose excitations propagate no faster than $c$ (e.g., the Coulomb potential). 

Still, generic quantum dynamics may also involve nonunitary quantum operations, such as measurements and outcome-dependent feedback. When the outcome-dependent operations are performed in the vicinity of the measurement, the entire process is captured by a Lindblad master equation involving only \emph{local} jump operators. Any dynamics captured by a local Lindbladian obeys the \emph{same} Lieb-Robinson bound \eqref{eq:LR bound} as local unitary time evolution alone \cite{Poulin}. 

However, when outcome-dependent feedback is applied \emph{nonlocally}---i.e., far from the location of measurement---the Lieb-Robinson bound \eqref{eq:LR bound} does \emph{not} apply \cite{SpeedLimit}. We allow for generic local quantum operations that are individually compatible with the usual Lieb-Robinson bound~\cite{Lieb1972, Poulin}, and further allow for the nonlocal communication of measurement outcomes; in the nonrelativistic limit where $c \to \infty$, this communication is effectively instantaneous. Using the Stinespring representation of Sec.~\ref{sec:Measurement formalism}, Ref.~\citenum{SpeedLimit} established that \emph{any} quantum task that achieves a useful task over some distance $x$ using measurements in $M$ regions obeys the \emph{generalized} bound,
\begin{equation}
    \label{eq:gen bound}
    x \, \leq \, 2 (M+1) vt \, ,~~
\end{equation}
where $t$ is the duration of unitary time evolution and $v$ is the Lieb-Robinson velocity associated with that evolution \eqref{eq:LR bound}. While the bound \eqref{eq:gen bound} was only proven explicitly for qubits on arbitrary graphs and interacting with their nearest neighbor only, it is highly likely that the generalized bound is compatible with any Lieb-Robinson bound for the purely unitary part \cite{SpeedLimit}. Proving this to be the case is an open direction for future work. We stress that the measurements need not be projective, and can be generalized to any quantum operations involving ancillas in $\Hilbert*{\text{ss}}$; however, any such operation \emph{not} accompanied by nonlocal outcome-dependent feedback provides no enhancement \cite{SpeedLimit}. 

We further comment that the generalized bound \eqref{eq:gen bound} does \emph{not} derive from commutator norms, as measuring $\PZ{i}$ and then communicating the outcome $n \in \{0,1\}$ and applying $\PX{j}$ if $n=1$ leads to $| [\PX{i} (0), \PZ{j} (t) | = 2$ even as $t \to 0$ and $x = d(i,j) \to \infty$. However, this sequence of operations does not transmit quantum information nor does it generate or meaningfully alter correlations or entanglement between $i$ and $j$. The commutator is nonzero solely because classical information was transmitted. Instead, the bound \eqref{eq:gen bound} derives from comparing the reduced density matrix $\rho^{\,}_{ij}$ between sites $i$ and $j$ separated by $x$ and showing that, if the bound is violated, it is arbitrarily close in trace distance to a state $\rho_{ij}'$ that contains no entanglement or correlations between $i$ and $j$. Because all useful quantum tasks inherently create entanglement and/or correlations over some distance $x$, the bound \eqref{eq:gen bound} constrains all useful quantum tasks \cite{SpeedLimit}.

Importantly, the generalized bound \eqref{eq:gen bound} implies that a finite number of measurements only leads to a \emph{finite} enhancement to the speed $v$ \eqref{eq:LR bound} of quantum information. This holds even in the nonrelativistic limit where $c \to \infty$ and communication of outcomes is \emph{instantaneous}. Intuitively, the bound reflects the fact that measurements can only be used to ``link up'' local regions---of a resource state, teleportation protocol, or similar. The most optimal protocols involve unitary evolution for time $t$ that generates local patches of the resource state in $M+1$ regions. Each patch grows to maximum size $\ell (t) = 2 v t$ \eqref{eq:LR bound}, and the $M$ measurements ``stitch'' these patches together to achieve a distance of $x  \leq (M+1) \, \ell(t) = 2 (M+1) vt$ \eqref{eq:gen bound}. In some cases, a protocol can achieve a useful quantum task starting from a product state (e.g., $\ket{00 \cdots 00}$) in as few as $t = 2$ layers of entangling unitary gates (i.e., depth two). In the context of teleportation, the outcomes of measurements adjacent to the initial site $i$ must be communicated to determine a feedback operation on the final site $f$, with $d(i,f)=x$; in the context of state preparation, measurements at the edges of the system are tied to feedback operations \emph{throughout} the system, with maximum linear size $x$ \cite{SpeedLimit}. 

Essentially, there are two mechanisms for transferring quantum information (or generating correlations and/or entanglement). The first is through unitary time evolution, which usually obeys a nonrelativistic Lieb-Robinson bound \eqref{eq:LR bound} and always obeys relativistic locality \cite{Lieb1972}. The second is through the communication of measurement outcomes and accompanying feedback, which obeys the generalized bound \eqref{eq:gen bound} whenever the unitary dynamics obey the Lieb-Robinson bound \eqref{eq:LR bound} and the number of measurement regions $M$ is finite \cite{SpeedLimit}. But when $M \to \infty$ or the unitary evolution does not obey Eq.~\ref{eq:LR bound}, there is no nonrelativistic notion of locality.

However, the \emph{relativistic} notion of locality always holds. We stress that the generalized bound \eqref{eq:gen bound} holds in the \emph{non}relativistic limit where $c \to \infty$; to demonstrate relativistic locality, we take $c$ to be finite but large. First, consider an optimal protocol with $M = \Order{x}$ measurements and unitary evolution with Lieb-Robinson velocity $v$ \eqref{eq:LR bound} for time $t_u$, which we minimize. Because the protocol is optimal, all $M \to \infty$ measurements occur after all unitary evolution \cite{SpeedLimit}, and we assume that they occur simultaneously in the rest frame of the system. Even in this limit, at least one measurement's outcome must be communicated a distance $x - 2 v t_u$ using a signal with speed $c$, where the unitary dynamics shorten this distance by $v t_u$ from both ``edges'' of the system. This protocol achieves a useful task over distance $x$ in time $t \geq x/c + (1 - 2 v / c) t_u$, i.e., 
\begin{equation}
    x \, \leq \, c (t - t_u) + 2 v t_u \, \leq \, 2  c  t,~~
\end{equation}
meaning that information travels at speed $2 v \ll c$ for time $t_u$ and speed $c$ for the remaining time; since $v \ll c$, this obeys relativistic locality. Even in the limit $v \to c$, we still have $x < 2 c t$, so relativistic locality is always satisfied. The reason is that the measurements themselves do not send information---that is achieved by the classical signals, which travel no faster than $c$.

Accordingly, \emph{all} quantum operations obey relativistic locality. A large fraction obey a stronger, nonrelativistic notion of locality, captured by the Lieb-Robinson bound \eqref{eq:LR bound} for unitary evolution and local quantum operations, and by the generalized bound \eqref{eq:gen bound} for local quantum operations combined with classical signaling and nonlocal feedback \cite{SpeedLimit, Poulin, Lieb1972}. Without the classical signals and the corresponding feedback, nonunitary operations provide no advantage over unitary dynamics \cite{SpeedLimit, Poulin, AaronMIPT}. The generation of correlations and/or entanglement and the transmission of quantum information involve combinations of unitary dynamics and classical signals, which individually and jointly obey relativistic locality. Hence, all quantum dynamics are local in the relativistic sense. 

This not only establishes that quantum mechanics is ``complete'' in the sense of EPR \cite{epr}, but also has extremely useful implications for quantum applications. For example, locality can be used to constrain quantum protocols, reveal useful tradeoffs (e.g., between unitary time evolution and measurements), identify optimal protocols tailored to particular hardware, and constrain the resource states compatible with a particular quantum task  \cite{AaronTeleport, SpeedLimit}, among other applications.

\section{Conclusion}

The description of quantum measurements as an entangling interaction between a physical system of interest and a measurement apparatus---as well as the unitary description \eqref{eq:Unitary Summary} of generic quantum operations in finite dimension---have long been known to the literature~\cite{VonNeumann, KrausBook, WheelerZurek, QC_book, Preskill_QI, AaronMIPT, SpeedLimit, decoherence, MargenauMeas, PeresMeas86, AaronTeleport, AaronJamesFuture, WolfNotes, Stinespring, ChoisThm}. Nonetheless, the analytical tools most commonly used in the literature to  describe measurements manifestly ignore details of the apparatus \cite{KrausMeas1969, KrausMeas1971, KrausMeas1981, KrausBook, lindblad1973entropy}, creating a disconnect between theory and experiment, leading to avoidable confusion about the nature of measurements, and complicating the description of adaptive protocols with midcircuit measurements~\cite{AaronMIPT, SpeedLimit, AaronTeleport}. 

In this work, we have given an overview of projective and destructive quantum measurements, their theoretical description, and their experimental implementation. In particular, we have highlighted the ``Stinespring'' representation of measurements in terms of a unitary $\umeas$ \eqref{eq:Stinespring Unitary General} acting on a dilated Hilbert space $\Hilbert{\text{dil}} = \hilbert \otimes \Hilbert{\text{ss}}$ \eqref{eq:Dilated Hilbert}, and how it follows logically from all axiomatic formulations of quantum mechanics~\cite{DiracQuantum, vonNeumannAxioms, HardyAxioms, FuchsAxioms, MackeyAxioms, WilceAxioms, MasanesAxioms, KapustinAxioms}. We have also shown how the unitary $\umeas$ relates to the standard Kraus representation \eqref{eq:Kraus Representation}~\cite{KrausMeas1969,  KrausMeas1971, KrausMeas1981, KrausBook},  von Neumann pointer Hamiltonian $H$ \eqref{eq:Pointer Hamiltonian}~\cite{VonNeumann, Preskill_QI}, and most importantly, to \emph{experiment}. Noting that a large number of experimental implementations culminate in one of a handful of measurement types, we consider the most prominent among these in Secs.~\ref{sec:photons} and \ref{sec:qubits}. In all cases, we find that the dilated unitary $\umeas$ \eqref{eq:Projective measurement gate} is not just a bookkeeping tool but \emph{physical}, capturing the time evolution of the system and measurement apparatus during the measurement process, in agreement with intuition from the literature. 

Generally speaking, projective and destructive measurements involve forming a maximally entangled state between the physical system and a classical measurement apparatus. Consider the measurement of an observable $\mobserv$ \eqref{eq:measured observable} with $\Noutcome$ unique eigenvalues (i.e., measurement outcomes) $\mEig*{m}$ with corresponding eigenprojectors $\proj{m}$ \eqref{eq:Spec Projecc}. In the \emph{minimal} Stinespring representation, the projective measurement outputs a state that is a sum over $m$ of the collapsed state $\proj{m} \ket{\psi} \in \hilbert$ of the system times the state $\ket{m}^{\,}_{\text{ss}} \in \Hilbert{\text{ss}}$ of the apparatus. Additionally, the measurement unitary $\umeas$ \eqref{eq:Projective measurement gate} is unique up to the choice of initial state $\ket{i}^{\,}_{\text{ss}} \in \Hilbert{\text{ss}}$ of the  apparatus. In the case of\emph{destructive} measurements---which are unitarily related to projective measurements---the final state of the physical system is not an eigenstate of the measured observable. In our consideration of physical implementations of measurements in Secs.~\ref{sec:photons} and \ref{sec:qubits}, we saw that the physical unitary corresponding to the time evolution of the system and apparatus generally realizes a \emph{nonminimal} Stinespring representation, in which $\DimOf*{\Hilbert{\text{ss}}} > \Noutcome$. However, upon ``binning'' distinct states of the apparatus corresponding to the same observed outcome---and ignoring states that do not encode information about the outcome---one recovers the minimal Stinespring representation \eqref{eq:Projective measurement gate}. 

In fact, one of the most important features of the Stinespring formulation of measurements on $\Hilbert{\rm dil}$ \eqref{eq:Dilated Hilbert} is its direct connection to experiment. In all of the examples of projective and destructive measurements considered in Secs.~\ref{sec:photons} and \ref{sec:qubits}, we found that the Stinespring measurement unitary $\umeas$ \eqref{eq:Projective measurement gate} corresponds \emph{exactly} to a simplified description of the actual, physical time evolution of the system and measurement apparatus during the measurement process. This direct correspondence suggests that $\umeas$ \eqref{eq:Projective measurement gate} is not only a valid bookkeeping tool, but the physical time evolution operator. This allows us to evolve operators in the Heisenberg picture through arbitrary combinations of measurements and outcome-dependent feedback \cite{SpeedLimit, AaronMIPT, AaronTeleport}. Thus, the minimal Stinespring representation not only encodes the minimal ingredients required for the experimental implementation of a given measurement, but is a powerful analytical tool for describing adaptive protocols with midcircuit measurements. 

Additionally, we have explained in Sec.~\ref{sec:Using Stinespring} how to extract standard results from the Stinespring formalism, and its application to quantum protocols. We explained in Sec.~\ref{subsec:extract} how to recover the familiar Born rule, expectation values, and other statistics from the \emph{post}measurement state $\varrho(t)$ \eqref{eq:dilated postmeas state}, using Gleason's theorem \cite{Gleason}. We showcased the utility of the Stinespring representation $\umeas$ \eqref{eq:Projective measurement gate} for adaptive protocols with measurements and outcome-dependent feedback in Sec.~\ref{subsec:adaptive}. These protocols can achieve tasks in much less time than their purely unitary counterparts \cite{SpeedLimit}. 

In Sec.~\ref{subsec:collapse} we derived the appearance of \emph{wavefunction collapse} via decoherence  \cite{decoherence}. Because the state of the measurement apparatus is readable to us, it must be stable to the quantum decoherence---i.e., the chaotic, highly entangling interactions with the environment---responsible for the classical behavior of the apparatus. Moreover, the states corresponding to distinct outcomes must be \emph{stable}; as is known from the quantum chaos literature \cite{RUCconTibor, U1FRUC, AaronMIPT}, this requires that these states have different ``charges'' under some symmetry (e.g., particle number), which also provides a ``classically preferred basis'' \cite{decoherence}. This decoherence effectively leads to a mixed state \eqref{eq:postmeasurement mixing}---i.e.,  a classical ensemble of ``collapsed'' pure quantum states corresponding to different outcomes. Despite being deterministic, we \emph{experience} this process as probabilistic because we only experience one of the pure states encoded in a mixed state, according to their probabilities. 

We have also explained using the Stinespring representation how \emph{all} quantum operations---including the combination of measurements and nonlocal feedback---obey relativistic locality. In Sec.~\ref{subsec:not spooky}, we explain the absence of ``spooky action at a distance'' in measurements of a well separated Bell state \cite{epr, Bell, SpeedLimit}. If Alice and Bob share an entangled Bell state, there is no measurement either can perform on their own qubit to learn anything about the other's. Alice's measurement of her qubit does not influence or in any way transmit information to Bob's qubit. Hence, there is simply no ``action at a distance'' of any kind, and ``quantum nonlocality'' \cite{NonlocalityAxiom, BellNonloc} is a misnomer. We also show how the act of measurement does not create new ``branches'' in the universal wavefunction $\ket{\Psi} \in \Hilbert*{\text{dil}}$. In fact, when the physical state $\ket{\psi} \in \hilbert$ is written in the basis of the observable to be measured, the projective measurement process does not even alter the structure of $\ket{\psi}$, but merely entangles the state of the apparatus into the existing superposition. Moreover, in Sec.~\ref{subsec:locality}, we summarized how the Stinespring representation was used in Ref.~\citenum{SpeedLimit} to establish a notion of locality in quantum dynamics with measurement, generalizing the Lieb-Robinson bound \cite{Lieb1972} to \emph{generic} quantum channels. Most importantly, we also showed how this bound implies that \emph{all quantum dynamics respect relativistic locality} (i.e., causality). 

In summary, we have shown how the unitary Stinespring representation of measurements on a dilated Hilbert space that includes the detector is both conceptually transparent and practically useful. While this representation is likely intuitive to many~\cite{QC_book}, it is not commonly used in the literature, despite its utility. In addition to providing a useful connection between theory and experiment, a powerful tool for describing protocols with midcircuit measurements, and insights into more foundational questions about quantum measurements, the Stinespring formalism also paves the way to entirely new types of results in the context of quantum computation, dynamics, and information processing. In particular, the formalism allows for the Heisenberg evolution of \emph{operators} in the presence of arbitrary quantum operations (e.g., measurements and nonlocal feedback), which allows for the systematic study of quantum protocols, resolving questions such as what resource states can be measured to achieve a given quantum task \cite{AaronTeleport}, and revealing new constraints and resource tradeoffs \cite{SpeedLimit}. We believe this formalism will be an invaluable tool in future studies of quantum protocols and near-term quantum technologies. In addition to forthcoming work proving the physicist's Stinespring theorem for infinite-dimensional systems~\cite{AaronJamesFuture}, we also plan to extend the Stinespring representation to generalized and weak measurements, and generic quantum operations. 

\section*{Acknowledgments}

We thank Josh Combes, Shawn Geller, Eric Song, Daniel Spiegel, James Woodcock, and Aaron Young for useful discussions and feedback on this manuscript.  DB is supported in part by NIST, the DOE Quantum Systems Accelerator (QSA) grant, ARO grant W911NF-19-1-0210, the Simons collaboration on Ultra-Quantum Matter (UQM), and NSF grants JILA-PFC PHY-2317149, QLCI-OMA-2016244, and PHY-2309135 via the Kavli Institute for Theoretical Phyiscs (KITP); DB also acknowledges the hospitality of the KITP while parts of this work were completed. AJF is supported in part by DOE grant DE-SC0024324.

\appendix

\renewcommand{\thesubsection}{\thesection.\arabic{subsection}}
\renewcommand{\thesubsubsection}{\thesubsection.\arabic{subsubsection}}
\renewcommand{\theequation}{\thesection.\arabic{equation}}

\section{Details of the Stern-Gerlach experiment}
\label{app:Stern Gerlach}

Here we elaborate on technical details relevant to the Stern-Gerlach experiment discussed in Sec.~\ref{subsec:SternGerlach}. In App.~\ref{app:SG general} we show that the evolution of the $y$ coordinate can be ignored; in App.~\ref{app:SG easy}, we derive the final wavefunction $\Psi_s (z,t)$ for the magnetic particle.

\subsection{Time evolution of operators in the general case}
\label{app:SG general}

Observables $\observ$ evolve in the Heisenberg picture under a Hamiltonian $H$ according to
\begin{equation}
    \label{eq:Heisenberg BCH}
    \observ (t) \, = \, \ee^{\ii H t/\hbar} \, \observ \, \ee^{-\ii H t/\hbar} \, = \, \sum\limits_{n=0}^{\infty} \frac{1}{n!} \, \left( \frac{t}{\ii \hbar} \right)^n \, \comm{\observ}{ H}^{\,}_n \, ,~~
\end{equation}
where the nested commutator satisfies $\comm{A}{B}^{\,}_{n+1} = \comm{\comm{A}{B}^{\,}_n}{B}$ with $\comm{A}{B}^{\,}_0 = A$. For the Stern-Gerlach Hamiltonian $\Ham_{\rm SG}$ \eqref{eq:Stern Gerlach full Hamiltonian}, the $j$th component of the position operator $\bvec{x}$ evolves via
\begin{align}
    x^{\vpp}_j (t) \, = \, \sum\limits_{n=0}^{\infty} \frac{1}{n!}  \, \comm{\observ}{ \left( -\frac{\hbar^2}{2M} \nabla^2 -  \mu^{\,}_B \,b\, y \, \Pauli{y}{} + \mu^{\,}_B \, ( B_0 + b \, z ) \, \Pauli{z}{} \right) \frac{t}{\ii \hbar} }^{\,}_n \, = \, \sum\limits_{n=0}^{\infty} \, \frac{1}{n!} \, x^{(n)}_j (t) \, , ~~
\end{align}
where we retain the time dependence of the $n$th term for convenience. The first several terms are
\begin{subequations}
\label{eq:SG nested commutator easy}
\begin{align}
    x^{(0)}_j (t) \, &= \, x^{\vpp}_j \label{eq:SG nested commutator 0} \\
    x^{(1)}_j (t) \, &= \, \frac{t}{\ii \hbar} \, \comm{x^{\vpp}_j}{H^{\,}_{\rm SG}} \, = \, \frac{\ii t \hbar}{2 M} \, \comm{x^{\vpp}_j}{\pdp{j}{2}} \, =  \, \frac{t \hbar}{\ii M} \pd{j} \, = \frac{t}{M} \, p^{\vpp}_j \label{eq:SG nested commutator 1} \\
    x^{(2)}_j (t) \, &= \,  \frac{t}{\ii \hbar} \, \comm{\frac{t \hbar}{\ii M} \pd{j}}{H^{\,}_{\rm SG}} \, 
    %\, \frac{\mu^{\,}_B \, t^2}{2 \, M} \comm{\pd{j}}{b \, y \, \Pauli{y}{} - (B_0 + b \, z ) \Pauli{z}{} } \notag \\
    = \, \frac{\mu^{\,}_B \, b \, t^2}{M} \, \left( \kron{j,y} \, \Pauli{y}{}  - \kron{j,z} \Pauli{z}{} \right) \label{eq:SG nested commutator 2} \, ,~~
\end{align}
\end{subequations}
and for convenience, we now treat the evolution of $y$ and $z$ separately. For $y(t)$, we have
\begin{subequations}
\label{eq:SG nested commutator Y}
\begin{align}
    y^{(3)} (t) \, &= \,  \frac{t}{\ii \hbar} \, \comm{\frac{\mu^{\,}_B \, b \, t^2}{M} \, \Pauli{y}{} }{H^{\,}_{\rm SG}} \, = \, \frac{\mu^{\,}_B \, b \, t^3}{\ii \hbar M} \, \comm{\Pauli{y}{}}{\mu^{\,}_B (B_0 + b \, z) \Pauli{z}{}}  \, = \, \frac{2 \, \mu^{2}_B \, b \, t^3}{\hbar M} \, \left( B_0 + b \, z \right) \, \Pauli{x}{} \label{eq:SG nested commutator Y 3} \\
    y^{(4)} (t) \, &= \, \frac{t}{\ii \hbar} \, \frac{2 \, \mu^{2}_B \, b \, t^3}{\hbar M} \, \comm{\left( B_0 + b \, z \right) \, \Pauli{x}{}}{H^{\,}_{\rm SG}} \notag \\
    &= \, \frac{2 \, \mu^{2}_B \, b \, t^4}{\ii \hbar^2 M} \left(  -\frac{b \, \hbar^2}{2M} \,   \comm{z}{\pdp{z}{2}} \, \Pauli{x}{}  - \mu^{\,}_B b y  \left( B_0 + b \, z \right) \comm{ \Pauli{x}{} }{\Pauli{y}{}}  + \mu^{\,}_B (B_0 + b \, z )^2 \comm{\Pauli{x}{}}{\Pauli{z}{}}  \right)  \notag \\
    &= \, \frac{2 \, \mu^{2}_B \, b \, t^4}{\ii \hbar^2 M} \left( \frac{b \,\hbar^2}{M} \, \Pauli{x}{} \, \pd{z} - 2 \, \ii \,  \mu^{\,}_B b y  \left( B_0 + b \, z \right) \, \Pauli{z}{} - 2 \, \ii \, \mu^{\,}_B (B_0 + b \, z )^2 \, \Pauli{y}{} \right)  \notag \\
    &= \, \frac{2 \, \mu_B^2 \, b^2 \, t^4}{\hbar \, M^2} \, p^{\vpp}_z \, \Pauli{x}{} - \frac{4 \, \mu_B^3 \, b \, t^4}{\hbar^2 \, M} \, \left( B_0 + b \, z \right)^2 \, \Pauli{y}{} - \frac{4 \, \mu_B^3 \, b^2 \, t^4}{ \hbar^2 \, M} \left( B_0 \, y + b \, y \, z \right) \, \Pauli{z}{} \, , ~~
     \label{eq:SG nested commutator Y 4} 
\end{align}
\end{subequations}
where the terms above are organized by the Pauli matrices. The analogous expressions for $z(t)$ are
\begin{subequations}
\label{eq:SG nested commutator Z}
\begin{align}
    z^{(3)} (t) \, &= \,  \frac{t}{\ii \hbar} \, \comm{-\frac{\mu^{\,}_B \, b \, t^2}{M} \, \Pauli{z}{} }{H^{\,}_{\rm SG}} \, = \, \frac{\ii \, \mu^{\,}_B \, b \, t^3}{\hbar M} \, \comm{\Pauli{z}{}}{-\mu^{\,}_B \, b \, y\,  \Pauli{y}{}} \, = - \frac{2 \, \mu^{2}_B \, b^2 \, t^3}{\hbar M} \, y \, \Pauli{x}{} \label{eq:SG nested commutator Z 3} \\
    z^{(4)} (t) \, &= \,  \frac{t}{\ii \hbar} \, \left( - \frac{2 \, \mu^{2}_B \, b^2 \, t^3}{\hbar M} \right) \, \comm{ y \, \Pauli{x}{} }{H^{\,}_{\rm SG}} \notag \\
    &= \, \frac{2 \, \ii \, \mu_B^2 \, b^2 \, t^4}{\hbar^2 \, M} \, \left( - \frac{\hbar^2}{2M} \, \comm{y}{\pdp{y}{2}} \, \Pauli{x}{} - \mu^{\,}_B \, b\, y^2 \, \comm{\Pauli{x}{}}{\Pauli{y}{}} + \mu^{\,}_B \, y \, \left( B_0 + b \, z \right) \, \comm{\Pauli{x}{}}{\Pauli{z}{}} \right) \notag \\
    &= \, \frac{2 \, \ii \, \mu_B^2 \, b^2 \, t^4}{\hbar^2 \, M} \,  \left( \frac{\hbar^2}{M} \, \pd{y} \, \Pauli{x}{} - 2 \, \ii \,\mu^{\,}_B \, b\, y^2 \, \Pauli{z}{} - 2 \, \ii \,  \mu^{\,}_B \, y \, \left( B_0 + b \, z \right) \, \Pauli{y}{}  \right) \notag \\
    &= \, - \frac{2 \, \mu_B^2 \, b^2 \, t^4}{\hbar \, M^2} \, p^{\vpp}_y \, \Pauli{x}{} + \frac{4 \, \mu_B^3 \, b^2 \, t^4}{ \hbar^2 \, M} \, \left( B_0 \, y + b \, y \, z \right) \, \Pauli{y}{} + \frac{4 \, \mu_B^3 \, b^3 \, t^4}{\hbar^2 \, M} \, y^2 \, \Pauli{z}{} \, , ~~
    \label{eq:SG nested commutator Z 4}
\end{align}
\end{subequations}
organized as before. Now, suppose we plan to measure $z$ at time $t$, and take the initial state to be
\begin{equation}
    \label{eq:Stern Gerlach initial state with y}
    \ket{\Psi(0))} \, = \, \iiint_{\Reals^3} \, \thed p_x \, \thed y \, \thed z \, \frac{\ee^{-(p_x - M \, v)^2/4 \, \delta_x^2}}{(2 \, \pi \, \delta_x^2)^{1/4}}  \, \frac{\ee^{-y^2/4 \, \delta_y^2}}{(2 \, \pi \, \delta_y^2)^{1/4}}  \, \frac{\ee^{-(z - z_0)^2/4 \, \delta_z^2}}{(2 \, \pi \, \delta_z^2)^{1/4}}  \, \ket{p_x,y,z,s} \, ,~~
\end{equation}
where $\Pauli{z}{} \, \ket{s} = s \, \ket{s}$. This initial is a Gaussian wave packet centered around $y=0$ and $z=z_0$ (in real space) and $p_x = M \, v$ (in momentum space), with spin  polarization $s$. To $\Order{t^4}$, we have
\begin{subequations}
    \label{eq:SG expval evolution}
\begin{align}
    \expval{x}(t) \, &= \, v \, t \label{eq:SG <x> evolution} \\
    \expval{y}(t) \, &= \, %\frac{\mu^{\,}_B \, b \, t^2}{M} \Im \left( c_{+}^* c_{-}^{\vps} \right)  + \frac{2\, \mu^2_B \, b \, t^3}{3 \, \hbar \, M} \left( B_0 + b \, z_0 \right) \Re \left( c_{+}^* c_{-}^{\vps} \right) - \frac{\mu_B^3 \, b \, t^4}{3 \, \hbar^2 \, M} \left( B_0^2 + 2 \, b \, B_0 \, z_0 + b^2 \, z_0^2 + b^2 \delta_z^2 \right) \Im \left( c_+^* c_{-}^{\vps} \right) 
    0 \label{eq:SG <y> evolution} \\
    \expval{z}(t) \, &= \, z_0 - \frac{\mu^{\,}_B \, b \, t^2}{2 M} \, s + \frac{\mu_B^3 \, b^3 \, t^4}{6 \, \hbar^2 \, M} \, \delta^2_y \, s \, ,~~\label{eq:SG <z> evolution}
\end{align}
\end{subequations}
meaning that the average correction to $z(t)$ due to the $y$ dynamics only appears at $\Order{t^4}$, and is proportional to the variance $\delta^{\,}_y$ in the initial $y$ position. By minimizing $\delta^{\,}_y$ and using particles with polarized $z$ spin $s$,  the $y$ dynamics can be neglected, as in Sec.~\ref{subsec:SternGerlach}.

\subsection{Time evolution of states in the simple case}
\label{app:SG easy}

We start by applying the rightmost operator in the  unitary $\umeas (t)$ \eqref{eq:Stern Gerlach nice unitary} to $\ket{\Psi(0)}$ \eqref{eq:Stern Gerlach initial state nice}, giving 
\begin{align}
    \label{eq:Stern Gerlach first unitary}
    \sum\limits_{s = \pm 1} \, c^{\vpp}_s \, \int \thed z \, \varphi_z (z) \, \ket{z,s} ~~\mapsto ~~\sum\limits_{s = \pm 1} \, c^{\vpp}_s \, \int \thed z \, \varphi (z - s \, t^2 \, \frac{b \, \mu^{\,}_B}{M} \, ) \, \ket{z,s} \, ,~~
\end{align}
via the same trick for exponentials of $\pd{z}$ used in the pointer-particle example \eqref{eq:Pointer State Time t} in Sec.~\ref{subsec:von Neumann}, where $\varphi (z) \propto \exp \left( - (z-z_0)^2 / 4 \, \delta^2 \right)$ \eqref{eq:Stern Gerlach initial state with y}. Applying the next factor in  $\umeas (t)$ \eqref{eq:Stern Gerlach nice unitary} leads to
\begin{align}
    \ket{\Psi(0)} \, &\mapsto \, \sum\limits_{s = \pm 1} \, c^{\vpp}_s \, \int \thed z \, \Bigg[ \,  \exp \left( \frac{\ii \, \hbar \, t}{2 M} \, \pdp{z}{2} \right) \,  \Bigg] \, \varphi (z - z_{s,t} ) \, \ket{z,s} \notag \\
    &= \, \sum\limits_{s = \pm 1} \, c^{\vpp}_s \, \int \thed z ~ \sum\limits_{n=0}^{\infty} \, \frac{1}{n!} \left( \frac{\hbar \, t}{2 \, \ii \, M} \right)^n ~ \pdp{z}{2n} \, \frac{\exp \left[ - \left( z-z_{s,t} \right)^2 / 4 \, \delta^2 \, \right]}{\left( 2 \, \pi \, \delta^2 \right)^{1/4}} ~ \ket{z,s} \notag \\
    &= \, \sum\limits_{s = \pm 1} \, c^{\vpp}_s \, \int \thed z ~ \sum\limits_{n=0}^{\infty} \, \frac{1}{n!} \left( \frac{\ii \, \hbar \, t}{8 \, \delta^2 \, M} \right)^n \, H^{\,}_{2n} \left(\frac{z-z_{s,t}}{2 \, \delta} \right)  ~\varphi (z - z_{s,t} ) \, \ket{z,s} \, ,~~
    \label{eq:Stern Gerlach second unitary part 1}
\end{align}
where $H_k(y)$ is the $k$th (physicist's) Hermite polynomial in $y$, and we have defined 
\begin{equation}
    \label{eq:Stern Gerlach zst}
    z_{s,t} \, \equiv \, z_0 + \frac{b \, \mu^{\,}_B}{M} \, s \, t^2 \, ,~~
\end{equation}
and we now simplify Eq.~\ref{eq:Stern Gerlach second unitary part 1} using a Taylor expansion of the Hermite polynomials. We define
\begin{align}
    F ( \lambda , \tau ) \, &= \, \sum\limits_{n=0}^{\infty} \, \frac{1}{n!} \left( \ii \, \tau  \right)^n \, H^{\vpp}_{2n} ( \lambda ) \, = \, \sum\limits_{n=0}^{\infty} \, \frac{1}{n!} \left( \ii \, \tau  \right)^n \,  \sum\limits_{k=0}^{\infty} \, \frac{\lambda^k}{k!} \, \left. \pdp{\lambda}{k} H^{\vpp}_{2n} (\lambda) \right|_{\lambda=0} \notag \\
    &= \, \sum\limits_{n=0}^{\infty} \, \frac{1}{n!} \left( \ii \, \tau  \right)^n \,  \sum\limits_{k=0}^{n} \, \frac{\lambda^k}{k!} \, \frac{2^k \, (2n)!}{(2n-k)!} \, H^{\vpp}_{2n-k} (0) \notag \\
    &= \, \sum\limits_{n=0}^{\infty} \, \frac{1}{n!} \left( \ii \, \tau  \right)^n \,  \sum\limits_{k=0}^{\infty} \, \frac{\lambda^{2k}}{(2k)!} \, \frac{4^k \, (2n)!}{(2n-2k)!} \, H^{\vpp}_{2m-2k} (0)  \, ,~~
\end{align}
where we used the fact that $H^{\vpp}_{2s+1}(0) = 0$, and we next switch the order of summation,
\begin{align}
    F(\lambda, \tau) \, &= \, \sum\limits_{k=0}^{\infty} \, \frac{(2\lambda)^{2k}}{(2k)!} \, \sum\limits_{n=k}^{\infty} \, \frac{\left( \ii \, \tau \right)^n}{n!} \, \frac{(2n)!}{(2n-2k)!} \, H^{\vpp}_{2(n-k)} ( 0 ) \, ,~~
\end{align}
and the sum over $n$ can be computed exactly to give
\begin{align}
    F(\lambda, \tau) \,  &= \, \sum\limits_{k=0}^{\infty} \, \frac{(2\lambda)^{2k}}{(2k)!} \, \frac{(2k)!}{k!} \frac{1}{\sqrt{1+4 \, \ii \, \tau}} \, \left( \frac{\ii \, \tau}{1 + 4 \, \ii \, \tau} \right)^n \, = \, \frac{1}{\sqrt{1+4 \, \ii \, \tau}} \, \exp \left( \frac{4 \, \ii \, \lambda^2 \, \tau}{1+4 \, \ii \, \tau} \right) \label{eq:Stern Gerlach F function} \, , ~~
\end{align}
so that the initial wavefunction \eqref{eq:Stern Gerlach initial state nice} evolves via $\umeas (t)$ \eqref{eq:Stern Gerlach nice unitary} into
\begin{equation*}
    \tag{\ref{eq:Stern Gerlach nice final wavefunction}}
    \Psi^{\vps}_s (z,t) \, = \, c^{\vpp}_s \, \frac{\ee^{\ii \, b^2 \, \mu^2_B \, t^3 / \hbar \, M} \, \ee^{- \ii \, s \, t \, \mu^{\,}_B (B_0 + b z)/\hbar}}{(2 \, \pi)^{1/4} \, ( \delta + \, \frac{\ii \, \delta \, t}{2 \delta})^{1/2}} \, \exp \left[ - \frac{\left( z - z_0 - \frac{b \, \mu^{\,}_B}{M} \, s \,  t^2 \right)^2}{4 \, \delta^2} \, \frac{1}{1+ \frac{\ii \, \hbar \, t}{2 \, M \, \delta^2}} \right] \, ,~~
\end{equation*}
as claimed in the main text. This simplifies upon evaluating $p_s(z,t)$ \eqref{eq:Stern Gerlach nice z distribution}.

\section{Photon absorption}
\label{app:photonabsorption}

Here we discuss details of the photon-counting measurements considered in Sec.~\ref{subsec:photon counting}. The physical system is a bosonic mode with %Hilbert space 
$\Hilbert{\text{ph}} = L^2(\Reals) = \ell^2 (\Nats) \cong \ell^2 (\Nats)$, and the apparatus consists of $\Noutcome \ll 1$ qubits with states $\ket{0}$ (default) and $\ket{1}$ (excited). We initialize the system in the state $\ket{\psi}$, the apparatus in the state $\ket{\bvec{0}}=\ket{0}^{\otimes \Noutcome}$, and evolve under a Hamiltonian $\Ham$ \eqref{eq:photon counting Hamiltonian}, leading to 
%that couples the system and apparatus with strength $g$, resulting in the postmeasurement state
\begin{equation*}
\tag{\ref{eq:photon count final state def}}
    \ket{\Psi} \, = \, \ee^{g\tau(\a{}\Pauli{+}{\Noutcome}-\adag{}\Pauli{-}{\Noutcome})} \cdots \ee^{g\tau(\a{}\Pauli{+}{1}-\adag{}\Pauli{-}{1})} \,  \ket{\psi}^{\vpp}_{\text{ph}} \otimes \ket{ \bvec{0}}^{\vpp}_{\text{ss}} \, ,~~
\end{equation*}
where $\tau$ is the duration and $g$ is the strength of the interaction between the light and the detector qubits, which are labeled according to the order in which they interact with the light. 

Next, consider the interaction between the electromagnetic mode and the $k$th detector qubit. Assuming that the former starts in the state $\ket{\phi}$ and taking $\Noutcome \gg 1$ and $\tau \ll 1$, we find that
\begin{align}
    \ee^{g \tau \left( \a{} \Pauli{+}{k} - \adag{} \Pauli{-}{k}\right) } \, \ket{\phi} \otimes \ket{0}^{\vpp}_k \, &= \, \ee^{g \tau \, \a{} \, \Pauli{+}{k}} \, \ee^{-g \tau \, \adag{} \, \Pauli{-}{k}} \, \ee^{g^2 \tau^2 \comm{\a{} \Pauli{+}{k}}{\adag{}\Pauli{-}{k}}/2} \left( 1 + \Order{g^3 \tau^3} \right) \, \ket{\phi} \otimes \ket{0}^{\vpp}_k \notag \\
    &=  \, \ee^{g \tau \, \a{} \, \Pauli{+}{k}} \, \ee^{-g \tau \, \adag{} \, \Pauli{-}{k}} \, \ee^{g^2 \tau^2 \left[ ( \adag{} \a{} + 1 ) \, \BKop{1}{1}^{\,}_k - \adag{} \a{} \,\BKop{0}{0}^{\,}_K  \right] /2} \, \ket{\phi} \otimes \ket{0}^{\vpp}_k + \Order{g^3 \tau^3} \notag \\
    &= \,  \ee^{g \tau \, \a{} \, \Pauli{+}{k}} \, \ee^{-g \tau \, \adag{} \, \Pauli{-}{k}} \, \ee^{- g^2 \tau^2 \adag{} \a{} /2} \, \ket{\phi} \otimes \ket{0}^{\vpp}_k + \Order{g^3 \tau^3} \notag \\
    &=  \, \ee^{g \tau \, \a{} \, \Pauli{+}{k}} \, \ee^{- g^2 \tau^2 \adag{} \a{} /2} \, \ket{\phi} \otimes \ket{0}^{\vpp}_k + \Order{g^3 \tau^3} \, , ~~\label{eq:effective photon qubit update}
\end{align}
so that the final state $\ket{\Psi}$ \eqref{eq:photon count final state def} is given by
\begin{equation}
\label{eq:photon count appendix Psi 2}
    \ket{\Psi} \, = \, \ee^{g\tau\a{}\Pauli{+}{\Noutcome}} \, \ee^{-g^2\tau^2\adag{}\a{}/2} \cdots \ee^{g\tau\a{}\Pauli{+}{1}} \, \ee^{-g^2\tau^2\adag{}\a{}/2} \ket{\psi}^{\vpp}_{\text{ph}} \otimes \ket{\bvec{0}}^{\vpp}_{\text{ss}} +\Order{\mathcal{N}g^3\tau^3} \, ,~~
\end{equation}
and we now reorder the product of operators above. We first note that
\begin{equation}
    \label{eq:BCH photon counting commutation}
    \exp \left( - g^2 \tau^2 \, \adag{} \a{}/2 \right) \, \exp \left( g \tau \, \a{} \Pauli{+}{k} \right) \, = \, \exp \left( \ee^{g^2 \tau^2 / 2} \, g \tau \, \a{} \Pauli{+}{k} \right) \, \exp \left( - g^2 \tau^2 \, \adag{} \a{}/2 \right)  \, ,~~
\end{equation}
which we use to commute all $\ee^{-g^2\tau^2 \adag{}\a{}/2}$ terms to the right of all others in Eq.~\ref{eq:photon count appendix Psi 2}, leading to
\begin{align}
    \ket{\Psi} \, &= \, \ee^{g\tau\a{}\Pauli{+}{\Noutcome}} \, \ee^{-g^2\tau^2\adag{}\a{}/2} \cdots \ee^{g\tau\a{}\Pauli{+}{1}} \, \ee^{-g^2\tau^2\adag{}\a{}/2} \ket{\psi}^{\vpp}_{\text{ph}} \otimes \ket{\bvec{0}}^{\vpp}_{\text{ss}} +\Order{\mathcal{N}g^3\tau^3} \notag \\
    &= \, \Bigg[ \prod\limits_{k=1}^{\Noutcome} \, \exp \Big( \ee^{g^2 \tau^2 (\Noutcome - k) /2}  g \tau \, \a{} \otimes \Pauli{+}{k} \,\Big) \, \Bigg]  \, \ee^{- \Noutcome \, g^2 \tau^2 \, \adag{} \a{} /2 } \, \ket{\psi}^{\vpp}_{\text{ph}} \otimes \ket{\bvec{0}}^{\vpp}_{\text{ss}} +\Order{\Noutcome g^3\tau^3} \, ,~\\
\intertext{where all terms in the product over qubits $k$ commute with one another, so that we can write}
    &= \, \exp \Bigg( g \tau \, \sum\limits_{k=1}^{\Noutcome} \, \ee^{g^2 \tau^2 (\Noutcome-k)/2} \, \a{} \otimes \Pauli{+}{k}  \Bigg) \, \exp \left( -\Noutcome \, g \tau \, \adag{} \a{} / 2 \right) \ket{\psi}^{\vpp}_{\text{ph}} \otimes \ket{\bvec{0}}^{\vpp}_{\text{ss}} +\Order{\Noutcome g^3\tau^3} \, ,~~ \label{eq:photon count appendix nice Psi}
\end{align}
which we now simplify using additional definitions. We first  define the \emph{attenuation parameter}
\begin{equation*}
    \tag{\ref{eq:zeta def}}
    \zeta \, \equiv \, \Noutcome \, g^2 \, \tau^2 \, , ~~
\end{equation*}
which remains finite as $\tau \to 0$ and $\Noutcome \to \infty$. We note that there are $\Noutcome$ leading corrections to Eq.~\ref{eq:photon count appendix nice Psi} with $\Order{g^3 \tau^3}$ coefficients, which may conspire to produce a correction no larger than $\Order{\Noutcome g^3 \tau^3} = \Order{ \zeta^{3/2} \Noutcome^{-1/2}}$, which still vanishes as $g \tau \to 0$ and $\Noutcome \to \infty$. We then define
%\begin{equation}
%\label{eq:big B appendix version 0}
    %g \tau \, \sum\limits_{k=1}^{\Noutcome} \, \ee^{g^2 \tau^2 (\Noutcome-k)/2} \, \a{} \otimes \Pauli{+}{k}  \, \equiv \sqrt{ \ee^{\zeta}-1} \, \a{} \otimes \Bdag{} \, ,~~
%\end{equation}
%which defines an effective bosonic operator
\begin{equation}
\label{eq:big B appendix version}
    \Bdag{} \, \equiv \, \sqrt{\frac{\zeta}{N \, (\ee^{\zeta} - 1 )}} \, \sum\limits_{k=1}^{\Noutcome} \, \ee^{\zeta(\Noutcome-k)/(2\Noutcome)} \, \Pauli{+}{k} \, ,~~
\end{equation}
as an effective bosonic operator. The postmeasurement state $\ket{\Psi}$ \eqref{eq:photon count appendix nice Psi} now becomes
\begin{equation}
\label{eqn:appPhotonAbsorption}
    \ket{\Psi} \, = \, \exp \bigg[ \left( \ee^{\zeta}-1 \right)^{1/2} \,  \a{} \otimes \Bdag{}\bigg] \, \exp \bigg[ -\frac{\zeta}{2} \, \adag{}\a{} \bigg] \, \ket{\psi}^{\vpp}_{\text{ph}} \otimes \ket{ \bvec{0}}^{\vpp}_{\text{ss}} +\Order{\zeta^{3/2}\Noutcome^{-1/2}} \, , ~~
\end{equation}
which reproduces Eq.~\ref{eqn: photodetection final state} upon reordering the two exponential terms above.

The lowering operator $B$ \eqref{eq:big B appendix version} only realizes a bosonic annihilation operator in the limit $\Noutcome \to \infty$ with $\zeta$ \eqref{eq:zeta def} finite. The standard commutation relation for arbitrary $\Noutcome$ and $\tau$ is given by
\begin{align}
    \comm{\B{}}{\Bdag{}} \, &= \, \frac{g^2 \, \tau^2}{\ee^{ \Noutcome g^2  \tau^2} - 1} \, \sum\limits_{k,k'=1}^{\Noutcome} \, \ee^{g^2 \tau^2 ( 2 \, \Noutcome - k - k')/2} \, \comm{\Pauli{-}{k}}{\Pauli{+}{k'}} \notag \\
    %
    %&= \, \frac{g^2 \, \tau^2}{\ee^{\Noutcome g^2 \tau^2} - 1} \, \sum\limits_{k=1}^{\Noutcome} \, \ee^{g^2 \tau^2 ( \Noutcome - k)} \, \comm{\Pauli{-}{k}}{\Pauli{+}{k}} \, 
    &= \, - \frac{g^2 \, \tau^2}{\ee^{\Noutcome g^2  \tau^2} - 1} \, \sum\limits_{k=1}^{\Noutcome} \, \ee^{g^2 \tau^2 ( \Noutcome - k)} \,\PZ{k} \, ,~~ \\
\intertext{and we rewrite $\PZ{k} = 2 \, n^{\,}_k - 1$, where $n^{\,}_k = \BKop{1}{1}^{\,}_k$. As discussed in Secs.~\ref{subsec:destructive} and \ref{subsec:photon counting}, all states with $m$ qubits in the state $\ket{1}$ reflect the same outcome, so we assume the qubits are excited in order of ascending $k$ without loss of generality. The above then becomes}
     \comm{\B{}}{\Bdag{}} \, &= \, \frac{g^2 \, \tau^2}{\ee^{\Noutcome \, g^2 \, \tau^2} - 1} \, \sum\limits_{k=1}^{\Noutcome} \, \ee^{g^2 \, \tau^2 ( \Noutcome - k)} \, \left( 1 - 2 \, n^{\vpp}_k \right) \notag \\
     &= \, \frac{g^2 \, \tau^2}{\ee^{\Noutcome \, g^2 \, \tau^2} - 1} \, \sum\limits_{n=0}^{\Noutcome} \left( \sum\limits_{k=1}^{\Noutcome} \, \ee^{g^2 \, \tau^2 ( \Noutcome - k)}  - 2 \sum\limits_{k=1}^{n}  \, \ee^{g^2 \, \tau^2 ( \Noutcome - k)}  \right) \, \BKop{n}{n}^{\vpp}_{\text{ss}} \notag \\
     &= \, \frac{g^2 \, \tau^2}{\ee^{g^2 \tau^2}-1} \, \sum\limits_{n=0}^{\Noutcome} \, \left( 1 - 2 \frac{1-\ee^{-n g^2 \tau^2}}{1-\ee^{-\Noutcome g^2 \tau^2}} \right) \, \BKop{n}{n}^{\vpp}_{\text{ss}}  \, , ~~
\end{align}
and taking the limits $\tau \to 0$ and $\Noutcome \to \infty$ in either order while keeping $\zeta$ \eqref{eq:zeta def} fixed  gives
\begin{align}
\label{eq:big B commutation verification}
    \comm{\B{}}{\Bdag{}} \, &= \, \sum\limits_{n=0}^{\infty} \, \BKop{n}{n}^{\vpp}_{\text{ss}} \, = \, \ident^{\vpp}_{\text{ss}} \, , ~~ 
\end{align}
so that $B$ is a bosonic annihilation operator in this limit. If we only take $\Noutcome \to \infty$, the above commutator is instead $g^2 \tau^2 (\ee^{g^2 \tau^2}-1)^{-1} \, \ident$; if we only take $\tau \to 0$, the commutator is instead $\sum_{n=0}^{\Noutcome} \, (1 - 2 n/\Noutcome) \, \BKop{n}{n}$. Hence, both limits are needed, which means that  $\zeta$ \eqref{eq:zeta def} must be \emph{finite}.

\section{Matrix element for homodyne detection}
\label{app:homdetect}

Here we calculate the matrix element  \ref{eqn:HomDetectMatrixElement} central to the discussion in Sec.~\ref{subsec:homodyne}. This calculation also appears in the literature (see, e.g., Ref.~\citenum{Combes2022}). The matrix element of interest is
\begin{equation}
\label{eq:homodyne matel}
    f(N,D) \, = \, \matel*{N,D}{U^{\vpd}_{\text{BS}}}{\Psi_0}
    %=\frac{ \ee^{\ii N\phi }}{\pi^{1/4}\abs{\beta}}\exp\bigg[-\frac{\big(N-\abs{\beta}^2\big)^2}{4 \abs{\beta}^2}\bigg]\,\psi_\phi\bigg(\frac{D \sqrt{2}}{\abs{\beta}}\bigg),
\end{equation}
where $\ket{\Psi_0}=\ket{\psi}_A\ket{\beta}_B$, with $\ket{\psi}$ arbitrary and $\ket{\beta}$ \eqref{eq:beta aux} a coherent state. The Fock states $\ket{N,D}$ correspond to the sum $N$ and difference $D$ of the occupation numbers of the A and B modes, i.e., 
\begin{equation*}
\tag{\ref{eq:homodyne sum diff basis}}
    \ket{N,D} \, = \, \frac{1}{\sqrt{\big(N/2+D\big)!}}\frac{1}{\sqrt{\big(N/2-D\big)!}}\big(\adag{}\big)^{N/2+D}\big(b^{\dagger} \big)^{N/2-D} \, \ket{00} \,  ,~~
\end{equation*}
and the unitary $U^{\vpd}_{\text{BS}}$ describing time evolution within the beam splitter (see Fig.~\ref{fig:homodyne detection}) is given by
\begin{equation*}
    \tag{\ref{eq:Beam Splitter U}}
    U^{\vpd}_{\text{BS}} \, = \, \exp \left( \frac{\pi}{4} (\adag{} b - \a{} b^{\dagger} )  \right) \, ,~~
\end{equation*}
so that the following properties hold for the annihilation operators for the two modes:
\begin{subequations}
\label{eq:Ubs annihilation relation}
\begin{align}
    \a{} \, U^{\vpd}_{\text{BS}} \, &= \, U^{\vpd}_{\text{BS}} \, \frac{(\a{}+b)}{\sqrt{2}} \label{eq:Ubs annihilation relation A} \\
    b \, U^{\vpd}_{\text{BS}} \, &= \,  U^{\vpd}_{\text{BS}} \frac{(b-\a{})}{\sqrt{2}} \, ,~~ \label{eq:Ubs annihilation relation B}
\end{align}
\end{subequations}
and using the fact that $b \, \ket{\beta}_B = \beta \, \ket{\beta}_B$, the matrix element $f(N,D)$ \eqref{eq:homodyne matel} takes the exact form
\begin{equation}
\label{eq:homodyne matel exact}
    f(N,D) \,  = \, \frac{\ee^{-\abs{\beta}^2/2} \, \matel*{0}{\, (\beta + \a{})^{N/2+D} \, (\beta - \a{})^{N/2-D} \, }{\psi}^{\vpp}_A}{\sqrt{2^N \, (N/2+D)! \, (N/2-D)!}} \, , ~~
\end{equation}
using Eq.~\ref{eq:beta aux} for $\inprod{0}{\beta}^{\,}_B$, the relation $\bra{0}^{\,}_A \bra{0}^{\,}_B \, U^{\vpp}_{\text{BS}} = \bra{0}^{\,}_A \bra{0}^{\,}_B$, and other relations above.

Assuming that the B-mode occupation $n_B \sim \abs{\beta}^2 \gg 1$ is large (the precise value must be determined self-consistently and depends on $n_A$), we can rewrite $f(N,D)$ \eqref{eq:homodyne matel exact} as
\begin{align}
\label{appeqn:expansion}
    &= \, \matel*{0}{\, (\beta+\a{})^{N/2+D} \, (\beta-\a{})^{N/2-D} \, }{\psi}^{\vpp}_A \, \notag  \\
    &=\, \abs{\beta}^N \, \ee^{-\ii \phi N} \, \matel*{0}{\, \exp \left[\left(\frac{N}{2}+D\right)\log\left(1+\frac{ \a{} \, \ee^{\ii \phi}}{\abs{\beta}}\right)+\left(\frac{N}{2}-D\right)\log\left(1-\frac{ \a{} \, \ee^{\ii \phi}}{\abs{\beta}}\right) \right] \, }{\psi}^{\vpp}_A \notag \\
    &\approx\, \abs{\beta}^N \, \ee^{-\ii \phi N} \, \matel*{0}{\, \exp \left(\frac{2 D}{\abs{\beta}} \ee^{\ii \phi} \, \a{} - \frac{N}{2 \abs{\beta}^2}(\ee^{\ii \phi} \, \a{})^2 + \dots \right) \, }{\psi}^{\vpp}_A \, , ~~
\end{align}
where we have ignored higher-order terms in the arguments of the logarithms and used the definition $\beta = \abs{\beta} \, \ee^{-\ii\phi}$. In the limit identified above, the matrix element $f(N,D)$ \eqref{eq:homodyne matel exact} becomes
\begin{equation}
    f(N,D) \, = \, \frac{ \abs{\beta}^N \, \ee^{ - \abs{\beta}^2/2 } \, \ee^{-\ii \phi N}}{\sqrt{2^N \, \left(N/2+D\right)! \, \left(N/2-D\right)!}} \, \matel*{0}{\, \exp \left( \frac{2D}{\beta} \, a - \frac{N}{2 \, \beta^2} \, a^2 \right) }{\psi}^{\vpp}_A \, , ~~
\end{equation}
which, as a function of $N$, is peaked around $N \sim \abs{\beta}^2$. Expanding about this value leads to
\begin{equation}
\label{eq:f mat el penultimate}
    f(N,D) \, = \, \frac{ \ee^{ \ii \phi N } \,  }{ \abs{\beta}\, \sqrt{\pi}} \, \ee^{ - D^2 /\abs{\beta}^2 } \, \ee^{\left(N-\abs{\beta}^2\right)^2 / 4 N}  \,  \underbrace{ \bra{0} \exp \left( -\frac{1}{2}( \ee^{\ii \phi} \a{} )^2 + \frac{2D}{\abs{\beta}}( \ee^{\ii \phi} \a{} ) \right)}_{\bra{P^{\,}_\phi}}\ket{\psi} \, ,~~
\end{equation}
and we now consider the properties of the state $\ket{P^{\,}_{\phi}}$ above, defined by
\begin{equation}
\label{eq:P phi def}
    \ket{P_\phi} \, = \, \exp \left[ -\frac{1}{2}(\ee^{-\ii \phi} \adag{})^2 + \frac{2 D}{\abs{\beta}}(\ee^{-\ii \phi} \adag{}) \right]\ket{0} \, \equiv \, \mathcal{G}\ket{0} \, ,~~
\end{equation}
where, for notational convenience, we have introduced the operator
\begin{equation}
    \mathcal{G} \,  \equiv \, \exp \left[ -\frac{1}{2}(\ee^{-\ii \phi} \adag{})^2 + \frac{2 D}{\abs{\beta}}(\ee^{-\ii \phi} \adag{}) \right] \, . ~~
\end{equation}
We first note that $\ket{P_\phi}$ is an eigenstate of the quadrature operator $x^{\,}_{\phi}$ \eqref{eq:quadrature X}, since 
\begin{equation}
    \mathcal{G}^{-1} \adag{} \mathcal{G} \, = \, \adag{}  ~~~~ \text{and}~~~~\mathcal{G}^{-1} \a{} \mathcal{G} \, = \, \a{}-\adag{} \ee^{-2 \ii \phi} + \frac{2D}{\abs{\beta}} \ee^{-\ii \phi} \, , ~~
\end{equation}
so that applying $x^{\,}_{\phi}$ \eqref{eq:quadrature X} to $\ket{P_\phi}$ \eqref{eq:P phi def} leads to
\begin{align}
    x^{\vpp}_{\phi} \, \ket{P^{\,}_\phi} \, &= \, \left( \frac{ \ee^{\ii \phi} \, \a{} + \ee^{-\ii \phi} \, \adag{}}{\sqrt{2}} \right) \, \mathcal{G}\, \ket{0} \, =  \, \mathcal{G}\bigg(\frac{\a{} \ee^{-\ii \phi}}{\sqrt{2}}+\frac{\sqrt{2} D}{\abs{\beta}}\bigg)\ket{0} \, = \, \bigg(\frac{D \sqrt{2}}{\abs{\beta}}\bigg)\ket{P_{\phi}} \,,~
\end{align}
meaning that $\ket{P_\phi}$ \eqref{eq:P phi def} is an eigenstate of $x^{\,}_{\phi}$ \eqref{eq:quadrature X} with eigenvalue $x^{\,}_{\phi} (D) = 2 D/\abs{\beta}$. Thus, $\ket{P_{\phi}}$ is proportional to an eigenstate $\ket{x^{\,}_{\phi} (D) } = \mathcal{N}^{-1} (D) \,  \ket{P_\phi}$ that is part of an orthonormal set, i.e.,
\begin{equation}
    \inprod{x^{\,}_\phi (D) }{x_\phi^{\,} (D')} \, = \, \DiracDelta{x^{\,}_{\phi} (D) - x^{\,}_{\phi} (D')} \, \propto \, \DiracDelta{D-D'} \, .~~
\end{equation}
We then obtain the ``normalization factor'' $\mathcal{N}$ by projecting $\ket{P_{\phi}} = \mathcal{N} \ket{x^{\,}_{\phi} (D)}$ onto $\ket{0}$,
\begin{equation}
    \inprod{0}{P_{\phi}} \, \overset{!}{=}  \, 1 \, = \, \mathcal{N} \inprod{0}{x^{\,}_{\phi} (D)} \, = \, \frac{\mathcal{N}}{\pi^{1/4}} \exp\bigg[-\frac{x_\phi^2 (D)}{2}\bigg]=\frac{\mathcal{N}}{\pi^{1/4}} \ee^{-D^2/\abs{\beta}^2} \, , ~~
\end{equation}
which we then insert into Eq.~\ref{eq:f mat el penultimate} to recover the result claimed in the main text,
\begin{align*}
    f(N,D) \, &= \, \frac{\ee^{\ii \phi N}}{\abs{\beta}\, \pi^{1/4}} \exp \bigg[\frac{\big(N-\abs{\beta}^2\big)^2}{4 N} \bigg] \, \inprod{x_\phi^{D}}{\psi} \\
    &= \, \frac{\ee^{\ii \phi N}}{\abs{\beta}\, \pi^{1/4}} \exp \bigg[\frac{\big(N-\abs{\beta}^2\big)^2}{4 N} \bigg] \, \psi^{\,}_\phi \left( \frac{D \sqrt{2}}{\abs{\beta}} \right) \, .~~ \tag{\ref{eqn:HomDetectMatrixElement}}
\end{align*}
To determine the conditions under which the foregoing derivation is valid, we consider the first two terms that we neglected in the argument of the exponential in Eq.~\ref{appeqn:expansion}, which are proportional to $D \, \abs{\beta}^{-3} \, (\a{})^3$ and $N \, \abs{\beta}^{-4} \, (\a{})^4$, respectively. However, we note that $N \approx  \abs{\beta}^2$ and, in typical states, $\a{} \sim \sqrt{\avg{\adag{}\a{}}}$ (up to a phase, where the average is taken with respect to the initial physical state $\ket{\psi}^{\,}_A$). Hence, we can safely neglect these higher-order terms when $\abs{\beta} \gg \matel{\psi}{\adag{}\a{}}{\psi}$ and $D \ll \abs{\beta}^3/\avg{\adag{}\a{}}^{3/2}$. Since $x_\phi^{\,} (D) \propto D/\abs{\beta}$ extends up to a distance in phase space of size $\sqrt{\avg{\adag{}\a{}}}$, the latter condition also implies that $\abs{\beta}\gg \matel{\psi}{\adag{}\a{}}{\psi}$, in agreement with Ref.~\citenum{Tyc_2004}.

\bibliography{thebib}

\begin{thebibliography}{131}%
\makeatletter
\providecommand \@ifxundefined [1]{%
 \@ifx{#1\undefined}
}%
\providecommand \@ifnum [1]{%
 \ifnum #1\expandafter \@firstoftwo
 \else \expandafter \@secondoftwo
 \fi
}%
\providecommand \@ifx [1]{%
 \ifx #1\expandafter \@firstoftwo
 \else \expandafter \@secondoftwo
 \fi
}%
\providecommand \natexlab [1]{#1}%
\providecommand \enquote  [1]{``#1''}%
\providecommand \bibnamefont  [1]{#1}%
\providecommand \bibfnamefont [1]{#1}%
\providecommand \citenamefont [1]{#1}%
\providecommand \href@noop [0]{\@secondoftwo}%
\providecommand \href [0]{\begingroup \@sanitize@url \@href}%
\providecommand \@href[1]{\@@startlink{#1}\@@href}%
\providecommand \@@href[1]{\endgroup#1\@@endlink}%
\providecommand \@sanitize@url [0]{\catcode `\\12\catcode `\$12\catcode
  `\&12\catcode `\#12\catcode `\^12\catcode `\_12\catcode `\%12\relax}%
\providecommand \@@startlink[1]{}%
\providecommand \@@endlink[0]{}%
\providecommand \url  [0]{\begingroup\@sanitize@url \@url }%
\providecommand \@url [1]{\endgroup\@href {#1}{\urlprefix }}%
\providecommand \urlprefix  [0]{URL }%
\providecommand \Eprint [0]{\href }%
\providecommand \doibase [0]{https://doi.org/}%
\providecommand \selectlanguage [0]{\@gobble}%
\providecommand \bibinfo  [0]{\@secondoftwo}%
\providecommand \bibfield  [0]{\@secondoftwo}%
\providecommand \translation [1]{[#1]}%
\providecommand \BibitemOpen [0]{}%
\providecommand \bibitemStop [0]{}%
\providecommand \bibitemNoStop [0]{.\EOS\space}%
\providecommand \EOS [0]{\spacefactor3000\relax}%
\providecommand \BibitemShut  [1]{\csname bibitem#1\endcsname}%
\let\auto@bib@innerbib\@empty
%</preamble>
\bibitem [{\citenamefont {Dirac}(1930)}]{DiracQuantum}%
  \BibitemOpen
  \bibfield  {author} {\bibinfo {author} {\bibfnamefont {P.~A.~M.}\
  \bibnamefont {Dirac}},\ }\href@noop {} {\emph {\bibinfo {title} {The
  Principles of Quantum Mechanics}}}\ (\bibinfo  {publisher} {Oxford University
  Press},\ \bibinfo {address} {Oxford, U.K.},\ \bibinfo {year}
  {1930})\BibitemShut {NoStop}%
\bibitem [{\citenamefont {Birkhoff}\ and\ \citenamefont
  {Neumann}(1936)}]{vonNeumannAxioms}%
  \BibitemOpen
  \bibfield  {author} {\bibinfo {author} {\bibfnamefont {G.}~\bibnamefont
  {Birkhoff}}\ and\ \bibinfo {author} {\bibfnamefont {J.~V.}\ \bibnamefont
  {Neumann}},\ }\bibfield  {title} {\bibinfo {title} {The logic of quantum
  mechanics},\ }\href {http://www.jstor.org/stable/1968621} {\bibfield
  {journal} {\bibinfo  {journal} {Ann. Math.}\ }\textbf {\bibinfo {volume}
  {37}},\ \bibinfo {pages} {823} (\bibinfo {year} {1936})}\BibitemShut
  {NoStop}%
\bibitem [{\citenamefont {Hardy}(2001)}]{HardyAxioms}%
  \BibitemOpen
  \bibfield  {author} {\bibinfo {author} {\bibfnamefont {L.}~\bibnamefont
  {Hardy}},\ }\href {https://doi.org/10.48550/arXiv.quant-ph/0101012} {\bibinfo
  {title} {Quantum theory from five reasonable axioms}} (\bibinfo {year}
  {2001}),\ \Eprint {https://arxiv.org/abs/quant-ph/0101012}
  {ar{X}iv:quant-ph/0101012 [quant-ph]} \BibitemShut {NoStop}%
\bibitem [{\citenamefont {Fuchs}(2002)}]{FuchsAxioms}%
  \BibitemOpen
  \bibfield  {author} {\bibinfo {author} {\bibfnamefont {C.~A.}\ \bibnamefont
  {Fuchs}},\ }\href {https://doi.org/10.48550/arXiv.quant-ph/0205039} {\bibinfo
  {title} {Quantum mechanics as quantum information (and only a little more)}}
  (\bibinfo {year} {2002}),\ \Eprint {https://arxiv.org/abs/quant-ph/0205039}
  {ar{X}iv:quant-ph/0205039 [quant-ph]} \BibitemShut {NoStop}%
\bibitem [{\citenamefont {Mackey}(2004)}]{MackeyAxioms}%
  \BibitemOpen
  \bibfield  {author} {\bibinfo {author} {\bibfnamefont {G.}~\bibnamefont
  {Mackey}},\ }\href {https://books.google.com/books?id=Gcsz0c39QzgC} {\emph
  {\bibinfo {title} {Mathematical Foundations of Quantum Mechanics}}}\
  (\bibinfo  {publisher} {Dover},\ \bibinfo {year} {2004})\BibitemShut
  {NoStop}%
\bibitem [{\citenamefont {Wilce}(2009)}]{WilceAxioms}%
  \BibitemOpen
  \bibfield  {author} {\bibinfo {author} {\bibfnamefont {A.}~\bibnamefont
  {Wilce}},\ }\href {https://doi.org/10.48550/arXiv.0912.5530} {\bibinfo
  {title} {Four and a half axioms for finite-dimensional quantum mechanics}}
  (\bibinfo {year} {2009}),\ \Eprint {https://arxiv.org/abs/0912.5530}
  {ar{X}iv:0912.5530 [quant-ph]} \BibitemShut {NoStop}%
\bibitem [{\citenamefont {Masanes}\ and\ \citenamefont
  {Müller}(2011)}]{MasanesAxioms}%
  \BibitemOpen
  \bibfield  {author} {\bibinfo {author} {\bibfnamefont {L.}~\bibnamefont
  {Masanes}}\ and\ \bibinfo {author} {\bibfnamefont {M.~P.}\ \bibnamefont
  {Müller}},\ }\bibfield  {title} {\bibinfo {title} {A derivation of quantum
  theory from physical requirements},\ }\href
  {https://doi.org/10.1088/1367-2630/13/6/063001} {\bibfield  {journal}
  {\bibinfo  {journal} {New J. Phys.}\ }\textbf {\bibinfo {volume} {13}},\
  \bibinfo {pages} {063001} (\bibinfo {year} {2011})}\BibitemShut {NoStop}%
\bibitem [{\citenamefont {Kapustin}(2013)}]{KapustinAxioms}%
  \BibitemOpen
  \bibfield  {author} {\bibinfo {author} {\bibfnamefont {A.}~\bibnamefont
  {Kapustin}},\ }\bibfield  {title} {\bibinfo {title} {{Is quantum mechanics
  exact?}},\ }\href {https://doi.org/10.1063/1.4811217} {\bibfield  {journal}
  {\bibinfo  {journal} {J. Math. Phys.}\ }\textbf {\bibinfo {volume} {54}},\
  \bibinfo {pages} {062107} (\bibinfo {year} {2013})}\BibitemShut {NoStop}%
\bibitem [{\citenamefont {Einstein}(1971)}]{nodice}%
  \BibitemOpen
  \bibfield  {author} {\bibinfo {author} {\bibfnamefont {A.}~\bibnamefont
  {Einstein}},\ }\bibfield  {title} {\bibinfo {title} {Letter to {M}ax {B}orn,
  1926},\ }in\ \href@noop {} {\emph {\bibinfo {booktitle} {{The Born-Einstein
  Letters}}}}\ (\bibinfo  {publisher} {Walker and Company},\ \bibinfo {address}
  {New York},\ \bibinfo {year} {1971})\ \bibinfo {note} {translated by {I}rene
  {B}orn}\BibitemShut {NoStop}%
\bibitem [{\citenamefont {Einstein}\ \emph {et~al.}(1935)\citenamefont
  {Einstein}, \citenamefont {Podolsky},\ and\ \citenamefont {Rosen}}]{epr}%
  \BibitemOpen
  \bibfield  {author} {\bibinfo {author} {\bibfnamefont {A.}~\bibnamefont
  {Einstein}}, \bibinfo {author} {\bibfnamefont {B.}~\bibnamefont {Podolsky}},\
  and\ \bibinfo {author} {\bibfnamefont {N.}~\bibnamefont {Rosen}},\ }\bibfield
   {title} {\bibinfo {title} {Can quantum-mechanical description of physical
  reality be considered complete?},\ }\href
  {https://doi.org/10.1103/PhysRev.47.777} {\bibfield  {journal} {\bibinfo
  {journal} {Phys. Rev.}\ }\textbf {\bibinfo {volume} {47}},\ \bibinfo {pages}
  {777} (\bibinfo {year} {1935})}\BibitemShut {NoStop}%
\bibitem [{\citenamefont {Wheeler}\ and\ \citenamefont
  {Zurek}(2014)}]{WheelerZurek}%
  \BibitemOpen
  \bibfield  {author} {\bibinfo {author} {\bibfnamefont {J.~A.}\ \bibnamefont
  {Wheeler}}\ and\ \bibinfo {author} {\bibfnamefont {W.~H.}\ \bibnamefont
  {Zurek}},\ }\href@noop {} {\emph {\bibinfo {title} {Quantum Theory and
  Measurement}}},\ Vol.~\bibinfo {volume} {53}\ (\bibinfo  {publisher}
  {Princeton University Press},\ \bibinfo {address} {US},\ \bibinfo {year}
  {2014})\BibitemShut {NoStop}%
\bibitem [{\citenamefont {Margenau}(1963)}]{MargenauMeas}%
  \BibitemOpen
  \bibfield  {author} {\bibinfo {author} {\bibfnamefont {H.}~\bibnamefont
  {Margenau}},\ }\bibfield  {title} {\bibinfo {title} {Measurements in quantum
  mechanics},\ }\href
  {https://doi.org/https://doi.org/10.1016/0003-4916(63)90264-1} {\bibfield
  {journal} {\bibinfo  {journal} {Ann. Phys.}\ }\textbf {\bibinfo {volume}
  {23}},\ \bibinfo {pages} {469} (\bibinfo {year} {1963})}\BibitemShut
  {NoStop}%
\bibitem [{\citenamefont {Peres}(1986)}]{PeresMeas86}%
  \BibitemOpen
  \bibfield  {author} {\bibinfo {author} {\bibfnamefont {A.}~\bibnamefont
  {Peres}},\ }\bibfield  {title} {\bibinfo {title} {When is a quantum
  measurement?},\ }\href {https://doi.org/10.1111/j.1749-6632.1986.tb12446.x}
  {\bibfield  {journal} {\bibinfo  {journal} {Amer. J. Phys.}\ }\textbf
  {\bibinfo {volume} {54}},\ \bibinfo {pages} {688} (\bibinfo {year}
  {1986})}\BibitemShut {NoStop}%
\bibitem [{\citenamefont {Lindblad}(1973)}]{lindblad1973entropy}%
  \BibitemOpen
  \bibfield  {author} {\bibinfo {author} {\bibfnamefont {G.}~\bibnamefont
  {Lindblad}},\ }\bibfield  {title} {\bibinfo {title} {Entropy, information and
  quantum measurements},\ }\href {https://doi.org/10.1007/BF01646743}
  {\bibfield  {journal} {\bibinfo  {journal} {Commun. Math. Phys.}\ }\textbf
  {\bibinfo {volume} {33}},\ \bibinfo {pages} {305} (\bibinfo {year}
  {1973})}\BibitemShut {NoStop}%
\bibitem [{\citenamefont {Schlosshauer}(2007)}]{decoherence}%
  \BibitemOpen
  \bibfield  {author} {\bibinfo {author} {\bibfnamefont {M.}~\bibnamefont
  {Schlosshauer}},\ }\href {https://doi.org/10.1007/978-3-540-35775-9} {\emph
  {\bibinfo {title} {Decoherence and the Quantum-To-Classical Transition}}},\
  Frontiers Collection\ (\bibinfo  {publisher} {Springer Berlin},\ \bibinfo
  {address} {Heidelberg, Germany},\ \bibinfo {year} {2007})\BibitemShut
  {NoStop}%
\bibitem [{\citenamefont {Brasil}\ \emph {et~al.}(2013)\citenamefont {Brasil},
  \citenamefont {Fanchini},\ and\ \citenamefont
  {Napolitano}}]{brasil2013simple}%
  \BibitemOpen
  \bibfield  {author} {\bibinfo {author} {\bibfnamefont {C.~A.}\ \bibnamefont
  {Brasil}}, \bibinfo {author} {\bibfnamefont {F.~F.}\ \bibnamefont
  {Fanchini}},\ and\ \bibinfo {author} {\bibfnamefont {R.~d.~J.}\ \bibnamefont
  {Napolitano}},\ }\bibfield  {title} {\bibinfo {title} {A simple derivation of
  the {L}indblad equation},\ }\href
  {https://doi.org/10.1590/S1806-11172013000100003} {\bibfield  {journal}
  {\bibinfo  {journal} {Revista Brasileira de Ensino de F{\'\i}sica}\ }\textbf
  {\bibinfo {volume} {35}},\ \bibinfo {pages} {01} (\bibinfo {year}
  {2013})}\BibitemShut {NoStop}%
\bibitem [{\citenamefont {Hellwig}\ and\ \citenamefont
  {Kraus}(1969)}]{KrausMeas1969}%
  \BibitemOpen
  \bibfield  {author} {\bibinfo {author} {\bibfnamefont {K.-E.}\ \bibnamefont
  {Hellwig}}\ and\ \bibinfo {author} {\bibfnamefont {K.}~\bibnamefont
  {Kraus}},\ }\bibfield  {title} {\bibinfo {title} {Pure operations and
  measurements},\ }\href {https://doi.org/10.1007/BF01645807} {\bibfield
  {journal} {\bibinfo  {journal} {Commun. Math. Phys.}\ }\textbf {\bibinfo
  {volume} {11}},\ \bibinfo {pages} {214} (\bibinfo {year} {1969})}\BibitemShut
  {NoStop}%
\bibitem [{\citenamefont {Kraus}(1971)}]{KrausMeas1971}%
  \BibitemOpen
  \bibfield  {author} {\bibinfo {author} {\bibfnamefont {K.}~\bibnamefont
  {Kraus}},\ }\bibfield  {title} {\bibinfo {title} {General state changes in
  quantum theory},\ }\href {https://doi.org/10.1016/0003-4916(71)90108-4}
  {\bibfield  {journal} {\bibinfo  {journal} {Ann. Phys.}\ }\textbf {\bibinfo
  {volume} {64}},\ \bibinfo {pages} {311} (\bibinfo {year} {1971})}\BibitemShut
  {NoStop}%
\bibitem [{\citenamefont {Kraus}(1981)}]{KrausMeas1981}%
  \BibitemOpen
  \bibfield  {author} {\bibinfo {author} {\bibfnamefont {K.}~\bibnamefont
  {Kraus}},\ }\bibfield  {title} {\bibinfo {title} {Measuring processes in
  quantum mechanics {I}. {C}ontinuous observation and the watchdog effect},\
  }\href {https://doi.org/10.1007/BF00726936} {\bibfield  {journal} {\bibinfo
  {journal} {Found. Phys.}\ }\textbf {\bibinfo {volume} {11}},\ \bibinfo
  {pages} {547} (\bibinfo {year} {1981})}\BibitemShut {NoStop}%
\bibitem [{\citenamefont {Kraus}(1983)}]{KrausBook}%
  \BibitemOpen
  \bibfield  {author} {\bibinfo {author} {\bibfnamefont {K.}~\bibnamefont
  {Kraus}},\ }\href {https://doi.org/https://doi.org/10.1007/3-540-12732-1}
  {\emph {\bibinfo {title} {{States, Effects, and Operations. Fundamental
  Notions of Quantum Theory.}}}},\ Lecture Notes in Physics\ (\bibinfo
  {publisher} {Springer Berlin},\ \bibinfo {address} {Heidelberg, Germany},\
  \bibinfo {year} {1983})\BibitemShut {NoStop}%
\bibitem [{\citenamefont {Friedman}\ \emph
  {et~al.}(2022{\natexlab{a}})\citenamefont {Friedman}, \citenamefont {Yin},
  \citenamefont {Hong},\ and\ \citenamefont {Lucas}}]{SpeedLimit}%
  \BibitemOpen
  \bibfield  {author} {\bibinfo {author} {\bibfnamefont {A.~J.}\ \bibnamefont
  {Friedman}}, \bibinfo {author} {\bibfnamefont {C.}~\bibnamefont {Yin}},
  \bibinfo {author} {\bibfnamefont {Y.}~\bibnamefont {Hong}},\ and\ \bibinfo
  {author} {\bibfnamefont {A.}~\bibnamefont {Lucas}},\ }\bibfield  {title}
  {\bibinfo {title} {Locality and error correction in quantum dynamics with
  measurement},\ }\bibfield  {journal} {\bibinfo  {journal} {arXiv}\ }\href
  {https://doi.org/https://doi.org/10.48550/arXiv.2206.09929}
  {https://doi.org/10.48550/arXiv.2206.09929} (\bibinfo {year}
  {2022}{\natexlab{a}}),\ \Eprint {https://arxiv.org/abs/2206.09929}
  {ar{X}iv:2206.09929 [quant-ph]} \BibitemShut {NoStop}%
\bibitem [{\citenamefont {Friedman}\ \emph
  {et~al.}(2022{\natexlab{b}})\citenamefont {Friedman}, \citenamefont {Hart},\
  and\ \citenamefont {Nandkishore}}]{AaronMIPT}%
  \BibitemOpen
  \bibfield  {author} {\bibinfo {author} {\bibfnamefont {A.~J.}\ \bibnamefont
  {Friedman}}, \bibinfo {author} {\bibfnamefont {O.}~\bibnamefont {Hart}},\
  and\ \bibinfo {author} {\bibfnamefont {R.}~\bibnamefont {Nandkishore}},\
  }\href {https://doi.org/10.48550/arXiv.2210.07256} {\bibinfo {title}
  {Measurement-induced phases of matter require feedback}} (\bibinfo {year}
  {2022}{\natexlab{b}}),\ \Eprint {https://arxiv.org/abs/2210.07256}
  {ar{X}iv:2210.07256 [quant-ph]} \BibitemShut {NoStop}%
\bibitem [{\citenamefont {Hong}\ \emph {et~al.}(2023)\citenamefont {Hong},
  \citenamefont {Stephen},\ and\ \citenamefont {Friedman}}]{AaronTeleport}%
  \BibitemOpen
  \bibfield  {author} {\bibinfo {author} {\bibfnamefont {Y.}~\bibnamefont
  {Hong}}, \bibinfo {author} {\bibfnamefont {D.~T.}\ \bibnamefont {Stephen}},\
  and\ \bibinfo {author} {\bibfnamefont {A.~J.}\ \bibnamefont {Friedman}},\
  }\href {https://doi.org/https://doi.org/10.48550/arXiv.2310.12227} {\bibinfo
  {title} {Quantum teleportation implies symmetry-protected topological order}}
  (\bibinfo {year} {2023}),\ \Eprint {https://arxiv.org/abs/2310.12227}
  {ar{X}iv:2310.12227 [quant-ph]} \BibitemShut {NoStop}%
\bibitem [{\citenamefont {Nielsen}\ and\ \citenamefont
  {Chuang}(2010)}]{QC_book}%
  \BibitemOpen
  \bibfield  {author} {\bibinfo {author} {\bibfnamefont {M.~A.}\ \bibnamefont
  {Nielsen}}\ and\ \bibinfo {author} {\bibfnamefont {I.~L.}\ \bibnamefont
  {Chuang}},\ }\href {https://doi.org/10.1017/CBO9780511976667} {\emph
  {\bibinfo {title} {{Quantum Computation and Quantum Information}}}}\
  (\bibinfo  {publisher} {Cambridge University Press},\ \bibinfo {address}
  {Cambridge, {UK}},\ \bibinfo {year} {2010})\BibitemShut {NoStop}%
\bibitem [{\citenamefont {Jozsa}(2005)}]{Jozsa-intro-MBQC}%
  \BibitemOpen
  \bibfield  {author} {\bibinfo {author} {\bibfnamefont {R.}~\bibnamefont
  {Jozsa}},\ }\href
  {https://doi.org/https://doi.org/10.48550/arXiv.quant-ph/0508124} {\bibinfo
  {title} {An introduction to measurement-based quantum computation}} (\bibinfo
  {year} {2005}),\ \Eprint {https://arxiv.org/abs/quant-ph/0508124}
  {ar{X}iv:quant-ph/0508124 [quant-ph]} \BibitemShut {NoStop}%
\bibitem [{\citenamefont {Paulsen}(2003)}]{PaulsenBook}%
  \BibitemOpen
  \bibfield  {author} {\bibinfo {author} {\bibfnamefont {V.}~\bibnamefont
  {Paulsen}},\ }\href {https://doi.org/10.1017/CBO9780511546631} {\emph
  {\bibinfo {title} {Completely Bounded Maps and Operator Algebras}}},\
  Cambridge Studies in Advanced Mathematics\ (\bibinfo  {publisher} {Cambridge
  University Press},\ \bibinfo {address} {Cambridge, {UK}},\ \bibinfo {year}
  {2003})\BibitemShut {NoStop}%
\bibitem [{\citenamefont {Takesaki}(2012)}]{Takesaki}%
  \BibitemOpen
  \bibfield  {author} {\bibinfo {author} {\bibfnamefont {M.}~\bibnamefont
  {Takesaki}},\ }\href {https://doi.org/10.1007/978-1-4612-6188-9} {\emph
  {\bibinfo {title} {Theory of Operator Algebras {I}}}}\ (\bibinfo  {publisher}
  {Springer New York},\ \bibinfo {address} {New York, {US}},\ \bibinfo {year}
  {2012})\BibitemShut {NoStop}%
\bibitem [{\citenamefont {Stinespring}(1955)}]{Stinespring}%
  \BibitemOpen
  \bibfield  {author} {\bibinfo {author} {\bibfnamefont {W.~F.}\ \bibnamefont
  {Stinespring}},\ }\bibfield  {title} {\bibinfo {title} {Positive functions on
  {${C}^*$}-algebras},\ }\href {https://doi.org/10.2307/2032342} {\bibfield
  {journal} {\bibinfo  {journal} {Proc. Amer. Math. Soc.}\ }\textbf {\bibinfo
  {volume} {6}},\ \bibinfo {pages} {211} (\bibinfo {year} {1955})}\BibitemShut
  {NoStop}%
\bibitem [{\citenamefont {Choi}(1975)}]{ChoisThm}%
  \BibitemOpen
  \bibfield  {author} {\bibinfo {author} {\bibfnamefont {M.-D.}\ \bibnamefont
  {Choi}},\ }\bibfield  {title} {\bibinfo {title} {Completely positive linear
  maps on complex matrices},\ }\href
  {https://doi.org/https://doi.org/10.1016/0024-3795(75)90075-0} {\bibfield
  {journal} {\bibinfo  {journal} {Lin. Alg. Appl.}\ }\textbf {\bibinfo {volume}
  {10}},\ \bibinfo {pages} {285} (\bibinfo {year} {1975})}\BibitemShut
  {NoStop}%
\bibitem [{\citenamefont {Wolf}(2012)}]{WolfNotes}%
  \BibitemOpen
  \bibfield  {author} {\bibinfo {author} {\bibfnamefont {M.~M.}\ \bibnamefont
  {Wolf}},\ }\href {https://mediatum.ub.tum.de/node?id=1701036} {\bibinfo
  {title} {{Quantum Channels and Operations --- Guided Tour}}} (\bibinfo {year}
  {2012}),\ \bibinfo {note} {graue Literatur}\BibitemShut {NoStop}%
\bibitem [{\citenamefont {Friedman}\ and\ \citenamefont
  {Woodcock}(2024)}]{AaronJamesFuture}%
  \BibitemOpen
  \bibfield  {author} {\bibinfo {author} {\bibfnamefont {A.~J.}\ \bibnamefont
  {Friedman}}\ and\ \bibinfo {author} {\bibfnamefont {J.}~\bibnamefont
  {Woodcock}},\ }\href@noop {} {\bibinfo {title} {Dilation theorems for quantum
  operations}} (\bibinfo {year} {2024}),\ \bibinfo {note}
  {to~appear}\BibitemShut {NoStop}%
\bibitem [{\citenamefont {von Neumann}(2018)}]{VonNeumann}%
  \BibitemOpen
  \bibfield  {author} {\bibinfo {author} {\bibfnamefont {J.}~\bibnamefont {von
  Neumann}},\ }\href {http://www.jstor.org/stable/j.ctt1wq8zhp} {\emph
  {\bibinfo {title} {Mathematical Foundations of Quantum Mechanics: New
  Edition}}},\ edited by\ \bibinfo {editor} {\bibfnamefont {R.~T.}\
  \bibnamefont {Beyer}}\ and\ \bibinfo {editor} {\bibfnamefont {N.~A.}\
  \bibnamefont {Wheeler}}\ (\bibinfo  {publisher} {Princeton University
  Press},\ \bibinfo {address} {US},\ \bibinfo {year} {2018})\BibitemShut
  {NoStop}%
\bibitem [{\citenamefont {Preskill}(2022)}]{Preskill_QI}%
  \BibitemOpen
  \bibfield  {author} {\bibinfo {author} {\bibfnamefont {J.}~\bibnamefont
  {Preskill}},\ }\href@noop {} {\bibinfo {title} {{The Physics of Quantum
  Information}}} (\bibinfo {year} {2022}),\ \Eprint
  {https://arxiv.org/abs/2208.08064} {ar{X}iv:2208.08064 [quant-ph]}
  \BibitemShut {NoStop}%
\bibitem [{\citenamefont {Davidson}(1996)}]{Davidson96}%
  \BibitemOpen
  \bibfield  {author} {\bibinfo {author} {\bibfnamefont {K.~R.}\ \bibnamefont
  {Davidson}},\ }\bibfield  {title} {\bibinfo {title} {{$C^*$-Algebras by
  Example}},\ }in\ \href {https://api.semanticscholar.org/CorpusID:117518955}
  {\emph {\bibinfo {booktitle} {{Fields Institute Monographs}}}},\
  Vol.~\bibinfo {volume} {6}\ (\bibinfo  {publisher} {AMS},\ \bibinfo {year}
  {1996})\BibitemShut {NoStop}%
\bibitem [{\citenamefont {Brown}(1982)}]{Brown}%
  \BibitemOpen
  \bibfield  {author} {\bibinfo {author} {\bibfnamefont {L.~G.}\ \bibnamefont
  {Brown}},\ }\bibfield  {title} {\bibinfo {title} {{Extensions of {AF}
  Algebras: The Projection Lifting Problem}},\ }in\ \href
  {https://bookstore.ams.org/pspum-38-1/} {\emph {\bibinfo {booktitle}
  {{Operator Algebras and Applications, Part 1}}}},\ \bibinfo {series}
  {{Proceedings of Symposia in Pure Mathematics}}, Vol.~\bibinfo {volume} {38}\
  (\bibinfo  {publisher} {AMS},\ \bibinfo {year} {1982})\BibitemShut {NoStop}%
\bibitem [{\citenamefont {Bratteli}(1972)}]{Bratteli}%
  \BibitemOpen
  \bibfield  {author} {\bibinfo {author} {\bibfnamefont {O.}~\bibnamefont
  {Bratteli}},\ }\bibfield  {title} {\bibinfo {title} {Inductive limits of
  finite-dimensional {$C^*$}-algebras},\ }\href
  {http://www.jstor.org/stable/1996380} {\bibfield  {journal} {\bibinfo
  {journal} {Trans. Amer. Math. Soc.}\ }\textbf {\bibinfo {volume} {171}},\
  \bibinfo {pages} {195} (\bibinfo {year} {1972})}\BibitemShut {NoStop}%
\bibitem [{\citenamefont {Drummond}\ and\ \citenamefont
  {Ficek}(2004)}]{SqueezeBook}%
  \BibitemOpen
  \bibfield  {author} {\bibinfo {author} {\bibfnamefont {P.~D.}\ \bibnamefont
  {Drummond}}\ and\ \bibinfo {author} {\bibfnamefont {Z.}~\bibnamefont
  {Ficek}},\ }\href {https://doi.org/10.1007/978-3-662-09645-1} {\emph
  {\bibinfo {title} {Quantum Squeezing}}}\ (\bibinfo  {publisher} {Springer
  Berlin},\ \bibinfo {address} {Heidelberg, Germany},\ \bibinfo {year}
  {2004})\BibitemShut {NoStop}%
\bibitem [{\citenamefont {Jacobs}(2014)}]{JacobsBook}%
  \BibitemOpen
  \bibfield  {author} {\bibinfo {author} {\bibfnamefont {K.}~\bibnamefont
  {Jacobs}},\ }\href {https://doi.org/0.1007/978-3-662-09645-1_3} {\emph
  {\bibinfo {title} {Quantum Measurement Theory and its Applications}}}\
  (\bibinfo  {publisher} {Cambridge University Press},\ \bibinfo {year}
  {2014})\BibitemShut {NoStop}%
\bibitem [{\citenamefont {Gerlach}\ and\ \citenamefont
  {Stern}(1922{\natexlab{a}})}]{SternGerlach1}%
  \BibitemOpen
  \bibfield  {author} {\bibinfo {author} {\bibfnamefont {W.}~\bibnamefont
  {Gerlach}}\ and\ \bibinfo {author} {\bibfnamefont {O.}~\bibnamefont
  {Stern}},\ }\bibfield  {title} {\bibinfo {title} {{Der experimentelle
  Nachweis des magnetischen Moments des Silberatoms}},\ }\href
  {https://doi.org/10.1007/BF01329580} {\bibfield  {journal} {\bibinfo
  {journal} {{Zeitschrift f{\"u}r Physik}}\ }\textbf {\bibinfo {volume} {8}},\
  \bibinfo {pages} {110} (\bibinfo {year} {1922}{\natexlab{a}})}\BibitemShut
  {NoStop}%
\bibitem [{\citenamefont {Gerlach}\ and\ \citenamefont
  {Stern}(1922{\natexlab{b}})}]{SternGerlach2}%
  \BibitemOpen
  \bibfield  {author} {\bibinfo {author} {\bibfnamefont {W.}~\bibnamefont
  {Gerlach}}\ and\ \bibinfo {author} {\bibfnamefont {O.}~\bibnamefont
  {Stern}},\ }\bibfield  {title} {\bibinfo {title} {{Das magnetische Moment des
  Silberatoms}},\ }\href {https://doi.org/10.1007/BF01326984} {\bibfield
  {journal} {\bibinfo  {journal} {{Zeitschrift f{\"u}r Physik}}\ }\textbf
  {\bibinfo {volume} {9}},\ \bibinfo {pages} {353} (\bibinfo {year}
  {1922}{\natexlab{b}})}\BibitemShut {NoStop}%
\bibitem [{\citenamefont {Gerlach}\ and\ \citenamefont
  {Stern}(1922{\natexlab{c}})}]{SternGerlach3}%
  \BibitemOpen
  \bibfield  {author} {\bibinfo {author} {\bibfnamefont {W.}~\bibnamefont
  {Gerlach}}\ and\ \bibinfo {author} {\bibfnamefont {O.}~\bibnamefont
  {Stern}},\ }\bibfield  {title} {\bibinfo {title} {{Der experimentelle
  Nachweis der Richtungsquantelung im Magnetfeld}},\ }\href
  {https://doi.org/10.1007/BF01326983} {\bibfield  {journal} {\bibinfo
  {journal} {{Zeitschrift f{\"u}r Physik}}\ }\textbf {\bibinfo {volume} {9}},\
  \bibinfo {pages} {349} (\bibinfo {year} {1922}{\natexlab{c}})}\BibitemShut
  {NoStop}%
\bibitem [{\citenamefont {Busch}(2010)}]{WignerBusch}%
  \BibitemOpen
  \bibfield  {author} {\bibinfo {author} {\bibfnamefont {P.}~\bibnamefont
  {Busch}},\ }\href@noop {} {\bibinfo {title} {Translation of ``{Die Messung
  quantenmechanischer Operatoren}'' by {E.P.~Wigner}}} (\bibinfo {year}
  {2010}),\ \Eprint {https://arxiv.org/abs/1012.4372} {ar{X}iv:1012.4372
  [quant-ph]} \BibitemShut {NoStop}%
\bibitem [{\citenamefont {Araki}\ and\ \citenamefont
  {Yanase}(1960)}]{ArakiYanase}%
  \BibitemOpen
  \bibfield  {author} {\bibinfo {author} {\bibfnamefont {H.}~\bibnamefont
  {Araki}}\ and\ \bibinfo {author} {\bibfnamefont {M.~M.}\ \bibnamefont
  {Yanase}},\ }\bibfield  {title} {\bibinfo {title} {Measurement of quantum
  mechanical operators},\ }\href {https://doi.org/10.1103/PhysRev.120.622}
  {\bibfield  {journal} {\bibinfo  {journal} {Phys. Rev.}\ }\textbf {\bibinfo
  {volume} {120}},\ \bibinfo {pages} {622} (\bibinfo {year}
  {1960})}\BibitemShut {NoStop}%
\bibitem [{\citenamefont {Le~Bellac}(2006)}]{LeBellac}%
  \BibitemOpen
  \bibfield  {author} {\bibinfo {author} {\bibfnamefont {M.}~\bibnamefont
  {Le~Bellac}},\ }\href {https://doi.org/10.1017/CBO9780511616471} {\emph
  {\bibinfo {title} {Quantum Physics}}},\ edited by\ \bibinfo {editor}
  {\bibfnamefont {P.}~\bibnamefont {Forcrand-Millard}}\ (\bibinfo  {publisher}
  {Cambridge University Press},\ \bibinfo {year} {2006})\BibitemShut {NoStop}%
\bibitem [{\citenamefont {Ehrenfest}(1927)}]{Ehrenfest}%
  \BibitemOpen
  \bibfield  {author} {\bibinfo {author} {\bibfnamefont {P.}~\bibnamefont
  {Ehrenfest}},\ }\bibfield  {title} {\bibinfo {title} {Bemerkung {\"u}ber die
  angen{\"a}herte {G}{\"u}ltigkeit der klassischen {M}echanik innerhalb der
  {Q}uantenmechanik},\ }\href {https://doi.org/10.1007/BF01329203} {\bibfield
  {journal} {\bibinfo  {journal} {Zeitschrift f{\"u}r {P}hysik}\ }\textbf
  {\bibinfo {volume} {45}},\ \bibinfo {pages} {455} (\bibinfo {year}
  {1927})}\BibitemShut {NoStop}%
\bibitem [{\citenamefont {Magnus}(1954)}]{MagnusBCH}%
  \BibitemOpen
  \bibfield  {author} {\bibinfo {author} {\bibfnamefont {W.}~\bibnamefont
  {Magnus}},\ }\bibfield  {title} {\bibinfo {title} {On the exponential
  solution of differential equations for a linear operator},\ }\href
  {https://doi.org/https://doi.org/10.1002/cpa.3160070404} {\bibfield
  {journal} {\bibinfo  {journal} {CPAM}\ }\textbf {\bibinfo {volume} {7}},\
  \bibinfo {pages} {649} (\bibinfo {year} {1954})}\BibitemShut {NoStop}%
\bibitem [{\citenamefont {Nogues}\ \emph {et~al.}(1999)\citenamefont {Nogues},
  \citenamefont {Rauschenbeutel}, \citenamefont {Osnaghi}, \citenamefont
  {Brune}, \citenamefont {Raimond},\ and\ \citenamefont
  {Haroche}}]{Nogues1999}%
  \BibitemOpen
  \bibfield  {author} {\bibinfo {author} {\bibfnamefont {G.}~\bibnamefont
  {Nogues}}, \bibinfo {author} {\bibfnamefont {A.}~\bibnamefont
  {Rauschenbeutel}}, \bibinfo {author} {\bibfnamefont {S.}~\bibnamefont
  {Osnaghi}}, \bibinfo {author} {\bibfnamefont {M.}~\bibnamefont {Brune}},
  \bibinfo {author} {\bibfnamefont {J.~M.}\ \bibnamefont {Raimond}},\ and\
  \bibinfo {author} {\bibfnamefont {S.}~\bibnamefont {Haroche}},\ }\bibfield
  {title} {\bibinfo {title} {Seeing a single photon without destroying it},\
  }\href {https://doi.org/10.1038/22275} {\bibfield  {journal} {\bibinfo
  {journal} {Nature}\ }\textbf {\bibinfo {volume} {400}},\ \bibinfo {pages}
  {239} (\bibinfo {year} {1999})}\BibitemShut {NoStop}%
\bibitem [{\citenamefont {Reiserer}\ \emph {et~al.}(2013)\citenamefont
  {Reiserer}, \citenamefont {Ritter},\ and\ \citenamefont
  {Rempe}}]{Reiserer2013}%
  \BibitemOpen
  \bibfield  {author} {\bibinfo {author} {\bibfnamefont {A.}~\bibnamefont
  {Reiserer}}, \bibinfo {author} {\bibfnamefont {S.}~\bibnamefont {Ritter}},\
  and\ \bibinfo {author} {\bibfnamefont {G.}~\bibnamefont {Rempe}},\ }\bibfield
   {title} {\bibinfo {title} {Nondestructive detection of an optical photon},\
  }\href {https://doi.org/10.1126/science.1246164} {\bibfield  {journal}
  {\bibinfo  {journal} {Science}\ }\textbf {\bibinfo {volume} {342}},\ \bibinfo
  {pages} {1349} (\bibinfo {year} {2013})}\BibitemShut {NoStop}%
\bibitem [{\citenamefont {Niemietz}\ \emph {et~al.}(2021)\citenamefont
  {Niemietz}, \citenamefont {Farrera}, \citenamefont {Langenfeld},\ and\
  \citenamefont {Rempe}}]{Niemietz2021}%
  \BibitemOpen
  \bibfield  {author} {\bibinfo {author} {\bibfnamefont {D.}~\bibnamefont
  {Niemietz}}, \bibinfo {author} {\bibfnamefont {P.}~\bibnamefont {Farrera}},
  \bibinfo {author} {\bibfnamefont {S.}~\bibnamefont {Langenfeld}},\ and\
  \bibinfo {author} {\bibfnamefont {G.}~\bibnamefont {Rempe}},\ }\bibfield
  {title} {\bibinfo {title} {Nondestructive detection of photonic qubits},\
  }\href {https://doi.org/10.1038/s41586-021-03290-z} {\bibfield  {journal}
  {\bibinfo  {journal} {Nature}\ }\textbf {\bibinfo {volume} {591}},\ \bibinfo
  {pages} {570} (\bibinfo {year} {2021})}\BibitemShut {NoStop}%
\bibitem [{\citenamefont {Glauber}(1963)}]{Glauber1963}%
  \BibitemOpen
  \bibfield  {author} {\bibinfo {author} {\bibfnamefont {R.~J.}\ \bibnamefont
  {Glauber}},\ }\bibfield  {title} {\bibinfo {title} {The quantum theory of
  optical coherence},\ }\href {https://doi.org/10.1103/PhysRev.130.2529}
  {\bibfield  {journal} {\bibinfo  {journal} {Phys. Rev.}\ }\textbf {\bibinfo
  {volume} {130}},\ \bibinfo {pages} {2529} (\bibinfo {year}
  {1963})}\BibitemShut {NoStop}%
\bibitem [{\citenamefont {Olivares}\ \emph {et~al.}(2019)\citenamefont
  {Olivares}, \citenamefont {Allevi}, \citenamefont {Caiazzo}, \citenamefont
  {Paris},\ and\ \citenamefont {Bondani}}]{Olivares_2019}%
  \BibitemOpen
  \bibfield  {author} {\bibinfo {author} {\bibfnamefont {S.}~\bibnamefont
  {Olivares}}, \bibinfo {author} {\bibfnamefont {A.}~\bibnamefont {Allevi}},
  \bibinfo {author} {\bibfnamefont {G.}~\bibnamefont {Caiazzo}}, \bibinfo
  {author} {\bibfnamefont {M.~G.~A.}\ \bibnamefont {Paris}},\ and\ \bibinfo
  {author} {\bibfnamefont {M.}~\bibnamefont {Bondani}},\ }\bibfield  {title}
  {\bibinfo {title} {Quantum tomography of light states by
  photon-number-resolving detectors},\ }\href
  {https://doi.org/10.1088/1367-2630/ab4afb} {\bibfield  {journal} {\bibinfo
  {journal} {New J. Phys.}\ }\textbf {\bibinfo {volume} {21}},\ \bibinfo
  {pages} {103045} (\bibinfo {year} {2019})}\BibitemShut {NoStop}%
\bibitem [{\citenamefont {Combes}\ and\ \citenamefont
  {Lund}(2022)}]{Combes2022}%
  \BibitemOpen
  \bibfield  {author} {\bibinfo {author} {\bibfnamefont {J.}~\bibnamefont
  {Combes}}\ and\ \bibinfo {author} {\bibfnamefont {A.~P.}\ \bibnamefont
  {Lund}},\ }\bibfield  {title} {\bibinfo {title} {Homodyne measurement with a
  {S}chr\"odinger cat state as a local oscillator},\ }\href
  {https://doi.org/10.1103/PhysRevA.106.063706} {\bibfield  {journal} {\bibinfo
   {journal} {Phys. Rev. A}\ }\textbf {\bibinfo {volume} {106}},\ \bibinfo
  {pages} {063706} (\bibinfo {year} {2022})}\BibitemShut {NoStop}%
\bibitem [{\citenamefont {Yuen}\ and\ \citenamefont
  {Shapiro}(1978)}]{Yuen1978}%
  \BibitemOpen
  \bibfield  {author} {\bibinfo {author} {\bibfnamefont {H.~P.}\ \bibnamefont
  {Yuen}}\ and\ \bibinfo {author} {\bibfnamefont {J.~H.}\ \bibnamefont
  {Shapiro}},\ }\bibfield  {title} {\bibinfo {title} {Quantum statistics of
  homodyne and heterodyne detection},\ }in\ \href
  {https://doi.org/10.1007/978-1-4757-0665-9_75} {\emph {\bibinfo {booktitle}
  {Coherence and Quantum Optics {IV}}}},\ \bibinfo {editor} {edited by\
  \bibinfo {editor} {\bibfnamefont {L.}~\bibnamefont {Mandel}}\ and\ \bibinfo
  {editor} {\bibfnamefont {E.}~\bibnamefont {Wolf}}}\ (\bibinfo  {publisher}
  {Springer},\ \bibinfo {address} {Boston, {US}},\ \bibinfo {year}
  {1978})\BibitemShut {NoStop}%
\bibitem [{\citenamefont {Collett}\ \emph {et~al.}(1987)\citenamefont
  {Collett}, \citenamefont {Loudon},\ and\ \citenamefont
  {Gardiner}}]{Collett1987}%
  \BibitemOpen
  \bibfield  {author} {\bibinfo {author} {\bibfnamefont {M.~J.}\ \bibnamefont
  {Collett}}, \bibinfo {author} {\bibfnamefont {R.}~\bibnamefont {Loudon}},\
  and\ \bibinfo {author} {\bibfnamefont {C.}~\bibnamefont {Gardiner}},\
  }\bibfield  {title} {\bibinfo {title} {Quantum theory of optical homodyne and
  heterodyne detection},\ }\href {https://doi.org/10.1080/09500348714550811}
  {\bibfield  {journal} {\bibinfo  {journal} {J. Mod. Opt.}\ }\textbf {\bibinfo
  {volume} {34}},\ \bibinfo {pages} {881} (\bibinfo {year} {1987})}\BibitemShut
  {NoStop}%
\bibitem [{\citenamefont {Barchielli}(1990)}]{Barchielli_1990}%
  \BibitemOpen
  \bibfield  {author} {\bibinfo {author} {\bibfnamefont {A.}~\bibnamefont
  {Barchielli}},\ }\bibfield  {title} {\bibinfo {title} {Direct and heterodyne
  detection and other applications of quantum stochastic calculus to quantum
  optics},\ }\href {https://doi.org/10.1088/0954-8998/2/6/002} {\bibfield
  {journal} {\bibinfo  {journal} {J. Eur. Opt. Soc. B Quant. Opt.}\ }\textbf
  {\bibinfo {volume} {2}},\ \bibinfo {pages} {423} (\bibinfo {year}
  {1990})}\BibitemShut {NoStop}%
\bibitem [{\citenamefont {Wiseman}\ and\ \citenamefont
  {Milburn}(1993)}]{Wiseman1993}%
  \BibitemOpen
  \bibfield  {author} {\bibinfo {author} {\bibfnamefont {H.~M.}\ \bibnamefont
  {Wiseman}}\ and\ \bibinfo {author} {\bibfnamefont {G.~J.}\ \bibnamefont
  {Milburn}},\ }\bibfield  {title} {\bibinfo {title} {Quantum theory of
  field-quadrature measurements},\ }\href
  {https://doi.org/10.1103/PhysRevA.47.642} {\bibfield  {journal} {\bibinfo
  {journal} {Phys. Rev. A}\ }\textbf {\bibinfo {volume} {47}},\ \bibinfo
  {pages} {642} (\bibinfo {year} {1993})}\BibitemShut {NoStop}%
\bibitem [{\citenamefont {Lvovsky}\ and\ \citenamefont
  {Raymer}(2009)}]{Lvovsky2009}%
  \BibitemOpen
  \bibfield  {author} {\bibinfo {author} {\bibfnamefont {A.~I.}\ \bibnamefont
  {Lvovsky}}\ and\ \bibinfo {author} {\bibfnamefont {M.~G.}\ \bibnamefont
  {Raymer}},\ }\bibfield  {title} {\bibinfo {title} {Continuous-variable
  optical quantum-state tomography},\ }\href
  {https://doi.org/10.1103/RevModPhys.81.299} {\bibfield  {journal} {\bibinfo
  {journal} {Rev. Mod. Phys.}\ }\textbf {\bibinfo {volume} {81}},\ \bibinfo
  {pages} {299} (\bibinfo {year} {2009})}\BibitemShut {NoStop}%
\bibitem [{\citenamefont {Dennis}(1986)}]{Dennis1986}%
  \BibitemOpen
  \bibfield  {author} {\bibinfo {author} {\bibfnamefont {P.~N.~J.}\
  \bibnamefont {Dennis}},\ }\href {https://doi.org/10.1007/978-1-4613-2171-2}
  {\emph {\bibinfo {title} {Photodetectors: {A}n Introduction to Current
  Technology}}}\ (\bibinfo  {publisher} {Springer US},\ \bibinfo {year}
  {1986})\BibitemShut {NoStop}%
\bibitem [{\citenamefont {Mollow}(1968)}]{Mollow1968}%
  \BibitemOpen
  \bibfield  {author} {\bibinfo {author} {\bibfnamefont {B.~R.}\ \bibnamefont
  {Mollow}},\ }\bibfield  {title} {\bibinfo {title} {Quantum theory of field
  attenuation},\ }\href {https://doi.org/10.1103/PhysRev.168.1896} {\bibfield
  {journal} {\bibinfo  {journal} {Phys. Rev.}\ }\textbf {\bibinfo {volume}
  {168}},\ \bibinfo {pages} {1896} (\bibinfo {year} {1968})}\BibitemShut
  {NoStop}%
\bibitem [{\citenamefont {Scully}\ and\ \citenamefont
  {Lamb}(1969)}]{Scully1969}%
  \BibitemOpen
  \bibfield  {author} {\bibinfo {author} {\bibfnamefont {M.~O.}\ \bibnamefont
  {Scully}}\ and\ \bibinfo {author} {\bibfnamefont {W.~E.}\ \bibnamefont
  {Lamb}},\ }\bibfield  {title} {\bibinfo {title} {Quantum theory of an optical
  maser. {III. T}heory of photoelectron counting statistics},\ }\href
  {https://doi.org/10.1103/PhysRev.179.368} {\bibfield  {journal} {\bibinfo
  {journal} {Phys. Rev.}\ }\textbf {\bibinfo {volume} {179}},\ \bibinfo {pages}
  {368} (\bibinfo {year} {1969})}\BibitemShut {NoStop}%
\bibitem [{\citenamefont {Srinivas}\ and\ \citenamefont
  {Davies}(1981)}]{Srinivas1981}%
  \BibitemOpen
  \bibfield  {author} {\bibinfo {author} {\bibfnamefont {M.}~\bibnamefont
  {Srinivas}}\ and\ \bibinfo {author} {\bibfnamefont {E.}~\bibnamefont
  {Davies}},\ }\bibfield  {title} {\bibinfo {title} {Photon counting
  probabilities in quantum optics},\ }\href {https://doi.org/10.1080/713820643}
  {\bibfield  {journal} {\bibinfo  {journal} {{Optica Acta: Int. J. Opt.}}\
  }\textbf {\bibinfo {volume} {28}},\ \bibinfo {pages} {981} (\bibinfo {year}
  {1981})}\BibitemShut {NoStop}%
\bibitem [{\citenamefont {Chmara}(1987)}]{Chmara1987}%
  \BibitemOpen
  \bibfield  {author} {\bibinfo {author} {\bibfnamefont {W.}~\bibnamefont
  {Chmara}},\ }\bibfield  {title} {\bibinfo {title} {A quantum open-systems
  theory approach to photodetection},\ }\href
  {https://doi.org/10.1080/09500348714550431} {\bibfield  {journal} {\bibinfo
  {journal} {J. Mod. Opt.}\ }\textbf {\bibinfo {volume} {34}},\ \bibinfo
  {pages} {455} (\bibinfo {year} {1987})}\BibitemShut {NoStop}%
\bibitem [{\citenamefont {Fleischhauer}\ and\ \citenamefont
  {Welsch}(1991)}]{Fleischhauer1991}%
  \BibitemOpen
  \bibfield  {author} {\bibinfo {author} {\bibfnamefont {M.}~\bibnamefont
  {Fleischhauer}}\ and\ \bibinfo {author} {\bibfnamefont {D.~G.}\ \bibnamefont
  {Welsch}},\ }\bibfield  {title} {\bibinfo {title} {Nonperturbative approach
  to multimode photodetection},\ }\href
  {https://doi.org/10.1103/PhysRevA.44.747} {\bibfield  {journal} {\bibinfo
  {journal} {Phys. Rev. A}\ }\textbf {\bibinfo {volume} {44}},\ \bibinfo
  {pages} {747} (\bibinfo {year} {1991})}\BibitemShut {NoStop}%
\bibitem [{\citenamefont {Gardiner}\ and\ \citenamefont
  {Zoller}(2004)}]{Gardiner2004-co}%
  \BibitemOpen
  \bibfield  {author} {\bibinfo {author} {\bibfnamefont {C.~W.}\ \bibnamefont
  {Gardiner}}\ and\ \bibinfo {author} {\bibfnamefont {P.}~\bibnamefont
  {Zoller}},\ }\href {https://link.springer.com/book/9783540223016} {\emph
  {\bibinfo {title} {Quantum Noise}}},\ \bibinfo {edition} {3rd}\ ed.,\
  {Springer Series in Synergetics}\ (\bibinfo  {publisher} {Springer},\
  \bibinfo {year} {2004})\BibitemShut {NoStop}%
\bibitem [{\citenamefont {Born}\ and\ \citenamefont {Wolf}(2020)}]{Born2020}%
  \BibitemOpen
  \bibfield  {author} {\bibinfo {author} {\bibfnamefont {M.}~\bibnamefont
  {Born}}\ and\ \bibinfo {author} {\bibfnamefont {E.}~\bibnamefont {Wolf}},\
  }\href {https://doi.org/10.1017/CBO9781139644181} {\emph {\bibinfo {title}
  {Principles of Optics}}},\ \bibinfo {edition} {7th}\ ed.\ (\bibinfo
  {publisher} {Cambridge University Press},\ \bibinfo {address} {{UK}},\
  \bibinfo {year} {2020})\BibitemShut {NoStop}%
\bibitem [{\citenamefont {Tyc}\ and\ \citenamefont {Sanders}(2004)}]{Tyc_2004}%
  \BibitemOpen
  \bibfield  {author} {\bibinfo {author} {\bibfnamefont {T.}~\bibnamefont
  {Tyc}}\ and\ \bibinfo {author} {\bibfnamefont {B.~C.}\ \bibnamefont
  {Sanders}},\ }\bibfield  {title} {\bibinfo {title} {Operational formulation
  of homodyne detection},\ }\href {https://doi.org/10.1088/0305-4470/37/29/010}
  {\bibfield  {journal} {\bibinfo  {journal} {J. Phys. A Math. Gen.}\ }\textbf
  {\bibinfo {volume} {37}},\ \bibinfo {pages} {7341} (\bibinfo {year}
  {2004})}\BibitemShut {NoStop}%
\bibitem [{\citenamefont {Yurke}\ \emph {et~al.}(1986)\citenamefont {Yurke},
  \citenamefont {McCall},\ and\ \citenamefont {Klauder}}]{Yurke1986}%
  \BibitemOpen
  \bibfield  {author} {\bibinfo {author} {\bibfnamefont {B.}~\bibnamefont
  {Yurke}}, \bibinfo {author} {\bibfnamefont {S.~L.}\ \bibnamefont {McCall}},\
  and\ \bibinfo {author} {\bibfnamefont {J.~R.}\ \bibnamefont {Klauder}},\
  }\bibfield  {title} {\bibinfo {title} {{SU(2) and SU(1,1) interferometers}},\
  }\href {https://doi.org/10.1103/PhysRevA.33.4033} {\bibfield  {journal}
  {\bibinfo  {journal} {Phys. Rev. A}\ }\textbf {\bibinfo {volume} {33}},\
  \bibinfo {pages} {4033} (\bibinfo {year} {1986})}\BibitemShut {NoStop}%
\bibitem [{\citenamefont {Prasad}\ \emph {et~al.}(1987)\citenamefont {Prasad},
  \citenamefont {Scully},\ and\ \citenamefont {Martienssen}}]{PRASAD1987}%
  \BibitemOpen
  \bibfield  {author} {\bibinfo {author} {\bibfnamefont {S.}~\bibnamefont
  {Prasad}}, \bibinfo {author} {\bibfnamefont {M.~O.}\ \bibnamefont {Scully}},\
  and\ \bibinfo {author} {\bibfnamefont {W.}~\bibnamefont {Martienssen}},\
  }\bibfield  {title} {\bibinfo {title} {A quantum description of the beam
  splitter},\ }\href
  {https://doi.org/https://doi.org/10.1016/0030-4018(87)90015-0} {\bibfield
  {journal} {\bibinfo  {journal} {Opt. Commun.}\ }\textbf {\bibinfo {volume}
  {62}},\ \bibinfo {pages} {139} (\bibinfo {year} {1987})}\BibitemShut
  {NoStop}%
\bibitem [{\citenamefont {Wallentowitz}\ and\ \citenamefont
  {Vogel}(1996)}]{Wallentowitz1996}%
  \BibitemOpen
  \bibfield  {author} {\bibinfo {author} {\bibfnamefont {S.}~\bibnamefont
  {Wallentowitz}}\ and\ \bibinfo {author} {\bibfnamefont {W.}~\bibnamefont
  {Vogel}},\ }\bibfield  {title} {\bibinfo {title} {Unbalanced homodyning for
  quantum state measurements},\ }\href
  {https://doi.org/10.1103/PhysRevA.53.4528} {\bibfield  {journal} {\bibinfo
  {journal} {Phys. Rev. A}\ }\textbf {\bibinfo {volume} {53}},\ \bibinfo
  {pages} {4528} (\bibinfo {year} {1996})}\BibitemShut {NoStop}%
\bibitem [{\citenamefont {Cives-Esclop}\ \emph {et~al.}(2000)\citenamefont
  {Cives-Esclop}, \citenamefont {Luis},\ and\ \citenamefont
  {Sánchez-Soto}}]{Cives2000}%
  \BibitemOpen
  \bibfield  {author} {\bibinfo {author} {\bibfnamefont {A.}~\bibnamefont
  {Cives-Esclop}}, \bibinfo {author} {\bibfnamefont {A.}~\bibnamefont {Luis}},\
  and\ \bibinfo {author} {\bibfnamefont {L.}~\bibnamefont {Sánchez-Soto}},\
  }\bibfield  {title} {\bibinfo {title} {Unbalanced homodyne detection with a
  weak local oscillator},\ }\href
  {https://doi.org/https://doi.org/10.1016/S0030-4018(99)00734-8} {\bibfield
  {journal} {\bibinfo  {journal} {Optics Communications}\ }\textbf {\bibinfo
  {volume} {175}},\ \bibinfo {pages} {153} (\bibinfo {year}
  {2000})}\BibitemShut {NoStop}%
\bibitem [{\citenamefont {K\"uhn}\ and\ \citenamefont
  {Vogel}(2016)}]{Kuhn2016}%
  \BibitemOpen
  \bibfield  {author} {\bibinfo {author} {\bibfnamefont {B.}~\bibnamefont
  {K\"uhn}}\ and\ \bibinfo {author} {\bibfnamefont {W.}~\bibnamefont {Vogel}},\
  }\bibfield  {title} {\bibinfo {title} {Unbalanced homodyne correlation
  measurements},\ }\href {https://doi.org/10.1103/PhysRevLett.116.163603}
  {\bibfield  {journal} {\bibinfo  {journal} {Phys. Rev. Lett.}\ }\textbf
  {\bibinfo {volume} {116}},\ \bibinfo {pages} {163603} (\bibinfo {year}
  {2016})}\BibitemShut {NoStop}%
\bibitem [{\citenamefont {Smithey}\ \emph {et~al.}(1993)\citenamefont
  {Smithey}, \citenamefont {Beck}, \citenamefont {Raymer},\ and\ \citenamefont
  {Faridani}}]{Smithey1993}%
  \BibitemOpen
  \bibfield  {author} {\bibinfo {author} {\bibfnamefont {D.~T.}\ \bibnamefont
  {Smithey}}, \bibinfo {author} {\bibfnamefont {M.}~\bibnamefont {Beck}},
  \bibinfo {author} {\bibfnamefont {M.~G.}\ \bibnamefont {Raymer}},\ and\
  \bibinfo {author} {\bibfnamefont {A.}~\bibnamefont {Faridani}},\ }\bibfield
  {title} {\bibinfo {title} {Measurement of the {W}igner distribution and the
  density matrix of a light mode using optical homodyne tomography:
  {A}pplication to squeezed states and the vacuum},\ }\href
  {https://doi.org/10.1103/PhysRevLett.70.1244} {\bibfield  {journal} {\bibinfo
   {journal} {Phys. Rev. Lett.}\ }\textbf {\bibinfo {volume} {70}},\ \bibinfo
  {pages} {1244} (\bibinfo {year} {1993})}\BibitemShut {NoStop}%
\bibitem [{\citenamefont {Shapiro}(2009)}]{Shapiro2009}%
  \BibitemOpen
  \bibfield  {author} {\bibinfo {author} {\bibfnamefont {J.~H.}\ \bibnamefont
  {Shapiro}},\ }\bibfield  {title} {\bibinfo {title} {The quantum theory of
  optical communications},\ }\href {https://doi.org/10.1109/JSTQE.2009.2024959}
  {\bibfield  {journal} {\bibinfo  {journal} {IEEE J. Quant. Electr.}\ }\textbf
  {\bibinfo {volume} {15}},\ \bibinfo {pages} {1547} (\bibinfo {year}
  {2009})}\BibitemShut {NoStop}%
\bibitem [{\citenamefont {Leibfried}\ \emph {et~al.}(2003)\citenamefont
  {Leibfried}, \citenamefont {Blatt}, \citenamefont {Monroe},\ and\
  \citenamefont {Wineland}}]{Didi2003}%
  \BibitemOpen
  \bibfield  {author} {\bibinfo {author} {\bibfnamefont {D.}~\bibnamefont
  {Leibfried}}, \bibinfo {author} {\bibfnamefont {R.}~\bibnamefont {Blatt}},
  \bibinfo {author} {\bibfnamefont {C.}~\bibnamefont {Monroe}},\ and\ \bibinfo
  {author} {\bibfnamefont {D.}~\bibnamefont {Wineland}},\ }\bibfield  {title}
  {\bibinfo {title} {Quantum dynamics of single trapped ions},\ }\href
  {https://doi.org/10.1103/RevModPhys.75.281} {\bibfield  {journal} {\bibinfo
  {journal} {Rev. Mod. Phys.}\ }\textbf {\bibinfo {volume} {75}},\ \bibinfo
  {pages} {281} (\bibinfo {year} {2003})}\BibitemShut {NoStop}%
\bibitem [{\citenamefont {Bakr}\ \emph {et~al.}(2010)\citenamefont {Bakr},
  \citenamefont {Peng}, \citenamefont {Tai}, \citenamefont {Ma}, \citenamefont
  {Simon}, \citenamefont {Gillen}, \citenamefont {Fölling}, \citenamefont
  {Pollet},\ and\ \citenamefont {Greiner}}]{Bakr2010}%
  \BibitemOpen
  \bibfield  {author} {\bibinfo {author} {\bibfnamefont {W.~S.}\ \bibnamefont
  {Bakr}}, \bibinfo {author} {\bibfnamefont {A.}~\bibnamefont {Peng}}, \bibinfo
  {author} {\bibfnamefont {M.~E.}\ \bibnamefont {Tai}}, \bibinfo {author}
  {\bibfnamefont {R.}~\bibnamefont {Ma}}, \bibinfo {author} {\bibfnamefont
  {J.}~\bibnamefont {Simon}}, \bibinfo {author} {\bibfnamefont {J.~I.}\
  \bibnamefont {Gillen}}, \bibinfo {author} {\bibfnamefont {S.}~\bibnamefont
  {Fölling}}, \bibinfo {author} {\bibfnamefont {L.}~\bibnamefont {Pollet}},\
  and\ \bibinfo {author} {\bibfnamefont {M.}~\bibnamefont {Greiner}},\
  }\bibfield  {title} {\bibinfo {title} {Probing the superfluid–to–{M}ott
  insulator transition at the single-atom level},\ }\href
  {https://doi.org/10.1126/science.1192368} {\bibfield  {journal} {\bibinfo
  {journal} {Science}\ }\textbf {\bibinfo {volume} {329}},\ \bibinfo {pages}
  {547} (\bibinfo {year} {2010})}\BibitemShut {NoStop}%
\bibitem [{\citenamefont {Sherson}\ \emph {et~al.}(2010)\citenamefont
  {Sherson}, \citenamefont {Weitenberg}, \citenamefont {Endres}, \citenamefont
  {Cheneau}, \citenamefont {Bloch},\ and\ \citenamefont {Kuhr}}]{Sherson2010}%
  \BibitemOpen
  \bibfield  {author} {\bibinfo {author} {\bibfnamefont {J.~F.}\ \bibnamefont
  {Sherson}}, \bibinfo {author} {\bibfnamefont {C.}~\bibnamefont {Weitenberg}},
  \bibinfo {author} {\bibfnamefont {M.}~\bibnamefont {Endres}}, \bibinfo
  {author} {\bibfnamefont {M.}~\bibnamefont {Cheneau}}, \bibinfo {author}
  {\bibfnamefont {I.}~\bibnamefont {Bloch}},\ and\ \bibinfo {author}
  {\bibfnamefont {S.}~\bibnamefont {Kuhr}},\ }\bibfield  {title} {\bibinfo
  {title} {Single-atom-resolved fluorescence imaging of an atomic {M}ott
  insulator},\ }\href {https://doi.org/10.1038/nature09378} {\bibfield
  {journal} {\bibinfo  {journal} {Nature}\ }\textbf {\bibinfo {volume} {467}},\
  \bibinfo {pages} {68} (\bibinfo {year} {2010})}\BibitemShut {NoStop}%
\bibitem [{\citenamefont {Bloch}\ \emph {et~al.}(2012)\citenamefont {Bloch},
  \citenamefont {Dalibard},\ and\ \citenamefont {Nascimb{\`e}ne}}]{Bloch2012}%
  \BibitemOpen
  \bibfield  {author} {\bibinfo {author} {\bibfnamefont {I.}~\bibnamefont
  {Bloch}}, \bibinfo {author} {\bibfnamefont {J.}~\bibnamefont {Dalibard}},\
  and\ \bibinfo {author} {\bibfnamefont {S.}~\bibnamefont {Nascimb{\`e}ne}},\
  }\bibfield  {title} {\bibinfo {title} {Quantum simulations with ultracold
  quantum gases},\ }\href {https://doi.org/10.1038/nphys2259} {\bibfield
  {journal} {\bibinfo  {journal} {Nature Phys.}\ }\textbf {\bibinfo {volume}
  {8}},\ \bibinfo {pages} {267} (\bibinfo {year} {2012})}\BibitemShut {NoStop}%
\bibitem [{\citenamefont {Ott}(2016)}]{Ott_2016}%
  \BibitemOpen
  \bibfield  {author} {\bibinfo {author} {\bibfnamefont {H.}~\bibnamefont
  {Ott}},\ }\bibfield  {title} {\bibinfo {title} {Single-atom detection in
  ultracold quantum gases: {A} review of current progress},\ }\href
  {https://doi.org/10.1088/0034-4885/79/5/054401} {\bibfield  {journal}
  {\bibinfo  {journal} {Rep. Prog. Phys.}\ }\textbf {\bibinfo {volume} {79}},\
  \bibinfo {pages} {054401} (\bibinfo {year} {2016})}\BibitemShut {NoStop}%
\bibitem [{\citenamefont {Doherty}\ \emph {et~al.}(2013)\citenamefont
  {Doherty}, \citenamefont {Manson}, \citenamefont {Delaney}, \citenamefont
  {Jelezko}, \citenamefont {Wrachtrup},\ and\ \citenamefont
  {Hollenberg}}]{DOHERTY20131}%
  \BibitemOpen
  \bibfield  {author} {\bibinfo {author} {\bibfnamefont {M.~W.}\ \bibnamefont
  {Doherty}}, \bibinfo {author} {\bibfnamefont {N.~B.}\ \bibnamefont {Manson}},
  \bibinfo {author} {\bibfnamefont {P.}~\bibnamefont {Delaney}}, \bibinfo
  {author} {\bibfnamefont {F.}~\bibnamefont {Jelezko}}, \bibinfo {author}
  {\bibfnamefont {J.}~\bibnamefont {Wrachtrup}},\ and\ \bibinfo {author}
  {\bibfnamefont {L.~C.}\ \bibnamefont {Hollenberg}},\ }\bibfield  {title}
  {\bibinfo {title} {The nitrogen-vacancy colour centre in diamond},\ }\href
  {https://doi.org/https://doi.org/10.1016/j.physrep.2013.02.001} {\bibfield
  {journal} {\bibinfo  {journal} {Physics Reports}\ }\textbf {\bibinfo {volume}
  {528}},\ \bibinfo {pages} {1} (\bibinfo {year} {2013})}\BibitemShut {NoStop}%
\bibitem [{\citenamefont {Barry}\ \emph {et~al.}(2020)\citenamefont {Barry},
  \citenamefont {Schloss}, \citenamefont {Bauch}, \citenamefont {Turner},
  \citenamefont {Hart}, \citenamefont {Pham},\ and\ \citenamefont
  {Walsworth}}]{Barry2020}%
  \BibitemOpen
  \bibfield  {author} {\bibinfo {author} {\bibfnamefont {J.~F.}\ \bibnamefont
  {Barry}}, \bibinfo {author} {\bibfnamefont {J.~M.}\ \bibnamefont {Schloss}},
  \bibinfo {author} {\bibfnamefont {E.}~\bibnamefont {Bauch}}, \bibinfo
  {author} {\bibfnamefont {M.~J.}\ \bibnamefont {Turner}}, \bibinfo {author}
  {\bibfnamefont {C.~A.}\ \bibnamefont {Hart}}, \bibinfo {author}
  {\bibfnamefont {L.~M.}\ \bibnamefont {Pham}},\ and\ \bibinfo {author}
  {\bibfnamefont {R.~L.}\ \bibnamefont {Walsworth}},\ }\bibfield  {title}
  {\bibinfo {title} {Sensitivity optimization for {NV}-diamond magnetometry},\
  }\href {https://doi.org/10.1103/RevModPhys.92.015004} {\bibfield  {journal}
  {\bibinfo  {journal} {Rev. Mod. Phys.}\ }\textbf {\bibinfo {volume} {92}},\
  \bibinfo {pages} {015004} (\bibinfo {year} {2020})}\BibitemShut {NoStop}%
\bibitem [{\citenamefont {Nagourney}\ \emph {et~al.}(1986)\citenamefont
  {Nagourney}, \citenamefont {Sandberg},\ and\ \citenamefont
  {Dehmelt}}]{Nagourney1986}%
  \BibitemOpen
  \bibfield  {author} {\bibinfo {author} {\bibfnamefont {W.}~\bibnamefont
  {Nagourney}}, \bibinfo {author} {\bibfnamefont {J.}~\bibnamefont
  {Sandberg}},\ and\ \bibinfo {author} {\bibfnamefont {H.}~\bibnamefont
  {Dehmelt}},\ }\bibfield  {title} {\bibinfo {title} {Shelved optical electron
  amplifier: Observation of quantum jumps},\ }\href
  {https://doi.org/10.1103/PhysRevLett.56.2797} {\bibfield  {journal} {\bibinfo
   {journal} {Phys. Rev. Lett.}\ }\textbf {\bibinfo {volume} {56}},\ \bibinfo
  {pages} {2797} (\bibinfo {year} {1986})}\BibitemShut {NoStop}%
\bibitem [{\citenamefont {Myerson}\ \emph {et~al.}(2008)\citenamefont
  {Myerson}, \citenamefont {Szwer}, \citenamefont {Webster}, \citenamefont
  {Allcock}, \citenamefont {Curtis}, \citenamefont {Imreh}, \citenamefont
  {Sherman}, \citenamefont {Stacey}, \citenamefont {Steane},\ and\
  \citenamefont {Lucas}}]{FluorescentMeas}%
  \BibitemOpen
  \bibfield  {author} {\bibinfo {author} {\bibfnamefont {A.~H.}\ \bibnamefont
  {Myerson}}, \bibinfo {author} {\bibfnamefont {D.~J.}\ \bibnamefont {Szwer}},
  \bibinfo {author} {\bibfnamefont {S.~C.}\ \bibnamefont {Webster}}, \bibinfo
  {author} {\bibfnamefont {D.~T.~C.}\ \bibnamefont {Allcock}}, \bibinfo
  {author} {\bibfnamefont {M.~J.}\ \bibnamefont {Curtis}}, \bibinfo {author}
  {\bibfnamefont {G.}~\bibnamefont {Imreh}}, \bibinfo {author} {\bibfnamefont
  {J.~A.}\ \bibnamefont {Sherman}}, \bibinfo {author} {\bibfnamefont {D.~N.}\
  \bibnamefont {Stacey}}, \bibinfo {author} {\bibfnamefont {A.~M.}\
  \bibnamefont {Steane}},\ and\ \bibinfo {author} {\bibfnamefont {D.~M.}\
  \bibnamefont {Lucas}},\ }\bibfield  {title} {\bibinfo {title} {High-fidelity
  readout of trapped-ion qubits},\ }\href
  {https://doi.org/10.1103/PhysRevLett.100.200502} {\bibfield  {journal}
  {\bibinfo  {journal} {Phys. Rev. Lett.}\ }\textbf {\bibinfo {volume} {100}},\
  \bibinfo {pages} {200502} (\bibinfo {year} {2008})}\BibitemShut {NoStop}%
\bibitem [{\citenamefont {Bohnet}\ \emph {et~al.}(2016)\citenamefont {Bohnet},
  \citenamefont {Sawyer}, \citenamefont {Britton}, \citenamefont {Wall},
  \citenamefont {Rey}, \citenamefont {Foss-Feig},\ and\ \citenamefont
  {Bollinger}}]{Bohnet2016}%
  \BibitemOpen
  \bibfield  {author} {\bibinfo {author} {\bibfnamefont {J.~G.}\ \bibnamefont
  {Bohnet}}, \bibinfo {author} {\bibfnamefont {B.~C.}\ \bibnamefont {Sawyer}},
  \bibinfo {author} {\bibfnamefont {J.~W.}\ \bibnamefont {Britton}}, \bibinfo
  {author} {\bibfnamefont {M.~L.}\ \bibnamefont {Wall}}, \bibinfo {author}
  {\bibfnamefont {A.~M.}\ \bibnamefont {Rey}}, \bibinfo {author} {\bibfnamefont
  {M.}~\bibnamefont {Foss-Feig}},\ and\ \bibinfo {author} {\bibfnamefont
  {J.~J.}\ \bibnamefont {Bollinger}},\ }\bibfield  {title} {\bibinfo {title}
  {Quantum spin dynamics and entanglement generation with hundreds of trapped
  ions},\ }\href {https://doi.org/10.1126/science.aad9958} {\bibfield
  {journal} {\bibinfo  {journal} {Science}\ }\textbf {\bibinfo {volume}
  {352}},\ \bibinfo {pages} {1297} (\bibinfo {year} {2016})}\BibitemShut
  {NoStop}%
\bibitem [{\citenamefont {de~Clercq}\ \emph {et~al.}(2016)\citenamefont
  {de~Clercq}, \citenamefont {Lo}, \citenamefont {Marinelli}, \citenamefont
  {Nadlinger}, \citenamefont {Oswald}, \citenamefont {Negnevitsky},
  \citenamefont {Kienzler}, \citenamefont {Keitch},\ and\ \citenamefont
  {Home}}]{deClercq2016}%
  \BibitemOpen
  \bibfield  {author} {\bibinfo {author} {\bibfnamefont {L.~E.}\ \bibnamefont
  {de~Clercq}}, \bibinfo {author} {\bibfnamefont {H.-Y.}\ \bibnamefont {Lo}},
  \bibinfo {author} {\bibfnamefont {M.}~\bibnamefont {Marinelli}}, \bibinfo
  {author} {\bibfnamefont {D.}~\bibnamefont {Nadlinger}}, \bibinfo {author}
  {\bibfnamefont {R.}~\bibnamefont {Oswald}}, \bibinfo {author} {\bibfnamefont
  {V.}~\bibnamefont {Negnevitsky}}, \bibinfo {author} {\bibfnamefont
  {D.}~\bibnamefont {Kienzler}}, \bibinfo {author} {\bibfnamefont
  {B.}~\bibnamefont {Keitch}},\ and\ \bibinfo {author} {\bibfnamefont {J.~P.}\
  \bibnamefont {Home}},\ }\bibfield  {title} {\bibinfo {title} {Parallel
  transport quantum logic gates with trapped ions},\ }\href
  {https://doi.org/10.1103/PhysRevLett.116.080502} {\bibfield  {journal}
  {\bibinfo  {journal} {Phys. Rev. Lett.}\ }\textbf {\bibinfo {volume} {116}},\
  \bibinfo {pages} {080502} (\bibinfo {year} {2016})}\BibitemShut {NoStop}%
\bibitem [{\citenamefont {Cui}\ \emph {et~al.}(2022)\citenamefont {Cui},
  \citenamefont {Valencia}, \citenamefont {Boyce}, \citenamefont {Clements},
  \citenamefont {Leibrandt},\ and\ \citenamefont {Hume}}]{Cui2022}%
  \BibitemOpen
  \bibfield  {author} {\bibinfo {author} {\bibfnamefont {K.}~\bibnamefont
  {Cui}}, \bibinfo {author} {\bibfnamefont {J.}~\bibnamefont {Valencia}},
  \bibinfo {author} {\bibfnamefont {K.~T.}\ \bibnamefont {Boyce}}, \bibinfo
  {author} {\bibfnamefont {E.~R.}\ \bibnamefont {Clements}}, \bibinfo {author}
  {\bibfnamefont {D.~R.}\ \bibnamefont {Leibrandt}},\ and\ \bibinfo {author}
  {\bibfnamefont {D.~B.}\ \bibnamefont {Hume}},\ }\bibfield  {title} {\bibinfo
  {title} {Scalable quantum logic spectroscopy},\ }\href
  {https://doi.org/10.1103/PhysRevLett.129.193603} {\bibfield  {journal}
  {\bibinfo  {journal} {Phys. Rev. Lett.}\ }\textbf {\bibinfo {volume} {129}},\
  \bibinfo {pages} {193603} (\bibinfo {year} {2022})}\BibitemShut {NoStop}%
\bibitem [{\citenamefont {Kwon}\ \emph {et~al.}(2017)\citenamefont {Kwon},
  \citenamefont {Ebert}, \citenamefont {Walker},\ and\ \citenamefont
  {Saffman}}]{Kwon2017}%
  \BibitemOpen
  \bibfield  {author} {\bibinfo {author} {\bibfnamefont {M.}~\bibnamefont
  {Kwon}}, \bibinfo {author} {\bibfnamefont {M.~F.}\ \bibnamefont {Ebert}},
  \bibinfo {author} {\bibfnamefont {T.~G.}\ \bibnamefont {Walker}},\ and\
  \bibinfo {author} {\bibfnamefont {M.}~\bibnamefont {Saffman}},\ }\bibfield
  {title} {\bibinfo {title} {Parallel low-loss measurement of multiple atomic
  qubits},\ }\href {https://doi.org/10.1103/PhysRevLett.119.180504} {\bibfield
  {journal} {\bibinfo  {journal} {Phys. Rev. Lett.}\ }\textbf {\bibinfo
  {volume} {119}},\ \bibinfo {pages} {180504} (\bibinfo {year}
  {2017})}\BibitemShut {NoStop}%
\bibitem [{\citenamefont {Martinez-Dorantes}\ \emph {et~al.}(2017)\citenamefont
  {Martinez-Dorantes}, \citenamefont {Alt}, \citenamefont {Gallego},
  \citenamefont {Ghosh}, \citenamefont {Ratschbacher}, \citenamefont
  {V\"olzke},\ and\ \citenamefont {Meschede}}]{Martinez2018b}%
  \BibitemOpen
  \bibfield  {author} {\bibinfo {author} {\bibfnamefont {M.}~\bibnamefont
  {Martinez-Dorantes}}, \bibinfo {author} {\bibfnamefont {W.}~\bibnamefont
  {Alt}}, \bibinfo {author} {\bibfnamefont {J.}~\bibnamefont {Gallego}},
  \bibinfo {author} {\bibfnamefont {S.}~\bibnamefont {Ghosh}}, \bibinfo
  {author} {\bibfnamefont {L.}~\bibnamefont {Ratschbacher}}, \bibinfo {author}
  {\bibfnamefont {Y.}~\bibnamefont {V\"olzke}},\ and\ \bibinfo {author}
  {\bibfnamefont {D.}~\bibnamefont {Meschede}},\ }\bibfield  {title} {\bibinfo
  {title} {Fast nondestructive parallel readout of neutral atom registers in
  optical potentials},\ }\href {https://doi.org/10.1103/PhysRevLett.119.180503}
  {\bibfield  {journal} {\bibinfo  {journal} {Phys. Rev. Lett.}\ }\textbf
  {\bibinfo {volume} {119}},\ \bibinfo {pages} {180503} (\bibinfo {year}
  {2017})}\BibitemShut {NoStop}%
\bibitem [{\citenamefont {Hines}\ \emph {et~al.}(2023)\citenamefont {Hines},
  \citenamefont {Rajagopal}, \citenamefont {Moreau}, \citenamefont {Wahrman},
  \citenamefont {Lewis}, \citenamefont {Markovi\ifmmode~\acute{c}\else
  \'{c}\fi{}},\ and\ \citenamefont {Schleier-Smith}}]{Hines2022}%
  \BibitemOpen
  \bibfield  {author} {\bibinfo {author} {\bibfnamefont {J.~A.}\ \bibnamefont
  {Hines}}, \bibinfo {author} {\bibfnamefont {S.~V.}\ \bibnamefont
  {Rajagopal}}, \bibinfo {author} {\bibfnamefont {G.~L.}\ \bibnamefont
  {Moreau}}, \bibinfo {author} {\bibfnamefont {M.~D.}\ \bibnamefont {Wahrman}},
  \bibinfo {author} {\bibfnamefont {N.~A.}\ \bibnamefont {Lewis}}, \bibinfo
  {author} {\bibfnamefont {O.}~\bibnamefont {Markovi\ifmmode~\acute{c}\else
  \'{c}\fi{}}},\ and\ \bibinfo {author} {\bibfnamefont {M.}~\bibnamefont
  {Schleier-Smith}},\ }\bibfield  {title} {\bibinfo {title} {Spin squeezing by
  {R}ydberg dressing in an array of atomic ensembles},\ }\href
  {https://doi.org/10.1103/PhysRevLett.131.063401} {\bibfield  {journal}
  {\bibinfo  {journal} {Phys. Rev. Lett.}\ }\textbf {\bibinfo {volume} {131}},\
  \bibinfo {pages} {063401} (\bibinfo {year} {2023})}\BibitemShut {NoStop}%
\bibitem [{\citenamefont {Martinez-Dorantes}\ \emph {et~al.}(2018)\citenamefont
  {Martinez-Dorantes}, \citenamefont {Alt}, \citenamefont {Gallego},
  \citenamefont {Ghosh}, \citenamefont {Ratschbacher},\ and\ \citenamefont
  {Meschede}}]{Martinez2018}%
  \BibitemOpen
  \bibfield  {author} {\bibinfo {author} {\bibfnamefont {M.}~\bibnamefont
  {Martinez-Dorantes}}, \bibinfo {author} {\bibfnamefont {W.}~\bibnamefont
  {Alt}}, \bibinfo {author} {\bibfnamefont {J.}~\bibnamefont {Gallego}},
  \bibinfo {author} {\bibfnamefont {S.}~\bibnamefont {Ghosh}}, \bibinfo
  {author} {\bibfnamefont {L.}~\bibnamefont {Ratschbacher}},\ and\ \bibinfo
  {author} {\bibfnamefont {D.}~\bibnamefont {Meschede}},\ }\bibfield  {title}
  {\bibinfo {title} {State-dependent fluorescence of neutral atoms in optical
  potentials},\ }\href {https://doi.org/10.1103/PhysRevA.97.023410} {\bibfield
  {journal} {\bibinfo  {journal} {Phys. Rev. A}\ }\textbf {\bibinfo {volume}
  {97}},\ \bibinfo {pages} {023410} (\bibinfo {year} {2018})}\BibitemShut
  {NoStop}%
\bibitem [{\citenamefont {Bochmann}\ \emph {et~al.}(2010)\citenamefont
  {Bochmann}, \citenamefont {M\"ucke}, \citenamefont {Guhl}, \citenamefont
  {Ritter}, \citenamefont {Rempe},\ and\ \citenamefont
  {Moehring}}]{Bochmann2010}%
  \BibitemOpen
  \bibfield  {author} {\bibinfo {author} {\bibfnamefont {J.}~\bibnamefont
  {Bochmann}}, \bibinfo {author} {\bibfnamefont {M.}~\bibnamefont {M\"ucke}},
  \bibinfo {author} {\bibfnamefont {C.}~\bibnamefont {Guhl}}, \bibinfo {author}
  {\bibfnamefont {S.}~\bibnamefont {Ritter}}, \bibinfo {author} {\bibfnamefont
  {G.}~\bibnamefont {Rempe}},\ and\ \bibinfo {author} {\bibfnamefont {D.~L.}\
  \bibnamefont {Moehring}},\ }\bibfield  {title} {\bibinfo {title} {Lossless
  state detection of single neutral atoms},\ }\href
  {https://doi.org/10.1103/PhysRevLett.104.203601} {\bibfield  {journal}
  {\bibinfo  {journal} {Phys. Rev. Lett.}\ }\textbf {\bibinfo {volume} {104}},\
  \bibinfo {pages} {203601} (\bibinfo {year} {2010})}\BibitemShut {NoStop}%
\bibitem [{\citenamefont {Terraciano}\ \emph {et~al.}(2009)\citenamefont
  {Terraciano}, \citenamefont {Olson~Knell}, \citenamefont {Norris},
  \citenamefont {Jing}, \citenamefont {Fern{\'a}ndez},\ and\ \citenamefont
  {Orozco}}]{Terraciano2009}%
  \BibitemOpen
  \bibfield  {author} {\bibinfo {author} {\bibfnamefont {M.~L.}\ \bibnamefont
  {Terraciano}}, \bibinfo {author} {\bibfnamefont {R.}~\bibnamefont
  {Olson~Knell}}, \bibinfo {author} {\bibfnamefont {D.~G.}\ \bibnamefont
  {Norris}}, \bibinfo {author} {\bibfnamefont {J.}~\bibnamefont {Jing}},
  \bibinfo {author} {\bibfnamefont {A.}~\bibnamefont {Fern{\'a}ndez}},\ and\
  \bibinfo {author} {\bibfnamefont {L.~A.}\ \bibnamefont {Orozco}},\ }\bibfield
   {title} {\bibinfo {title} {Photon burst detection of single atoms in an
  optical cavity},\ }\href {https://doi.org/10.1038/nphys1282} {\bibfield
  {journal} {\bibinfo  {journal} {Nature Physics}\ }\textbf {\bibinfo {volume}
  {5}},\ \bibinfo {pages} {480} (\bibinfo {year} {2009})}\BibitemShut {NoStop}%
\bibitem [{\citenamefont {Bakr}\ \emph {et~al.}(2009)\citenamefont {Bakr},
  \citenamefont {Gillen}, \citenamefont {Peng}, \citenamefont {F{\"o}lling},\
  and\ \citenamefont {Greiner}}]{Bakr2009}%
  \BibitemOpen
  \bibfield  {author} {\bibinfo {author} {\bibfnamefont {W.~S.}\ \bibnamefont
  {Bakr}}, \bibinfo {author} {\bibfnamefont {J.~I.}\ \bibnamefont {Gillen}},
  \bibinfo {author} {\bibfnamefont {A.}~\bibnamefont {Peng}}, \bibinfo {author}
  {\bibfnamefont {S.}~\bibnamefont {F{\"o}lling}},\ and\ \bibinfo {author}
  {\bibfnamefont {M.}~\bibnamefont {Greiner}},\ }\bibfield  {title} {\bibinfo
  {title} {A quantum-gas microscope for detecting single atoms in a
  {H}ubbard-regime optical lattice},\ }\href
  {https://doi.org/10.1038/nature08482} {\bibfield  {journal} {\bibinfo
  {journal} {Nature}\ }\textbf {\bibinfo {volume} {462}},\ \bibinfo {pages}
  {74} (\bibinfo {year} {2009})}\BibitemShut {NoStop}%
\bibitem [{\citenamefont {Endres}\ \emph {et~al.}(2013)\citenamefont {Endres},
  \citenamefont {Cheneau}, \citenamefont {Fukuhara}, \citenamefont
  {Weitenberg}, \citenamefont {Schau{\ss}}, \citenamefont {Gross},
  \citenamefont {Mazza}, \citenamefont {Ba{\~{n}}uls}, \citenamefont {Pollet},
  \citenamefont {Bloch},\ and\ \citenamefont {Kuhr}}]{Endres2013}%
  \BibitemOpen
  \bibfield  {author} {\bibinfo {author} {\bibfnamefont {M.}~\bibnamefont
  {Endres}}, \bibinfo {author} {\bibfnamefont {M.}~\bibnamefont {Cheneau}},
  \bibinfo {author} {\bibfnamefont {T.}~\bibnamefont {Fukuhara}}, \bibinfo
  {author} {\bibfnamefont {C.}~\bibnamefont {Weitenberg}}, \bibinfo {author}
  {\bibfnamefont {P.}~\bibnamefont {Schau{\ss}}}, \bibinfo {author}
  {\bibfnamefont {C.}~\bibnamefont {Gross}}, \bibinfo {author} {\bibfnamefont
  {L.}~\bibnamefont {Mazza}}, \bibinfo {author} {\bibfnamefont {M.~C.}\
  \bibnamefont {Ba{\~{n}}uls}}, \bibinfo {author} {\bibfnamefont
  {L.}~\bibnamefont {Pollet}}, \bibinfo {author} {\bibfnamefont
  {I.}~\bibnamefont {Bloch}},\ and\ \bibinfo {author} {\bibfnamefont
  {S.}~\bibnamefont {Kuhr}},\ }\bibfield  {title} {\bibinfo {title}
  {Single-site- and single-atom-resolved measurement of correlation
  functions},\ }\href {https://doi.org/10.1007/s00340-013-5552-9} {\bibfield
  {journal} {\bibinfo  {journal} {Appl. Phys. B}\ }\textbf {\bibinfo {volume}
  {113}},\ \bibinfo {pages} {27} (\bibinfo {year} {2013})}\BibitemShut
  {NoStop}%
\bibitem [{\citenamefont {Schlosser}\ \emph {et~al.}(2001)\citenamefont
  {Schlosser}, \citenamefont {Reymond}, \citenamefont {Protsenko},\ and\
  \citenamefont {Grangier}}]{Schlosser2001}%
  \BibitemOpen
  \bibfield  {author} {\bibinfo {author} {\bibfnamefont {N.}~\bibnamefont
  {Schlosser}}, \bibinfo {author} {\bibfnamefont {G.}~\bibnamefont {Reymond}},
  \bibinfo {author} {\bibfnamefont {I.}~\bibnamefont {Protsenko}},\ and\
  \bibinfo {author} {\bibfnamefont {P.}~\bibnamefont {Grangier}},\ }\bibfield
  {title} {\bibinfo {title} {Sub-{P}oissonian loading of single atoms in a
  microscopic dipole trap},\ }\href {https://doi.org/10.1038/35082512}
  {\bibfield  {journal} {\bibinfo  {journal} {Nature}\ }\textbf {\bibinfo
  {volume} {411}},\ \bibinfo {pages} {1024} (\bibinfo {year}
  {2001})}\BibitemShut {NoStop}%
\bibitem [{\citenamefont {Evered}\ \emph {et~al.}(2023)\citenamefont {Evered},
  \citenamefont {Bluvstein}, \citenamefont {Kalinowski}, \citenamefont {Ebadi},
  \citenamefont {Manovitz}, \citenamefont {Zhou}, \citenamefont {Li},
  \citenamefont {Geim}, \citenamefont {Wang}, \citenamefont {Maskara},
  \citenamefont {Levine}, \citenamefont {Semeghini}, \citenamefont {Greiner},
  \citenamefont {Vuleti{\'{c}}},\ and\ \citenamefont {Lukin}}]{Evered2023}%
  \BibitemOpen
  \bibfield  {author} {\bibinfo {author} {\bibfnamefont {S.~J.}\ \bibnamefont
  {Evered}}, \bibinfo {author} {\bibfnamefont {D.}~\bibnamefont {Bluvstein}},
  \bibinfo {author} {\bibfnamefont {M.}~\bibnamefont {Kalinowski}}, \bibinfo
  {author} {\bibfnamefont {S.}~\bibnamefont {Ebadi}}, \bibinfo {author}
  {\bibfnamefont {T.}~\bibnamefont {Manovitz}}, \bibinfo {author}
  {\bibfnamefont {H.}~\bibnamefont {Zhou}}, \bibinfo {author} {\bibfnamefont
  {S.~H.}\ \bibnamefont {Li}}, \bibinfo {author} {\bibfnamefont {A.~A.}\
  \bibnamefont {Geim}}, \bibinfo {author} {\bibfnamefont {T.~T.}\ \bibnamefont
  {Wang}}, \bibinfo {author} {\bibfnamefont {N.}~\bibnamefont {Maskara}},
  \bibinfo {author} {\bibfnamefont {H.}~\bibnamefont {Levine}}, \bibinfo
  {author} {\bibfnamefont {G.}~\bibnamefont {Semeghini}}, \bibinfo {author}
  {\bibfnamefont {M.}~\bibnamefont {Greiner}}, \bibinfo {author} {\bibfnamefont
  {V.}~\bibnamefont {Vuleti{\'{c}}}},\ and\ \bibinfo {author} {\bibfnamefont
  {M.~D.}\ \bibnamefont {Lukin}},\ }\bibfield  {title} {\bibinfo {title}
  {High-fidelity parallel entangling gates on a neutral-atom quantum
  computer},\ }\href {https://doi.org/10.1038/s41586-023-06481-y} {\bibfield
  {journal} {\bibinfo  {journal} {Nature}\ }\textbf {\bibinfo {volume} {622}},\
  \bibinfo {pages} {268} (\bibinfo {year} {2023})}\BibitemShut {NoStop}%
\bibitem [{\citenamefont {Young}\ \emph {et~al.}(2020)\citenamefont {Young},
  \citenamefont {Eckner}, \citenamefont {Milner}, \citenamefont {Kedar},
  \citenamefont {Norcia}, \citenamefont {Oelker}, \citenamefont {Schine},
  \citenamefont {Ye},\ and\ \citenamefont {Kaufman}}]{Young2020}%
  \BibitemOpen
  \bibfield  {author} {\bibinfo {author} {\bibfnamefont {A.~W.}\ \bibnamefont
  {Young}}, \bibinfo {author} {\bibfnamefont {W.~J.}\ \bibnamefont {Eckner}},
  \bibinfo {author} {\bibfnamefont {W.~R.}\ \bibnamefont {Milner}}, \bibinfo
  {author} {\bibfnamefont {D.}~\bibnamefont {Kedar}}, \bibinfo {author}
  {\bibfnamefont {M.~A.}\ \bibnamefont {Norcia}}, \bibinfo {author}
  {\bibfnamefont {E.}~\bibnamefont {Oelker}}, \bibinfo {author} {\bibfnamefont
  {N.}~\bibnamefont {Schine}}, \bibinfo {author} {\bibfnamefont
  {J.}~\bibnamefont {Ye}},\ and\ \bibinfo {author} {\bibfnamefont {A.~M.}\
  \bibnamefont {Kaufman}},\ }\bibfield  {title} {\bibinfo {title}
  {Half-minute-scale atomic coherence and high relative stability in a tweezer
  clock},\ }\href {https://doi.org/10.1038/s41586-020-3009-y} {\bibfield
  {journal} {\bibinfo  {journal} {Nature}\ }\textbf {\bibinfo {volume} {588}},\
  \bibinfo {pages} {408} (\bibinfo {year} {2020})}\BibitemShut {NoStop}%
\bibitem [{\citenamefont {Barredo}\ \emph {et~al.}(2018)\citenamefont
  {Barredo}, \citenamefont {Lienhard}, \citenamefont {de~L{\'e}s{\'e}leuc},
  \citenamefont {Lahaye},\ and\ \citenamefont {Browaeys}}]{Barredo2018}%
  \BibitemOpen
  \bibfield  {author} {\bibinfo {author} {\bibfnamefont {D.}~\bibnamefont
  {Barredo}}, \bibinfo {author} {\bibfnamefont {V.}~\bibnamefont {Lienhard}},
  \bibinfo {author} {\bibfnamefont {S.}~\bibnamefont {de~L{\'e}s{\'e}leuc}},
  \bibinfo {author} {\bibfnamefont {T.}~\bibnamefont {Lahaye}},\ and\ \bibinfo
  {author} {\bibfnamefont {A.}~\bibnamefont {Browaeys}},\ }\bibfield  {title}
  {\bibinfo {title} {Synthetic three-dimensional atomic structures assembled
  atom by atom},\ }\href {https://doi.org/10.1038/s41586-018-0450-2} {\bibfield
   {journal} {\bibinfo  {journal} {Nature}\ }\textbf {\bibinfo {volume}
  {561}},\ \bibinfo {pages} {79} (\bibinfo {year} {2018})}\BibitemShut
  {NoStop}%
\bibitem [{\citenamefont {Blais}\ \emph {et~al.}(2021)\citenamefont {Blais},
  \citenamefont {Grimsmo}, \citenamefont {Girvin},\ and\ \citenamefont
  {Wallraff}}]{Blais2021}%
  \BibitemOpen
  \bibfield  {author} {\bibinfo {author} {\bibfnamefont {A.}~\bibnamefont
  {Blais}}, \bibinfo {author} {\bibfnamefont {A.~L.}\ \bibnamefont {Grimsmo}},
  \bibinfo {author} {\bibfnamefont {S.~M.}\ \bibnamefont {Girvin}},\ and\
  \bibinfo {author} {\bibfnamefont {A.}~\bibnamefont {Wallraff}},\ }\bibfield
  {title} {\bibinfo {title} {Circuit quantum electrodynamics},\ }\href
  {https://doi.org/10.1103/RevModPhys.93.025005} {\bibfield  {journal}
  {\bibinfo  {journal} {Rev. Mod. Phys.}\ }\textbf {\bibinfo {volume} {93}},\
  \bibinfo {pages} {025005} (\bibinfo {year} {2021})}\BibitemShut {NoStop}%
\bibitem [{\citenamefont {Jaynes}\ and\ \citenamefont
  {Cummings}(1963)}]{Jaynes1963}%
  \BibitemOpen
  \bibfield  {author} {\bibinfo {author} {\bibfnamefont {E.}~\bibnamefont
  {Jaynes}}\ and\ \bibinfo {author} {\bibfnamefont {F.}~\bibnamefont
  {Cummings}},\ }\bibfield  {title} {\bibinfo {title} {Comparison of quantum
  and semiclassical radiation theories with application to the beam maser},\
  }\href {https://doi.org/10.1109/PROC.1963.1664} {\bibfield  {journal}
  {\bibinfo  {journal} {Proceedings of the IEEE}\ }\textbf {\bibinfo {volume}
  {51}},\ \bibinfo {pages} {89} (\bibinfo {year} {1963})}\BibitemShut {NoStop}%
\bibitem [{\citenamefont {Puppe}\ \emph {et~al.}(2007)\citenamefont {Puppe},
  \citenamefont {Schuster}, \citenamefont {Grothe}, \citenamefont {Kubanek},
  \citenamefont {Murr}, \citenamefont {Pinkse},\ and\ \citenamefont
  {Rempe}}]{Puppe2007}%
  \BibitemOpen
  \bibfield  {author} {\bibinfo {author} {\bibfnamefont {T.}~\bibnamefont
  {Puppe}}, \bibinfo {author} {\bibfnamefont {I.}~\bibnamefont {Schuster}},
  \bibinfo {author} {\bibfnamefont {A.}~\bibnamefont {Grothe}}, \bibinfo
  {author} {\bibfnamefont {A.}~\bibnamefont {Kubanek}}, \bibinfo {author}
  {\bibfnamefont {K.}~\bibnamefont {Murr}}, \bibinfo {author} {\bibfnamefont
  {P.~W.~H.}\ \bibnamefont {Pinkse}},\ and\ \bibinfo {author} {\bibfnamefont
  {G.}~\bibnamefont {Rempe}},\ }\bibfield  {title} {\bibinfo {title} {Trapping
  and observing single atoms in a blue-detuned intracavity dipole trap},\
  }\href {https://doi.org/10.1103/PhysRevLett.99.013002} {\bibfield  {journal}
  {\bibinfo  {journal} {Phys. Rev. Lett.}\ }\textbf {\bibinfo {volume} {99}},\
  \bibinfo {pages} {013002} (\bibinfo {year} {2007})}\BibitemShut {NoStop}%
\bibitem [{\citenamefont {Schleier-Smith}\ \emph {et~al.}(2010)\citenamefont
  {Schleier-Smith}, \citenamefont {Leroux},\ and\ \citenamefont
  {Vuleti\ifmmode~\acute{c}\else \'{c}\fi{}}}]{SchleierSmith2010}%
  \BibitemOpen
  \bibfield  {author} {\bibinfo {author} {\bibfnamefont {M.~H.}\ \bibnamefont
  {Schleier-Smith}}, \bibinfo {author} {\bibfnamefont {I.~D.}\ \bibnamefont
  {Leroux}},\ and\ \bibinfo {author} {\bibfnamefont {V.}~\bibnamefont
  {Vuleti\ifmmode~\acute{c}\else \'{c}\fi{}}},\ }\bibfield  {title} {\bibinfo
  {title} {States of an ensemble of two-level atoms with reduced quantum
  uncertainty},\ }\href {https://doi.org/10.1103/PhysRevLett.104.073604}
  {\bibfield  {journal} {\bibinfo  {journal} {Phys. Rev. Lett.}\ }\textbf
  {\bibinfo {volume} {104}},\ \bibinfo {pages} {073604} (\bibinfo {year}
  {2010})}\BibitemShut {NoStop}%
\bibitem [{\citenamefont {Cox}\ \emph {et~al.}(2016)\citenamefont {Cox},
  \citenamefont {Greve}, \citenamefont {Weiner},\ and\ \citenamefont
  {Thompson}}]{Cox2016}%
  \BibitemOpen
  \bibfield  {author} {\bibinfo {author} {\bibfnamefont {K.~C.}\ \bibnamefont
  {Cox}}, \bibinfo {author} {\bibfnamefont {G.~P.}\ \bibnamefont {Greve}},
  \bibinfo {author} {\bibfnamefont {J.~M.}\ \bibnamefont {Weiner}},\ and\
  \bibinfo {author} {\bibfnamefont {J.~K.}\ \bibnamefont {Thompson}},\
  }\bibfield  {title} {\bibinfo {title} {Deterministic squeezed states with
  collective measurements and feedback},\ }\href
  {https://doi.org/10.1103/PhysRevLett.116.093602} {\bibfield  {journal}
  {\bibinfo  {journal} {Phys. Rev. Lett.}\ }\textbf {\bibinfo {volume} {116}},\
  \bibinfo {pages} {093602} (\bibinfo {year} {2016})}\BibitemShut {NoStop}%
\bibitem [{\citenamefont {Hosten}\ \emph {et~al.}(2016)\citenamefont {Hosten},
  \citenamefont {Engelsen}, \citenamefont {Krishnakumar},\ and\ \citenamefont
  {Kasevich}}]{Hosten2016}%
  \BibitemOpen
  \bibfield  {author} {\bibinfo {author} {\bibfnamefont {O.}~\bibnamefont
  {Hosten}}, \bibinfo {author} {\bibfnamefont {N.~J.}\ \bibnamefont
  {Engelsen}}, \bibinfo {author} {\bibfnamefont {R.}~\bibnamefont
  {Krishnakumar}},\ and\ \bibinfo {author} {\bibfnamefont {M.~A.}\ \bibnamefont
  {Kasevich}},\ }\bibfield  {title} {\bibinfo {title} {Measurement noise 100
  times lower than the quantum-projection limit using entangled atoms},\ }\href
  {https://doi.org/10.1038/nature16176} {\bibfield  {journal} {\bibinfo
  {journal} {Nature}\ }\textbf {\bibinfo {volume} {529}},\ \bibinfo {pages}
  {505} (\bibinfo {year} {2016})}\BibitemShut {NoStop}%
\bibitem [{\citenamefont {Robinson}\ \emph {et~al.}(2022)\citenamefont
  {Robinson}, \citenamefont {Miklos}, \citenamefont {Tso}, \citenamefont
  {Kennedy}, \citenamefont {Bothwell}, \citenamefont {Kedar}, \citenamefont
  {Thompson},\ and\ \citenamefont {Ye}}]{robinson2022direct}%
  \BibitemOpen
  \bibfield  {author} {\bibinfo {author} {\bibfnamefont {J.~M.}\ \bibnamefont
  {Robinson}}, \bibinfo {author} {\bibfnamefont {M.}~\bibnamefont {Miklos}},
  \bibinfo {author} {\bibfnamefont {Y.~M.}\ \bibnamefont {Tso}}, \bibinfo
  {author} {\bibfnamefont {C.~J.}\ \bibnamefont {Kennedy}}, \bibinfo {author}
  {\bibfnamefont {T.}~\bibnamefont {Bothwell}}, \bibinfo {author}
  {\bibfnamefont {D.}~\bibnamefont {Kedar}}, \bibinfo {author} {\bibfnamefont
  {J.~K.}\ \bibnamefont {Thompson}},\ and\ \bibinfo {author} {\bibfnamefont
  {J.}~\bibnamefont {Ye}},\ }\href@noop {} {\bibinfo {title} {Direct comparison
  of two spin-squeezed optical clocks below the quantum projection noise
  limit}} (\bibinfo {year} {2022}),\ \Eprint {https://arxiv.org/abs/2211.08621}
  {ar{X}iv:2211.08621 [quant-ph]} \BibitemShut {NoStop}%
\bibitem [{\citenamefont {Volz}\ \emph {et~al.}(2011)\citenamefont {Volz},
  \citenamefont {Gehr}, \citenamefont {Dubois}, \citenamefont {Est{\`e}ve},\
  and\ \citenamefont {Reichel}}]{Volz2011}%
  \BibitemOpen
  \bibfield  {author} {\bibinfo {author} {\bibfnamefont {J.}~\bibnamefont
  {Volz}}, \bibinfo {author} {\bibfnamefont {R.}~\bibnamefont {Gehr}}, \bibinfo
  {author} {\bibfnamefont {G.}~\bibnamefont {Dubois}}, \bibinfo {author}
  {\bibfnamefont {J.}~\bibnamefont {Est{\`e}ve}},\ and\ \bibinfo {author}
  {\bibfnamefont {J.}~\bibnamefont {Reichel}},\ }\bibfield  {title} {\bibinfo
  {title} {Measurement of the internal state of a single atom without energy
  exchange},\ }\href {https://doi.org/10.1038/nature10225} {\bibfield
  {journal} {\bibinfo  {journal} {Nature}\ }\textbf {\bibinfo {volume} {475}},\
  \bibinfo {pages} {210} (\bibinfo {year} {2011})}\BibitemShut {NoStop}%
\bibitem [{\citenamefont {Windpassinger}\ \emph {et~al.}(2008)\citenamefont
  {Windpassinger}, \citenamefont {Oblak}, \citenamefont {Petrov}, \citenamefont
  {Kubasik}, \citenamefont {Saffman}, \citenamefont {Alzar}, \citenamefont
  {Appel}, \citenamefont {M\"uller}, \citenamefont {Kj\ae{}rgaard},\ and\
  \citenamefont {Polzik}}]{Windpassinger2008}%
  \BibitemOpen
  \bibfield  {author} {\bibinfo {author} {\bibfnamefont {P.~J.}\ \bibnamefont
  {Windpassinger}}, \bibinfo {author} {\bibfnamefont {D.}~\bibnamefont
  {Oblak}}, \bibinfo {author} {\bibfnamefont {P.~G.}\ \bibnamefont {Petrov}},
  \bibinfo {author} {\bibfnamefont {M.}~\bibnamefont {Kubasik}}, \bibinfo
  {author} {\bibfnamefont {M.}~\bibnamefont {Saffman}}, \bibinfo {author}
  {\bibfnamefont {C.~L.~G.}\ \bibnamefont {Alzar}}, \bibinfo {author}
  {\bibfnamefont {J.}~\bibnamefont {Appel}}, \bibinfo {author} {\bibfnamefont
  {J.~H.}\ \bibnamefont {M\"uller}}, \bibinfo {author} {\bibfnamefont
  {N.}~\bibnamefont {Kj\ae{}rgaard}},\ and\ \bibinfo {author} {\bibfnamefont
  {E.~S.}\ \bibnamefont {Polzik}},\ }\bibfield  {title} {\bibinfo {title}
  {Nondestructive probing of {R}abi oscillations on the cesium clock transition
  near the standard quantum limit},\ }\href
  {https://doi.org/10.1103/PhysRevLett.100.103601} {\bibfield  {journal}
  {\bibinfo  {journal} {Phys. Rev. Lett.}\ }\textbf {\bibinfo {volume} {100}},\
  \bibinfo {pages} {103601} (\bibinfo {year} {2008})}\BibitemShut {NoStop}%
\bibitem [{\citenamefont {Andrews}\ \emph {et~al.}(1996)\citenamefont
  {Andrews}, \citenamefont {Mewes}, \citenamefont {van Druten}, \citenamefont
  {Durfee}, \citenamefont {Kurn},\ and\ \citenamefont
  {Ketterle}}]{Andrews1996}%
  \BibitemOpen
  \bibfield  {author} {\bibinfo {author} {\bibfnamefont {M.~R.}\ \bibnamefont
  {Andrews}}, \bibinfo {author} {\bibfnamefont {M.-O.}\ \bibnamefont {Mewes}},
  \bibinfo {author} {\bibfnamefont {N.~J.}\ \bibnamefont {van Druten}},
  \bibinfo {author} {\bibfnamefont {D.~S.}\ \bibnamefont {Durfee}}, \bibinfo
  {author} {\bibfnamefont {D.~M.}\ \bibnamefont {Kurn}},\ and\ \bibinfo
  {author} {\bibfnamefont {W.}~\bibnamefont {Ketterle}},\ }\bibfield  {title}
  {\bibinfo {title} {Direct, nondestructive observation of a {B}ose
  condensate},\ }\href {https://doi.org/10.1126/science.273.5271.84} {\bibfield
   {journal} {\bibinfo  {journal} {Science}\ }\textbf {\bibinfo {volume}
  {273}},\ \bibinfo {pages} {84} (\bibinfo {year} {1996})}\BibitemShut
  {NoStop}%
\bibitem [{\citenamefont {Bradley}\ \emph {et~al.}(1997)\citenamefont
  {Bradley}, \citenamefont {Sackett},\ and\ \citenamefont
  {Hulet}}]{Bradley1997}%
  \BibitemOpen
  \bibfield  {author} {\bibinfo {author} {\bibfnamefont {C.~C.}\ \bibnamefont
  {Bradley}}, \bibinfo {author} {\bibfnamefont {C.~A.}\ \bibnamefont
  {Sackett}},\ and\ \bibinfo {author} {\bibfnamefont {R.~G.}\ \bibnamefont
  {Hulet}},\ }\bibfield  {title} {\bibinfo {title} {{B}ose-{E}instein
  condensation of lithium: {O}bservation of limited condensate number},\ }\href
  {https://doi.org/10.1103/PhysRevLett.78.985} {\bibfield  {journal} {\bibinfo
  {journal} {Phys. Rev. Lett.}\ }\textbf {\bibinfo {volume} {78}},\ \bibinfo
  {pages} {985} (\bibinfo {year} {1997})}\BibitemShut {NoStop}%
\bibitem [{\citenamefont {Meppelink}\ \emph {et~al.}(2010)\citenamefont
  {Meppelink}, \citenamefont {Rozendaal}, \citenamefont {Koller}, \citenamefont
  {Vogels},\ and\ \citenamefont {van~der Straten}}]{Meppelink2010}%
  \BibitemOpen
  \bibfield  {author} {\bibinfo {author} {\bibfnamefont {R.}~\bibnamefont
  {Meppelink}}, \bibinfo {author} {\bibfnamefont {R.~A.}\ \bibnamefont
  {Rozendaal}}, \bibinfo {author} {\bibfnamefont {S.~B.}\ \bibnamefont
  {Koller}}, \bibinfo {author} {\bibfnamefont {J.~M.}\ \bibnamefont {Vogels}},\
  and\ \bibinfo {author} {\bibfnamefont {P.}~\bibnamefont {van~der Straten}},\
  }\bibfield  {title} {\bibinfo {title} {Thermodynamics of
  {B}ose-{E}instein-condensed clouds using phase-contrast imaging},\ }\href
  {https://doi.org/10.1103/PhysRevA.81.053632} {\bibfield  {journal} {\bibinfo
  {journal} {Phys. Rev. A}\ }\textbf {\bibinfo {volume} {81}},\ \bibinfo
  {pages} {053632} (\bibinfo {year} {2010})}\BibitemShut {NoStop}%
\bibitem [{\citenamefont {Gleason}(1957)}]{Gleason}%
  \BibitemOpen
  \bibfield  {author} {\bibinfo {author} {\bibfnamefont {A.}~\bibnamefont
  {Gleason}},\ }\bibfield  {title} {\bibinfo {title} {Measures on the closed
  subspaces of a {H}ilbert space},\ }\href
  {http://www.iumj.indiana.edu/IUMJ/fulltext.php?artid=56050&year=1957&volume=6}
  {\bibfield  {journal} {\bibinfo  {journal} {Indiana Univ. Math. J.}\ }\textbf
  {\bibinfo {volume} {6}},\ \bibinfo {pages} {885} (\bibinfo {year}
  {1957})}\BibitemShut {NoStop}%
\bibitem [{\citenamefont {Lieb}\ and\ \citenamefont
  {Robinson}(1972)}]{Lieb1972}%
  \BibitemOpen
  \bibfield  {author} {\bibinfo {author} {\bibfnamefont {E.~H.}\ \bibnamefont
  {Lieb}}\ and\ \bibinfo {author} {\bibfnamefont {D.~W.}\ \bibnamefont
  {Robinson}},\ }\bibfield  {title} {\bibinfo {title} {The finite group
  velocity of quantum spin systems},\ }\href
  {https://doi.org/10.1007/BF01645779} {\bibfield  {journal} {\bibinfo
  {journal} {Commun. Math. Phys.}\ }\textbf {\bibinfo {volume} {28}},\ \bibinfo
  {pages} {251} (\bibinfo {year} {1972})}\BibitemShut {NoStop}%
\bibitem [{\citenamefont {Poulin}(2010)}]{Poulin}%
  \BibitemOpen
  \bibfield  {author} {\bibinfo {author} {\bibfnamefont {D.}~\bibnamefont
  {Poulin}},\ }\bibfield  {title} {\bibinfo {title} {{L}ieb-{R}obinson bound
  and locality for general {M}arkovian quantum dynamics},\ }\href
  {https://doi.org/10.1103/PhysRevLett.104.190401} {\bibfield  {journal}
  {\bibinfo  {journal} {Phys. Rev. Lett.}\ }\textbf {\bibinfo {volume} {104}},\
  \bibinfo {pages} {190401} (\bibinfo {year} {2010})}\BibitemShut {NoStop}%
\bibitem [{\citenamefont {Deutsch}(1991)}]{ETH1}%
  \BibitemOpen
  \bibfield  {author} {\bibinfo {author} {\bibfnamefont {J.~M.}\ \bibnamefont
  {Deutsch}},\ }\bibfield  {title} {\bibinfo {title} {Quantum statistical
  mechanics in a closed system},\ }\href
  {https://doi.org/10.1103/PhysRevA.43.2046} {\bibfield  {journal} {\bibinfo
  {journal} {Phys. Rev. A}\ }\textbf {\bibinfo {volume} {43}},\ \bibinfo
  {pages} {2046} (\bibinfo {year} {1991})}\BibitemShut {NoStop}%
\bibitem [{\citenamefont {Srednicki}(1994)}]{ETH2}%
  \BibitemOpen
  \bibfield  {author} {\bibinfo {author} {\bibfnamefont {M.}~\bibnamefont
  {Srednicki}},\ }\bibfield  {title} {\bibinfo {title} {Chaos and quantum
  thermalization},\ }\href {https://doi.org/10.1103/PhysRevE.50.888} {\bibfield
   {journal} {\bibinfo  {journal} {Phys. Rev. E}\ }\textbf {\bibinfo {volume}
  {50}},\ \bibinfo {pages} {888} (\bibinfo {year} {1994})}\BibitemShut
  {NoStop}%
\bibitem [{\citenamefont {Rigol}\ \emph {et~al.}(2008)\citenamefont {Rigol},
  \citenamefont {Dunjko},\ and\ \citenamefont {Olshanii}}]{ETH3}%
  \BibitemOpen
  \bibfield  {author} {\bibinfo {author} {\bibfnamefont {M.}~\bibnamefont
  {Rigol}}, \bibinfo {author} {\bibfnamefont {V.}~\bibnamefont {Dunjko}},\ and\
  \bibinfo {author} {\bibfnamefont {M.}~\bibnamefont {Olshanii}},\ }\bibfield
  {title} {\bibinfo {title} {Thermalization and its mechanism for generic
  isolated quantum systems},\ }\href {http://dx.doi.org/10.1038/nature06838}
  {\bibfield  {journal} {\bibinfo  {journal} {Nature}\ }\textbf {\bibinfo
  {volume} {452}},\ \bibinfo {pages} {854} (\bibinfo {year}
  {2008})}\BibitemShut {NoStop}%
\bibitem [{\citenamefont {Doyon}(2017)}]{DoyonETH}%
  \BibitemOpen
  \bibfield  {author} {\bibinfo {author} {\bibfnamefont {B.}~\bibnamefont
  {Doyon}},\ }\bibfield  {title} {\bibinfo {title} {Thermalization and
  pseudolocality in extended quantum systems},\ }\href
  {https://doi.org/10.1007/s00220-017-2836-7} {\bibfield  {journal} {\bibinfo
  {journal} {Commun. Math. Phys.}\ }\textbf {\bibinfo {volume} {351}},\
  \bibinfo {pages} {155} (\bibinfo {year} {2017})}\BibitemShut {NoStop}%
\bibitem [{\citenamefont {Bohigas}\ \emph {et~al.}(1984)\citenamefont
  {Bohigas}, \citenamefont {Giannoni},\ and\ \citenamefont
  {Schmit}}]{BohigasChaos}%
  \BibitemOpen
  \bibfield  {author} {\bibinfo {author} {\bibfnamefont {O.}~\bibnamefont
  {Bohigas}}, \bibinfo {author} {\bibfnamefont {M.~J.}\ \bibnamefont
  {Giannoni}},\ and\ \bibinfo {author} {\bibfnamefont {C.}~\bibnamefont
  {Schmit}},\ }\bibfield  {title} {\bibinfo {title} {Characterization of
  chaotic quantum spectra and universality of level fluctuation laws},\ }\href
  {https://doi.org/10.1103/PhysRevLett.52.1} {\bibfield  {journal} {\bibinfo
  {journal} {Phys. Rev. Lett.}\ }\textbf {\bibinfo {volume} {52}},\ \bibinfo
  {pages} {1} (\bibinfo {year} {1984})}\BibitemShut {NoStop}%
\bibitem [{\citenamefont {Roberts}\ and\ \citenamefont
  {Yoshida}(2017)}]{YoshidaCBD}%
  \BibitemOpen
  \bibfield  {author} {\bibinfo {author} {\bibfnamefont {D.~A.}\ \bibnamefont
  {Roberts}}\ and\ \bibinfo {author} {\bibfnamefont {B.}~\bibnamefont
  {Yoshida}},\ }\bibfield  {title} {\bibinfo {title} {Chaos and complexity by
  design},\ }\href {https://doi.org/10.1007/jhep04(2017)121} {\bibfield
  {journal} {\bibinfo  {journal} {JHEP}\ }\textbf {\bibinfo {volume}
  {2017}}\bibinfo  {number} { (4)}}\BibitemShut {NoStop}%
\bibitem [{\citenamefont {Nahum}\ \emph {et~al.}(2017)\citenamefont {Nahum},
  \citenamefont {Vijay},\ and\ \citenamefont {Haah}}]{NahumOperator}%
  \BibitemOpen
\bibfield  {number} {  }\bibfield  {author} {\bibinfo {author} {\bibfnamefont
  {A.}~\bibnamefont {Nahum}}, \bibinfo {author} {\bibfnamefont
  {S.}~\bibnamefont {Vijay}},\ and\ \bibinfo {author} {\bibfnamefont
  {J.}~\bibnamefont {Haah}},\ }\href
  {https://doi.org/10.48550/arXiv.1705.08975} {\bibinfo {title} {Operator
  spreading in random unitary circuits}} (\bibinfo {year} {2017}),\ \Eprint
  {https://arxiv.org/abs/1705.08975} {ar{X}iv:1705.08975 [cond-mat.str-el]}
  \BibitemShut {NoStop}%
\bibitem [{\citenamefont {Rakovszky}\ \emph {et~al.}(2017)\citenamefont
  {Rakovszky}, \citenamefont {Pollmann},\ and\ \citenamefont {von
  Keyserlingk}}]{RUCconTibor}%
  \BibitemOpen
  \bibfield  {author} {\bibinfo {author} {\bibfnamefont {T.}~\bibnamefont
  {Rakovszky}}, \bibinfo {author} {\bibfnamefont {F.}~\bibnamefont
  {Pollmann}},\ and\ \bibinfo {author} {\bibfnamefont {C.~W.}\ \bibnamefont
  {von Keyserlingk}},\ }\href {https://doi.org/10.48550/arXiv.1710.09827}
  {\bibinfo {title} {Diffusive hydrodynamics of out-of-time-ordered correlators
  with charge conservation}} (\bibinfo {year} {2017}),\ \Eprint
  {https://arxiv.org/abs/1710.09827} {ar{X}iv:1710.09827 [cond-mat.stat-mech]}
  \BibitemShut {NoStop}%
\bibitem [{\citenamefont {Friedman}\ \emph {et~al.}(2019)\citenamefont
  {Friedman}, \citenamefont {Chan}, \citenamefont {De~Luca},\ and\
  \citenamefont {Chalker}}]{U1FRUC}%
  \BibitemOpen
  \bibfield  {author} {\bibinfo {author} {\bibfnamefont {A.~J.}\ \bibnamefont
  {Friedman}}, \bibinfo {author} {\bibfnamefont {A.}~\bibnamefont {Chan}},
  \bibinfo {author} {\bibfnamefont {A.}~\bibnamefont {De~Luca}},\ and\ \bibinfo
  {author} {\bibfnamefont {J.~T.}\ \bibnamefont {Chalker}},\ }\bibfield
  {title} {\bibinfo {title} {Spectral statistics and many-body quantum chaos
  with conserved charge},\ }\href
  {https://doi.org/10.1103/PhysRevLett.123.210603} {\bibfield  {journal}
  {\bibinfo  {journal} {Phys. Rev. Lett.}\ }\textbf {\bibinfo {volume} {123}},\
  \bibinfo {pages} {210603} (\bibinfo {year} {2019})}\BibitemShut {NoStop}%
\bibitem [{\citenamefont {Singh}\ \emph {et~al.}(2021)\citenamefont {Singh},
  \citenamefont {Ware}, \citenamefont {Vasseur},\ and\ \citenamefont
  {Friedman}}]{ConstrainedRUC}%
  \BibitemOpen
  \bibfield  {author} {\bibinfo {author} {\bibfnamefont {H.}~\bibnamefont
  {Singh}}, \bibinfo {author} {\bibfnamefont {B.~A.}\ \bibnamefont {Ware}},
  \bibinfo {author} {\bibfnamefont {R.}~\bibnamefont {Vasseur}},\ and\ \bibinfo
  {author} {\bibfnamefont {A.~J.}\ \bibnamefont {Friedman}},\ }\bibfield
  {title} {\bibinfo {title} {Subdiffusion and many-body quantum chaos with
  kinetic constraints},\ }\href
  {https://doi.org/10.1103/PhysRevLett.127.230602} {\bibfield  {journal}
  {\bibinfo  {journal} {Phys. Rev. Lett.}\ }\textbf {\bibinfo {volume} {127}},\
  \bibinfo {pages} {230602} (\bibinfo {year} {2021})}\BibitemShut {NoStop}%
\bibitem [{\citenamefont {Bell}(1964)}]{Bell}%
  \BibitemOpen
  \bibfield  {author} {\bibinfo {author} {\bibfnamefont {J.~S.}\ \bibnamefont
  {Bell}},\ }\bibfield  {title} {\bibinfo {title} {On the
  {E}instein-{P}odolsky-{R}osen paradox},\ }\href
  {https://doi.org/10.1103/PhysicsPhysiqueFizika.1.195} {\bibfield  {journal}
  {\bibinfo  {journal} {Physics Physique Fizika}\ }\textbf {\bibinfo {volume}
  {1}},\ \bibinfo {pages} {195} (\bibinfo {year} {1964})}\BibitemShut {NoStop}%
\bibitem [{\citenamefont {Greenberger}\ \emph {et~al.}(1989)\citenamefont
  {Greenberger}, \citenamefont {Horne},\ and\ \citenamefont
  {Zeilinger}}]{GHZ89}%
  \BibitemOpen
  \bibfield  {author} {\bibinfo {author} {\bibfnamefont {D.~M.}\ \bibnamefont
  {Greenberger}}, \bibinfo {author} {\bibfnamefont {M.~A.}\ \bibnamefont
  {Horne}},\ and\ \bibinfo {author} {\bibfnamefont {A.}~\bibnamefont
  {Zeilinger}},\ }\bibfield  {title} {\bibinfo {title} {Going beyond {B}ell’s
  theorem},\ }in\ \href {https://doi.org/10.1007/978-94-017-0849-4_10} {\emph
  {\bibinfo {booktitle} {Bell’s theorem, quantum theory and conceptions of
  the universe}}}\ (\bibinfo  {publisher} {Springer},\ \bibinfo {year} {1989})\
  pp.\ \bibinfo {pages} {69--72}\BibitemShut {NoStop}%
\bibitem [{\citenamefont {Hastings}(2010)}]{hastings_rev}%
  \BibitemOpen
  \bibfield  {author} {\bibinfo {author} {\bibfnamefont {M.~B.}\ \bibnamefont
  {Hastings}},\ }\href {https://doi.org/10.48550/arxiv.1008.5137} {\bibinfo
  {title} {Locality in quantum systems}} (\bibinfo {year} {2010}),\ \Eprint
  {https://arxiv.org/abs/1008.5137} {ar{X}iv:1008.5137 [math-ph]} \BibitemShut
  {NoStop}%
\bibitem [{\citenamefont {Popescu}\ and\ \citenamefont
  {Rohrlich}(1994)}]{NonlocalityAxiom}%
  \BibitemOpen
  \bibfield  {author} {\bibinfo {author} {\bibfnamefont {S.}~\bibnamefont
  {Popescu}}\ and\ \bibinfo {author} {\bibfnamefont {D.}~\bibnamefont
  {Rohrlich}},\ }\bibfield  {title} {\bibinfo {title} {Quantum nonlocality as
  an axiom},\ }\href {https://doi.org/10.1007/BF02058098} {\bibfield  {journal}
  {\bibinfo  {journal} {Found. Phys.}\ }\textbf {\bibinfo {volume} {24}},\
  \bibinfo {pages} {379} (\bibinfo {year} {1994})}\BibitemShut {NoStop}%
\bibitem [{\citenamefont {Brunner}\ \emph {et~al.}(2014)\citenamefont
  {Brunner}, \citenamefont {Cavalcanti}, \citenamefont {Pironio}, \citenamefont
  {Scarani},\ and\ \citenamefont {Wehner}}]{BellNonloc}%
  \BibitemOpen
  \bibfield  {author} {\bibinfo {author} {\bibfnamefont {N.}~\bibnamefont
  {Brunner}}, \bibinfo {author} {\bibfnamefont {D.}~\bibnamefont {Cavalcanti}},
  \bibinfo {author} {\bibfnamefont {S.}~\bibnamefont {Pironio}}, \bibinfo
  {author} {\bibfnamefont {V.}~\bibnamefont {Scarani}},\ and\ \bibinfo {author}
  {\bibfnamefont {S.}~\bibnamefont {Wehner}},\ }\bibfield  {title} {\bibinfo
  {title} {Bell nonlocality},\ }\href
  {https://doi.org/10.1103/RevModPhys.86.419} {\bibfield  {journal} {\bibinfo
  {journal} {Rev. Mod. Phys.}\ }\textbf {\bibinfo {volume} {86}},\ \bibinfo
  {pages} {419} (\bibinfo {year} {2014})}\BibitemShut {NoStop}%
\bibitem [{\citenamefont {Foss-Feig}\ \emph {et~al.}(2015)\citenamefont
  {Foss-Feig}, \citenamefont {Gong}, \citenamefont {Clark},\ and\ \citenamefont
  {Gorshkov}}]{LRfossfeig}%
  \BibitemOpen
  \bibfield  {author} {\bibinfo {author} {\bibfnamefont {M.}~\bibnamefont
  {Foss-Feig}}, \bibinfo {author} {\bibfnamefont {Z.-X.}\ \bibnamefont {Gong}},
  \bibinfo {author} {\bibfnamefont {C.~W.}\ \bibnamefont {Clark}},\ and\
  \bibinfo {author} {\bibfnamefont {A.~V.}\ \bibnamefont {Gorshkov}},\
  }\bibfield  {title} {\bibinfo {title} {Nearly linear light cones in
  long-range interacting quantum systems},\ }\href
  {https://doi.org/10.1103/PhysRevLett.114.157201} {\bibfield  {journal}
  {\bibinfo  {journal} {Phys. Rev. Lett.}\ }\textbf {\bibinfo {volume} {114}},\
  \bibinfo {pages} {157201} (\bibinfo {year} {2015})}\BibitemShut {NoStop}%
\bibitem [{\citenamefont {Chen}\ and\ \citenamefont
  {Lucas}(2019)}]{chen2019finite}%
  \BibitemOpen
  \bibfield  {author} {\bibinfo {author} {\bibfnamefont {C.-F.}\ \bibnamefont
  {Chen}}\ and\ \bibinfo {author} {\bibfnamefont {A.}~\bibnamefont {Lucas}},\
  }\bibfield  {title} {\bibinfo {title} {{Finite speed of quantum scrambling
  with long-range interactions}},\ }\href
  {https://doi.org/10.1103/PhysRevLett.123.250605} {\bibfield  {journal}
  {\bibinfo  {journal} {Phys. Rev. Lett.}\ }\textbf {\bibinfo {volume} {123}},\
  \bibinfo {pages} {250605} (\bibinfo {year} {2019})}\BibitemShut {NoStop}%
\bibitem [{\citenamefont {Tran}\ \emph {et~al.}(2020)\citenamefont {Tran},
  \citenamefont {Chen}, \citenamefont {Ehrenberg}, \citenamefont {Guo},
  \citenamefont {Deshpande}, \citenamefont {Hong}, \citenamefont {Gong},
  \citenamefont {Gorshkov},\ and\ \citenamefont {Lucas}}]{Tran:2020xpc}%
  \BibitemOpen
  \bibfield  {author} {\bibinfo {author} {\bibfnamefont {M.~C.}\ \bibnamefont
  {Tran}}, \bibinfo {author} {\bibfnamefont {C.-F.}\ \bibnamefont {Chen}},
  \bibinfo {author} {\bibfnamefont {A.}~\bibnamefont {Ehrenberg}}, \bibinfo
  {author} {\bibfnamefont {A.~Y.}\ \bibnamefont {Guo}}, \bibinfo {author}
  {\bibfnamefont {A.}~\bibnamefont {Deshpande}}, \bibinfo {author}
  {\bibfnamefont {Y.}~\bibnamefont {Hong}}, \bibinfo {author} {\bibfnamefont
  {Z.-X.}\ \bibnamefont {Gong}}, \bibinfo {author} {\bibfnamefont {A.~V.}\
  \bibnamefont {Gorshkov}},\ and\ \bibinfo {author} {\bibfnamefont
  {A.}~\bibnamefont {Lucas}},\ }\href
  {https://doi.org/10.48550/arXiv.2001.11509} {\bibinfo {title} {Hierarchy of
  linear light cones with long-range interactions}} (\bibinfo {year} {2020}),\
  \Eprint {https://arxiv.org/abs/2001.11509} {ar{X}iv:2001.11509 [quant-ph]}
  \BibitemShut {NoStop}%
\bibitem [{\citenamefont {Chen}\ \emph {et~al.}(2023)\citenamefont {Chen},
  \citenamefont {Lucas},\ and\ \citenamefont {Yin}}]{CLYreview}%
  \BibitemOpen
  \bibfield  {author} {\bibinfo {author} {\bibfnamefont {C.-F.}\ \bibnamefont
  {Chen}}, \bibinfo {author} {\bibfnamefont {A.}~\bibnamefont {Lucas}},\ and\
  \bibinfo {author} {\bibfnamefont {C.}~\bibnamefont {Yin}},\ }\bibfield
  {title} {\bibinfo {title} {Speed limits and locality in many-body quantum
  dynamics},\ }\href {https://doi.org/10.1088/1361-6633/acfaae} {\bibfield
  {journal} {\bibinfo  {journal} {Rep. Prog. Phys.}\ }\textbf {\bibinfo
  {volume} {86}},\ \bibinfo {pages} {116001} (\bibinfo {year}
  {2023})}\BibitemShut {NoStop}%
\end{thebibliography}%

\end{document}